%% file: 0main.tex
\documentclass[preprint]{aastex}

\pdfoutput=1
\usepackage{hyperref}

\usepackage{amsmath}
\usepackage{amsfonts}
\usepackage{dsfont}
\usepackage{amsxtra}
\usepackage{amssymb}
\usepackage{upgreek}

\usepackage{multirow}
\usepackage{url}

\usepackage{graphicx,epsfig}
\usepackage{bm}

\usepackage{xspace} 

\newcommand{\be}{\begin{equation}}
\newcommand{\ee}{\end{equation}}
\newcommand{\bea}{\begin{eqnarray}}
\newcommand{\eea}{\end{eqnarray}}
\newcommand{\beaa}{\begin{eqnarray*}}
\newcommand{\eeaa}{\end{eqnarray*}}
\newcommand{\ba}{\begin{array}}
\newcommand{\ea}{\end{array}}
\newcommand{\bi}{\begin{itemize}}
\newcommand{\ei}{\end{itemize}}
\newcommand{\ben}{\begin{enumerate}}
\newcommand{\een}{\end{enumerate}}

\newcommand{\ra}{\rightarrow}

\newcommand{\td}{\tilde}

\newcommand{\lb}{\label}
\newcommand{\g}{\gamma}
\newcommand{\G}{\Gamma}

\newcommand{\al}{\alpha}

\newcommand{\ld}{\lambda}

\newcommand{\vp}{\varphi}

\newcommand{\Om}{\Omega}

\newcommand{\sm}{\sigma}

\newcommand{\Fermi}{\textsl{Fermi}\xspace}

\newcommand{\WMAP}{\textsl{WMAP}\xspace}
\newcommand{\Planck}{\textsl{Planck}\xspace}
\newcommand{\Suzaku}{\textsl{Suzaku}\xspace}
\newcommand{\PAMELA}{\textsl{PAMELA}\xspace}

\newcommand{\onepic}{0.45}
\newcommand{\twopic}{0.4}

\newcommand{\threepic}{0.36}
\newcommand{\fourpic}{0.28}

\shorttitle{The {\Fermi} Bubbles}

\begin{document}

\title{The Spectrum and Morphology of the {\Fermi} Bubbles}

\input{authorlist}

\begin{abstract}
The {\Fermi} bubbles are two large structures in the gamma-ray sky extending to $55^\circ$ above and below the Galactic center.
We analyze 50 months of \hbox{\Fermi} Large Area Telescope data between 100 MeV and 500 GeV
above $10^\circ$ in Galactic latitude to derive the spectrum and morphology of the {\Fermi} bubbles.
We thoroughly explore the systematic uncertainties
that arise when modeling the
Galactic diffuse emission through two separate approaches.
The gamma-ray spectrum is well described by either a log parabola
or a power law with an exponential cutoff.
We exclude a simple power law with more than 7$\sigma$ significance.
The power law with an exponential cutoff has an index of $1.9 \pm 0.2$ and a cutoff energy of
\hbox{$110\pm 50$ GeV}.
We find that the gamma-ray luminosity of the bubbles is
\hbox{$4.4^{+2.4}_{-0.9} \times 10^{37}$ erg s$^{-1}$}.
We confirm a significant enhancement of gamma-ray emission in the south-eastern part of the bubbles,
but we do not find significant evidence for a jet.
No significant variation of the spectrum across the bubbles is detected.
The width of the boundary of the bubbles is estimated to be $3.4^{+3.7}_{-2.6}$ deg.
Both inverse Compton (IC) models and hadronic models including IC emission from secondary leptons fit the gamma-ray data well.
In the IC scenario, 
the synchrotron emission from the same population of electrons 
can also explain the \textsl{WMAP} and \Planck microwave haze
with a magnetic field between 5 and 20 $\upmu$G.

\end{abstract}

\keywords{
astroparticle physics --- 
cosmic rays ---
Galaxy: general ---
Galaxy: halo ---
gamma rays: diffuse background ---
methods: data analysis
}


\input{1intro}
\input{3galprop}

\input{4local}

\input{5comparison}

\input{6morphology}

\input{7interpretation}
\input{8conclusion}

\input{acknowledgement}
\newpage
\appendix
\input{9bias}

\input{9models}

\newpage
\bibliography{0main}        

\end{document}

%% file: authorlist.tex
\author{
M.~Ackermann\altaffilmark{1}, 
A.~Albert\altaffilmark{2}, 
W.~B.~Atwood\altaffilmark{3}, 
L.~Baldini\altaffilmark{4}, 
J.~Ballet\altaffilmark{5}, 
G.~Barbiellini\altaffilmark{6,7}, 
D.~Bastieri\altaffilmark{8,9}, 
R.~Bellazzini\altaffilmark{4}, 
E.~Bissaldi\altaffilmark{10}, 
R.~D.~Blandford\altaffilmark{2}
% ...
}
\altaffiltext{1}{Deutsches Elektronen Synchrotron DESY, D-15738 Zeuthen, Germany}
\altaffiltext{2}{W. W. Hansen Experimental Physics Laboratory, Kavli Institute for Particle Astrophysics and Cosmology, Department of Physics and SLAC National Accelerator Laboratory, Stanford University, Stanford, CA 94305, USA}
\altaffiltext{3}{Santa Cruz Institute for Particle Physics, Department of Physics and Department of Astronomy and Astrophysics, University of California at Santa Cruz, Santa Cruz, CA 95064, USA}
\altaffiltext{4}{Istituto Nazionale di Fisica Nucleare, Sezione di Pisa, I-56127 Pisa, Italy}
\altaffiltext{5}{Laboratoire AIM, CEA-IRFU/CNRS/Universit\'e Paris Diderot, Service d'Astrophysique, CEA Saclay, 91191 Gif sur Yvette, France}
\altaffiltext{6}{Istituto Nazionale di Fisica Nucleare, Sezione di Trieste, I-34127 Trieste, Italy}
\altaffiltext{7}{Dipartimento di Fisica, Universit\`a di Trieste, I-34127 Trieste, Italy}
\altaffiltext{8}{Istituto Nazionale di Fisica Nucleare, Sezione di Padova, I-35131 Padova, Italy}
\altaffiltext{9}{Dipartimento di Fisica e Astronomia ``G. Galilei'', Universit\`a di Padova, I-35131 Padova, Italy}
\altaffiltext{10}{Istituto Nazionale di Fisica Nucleare, Sezione di Trieste, and Universit\`a di Trieste, I-34127 Trieste, Italy}
% ...

%% file: 1intro.tex
\section{Introduction}

Radio and X-ray lobes are often observed in galaxies 
with significant accretion onto the central supermassive black hole
or with starburst activity in the vicinity of the galactic nucleus.
Similar features might therefore be expected in our own Galaxy.

Gamma-ray lobes, called the {\Fermi} bubbles, were 
discovered \citep{2010ApJ...717..825D, 2010ApJ...724.1044S, 2012ApJ...753...61S} 
in a search for a gamma-ray counterpart to the Wilkinson Microwave Anisotropy Probe ({\WMAP}) haze \citep{2004ApJ...614..186F},
which is residual microwave emission around the Galactic center that
remains after subtracting synchrotron, free-free, thermal dust,
and cosmic microwave background components from the {\WMAP} data.
The {\Fermi} bubbles are two large structures that extend to $55^\circ$ above and below  
the Galactic center. 
They were reported to have an approximately $E^{-2}$ gamma-ray spectrum between 
1 GeV and 100 GeV and
well defined edges \citep{2010ApJ...724.1044S}.
Further analysis revealed an enhanced gamma-ray emission in the south-east side of the bubbles
with a cocoon-like shape and a tentative 
identification of jet-like structures~\citep{2012ApJ...753...61S}. 

Soon after the discovery of the {\Fermi} bubbles, several models of their formation as
well as the acceleration of particles and gamma-ray production were proposed.
The formation of the bubbles can be modeled by emission of a jet from the black hole
\citep{2012ApJ...756..181G, 2012ApJ...756..182G, 2012ApJ...761..185Y},
a spherical outflow from the black hole
\citep{2011MNRAS.415L..21Z}, 
a wind from supernova explosions \citep{2011PhRvL.106j1102C}, 
or a sequence of shocks from several accretion events onto the black hole
\citep{2011ApJ...731L..17C}.
Observations of anomalously high ionization in the Magellanic Stream can be interpreted as due to active galactic nucleus (AGN) activity in the Milky Way a few million years ago, which may have caused the formation of the bubbles
\citep{2013ApJ...778...58B}.
The gamma-ray emission could be explained by hadronic production through collisions
of cosmic-ray (CR) protons with diffuse gas in the bubbles \citep{2011PhRvL.106j1102C}
or through inverse Compton (IC) scattering  
of high-energy electrons on radiation fields \citep{2010ApJ...724.1044S, 2011PhRvL.107i1101M}.

Significant effort has gone into searching for counterparts of the bubbles in X-rays, radio emission, and very high energy gamma rays.
Radio and microwave emission is expected in leptonic models of the {\Fermi} bubbles due to synchrotron radiation.
The presence of the microwave haze was confirmed with 7 years of the \WMAP data 
\citep{Pietrobon:2011hh, 2012ApJ...750...17D}
and by the \Planck collaboration 
\citep{2013A&A...554A.139P}.
There is a tentative association of the {\Fermi} bubbles with some 
features in the S-band Polarization All Sky Survey (S-PASS) radio data \citep{2013Natur.493...66C}
and in \WMAP polarization maps \citep{2012ApJ...747L..12J}.
In X-rays, one expects to see lower-density hot gas inside the bubbles and higher-density colder
gas outside.
There are features possibly associated with the {\Fermi} bubbles in {\it R\"ontgensatellit} ({\it ROSAT}) data \citep{2010ApJ...724.1044S}
and in \Suzaku data \citep{2013ApJ...779...57K}.

Although the {\Fermi} bubbles appear to be aligned transverse to the plane of the Galaxy and emanating from the region near the GC, it is not certain
that they are associated with the 
GC region rather than from a region closer to the Earth.
There are nevertheless several indirect arguments that the {\Fermi} bubbles were created
by a phenomenon in or around the Galactic center.
First, a symmetry argument: the bubbles appear to be directly above and below the GC.
Second, the hard energy spectrum and the sharp edges favor a transient nature for the bubbles.
Locally the bubbles can be produced by a SN explosion,
but in this case one expects to find strong synchrotron emission,
while the bubbles, assuming that the association with the {\WMAP} and {\Planck} haze is correct, 
have very weak synchrotron emission at high latitudes and relatively strong 
synchrotron emission at lower latitudes, which can be naturally explained by a decreasing magnetic field 
at larger distances from the Galactic plane.
Third, a series of \Suzaku X-ray observations across the edge of the bubbles \citep{2013ApJ...779...57K} 
reveal a spectral component that has a drop in emission measure
across the edge and a characteristic absorption at lower X-ray energies that favors a large distance
to the X-ray emitting region.

The {\Fermi}-LAT gamma-ray data are crucial for understanding
the physics of the bubbles
and to guide future multiwavelength observations.
We use 50 months of {\Fermi}-LAT data to study the details of the energy spectrum and
the morphology of the bubbles.
One of the main challenges is the spatial overlap of the {\Fermi} bubbles with the other components
of Galactic gamma-ray emission.
We pay special attention to systematic uncertainties associated with the modeling of the 
Galactic diffuse gamma-ray emission and the definition of the spatial extent of the bubbles. 

In Section \ref{sect:data}, we describe the gamma-ray data 
and our general analysis strategy.
In this paper, we use the method of template fitting, where the emission components
are modeled by their distributions on the sky.
In Section \ref{sect:galprop}, we model the Galactic foreground emission components
by using maps generated with the GALPROP\footnote{\url{http://galprop.stanford.edu}} \citep{Moskalenko:1997gh, Strong:1998fr, Strong:2004de, Ptuskin:2005ax, 2007ARNPS..57..285S, Porter:2008ve,Vladimirov:2010aq} 
CR propagation and interactions code as templates.
In Section \ref{sect:local} we present an alternative approach to model the Galactic foreground emission
that does not rely on 
the GALPROP calculation of the CR distribution in the Milky Way.
In both Sections \ref{sect:galprop} and \ref{sect:local},
we define the template of the {\Fermi} bubbles by applying a significance cut
in the residual maps obtained after subtraction of the other components of gamma-ray emission
from the data.
The energy spectra of the components are found by simultaneous fits of all the spatial templates to the data.
In Section \ref{sect:comparison}, we fit different functions to the energy spectrum of the bubbles
and estimate the statistical and systematic uncertainties in the fit parameters.
In Section \ref{sect:morphology}, we address several questions on the morphology
and spectral variation across the projected area of the bubbles.
In Section \ref{sect:interp}, we fit the spectrum of the {\Fermi} bubbles using hadronic and IC models of gamma-ray production.
We compare the synchrotron radiation from the electrons in the IC scenario and from the secondary electrons
and positrons in the hadronic scenario with the \WMAP and \Planck haze data.
We present our conclusions in Section \ref{sect:conclusions}.
Appendix \ref{sect:bias} has technical details on the fitting procedure, and
Appendix \ref{sec:IC_h_models} contains details on the IC and hadronic models of the bubbles.

%% file: 3galprop.tex
\section{Data set and analysis strategy}
\label{sect:data}

\begin{figure}[t] 
\begin{center}
\includegraphics[scale=\threepic]{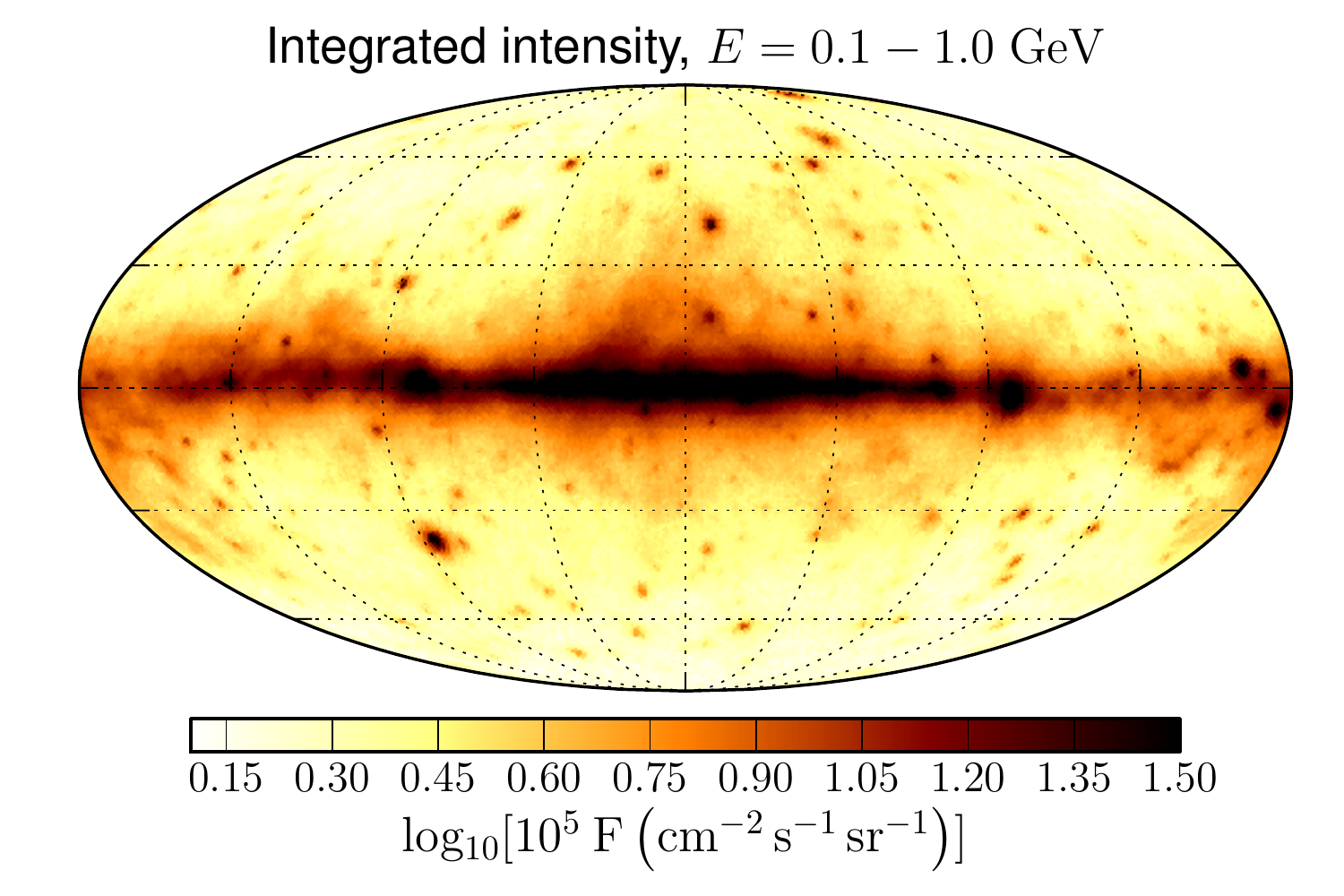}
\includegraphics[scale= \threepic]{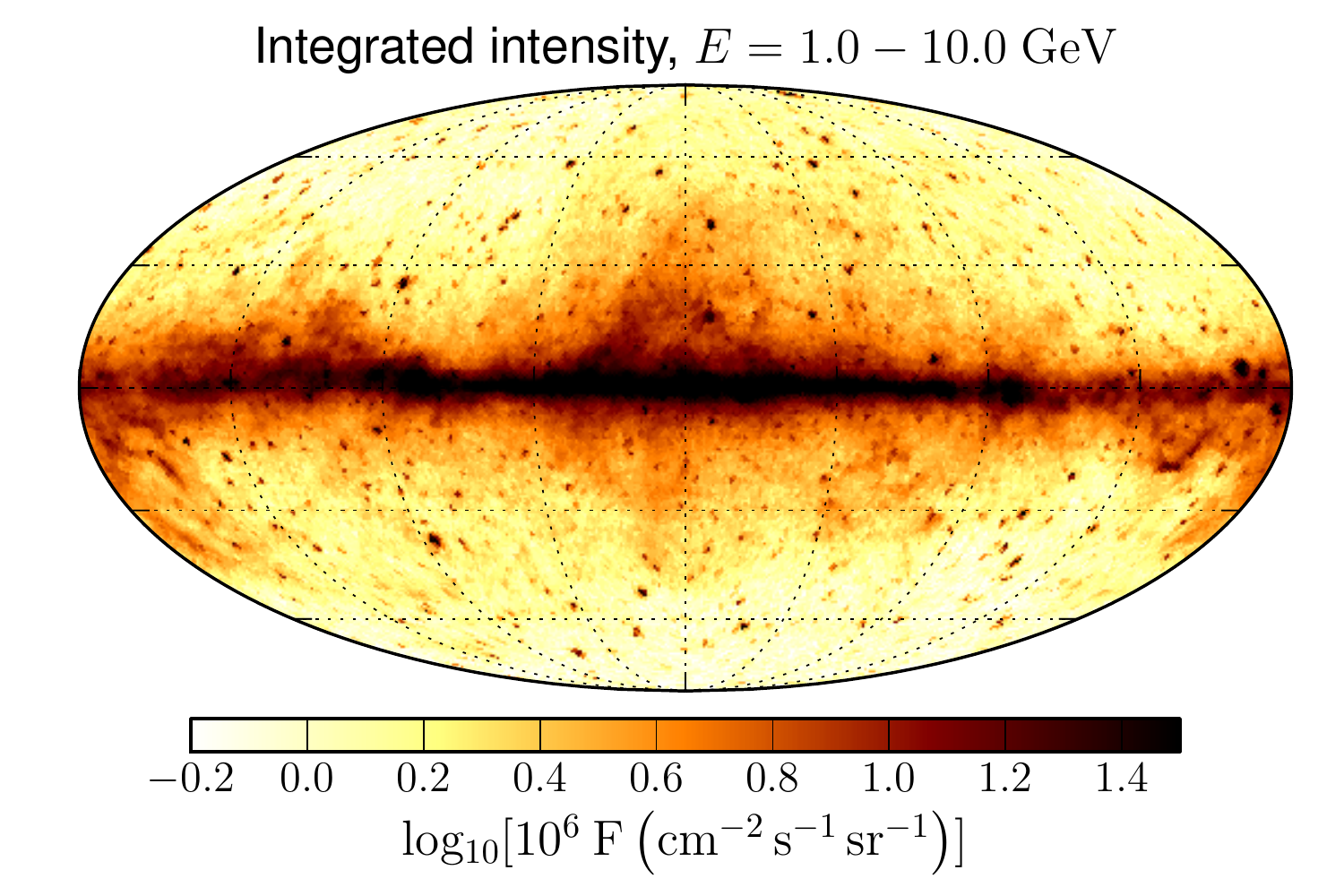}
\includegraphics[scale= \threepic]{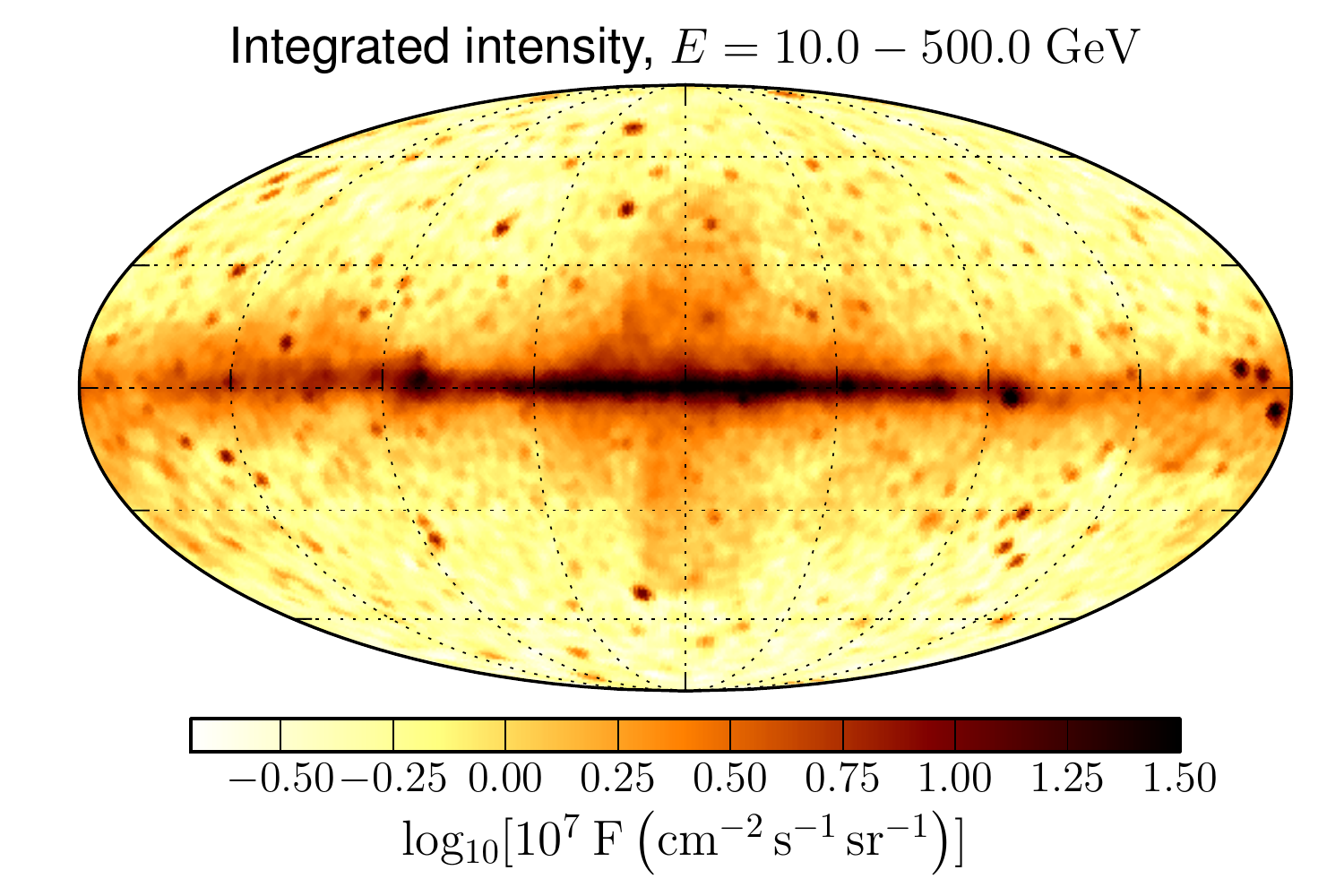}
\noindent
\caption{\small 
Gamma-ray intensity maps integrated in three large energy bins for the data set used in this paper. 
Throughout this paper we show skymaps in Galactic coordinates centered on the Galactic center using the Mollweide projection. 
The pixel size is $0.9^{\circ}$. The map on the right is smoothed with a $\sigma = 1^{\circ}$ Gaussian kernel.
The smoothing is for presentation only; we do not smooth the data maps when fitting the models.
}
\label{fig:data_big}
\end{center}
\vspace{1mm}
\end{figure}

In this analysis we use 50 months of  {\Fermi} LAT \citep{2009ApJ...697.1071A}
data recorded between 2008 August 4 and 2012 October 7
({\Fermi} Mission Elapsed Time 239557448\,s - 371262668\,s),  restricted to the Pass~7 reprocessed UltraClean class. We select the standard good-time intervals, e.g., when the satellite is not passing through the South Atlantic Anomaly.
The UltraClean class provides the cleanest standard gamma-ray sample with respect to the contamination from misclassified charged particle interactions in the {\Fermi} LAT~\citep{Ackermann:2012kna}.
The Pass~7 reprocessed data\footnote{\url{http://fermi.gsfc.nasa.gov/ssc/data/analysis/documentation/Pass7REP_usage.html}} benefits from an updated calibration that improves the energy measurement and event-direction reconstruction accuracy at energies above 1 GeV. 
To minimize the contamination from the Earth-limb emission,
we select events with a zenith angle $< 90^{\circ}$.
In addition we require that the angle of the event with respect to the instrument axis is $< 72^{\circ}$, because there is increased CR background leakage for highly inclined events~\citep{2009ApJ...697.1071A}. 
The exposure and the effective point-spread function (PSF), which are functions of the position in the sky and measured energy as well as the pointing history of the observations, were generated using the standard {\Fermi} LAT ScienceTools package version 9-28-00 available from the {\Fermi} Science Support Center\footnote{\url{http://fermi.gsfc.nasa.gov/ssc/data/analysis/}} using the P7REP\_ULTRACLEAN\_V15 instrument response functions. 
We mask the Galactic plane within $|b| < 10^\circ$,
use events with energies between 100 MeV and 500 GeV separated in 25 logarithmic energy bins, and combine front and back-converting events.
Spatial binning is performed using HEALPix\footnote{\url{http://sourceforge.net/projects/healpix/}} \citep{Gorski:2004by} with a pixelization of order 6 ($\sim 0.9^{\circ}$ pixel size). 
The gamma-ray intensity integrated in three broad energy bins is shown in Figure \ref{fig:data_big}. The {\Fermi} bubbles are visible at  energies $>10$ GeV without any detailed analysis. To calculate the spectrum of the bubbles, the subtraction of the foreground emission components is required.

The observed gamma-ray emission can be divided into resolved point sources (PS) and diffuse emission.
Most of the diffuse gamma-ray emission comes from interactions of CR nuclei with interstellar gas.
Another important component of diffuse emission at high energies is due to IC emission from leptonic CRs interacting with the low-energy interstellar radiation field (ISRF).
At energies below approximately 10 GeV, bremsstrahlung emission from electrons and positrons interacting with interstellar gas is important as well.
These three components have a characteristic distribution on the sky that peaks near the Galactic plane.
An additional isotropic gamma-ray component is made of several contributions,
including residual CR contamination, unresolved extragalactic point sources, and extragalactic diffuse background.

Another diffuse emission component is represented by Loop~I, a giant radio loop spanning $100^{\circ}$ on the 
sky~\citep{1962MNRAS.124..405L}, 
which is also visible in the gamma-ray sky~\citep{Casandjian:2009wq}. The origin of Loop~I is an open question. It could be of local origin, produced either by a nearby supernova explosion or by the wind activity of the Scorpio-Centaurus OB association at a distance of 170 pc~\citep{Wolleben:2007pq}. Alternatively, it could be interpreted as a large-scale outflow from the Galactic center~\citep{2013ApJ...779...57K}. In this paper we consider Loop~I only as a foreground for the bubbles. A dedicated study of this feature is beyond the scope of this paper and left for future work.

The hadronic, IC, bremsstrahlung, isotropic and Loop~I components comprise the most important emission components for an analysis of additional large scale gamma-ray structures, 
such as the {\Fermi} bubbles.
The general analysis strategy in the evaluation of the gamma-ray emission from the {\Fermi} bubbles in this paper
can be divided into the following steps:
\vspace{-5mm}
\ben
\item
Construct a foreground emission model that includes known components\footnote{Note that the diffuse model provided by the Fermi-LAT collaboration through the Fermi Science Support Center cannot be used in this analysis, 
because it was developed for studies of point-like and small objects and, in particular, it already includes a simple model for the Fermi bubbles.}, namely
point sources, hadronic emission from interactions of CR nuclei and ions with interstellar gas,
IC emission and bremsstrahlung from CR electrons, isotropic emission and emission from Loop~I (Section~\ref{sect:galprop}).
\item
Use the residual maps obtained by fitting and subtracting the known components from the data
to find a template for the {\Fermi} bubbles (Section~\ref{subsec:BubbleTemp}).
\item
Find the spectrum of the {\Fermi} bubbles using the template derived in the previous step together with the templates for the other components (Section~\ref{sec:gadgetSys}).
\een

\section{Characterization of the {\Fermi} bubbles using GALPROP templates}
\lb{sect:galprop}

In this section we use the GALPROP package v54.1 to generate templates for the Galactic IC emission component and
for the hadronic and the bremsstrahlung gamma-ray components. The latter ones are correlated  with the
distribution of interstellar gas. In the following, these components will be referred to as  ``gas-correlated".
GALPROP calculates the propagation and interaction of CRs in the Galaxy by numerically solving the diffusion equation for a specified model of the CR source distribution, a
CR injection spectrum, and a model of the transport in the Galaxy. Parameters of the model are constrained by reproducing CR observables, including CR secondary abundances and spectra obtained from direct measurements in the solar system, and diffuse gamma-ray and synchrotron emission. We assume diffusive reacceleration with a Kolmogorov spectrum of interstellar turbulence and no convection. 
The diffusion coefficient is assumed to be isotropic.
In the current version of GALPROP, all calculations assume azimuthal symmetry of the CR density with respect to the GC.
Surveys of the 21-cm line of H~I and the 2.6-mm line of CO (a tracer of H$_2$) are used to evaluate the distribution of the target gas.
The model of the ionized gas 
is based on observations of
the pulsar dispersion measures and H$\alpha$ emission
\citep{2008PASA...25..184G}.
Dust maps are used to correct for the dark gas distribution that refers to neutral interstellar gas unaccounted for by the H~I and CO surveys~\citep{2005Sci...307.1292G,FermiLAT:2012aa}.

\begin{figure}[htbp] 
\begin{center}
\includegraphics[scale=\twopic]{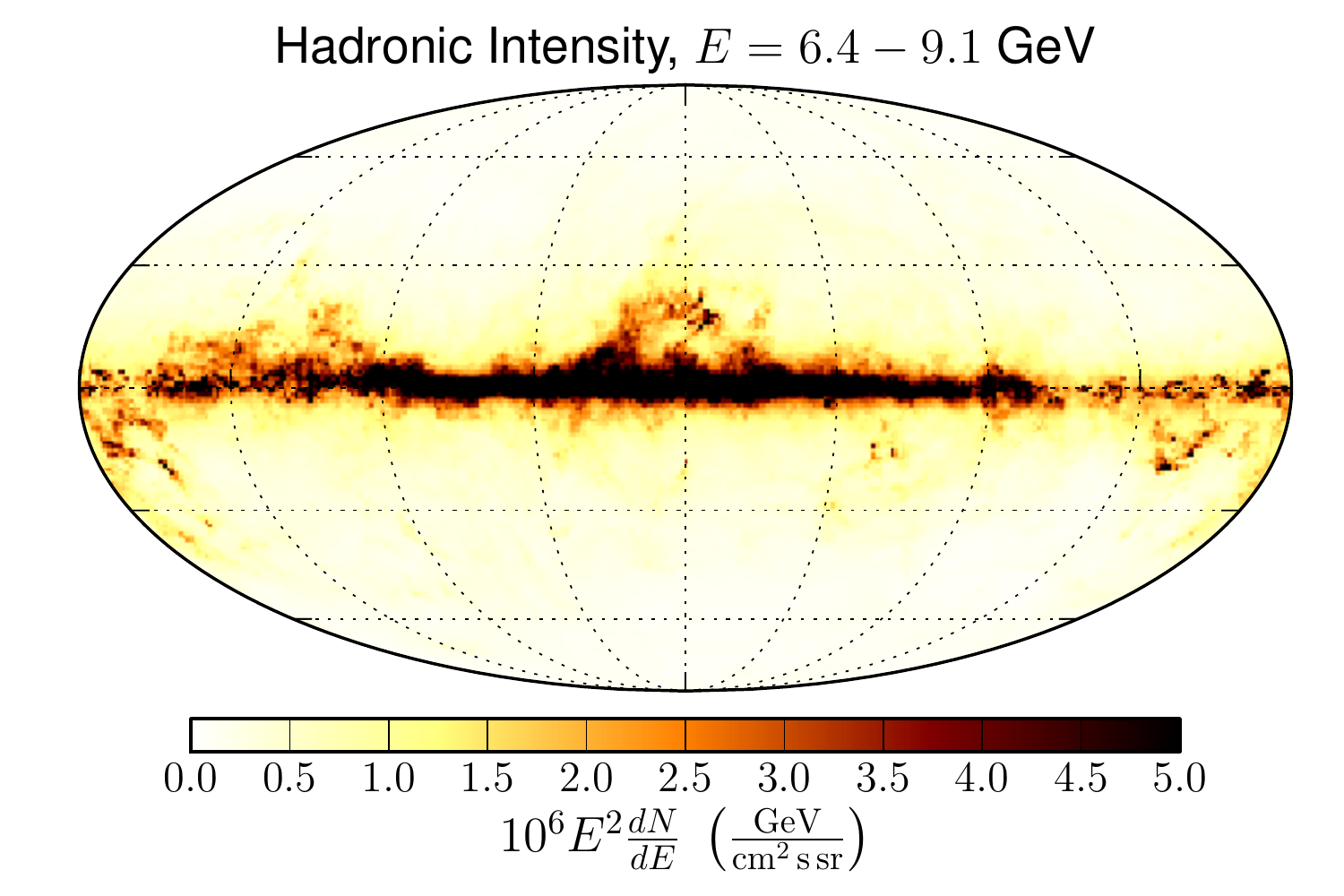}
\includegraphics[scale=\twopic]{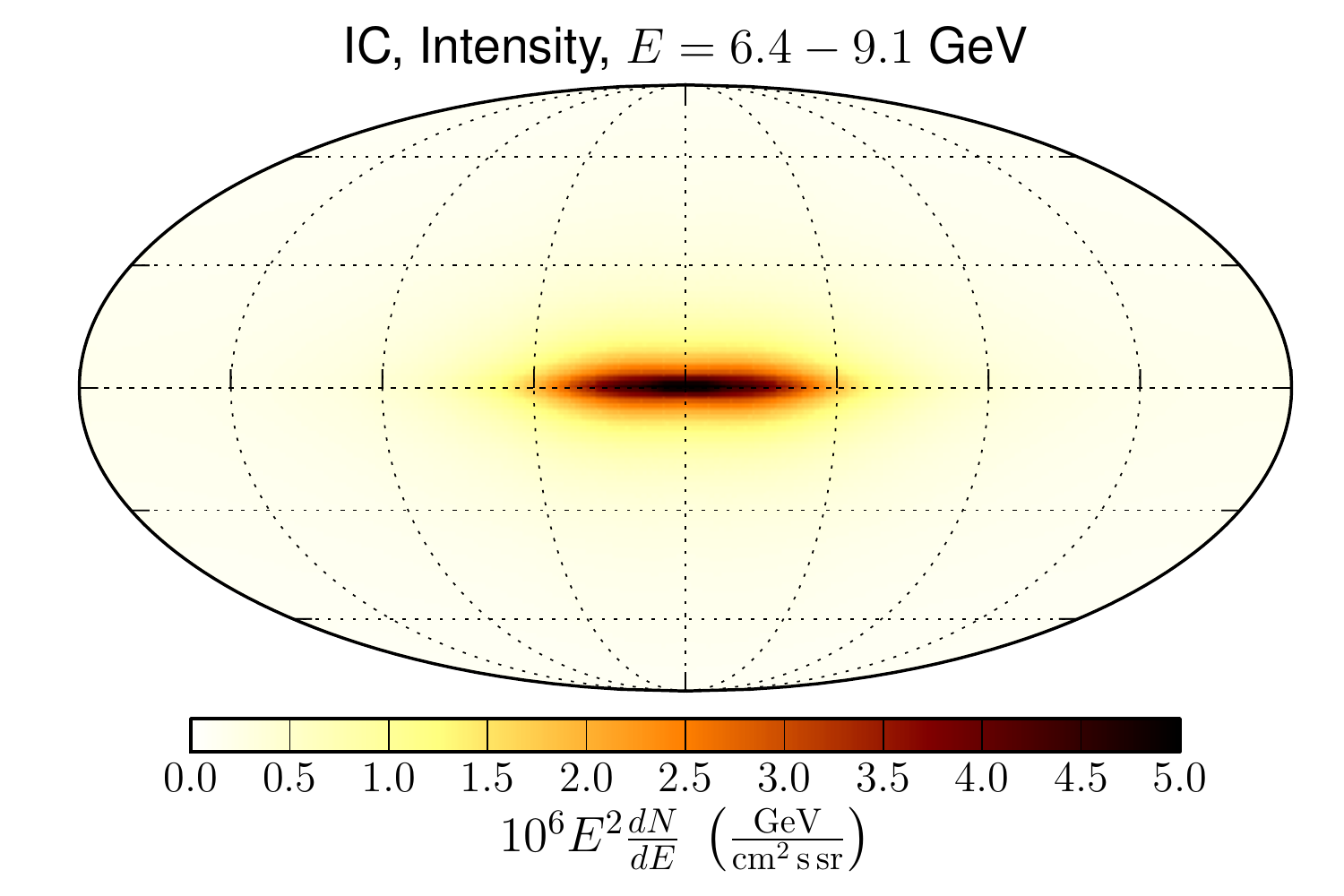}
\includegraphics[scale=\twopic]{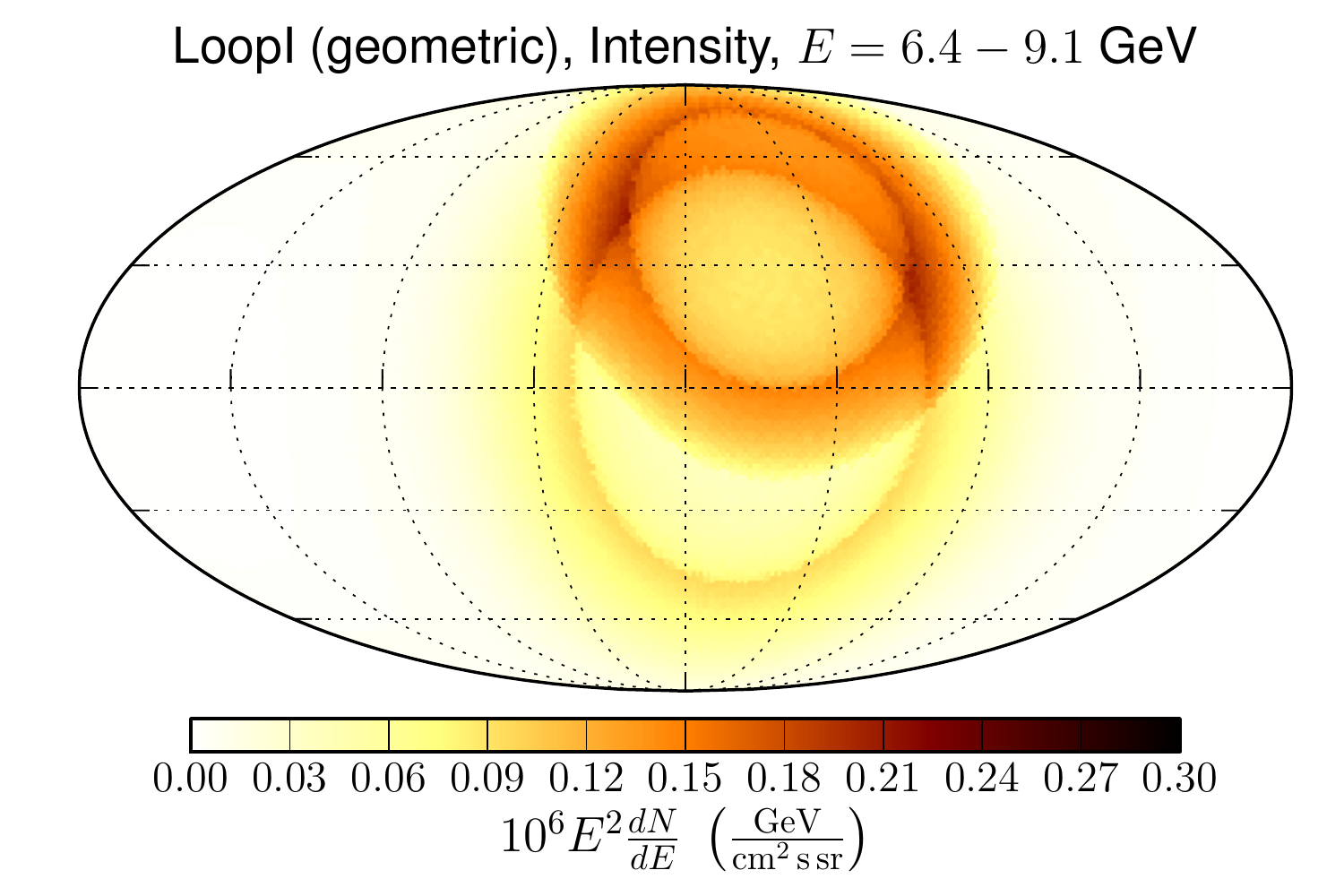}
\includegraphics[scale=\twopic]{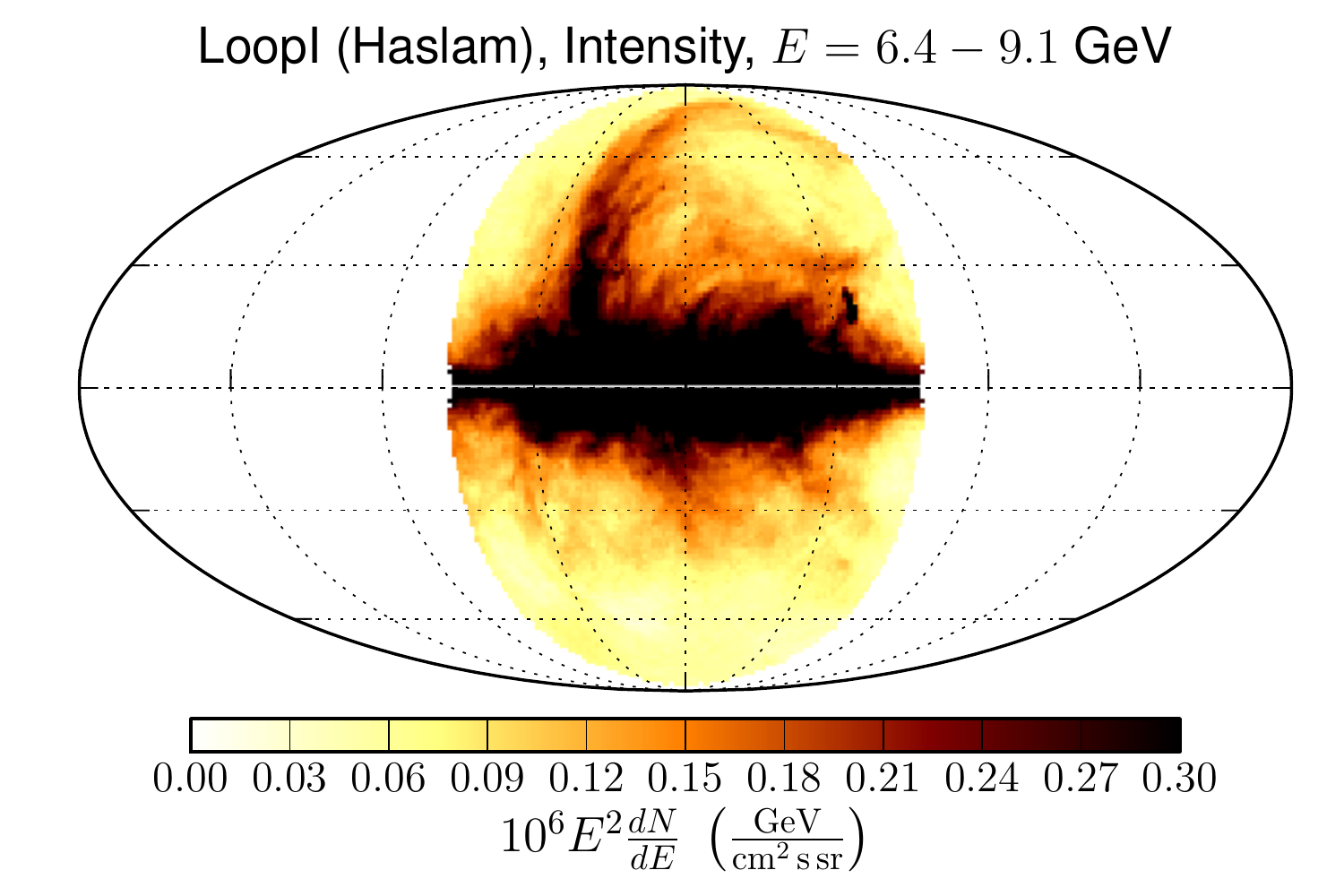}
\noindent
\caption{\small 
Template intensities in the energy bin $E=6.4-9.1$ GeV. Top left: gas-correlated template (sum of hadronic and bremsstrahlung 
for neutral and ionized atomic and molecular hydrogen) obtained from GALPROP.
Top right: IC map obtained from GALPROP. 
Bottom:  Loop~I template based on geometrical model~\citep{Wolleben:2007pq} (left)
and on the Haslam map (right). The Loop~I template normalizations are obtained by fitting to the {\Fermi} LAT data. 
}
\label{fig:LoopIGulli}
\end{center}
\vspace{1mm}
\end{figure}

From the GALPROP calculation, we obtain gamma-ray emissivities corresponding to bremsstrahlung, and hadronic interactions with neutral (H~I), ionized (H~II), and molecular (H$_2$) hydrogen in different Galactocentric rings, and IC emission.
Since the bremsstrahlung is correlated with the distribution of gas, it is combined with the template of gamma rays from hadronic interactions.
In the following, we also combine H~I and H~II rings, while we keep H$_2$ rings independent.
The gamma-ray intensity is proportional to a product of the gas and CR densities.
The integral CR density is not well constrained a priori.
To reduce the uncertainty related to the CR density, we use the gas-correlated templates in Galactocentric rings. The local ring template (r$=$8--10 kpc from the Galactic center) for H~I 
accounts for the vast majority of the gamma rays at latitudes $|b|>10^{\circ}$, and therefore its normalization is kept free in the fit while we fix the other rings.
The molecular hydrogen gas is mainly concentrated in isolated clouds at low latitudes. Since this contribution is small at high latitudes compared to the atomic hydrogen contribution, it is fixed in the fit. 
An IC template is created that takes into account 
the anisotropy of the ISRF due to an anisotropic flux of photons from the Galactic plane%
\footnote{We correct for the anisotropy of the cross-section by multiplying the generated IC maps with a map of the ratio between the predicted IC emission from a full anisotropic calculation, and the prediction assuming an isotropic 
cross-section~\citep{FutureEGB}.}. 
Examples of hadronic and IC templates are shown in Figure \ref{fig:LoopIGulli}.
A detailed study of different GALPROP models and comparisons with {\Fermi} LAT data is presented in~\cite{FermiLAT:2012aa}. For this paper, we consider a subset of models used in~\cite{FermiLAT:2012aa} with updated H~I gas maps. 
Compared to the gas maps used in \cite{FermiLAT:2012aa}, the maps used here exclude the large and small Magellanic clouds, M31, and M33 as well as the Magellanic Stream and other high velocity clouds. 
The choice of a specific GALPROP model is a possible source of systematic uncertainties in our results. 
We address the question of systematic uncertainties in Section~\ref{sec:gadgetSys}.

The following parameters describe our baseline GALPROP model\footnote{GALDEF file: galdef\_54\_Lorimer\_z10kpc\_R20kpc\_Ts100000K\_EBV5mag}:
the CR population is traced by the measured pulsar distribution \citep{Lorimer:2006qs}, the CR confinement volume has a height of 10 kpc and a radius of 20 kpc, 
and H~I column densities are derived from the 21-cm line intensities in the approximation of an optically thin medium,
which is formally modeled by setting the spin temperature to 100,000 K.

The emission of Loop~I is not modeled by GALPROP.  It is a very important contribution because it overlaps the bubbles, especially in the Northern Galactic hemisphere (see Figure \ref{fig:data_big}). Here we use two different approaches to model the emission of Loop~I. 
In the first approach (used for the baseline model), we take a large elliptical region from the \citet{Haslam:1982zz} map at 408 MHz around the Galactic center as a template of Loop~I (Figure~\ref{fig:LoopIGulli}, bottom right).
This approach is based on the assumption that the features in the Haslam map are produced by the synchrotron radiation from
the same population of electrons that emit IC gamma rays.
As an alternative way to model Loop~I we use a geometric template based on a polarization survey at 1.4 GHz~\citep{Wolleben:2007pq}. The geometric Loop~I model assumes synchrotron emission from two shells. Each shell is described by 5 parameters: the center coordinates $\ell$, $b$; the distance to the center $d$;  and the inner ($r_{in}$) and outer ($r_{out}$) radius of the shell. 
The parameters are set to: $\ell_{1} = 341^{\circ}$, $b_{1} = 3^{\circ}$, $d_{1} = 78$\;pc, $r_{in,1} = 62$\;pc, $r_{out,1} = 81$\;pc, $\ell_{2} = 332^{\circ}$, $b_{2} = 37^{\circ}$, $d_{2}=95$\;pc, $r_{in,2} = 58$\;pc, $r_{out,2} = 82$\;pc (Figure~\ref{fig:LoopIGulli}, bottom left). Due to line of sight integration, the two uniform intensity spherical shells appear non-uniform with diffuse edges.
The geometric Loop~I template is included in the derivation of the systematic uncertainties (see Section~\ref{sec:gadgetSys}). 
An isotropic template accounts for the extragalactic diffuse emission and the residual CR contamination.

A template for the bubbles is defined from the residual maps in Section \ref{subsec:BubbleTemp}. In the definition of the templates of the bubbles, 
point sources from the 2FGL catalog \citep{2012ApJS..199...31N} 
with a test statistic $>25$ are masked with a radius of $1.48^{\circ}$, which corresponds to the $95\%$ containment region of the PSF at 1 GeV. 
The test statistic is defined as TS = 2 $\Delta \log \mathcal{L}$, where $\mathcal{L}$ is the likelihood. TS $=25$ corresponds to a significance of  just over $4\sigma$. The twelve extended sources in the 2FGL catalog are masked conservatively within a circle of radius equal to the sum of the major semiaxis of the source template and the $95\%$ PSF containment radius. The source mask is displayed in Figure~\ref{fig:mask} on the right.
Since the definition of the bubbles relies only on an analysis at high energies ($>6$ GeV), where the PSF core is narrow ($<0.2^\circ$),
the masked point sources cover about $18\%$ of the area of the bubbles at $|b|>10^{\circ}$.
The spectrum of the bubbles is calculated over a broad energy range extending from $100$ MeV to $500$ GeV. The broad PSF at low energies does not allow sufficient masking of the sources without masking much of the area of the bubbles, and therefore requires a more accurate modeling. 
Since the 2FGL catalog was obtained with only 2 years of data
and this analysis uses more than 4 years, 
we fit 472 bright point sources 
in the 2FGL catalog with TS $> 200$ using the full 50 months data set 
in order to account for flares happening outside of the time window of the 2FGL analysis. For each source, we fit the normalization in each energy bin. The remaining less significant sources are merged into a single template with fluxes from the 2FGL catalog. 
The overall normalization of this template is free in the fit. Individual sources fainter than the 2FGL limit are effectively part of the isotropic background. Their effect on the overall fit would be in the isotropic component and they would not impact the results for the bubbles.

The templates listed above are fit to the data (see Table~\ref{tab:templateList} for a summary of the templates used in the derivation of the spectra). The fit is performed in each energy bin individually, i.e., if a template is kept free in the fit, its normalization in each energy bin is a free parameter in the fit. 

\clearpage

\clearpage

\begin{deluxetable}{cc}
\tabletypesize{\scriptsize}
\tablecaption{Template maps used in all-sky fit for derivation of the spectra.\label{tab:templateList}}
\tablewidth{0pt}

\tablehead{\colhead{Template} & \colhead{Description}}

\startdata
Neutral and ionized atomic hydrogen   &   GALPROP: bremsstrahlung and hadronic production \\
 (sum of H~I and H~II) & local ring:   8 -  10 kpc  (free)\\
	&    non-local component: 0 - 8 kpc and 10 - 50 kpc (fixed)\\
	\tableline
Molecular hydrogen (H$_2$)   &   GALPROP: bremsstrahlung and hadronic production \\
& all Galactocentric rings combined (fixed)\\
\tableline
Inverse Compton radiation & GALPROP (free)\\ 
\tableline
Bright 2FGL sources & ${\rm TS} > 200$, 472 sources: each fitted individually \\ 
\tableline
Weak 2FGL sources & one template obtained by adding 2FGL fluxes (free) \\
\tableline
Isotropic & extragalactic diffuse and  \\
                & residual CR contamination (free) \\
\tableline
Loop~I    & Haslam map or geometric template (free) \\ 
\tableline
Bubble    & template obtained from residuals (Section~\ref{subsec:BubbleTemp}) (free)\\ 

\enddata

\end{deluxetable}


%

\subsection{Fitting algorithm}

In order to extract the {\Fermi} bubbles' morphology and spectrum from the {\Fermi} LAT data we use the
all-sky fitting tool GaDGET \citep{Ackermann:2009zz}, which 
simultaneously fits the different components of the
diffuse Galactic emission to the {\Fermi} LAT data in a maximum likelihood procedure
based on Poisson statistics.
MINUIT~\citep{James:1975dr} was used as the optimizer with a tolerance of $1.0\times10^{-4}$. 

The model is a linear combination of templates
\begin{equation}
\mu_{ij} (f) = \sum_{m} f_{im} t_{ij}^{m},
\end{equation}
where $t_{ij}^m$ is the template of the component $m$ in energy bin $i$ and in pixel $j$. Coefficients $f_{im}$ are the fitting parameters.
The model maps, which are in flux units, are 
multiplied with the exposure and convolved with the instrument PSF to obtain the corresponding count maps for comparison with the data count maps.

\subsection{Defining a template for the {\Fermi} Bubbles}
\label{subsec:BubbleTemp}
\begin{figure}[htbp]
\begin{center}
\includegraphics[scale=\twopic]{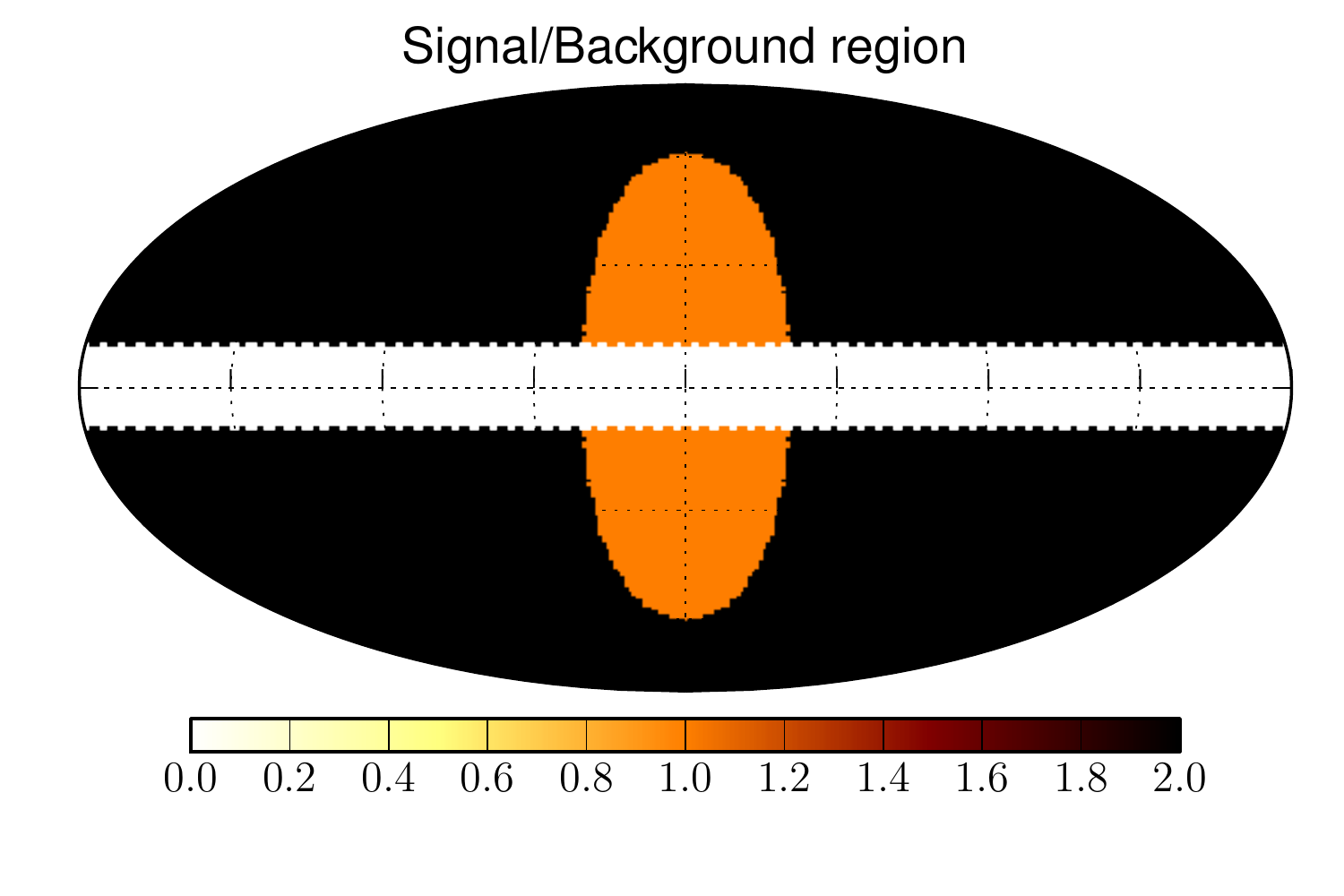}
\includegraphics[scale=\twopic]{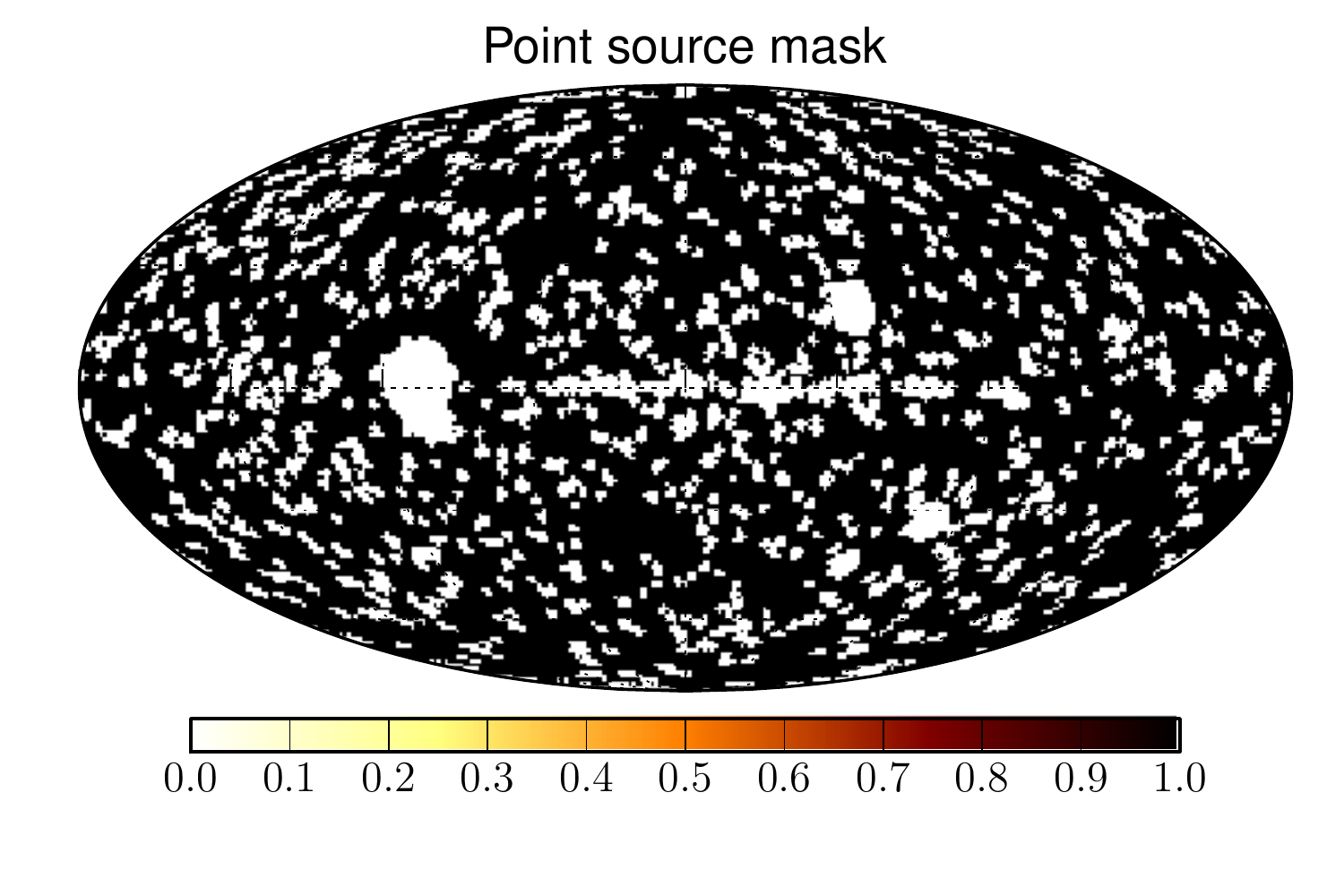}
\noindent
\caption{\small 
Left: elliptical region that covers the bubbles (orange) and background region (black). 
The Galactic plane is masked at $|b|<10^{\circ}$.
Right: mask for point sources (TS$>25$) and extended sources from the 2FGL catalog \citep{2012ApJS..199...31N}.
}
\label{fig:mask}
\end{center}
\vspace{1mm}
\end{figure}

In order to define a template for the {\Fermi} bubbles, we perform an all-sky fit with the templates listed above except, of course, a template of the bubbles. We define the template of the bubbles based on the residuals at energies above $6.4$ GeV where the flux of the bubbles 
becomes readily apparent\footnote{Between 5 and 10 GeV the bubbles become clearly visible in the residuals and the exact choice of the threshold 
does not affect the results.}.
Since no template accounts for the bubbles' flux, the coefficients of the fit will partially compensate for it, introducing a bias.
To avoid this bias we mask an elliptical region
that approximately covers the {\Fermi} bubbles (Figure~\ref{fig:mask} on the left).
Point sources are also masked (see above for details).
The PS mask is shown in Figure~\ref{fig:mask} on the right.
In the resulting residual map, the masked pixels are filled with the average of the neighboring not-masked pixels. 

\begin{figure}[htbp] 
\begin{center}
\includegraphics[scale=\twopic]{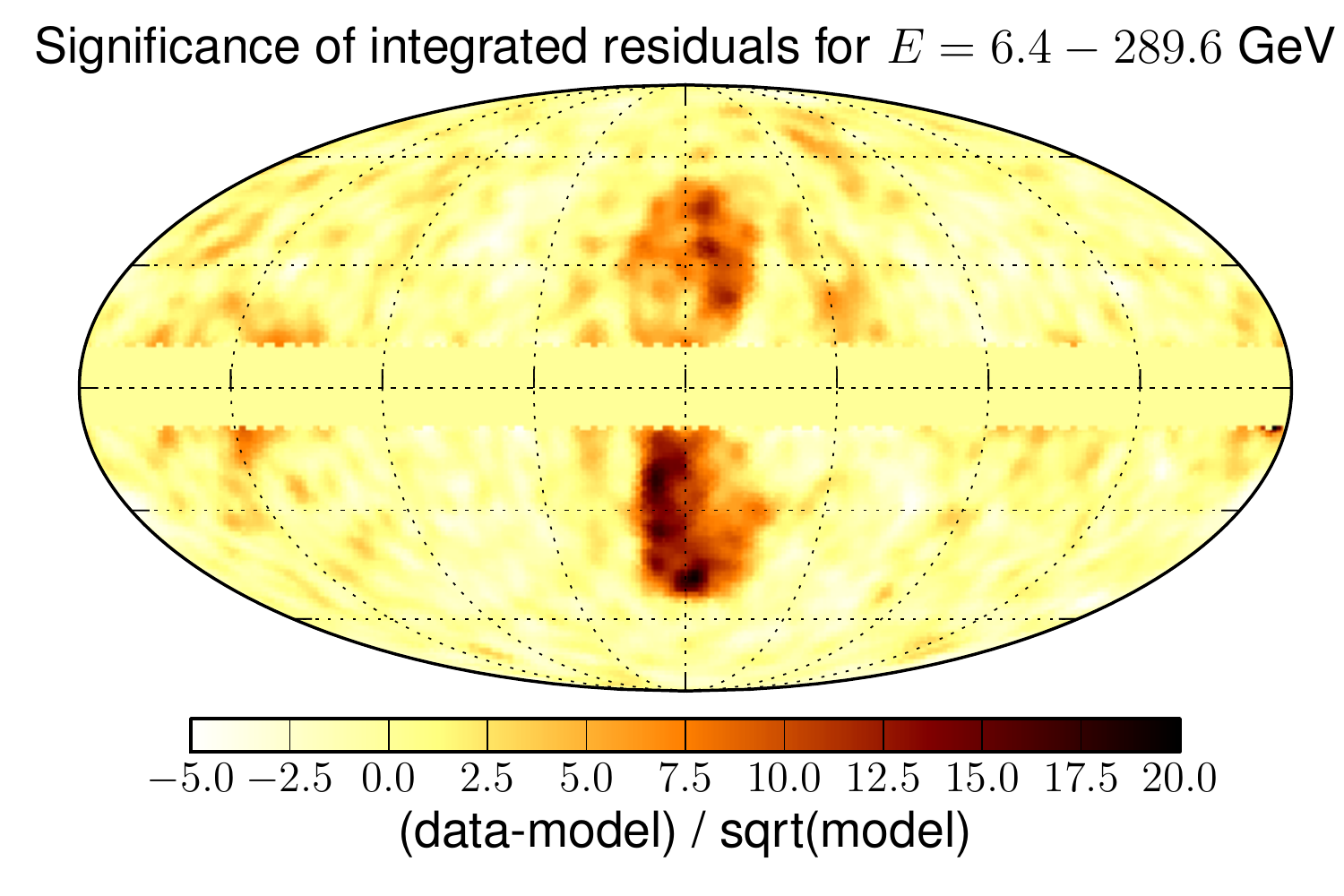}
\includegraphics[scale=\twopic]{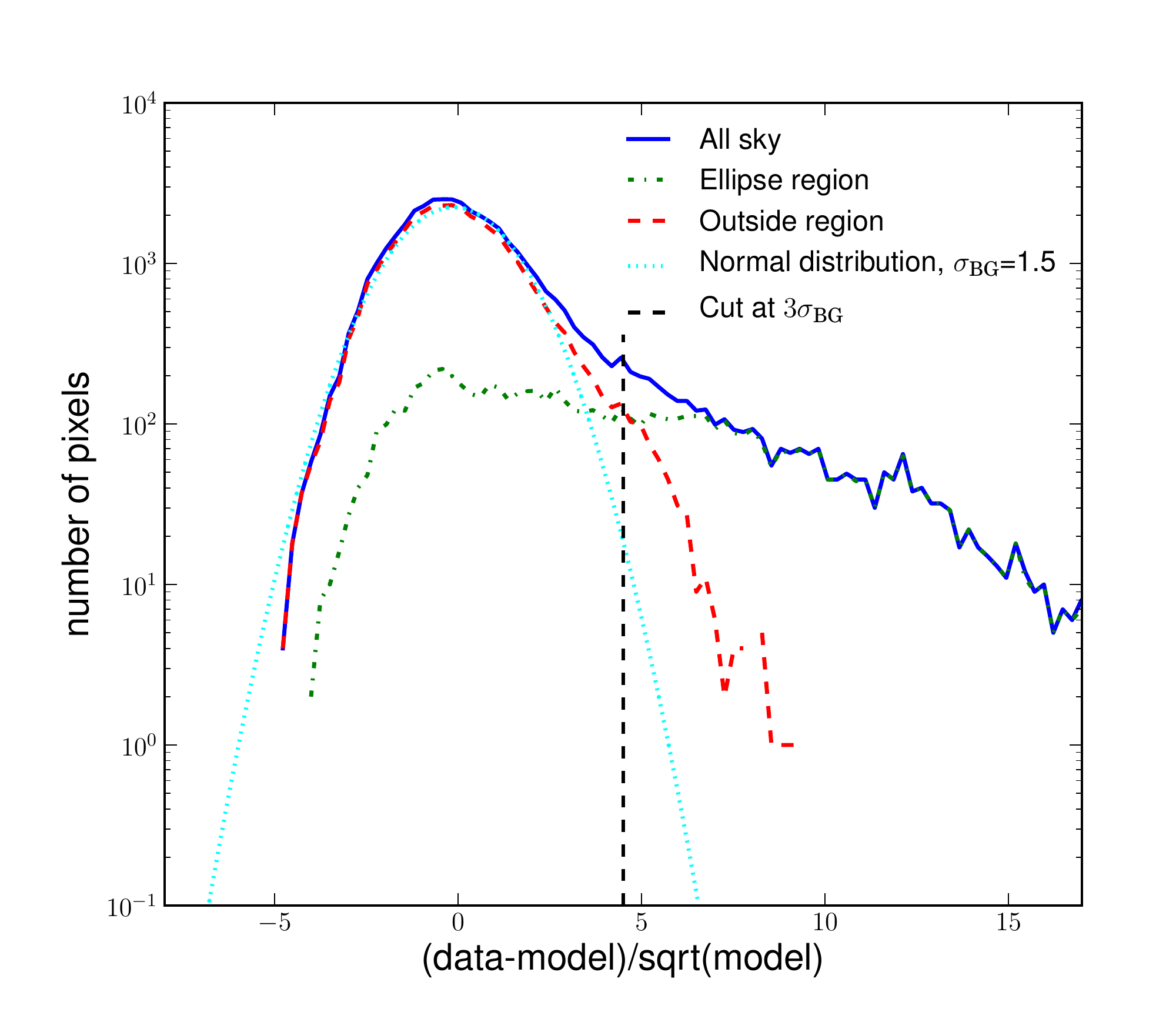}
\noindent
\caption{\small 
Left: significance of integrated residual map at energies $6.4 \, {\rm GeV} < E < 290 \, {\rm GeV}$, defined as $(\rm{data-model})/\sqrt{\rm{model}}$, 
smoothed with a $2^\circ$ Gaussian kernel. 
The large-scale residuals outside of the bubbles 
are due to imperfect modeling of Loop~I and the local gas.
Right: histogram of values in the smoothed residual significance map. 
Dashed (red): background region (Figure \ref{fig:mask}). 
Dash-dotted (green): the region of interest. 
Solid (blue): all sky. 
Dotted (cyan): Gaussian fit to the background distribution, the width is $\sigma_{\rm{BG}} = 1.5$.
The threshold in the definition of the bubbles' template is set to $3\sigma_{\rm{BG}}$ 
and is shown as a vertical dashed black line.
All pixels inside the elliptical masking region and above $|b| = 10^{\circ}$ with the level of residual flux larger than the threshold
are included in the template of the bubbles (Figure \ref{fig:structTemp}). 
}
\label{fig:allSkyRes}
\end{center}
\vspace{1mm}
\end{figure}

\begin{figure}[htbp] 
\begin{center}
\includegraphics[scale=0.4]{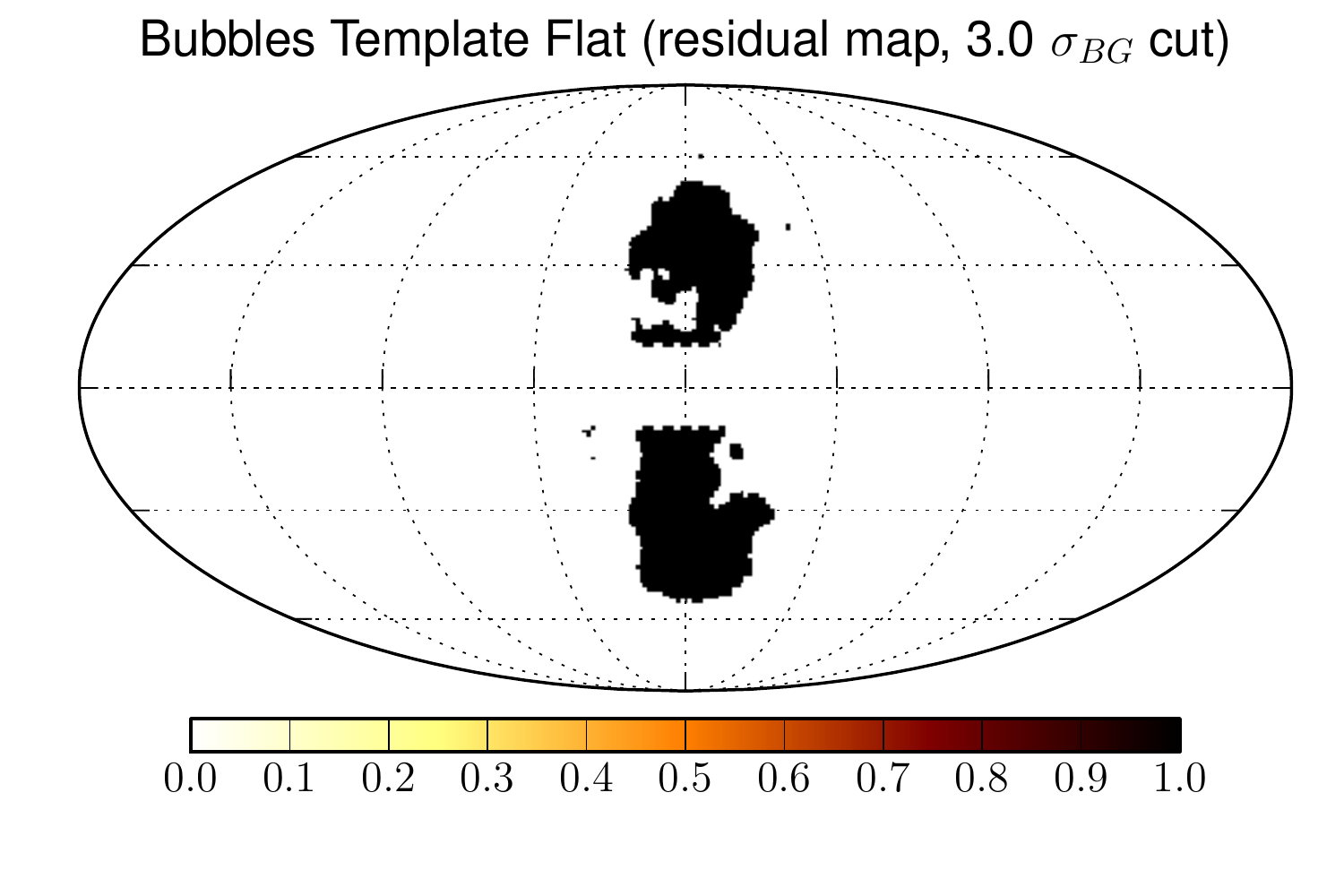}
\includegraphics[scale=0.4]{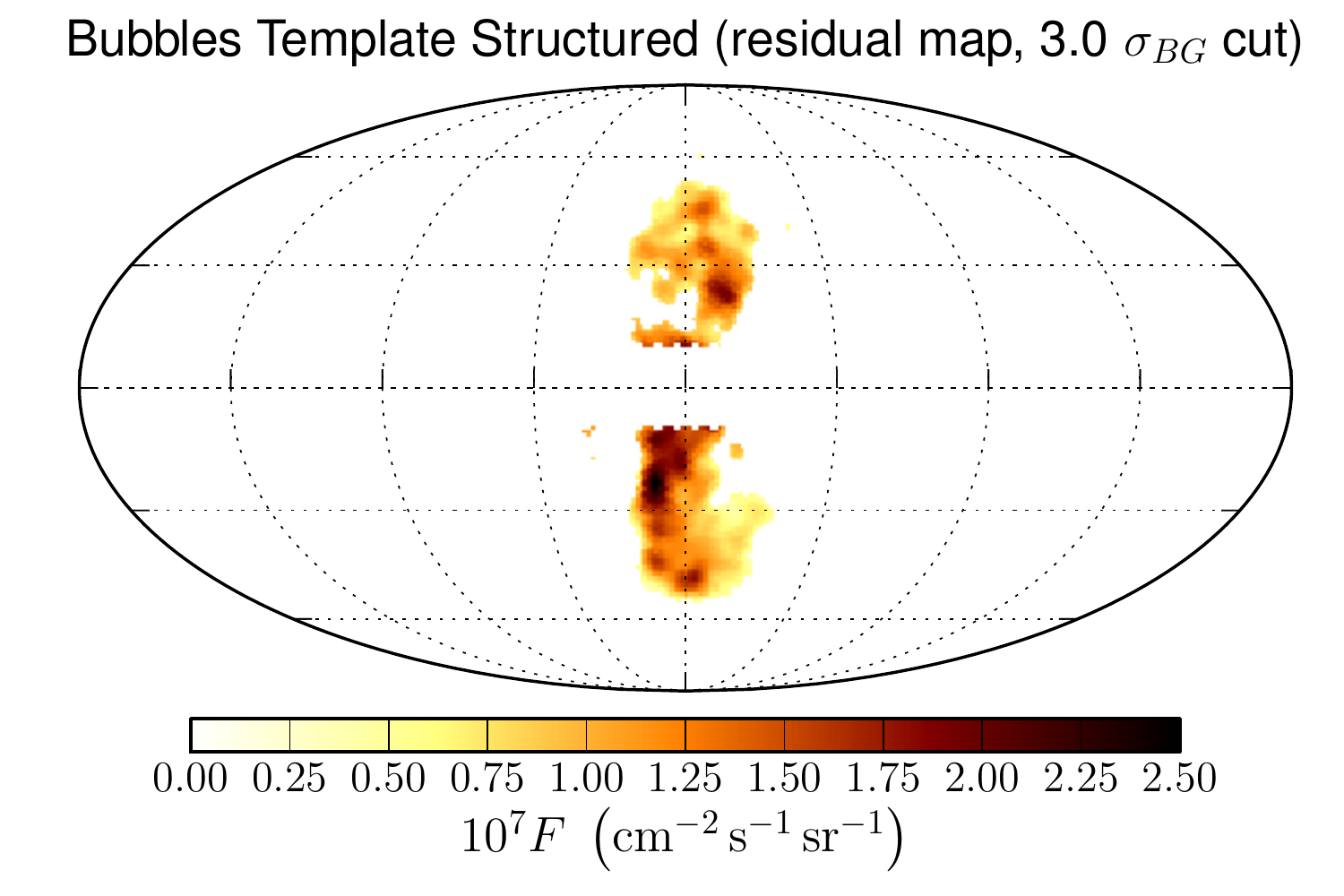}
\noindent
\caption{\small 
Templates of the bubbles defined from the residual significance map using a threshold of $3\sigma_{\rm BG}$,
where $\sigma_{\rm BG}$ is defined in Figure \ref{fig:allSkyRes}.
Left: flat 0-1 template (the value is 1 if the significance of the residual is more than $3\sigma_{\rm BG}$
and 0 otherwise).
Right: structured template proportional to the residual flux.
}
\label{fig:structTemp}
\end{center}
\vspace{1mm}
\end{figure}

The significance map of the residuals  (defined as (data$-$model)/$\sqrt{\rm{model}}$) for the baseline model defined in Section \ref{sect:galprop}
integrated over energies $6.4\, {\rm GeV} < E < 290 \, {\rm GeV}$
is shown in Figure~\ref{fig:allSkyRes} (left). 
The template of the bubbles is determined by applying a threshold in the smoothed significance map. 
To find the threshold, we create a histogram (Figure~\ref{fig:allSkyRes}, right) with the smoothed significance in each pixel in the region of interest and outside the region of interest (orange and black respectively in Figure~\ref{fig:mask}, left). 
We consider the outside region as the background region and fit a Gaussian to the histogram. 
The width of the Gaussian is denoted as $\sigma_{\rm{BG}}$.
The threshold in the definition of the template of the bubbles is set to $3\sigma_{\rm{BG}}$. The 
results for a different threshold of $4\sigma_{\rm{BG}}$ are included in the systematic uncertainties (see Section~\ref{sec:gadgetSys}). 
Figure~\ref{fig:structTemp} shows the resulting templates of the bubbles. 
We distinguish between \textit{flat} and \textit{structured} templates. In the flat template the value is 1 if the significance of the residual is more than the threshold
and 0 otherwise, while the structured template is equal to the residual flux if the significance of the residual is greater than the threshold and 0 otherwise. 
In the baseline model, the bubbles are modeled with a structured template created 
with a significance threshold of $3\sigma_{\rm{BG}}$.

\begin{figure}[htbp] 
\begin{center}
\includegraphics[scale=0.4]{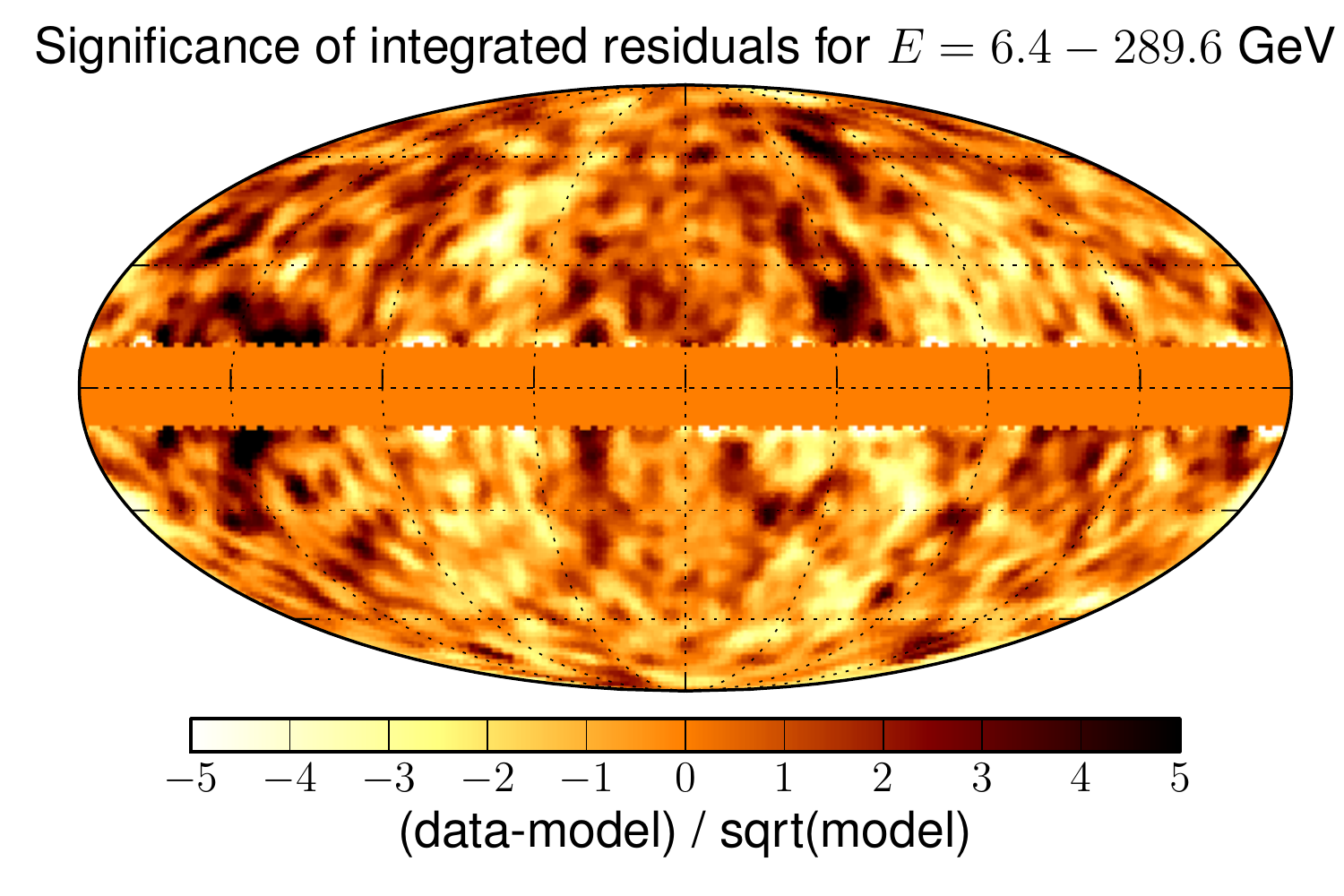}
\includegraphics[scale=0.3]{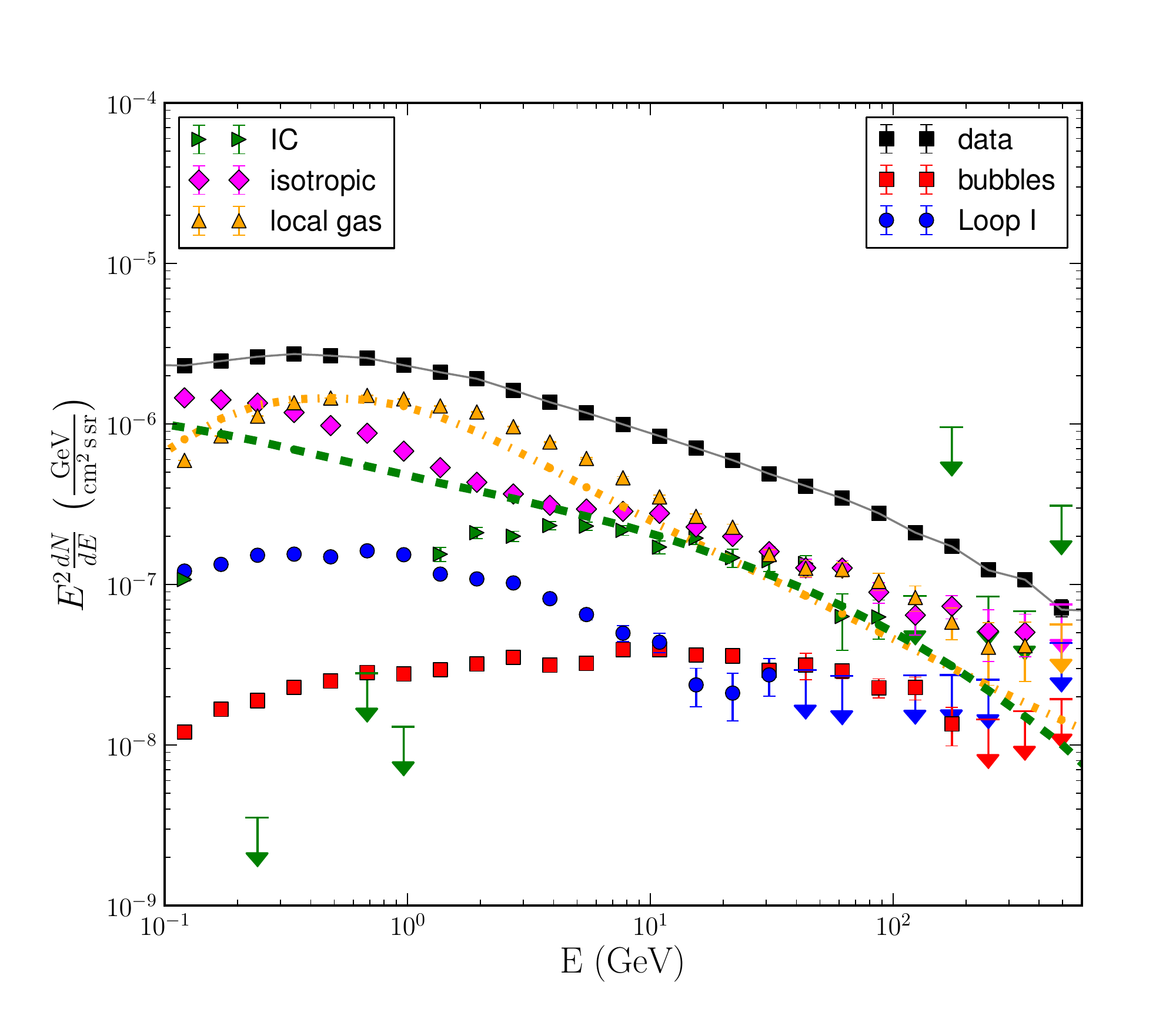}
\noindent
\caption{\small 
Left: residual map after including the structured template of the bubbles in the fit integrated over energies $6.4 \, {\rm GeV} < E < 290 \, {\rm GeV}$ smoothed with a $2^\circ$ Gaussian kernel. 
Remaining residuals to the northeast of the bubbles indicate an imperfect modeling of Loop~I.
Right: spectra of the bubbles and the other components obtained from the fit. The arrows correspond to $2\sigma$ upper limits.
The spectrum of the bubbles is computed as the mean over the points inside the bubbles template.
The lines show the spectra predicted by GALPROP. The drop in the extracted IC spectrum is due to a correlation between the IC template and Haslam map. The Haslam map contains synchrotron radiation emitted by the same population of electrons that is emitting the IC emission.
}
\label{fig:fitWithBubble}
\end{center}
\vspace{1mm}
\end{figure}

In the next step the template of the bubbles is included in the all-sky fit. This time, since no PS mask is applied, the point sources are included in the fit and not masked as in the derivation of the {\Fermi} bubbles' template. 
The integrated residual map after including the structured template of the bubbles in the fit and fitting the point sources is shown in Figure~\ref{fig:fitWithBubble} (left).
The spectra for the different components are presented
in Figure~\ref{fig:fitWithBubble} (right).

In the rest of the paper, the model of the foreground emission components and the {\Fermi} bubbles
presented in this section will be referred to as the baseline model.

\newpage

\subsection{Systematic uncertainties}
\label{sec:gadgetSys}

To estimate the systematic uncertainties in the spectrum of the {\Fermi} bubbles due to uncertainty in the modeling of the diffuse foregrounds and the bubbles,
we study the variation of our results when using
different GALPROP configurations and 
definitions of the templates of Loop~I and the bubbles. 

\begin{figure}[htbp] 
\begin{center}
\includegraphics[scale=0.4]{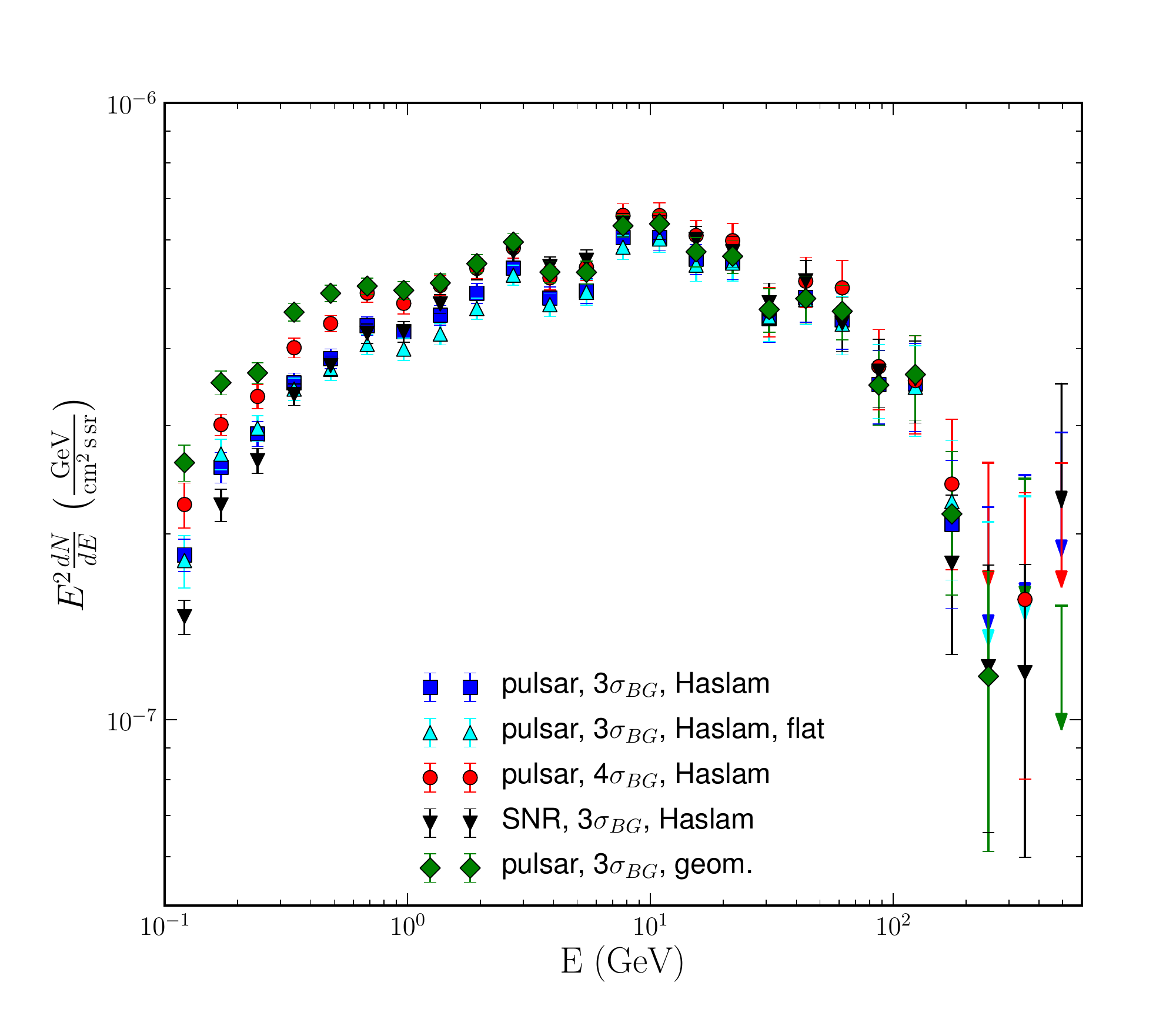}
\noindent
\caption{\small 
Spectra of the bubbles for different model configurations. The cases differ by significance threshold used to define the template of the bubbles 
($3\sigma_{\rm BG}$ or $4\sigma_{\rm BG}$), whether the CR source distribution is traced 
by pulsar or SNR distributions, and the use of the geometric Loop~I template or the Haslam map. 
}
\label{fig:sysErrorsSpec}
\end{center}
\vspace{1mm}
\end{figure}

We tested two different tracers of the CR source distributions: the measured distribution of supernova remnants (SNR)~\citep{Case:1998qg} and the measured pulsar distribution \citep{Lorimer:2006qs}. In addition, we varied the size of the CR confinement volume. 
In GALPROP, the diffusion zone is a cylinder with radius $R_h$ and height $z_{h}$ above the Galactic plane.
$R_h = 20$~kpc and $30$~kpc and $z_h=4$ and 10~kpc have been tested. 
Furthermore, two different spin temperature values ($T_{\rm S} = 150$~K and $10^5$~K) 
are used to correct for the H~I opacity in order to derive the H~I column density.
$T_{\rm S} = 10^5$~K corresponds to the optically thin approximation.
Loop~I is either modeled by the Haslam map or by the geometric template.
We also use a flat template of the bubbles instead of the structured template 
and vary the significance threshold used to define the bubbles ($3\sigma_{\rm{BG}}$ and $4\sigma_{\rm{BG}}$).
Figure~\ref{fig:sysErrorsSpec} shows some examples of the bubbles' spectra for different parameters of the model.
Each case represents the change of one parameter relative to the baseline model configuration.

The all-sky fit and the extraction of the bubble spectrum was repeated with the different GALPROP configurations, Loop~I models, and templates for the bubbles. 
The envelope of all tested models defines the systematic error band. 
The final result is shown in Section \ref{sect:comparison} where it is also compared to the spectrum obtained from the local template analysis (Section \ref{sect:local}).

%% file: 4local.tex
\section{Characterization of the bubbles using local templates analysis}
\lb{sect:local}

In this section we present an alternative approach for modeling the Galactic foreground emission
which does not rely on 
the GALPROP modeling of CR propagation and interactions in the Milky Way.
For example, one of the assumptions in the analysis presented in Section~\ref{sect:galprop} is that the CR spectrum 
is independent of the azimuthal direction from the Galactic center. 
In general, this assumption may be violated, e.g., in the spiral arms.
In this section, we relax this assumption by fitting templates in small regions on the sky (in the following called patches).
Instead of using the gamma-ray emission maps provided by GALPROP, 
we directly use gas maps to trace the intensities of gamma-ray emission:
H~I and CO surveys uncorrected for absorption and dark gas 
\citep[for a description see][]{FermiLAT:2012aa} together with
the Schlegel, Finkbeiner and Davis (SFD) dust map \citep{Schlegel:1997yv} 
to account for gas not traced by the H~I or CO lines.
The IC component is modeled by a bivariate Gaussian; 
the parameters of the Gaussians are found from fitting the model to the data (more details in Section \ref{sect:IC_comp}).

Models for the gamma-ray emission components are derived one at a time.
We start with the component with the brightest integrated flux, namely the gas-correlated emission.
After that we subtract the gas-correlated emission from the data and 
define a template for the IC emission together with the isotropic component.
Then we additionally subtract the IC and the isotropic components 
and determine the Loop~I and the bubbles' templates from the residuals.
At each step the components that have not yet been determined are represented 
by proxy templates in order to avoid a bias in the fluxes.
In the end, all templates are fit to the data simultaneously to determine the spectrum of the emission components.

In this analysis,
we subtract the 2FGL point sources from the data using the 2FGL fluxes.
In addition we mask the cores of the bright point sources
with fluxes above 1 GeV greater than $2 \times 10^{-9} \; \textrm{ph cm}^{-2} {\rm s^{-1}}$ (556 sources) 
within $1^\circ$, which corresponds to the 68\% radius of the PSF at approximately 700 MeV.
We test the influence of point sources on the determination of the spectrum of the {\Fermi} bubbles by refitting the 
point sources with TS $>$ 200 to take into account flaring outside the 2FGL time window in 
Section \ref{subsec:SysErrorLocal}.

\subsection{Gas-correlated components via local template fitting}
\lb{sect:gas-cor}

Gamma-ray intensities from a given direction that arise from hadronic interactions and bremsstrahlung are
proportional to the column density of gas.
The normalization coefficient is the 
emissivity, which can be given in terms of the gamma-ray emission rate per atom.
In this section, we assume that in a limited region of the high-latitude sky (dubbed hereafter a patch),
the CR density is approximately constant.
In this case, we can determine the emissivity directly as the proportionality constant between gas column densities and gamma-ray intensities from a fit to the data in each patch of the sky.

The other emission components, i.e., IC, isotropic, Loop~I, and the {\Fermi}-bubbles, have to be modeled simultaneously with the gas-correlated component.
Otherwise the gas-correlated contribution may be overestimated.
We assume that the other components either are sufficiently smooth, or have features uncorrelated
with the gas distribution,
so that they can be modeled by a combination of some smooth functions in each patch.
We choose two-dimensional polynomials defined locally on each patch as a linear basis for the smooth functions
(determined below).

\begin{figure}[htbp] 
\vspace{-1mm}
\begin{center}
\epsfig{figure = 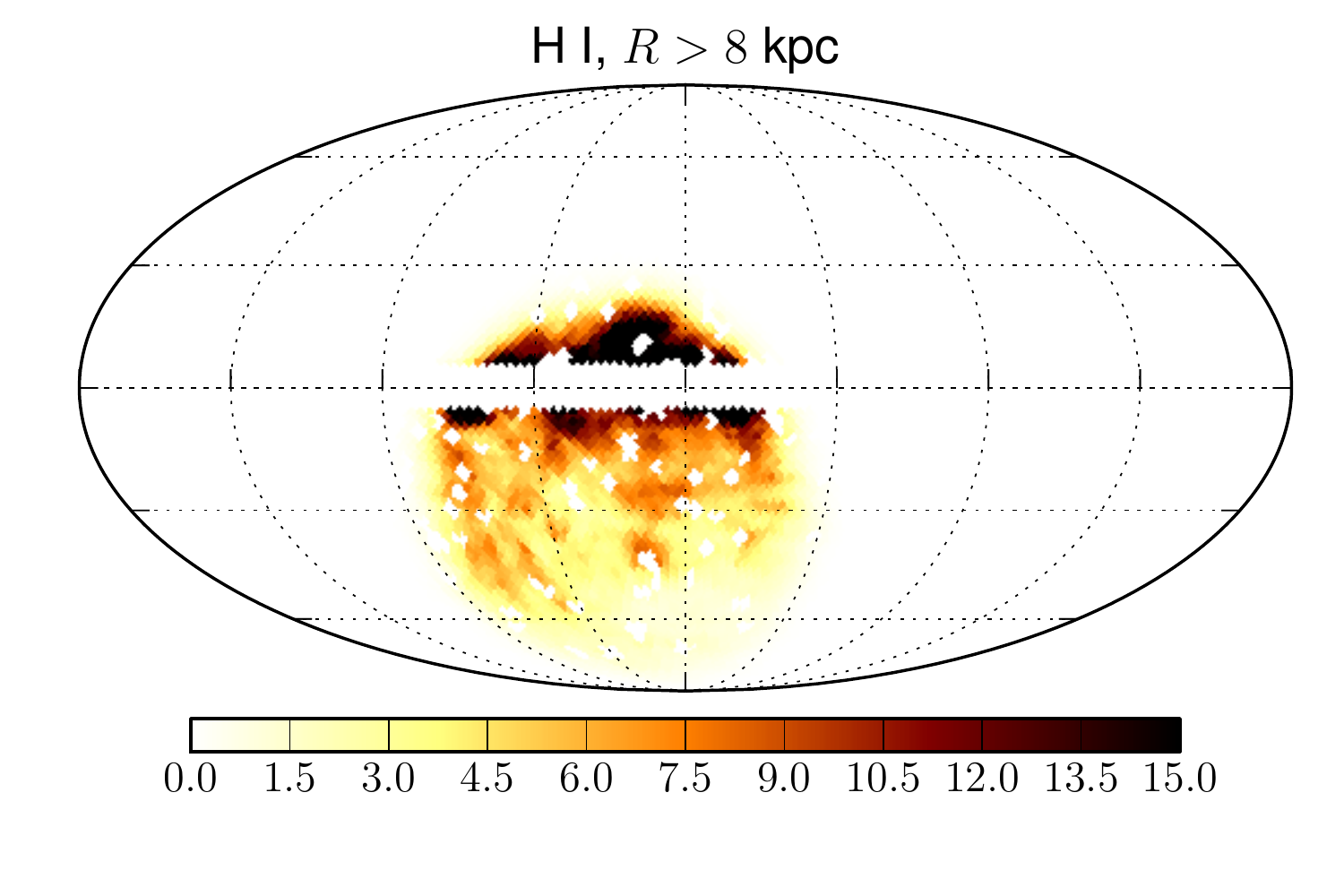, scale=\twopic}
\epsfig{figure = 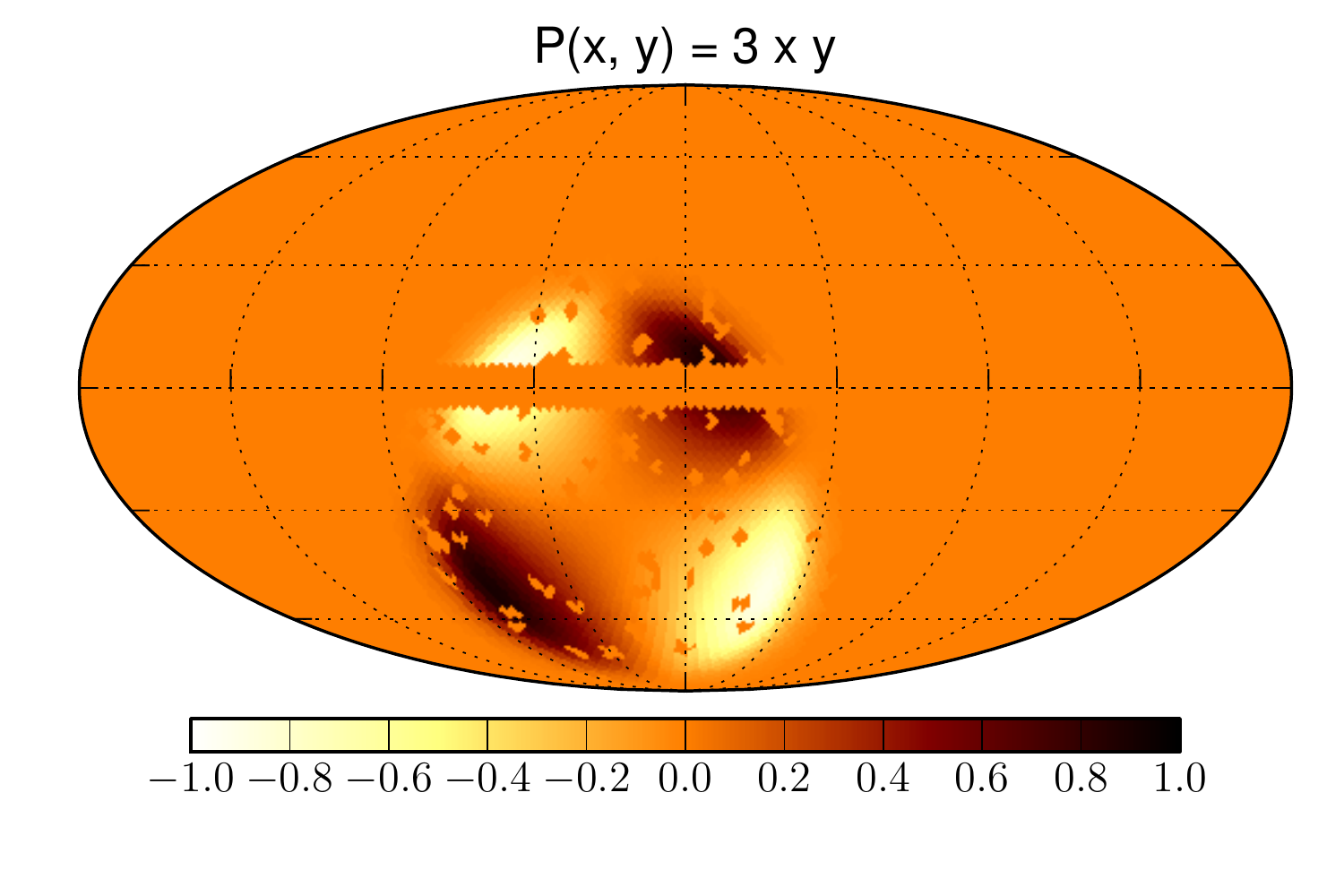, scale=\twopic}
\noindent
\caption{\small 
An example of the gas map (left) and a quadratic polynomial (right) in a local patch on the sky.
The center of the patch is at $b = -30^\circ,\,\ell = 22.5^\circ$,
the radius is $50^\circ$ and the width of the boundary is ${2}/{\lambda} = 10^\circ$.
The left map is proportional to the H~I column density integrated for $R > 8$ kpc,
where $R$ is the Galactocentric radius.
The data are fitted with a linear combination of gas maps and polynomials of different degrees.
}
\label{fig:42local_patch}
\end{center}
\vspace{1mm}
\end{figure}

In order to avoid sharp edges for the patches, which may lead to artificial features in the residuals, 
we use a hyperbolic tangent function to 
obtain patches with smooth boundaries.
The all-sky data and model maps are restricted to the patch by multiplication with the following weight function:
\be
\lb{eq:weight_def}
w(\theta_j) = \frac{1 + \tanh \ld (\theta_0 - \theta_j)}{2},
\ee
where $\theta_j$ is the angle from the center of the patch to the center of
pixel $j$, the radius of the patch is $\theta_0$ and 
the width of the boundary is $\Delta \theta = 2 / \ld$. 
This function smoothly interpolates between $w(\theta) \approx 1$ for $\theta \ll \theta_0$ and 
$w(\theta) \approx 0$ for $\theta \gg \theta_0$,
where the width of the edge is assumed to be much smaller than the patch radius, i.e., $\ld \theta_0 \gg 1$.
In Figure \ref{fig:42local_patch} we show an example of a patch with a gas template
and a local polynomial template multiplied with the weight function.

In the analysis, we separate the sky into 24 patches.
Each patch has a $50^\circ$ radius and a $10^\circ$ edge width.
There are 16 patches centered at $b = \pm 30^\circ$ and at $\ell = 22.5^\circ + 45^\circ  n$, where $n = 0, 1, 2, \ldots, 7$.
There are 8 patches centered at $b = \pm 60^\circ$ and at $\ell = 45^\circ + 90^\circ n$, where $n = 0, 1, 2, 3$.
The radius of the patches is chosen to be rather large so that the template fitting procedure converges well.
The centers of the patches are chosen to cover the sky approximately  uniformly with a significant overlap
among the patches.

The data in each patch are modeled by a combination of gas templates and a combination of
local polynomials up to a maximal degree $k_{\rm max}$ which is determined in each patch from the convergence 
of the fit (more details below).
The model in the patch $\al$ is
\be
\mu_{ij}^\al = \sum_m f^\al_{im} T^m_j + \sum_{n} k^\al_{in} P_n(x_j, y_j),
\ee
where $i$ labels the energy bins, $j$ labels the pixels in the patch, 
$n$ labels the polynomials, 
and $m$ labels the gas templates, $T_j^m$.
In this analysis, we use four gas templates: 
H~I summed over Galactocentric rings with $R < 8$ kpc,
H~I summed over rings with $R > 8$ kpc,
${\rm H}_2$ summed over all rings,
and the SFD dust template \citep{Schlegel:1997yv}.
We neglect the contribution from ionized hydrogen (H~II) in this analysis.
(If the corresponding contribution is smooth, it becomes a part of the local polynomial term).
The scaling coefficient
$f^\al_m$ is proportional to the emissivity, corresponding 
to template $m$ as fitted in patch $\al$.
Some of the patches at high latitudes 
may have little contribution from some gas templates,
such as H~I in the inner Galaxy rings or ${\rm H}_2$.
The template is included in the fit if 
the ``scalar product'' between the template and the weight function that determines the patch is sufficiently large:
$\sum_j^{N_{\rm pix}} T_j^m w^\al_j > 0.01 |T^m| |w^\al|$ 
where the norm is defined as the square root of the sum of the map values squared.
The total number of gas templates in each patch is $\leq4$. 

The second term in the model describes components that are not correlated with the gas distribution 
(isotropic, IC, Loop~I, bubbles).
These components are modeled by a combination of polynomials in local coordinates on the patch.
We define the local polynomials on the sphere by taking a polynomial function on the plane tangent to the center of the patch
and projecting the values of the function from the plane to the sphere using a stereographic projection.
In terms of local coordinates $x$ and $y$ on the tangent plane,
the polynomials are $P_n = 1, x, y, x^2, xy, y^2,\ldots$.
If $k$ is the maximal degree, then the total number of polynomials is
$n_{\rm max} = 1 + 2 + 3 + \ldots + (k + 1) = \frac{(k + 1)(k + 2)}{2}$.
The maximal degree of the polynomials $k_{\rm max}$ depends on the position of the patch 
and on the energy range.
We specify $k_{\rm max}$ at the end of this section in the discussion of the stability of the fit.

In order to speed up the calculation, we use the quadratic approximation to the log likelihood,
which for all-sky fits takes the form
\be
\lb{eq:loglike}
- 2 \log \mathcal{L} 
\approx \sum_i^{\rm E\, bins} \sum_j^{\rm pixels} \frac{(d_{ij} - \mu_{ij})^2}{\sm^2_{ij}},
\ee
where $d_{ij}$ and $\mu_{ij}$ are the number of gamma rays in the data and the model prediction
for energy bin $i$ in pixel $j$.
We use the smoothed counts map $\td{d}_{ij}$ as an estimator for the standard deviation $\sm^2_{ij}$, where the smoothing radius $R_i$ depends on energy bin $i$.
For each energy bin, the radius is chosen such that there are on average at least 100 photons inside the circle 
of radius $R_i$.
The minimal value of the radius is $2^\circ$, which corresponds to an average over about 15 pixels.
As we discuss in Appendix \ref{sect:bias},
the choices $\sm^2_{ij} = d_{ij}$ or $\sm^2_{ij} = \mu_{ij}$ result in biased $\chi^2$ fitting,
while $\sm^2_{ij} = \td{d}_{ij}$ reduces the bias.

The data fitting in a local patch $\al$ is performed with the following weighted $\chi^2$
\be
\lb{eq:logLL}
\chi^2 = \sum_i^{\rm E\, bins} \sum_j^{\rm pixels} {w^\al_j}^2 \frac{(d_{ij} - \mu^\al_{ij})^2}{\sm^2_{ij}}.
\ee
Notice that the weighted $\chi^2$ is equivalent to a multiplication of the data $d_{ij}$ and the model $\mu^\al_{ij}$ 
with the weight $w^\al_j$ (without changing $\sm_{ij}$).
The best-fit parameters and, hence, the model $\mu^\al_{ij}$ depend on the patch.
In this part of the analysis we mask $|b| < 5^\circ$ which is different from the Galactic plane 
mask $|b| < 10^\circ$ adopted in the rest of the paper.
The reason is that the regions closer to the Galactic plane have more features in the gas maps,
especially for the inner Galaxy H~I template and the H$_2$ template. Including these regions improves the convergence of the fits.

We use different fitting strategies at low and high energies.
At low energies (0.1 - 10 GeV), the statistics are high and the 
normalization of the templates can be determined 
in every energy bin independently.
At high energies the statistics are low.
In order to avoid high statistical fluctuations among the energy bins,
we assume that a single power law is a good approximation for the gas-correlated 
emission and find the best-fit power-law index for the energy bins between 3 and 500 GeV.
In the intermediate region, i.e., between 3 and 10 GeV, we take the mean of the spectra. 
The best-fit combinations of polynomial templates are determined independently in each patch and in each energy bin
both at low and at high energies.

The resulting contribution of the gas-correlated components depends on the degree of the polynomials
used to model the other components.
If the degree of the polynomial is too small, then the contribution of gas-correlated components will be overestimated
due to absorption of the flux from the other components.
For imperfect gas templates, a large degree of the polynomials leads to more of the gas-correlated gamma-ray emission being attributed to the polynomial templates.
For a sufficiently high degree, the polynomial templates will start to respond to the noise, thus, over-fitting the data.
Different portions of the sky require a different maximal degree of the polynomials.
Near the Galactic plane and close to local features, such as the {\Fermi} bubbles,
the degree should be larger than at high latitudes,
where the isotropic and the local IC emission can be described by
smoother templates.

In the following we describe the determination of the maximal degree of the polynomials for a particular patch.
The convergence of $\chi^2$ is not a good indicator: the $\chi^2$ value decreases 
with larger degree polynomials as more and more features are included in the model,
but it does not reach the level of statistical noise up until a very high degree, where we are already likely to include some of the gas-correlated emission into the local polynomials.
One of the characteristics of a good model of gas-correlated emission is the stability of the corresponding
spectrum as the degree increases:
we stop increasing the degree of the polynomials $k$ when the differences between the 
photon spectra of the gas-correlated component become smaller than a certain threshold.
In particular, we calculate the difference of the spectra between $k$ and $k' = k-1,\, k-2$,
i.e., we compare the last spectrum with the previous two spectra
\be
t^{(k')} = \frac{1}{N_{\rm bins}} \sum_{i}^{\rm E\, bins}  \frac{(F^{(k)}_i - F^{(k')}_i)^2}{{\sigma_i^{(k)}}^2},
\ee
where $F^{(k)}_i$ is the gas-correlated intensity for $|b| > 5^\circ$ in an energy bin $i$
derived with the maximal degree of polynomials $k$,
${\sigma_i^{(k)}}$ is the statistical uncertainty of $F^{(k)}_i$.
At energies below 10 GeV, we use the threshold $t^{(k')} < 20$.
This is much larger than the expected random fluctuations,
but at low energies the differences are dominated by the systematic uncertainties and
this level gives a good fit in Monte Carlo simulations.
At energies above 3 GeV, the stability condition is $t^{(k')} < 1.5$,
which is comparable to the statistical noise.
In each case, if the stability condition is not satisfied, we stop at a maximal degree of 12,
which corresponds to an angular scale of about~$9^\circ$.

\begin{figure}[htbp] 
\vspace{-1mm}
\begin{center}
\epsfig{figure = 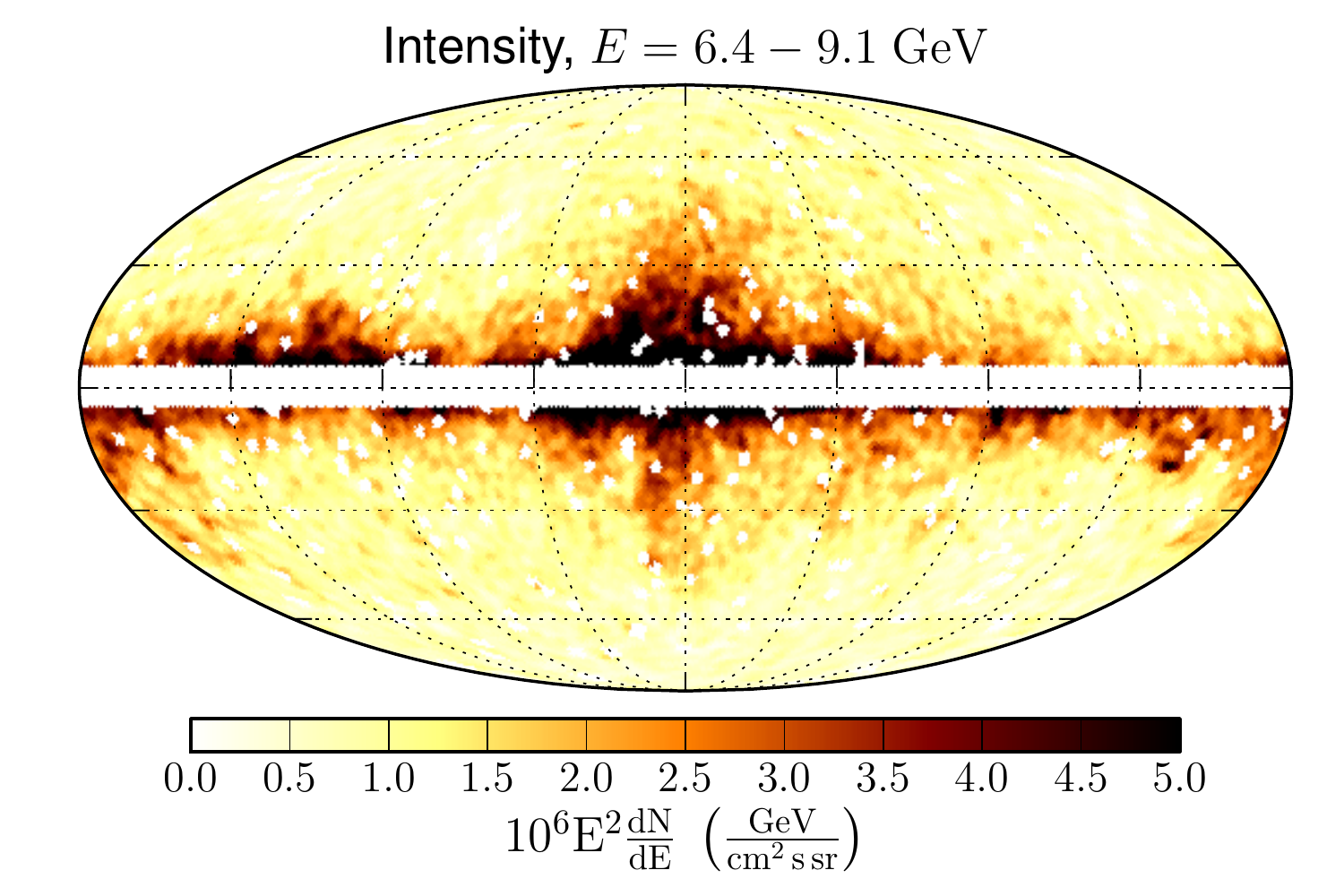, scale=\twopic}
\epsfig{figure = 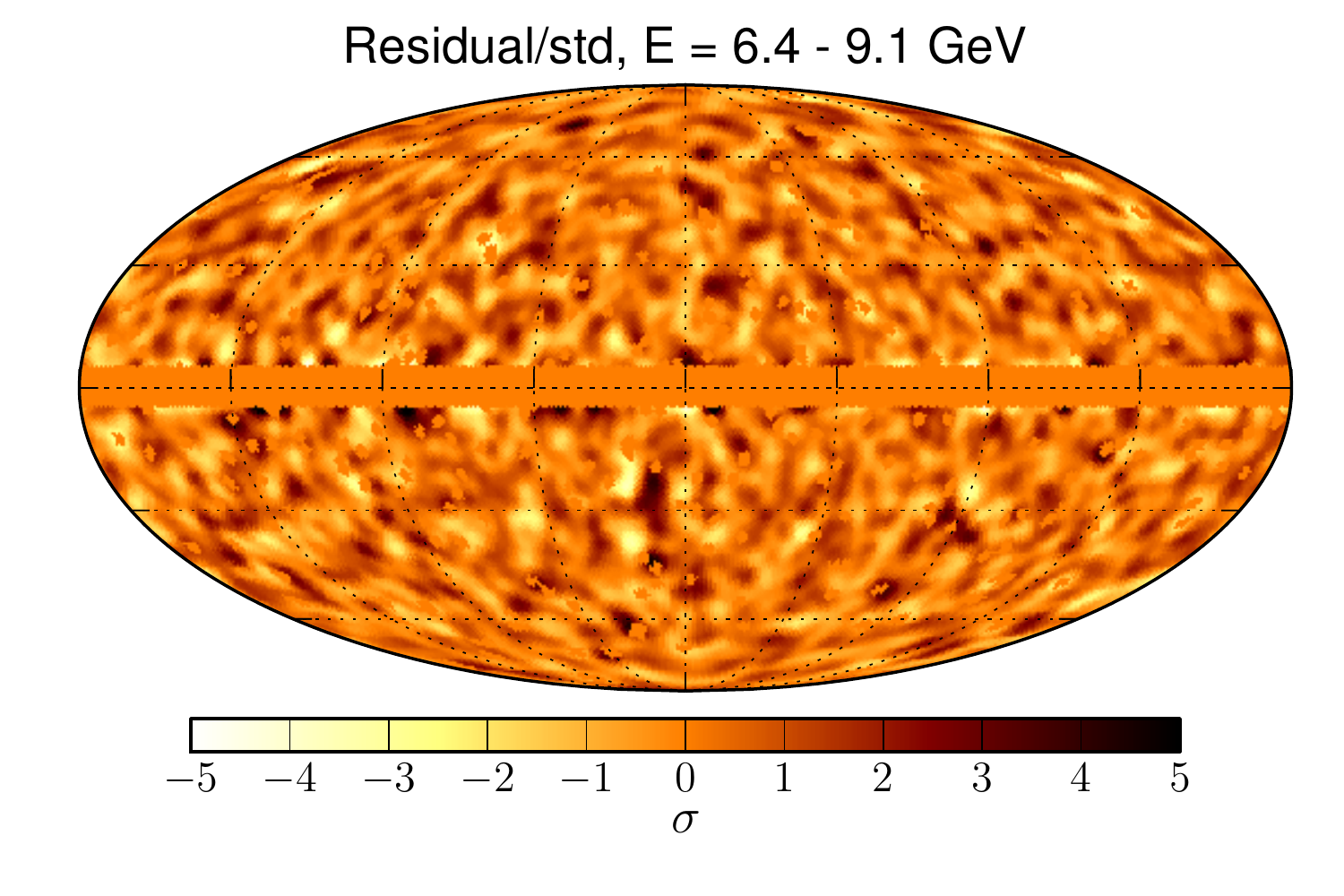, scale=\twopic} \\
\epsfig{figure = 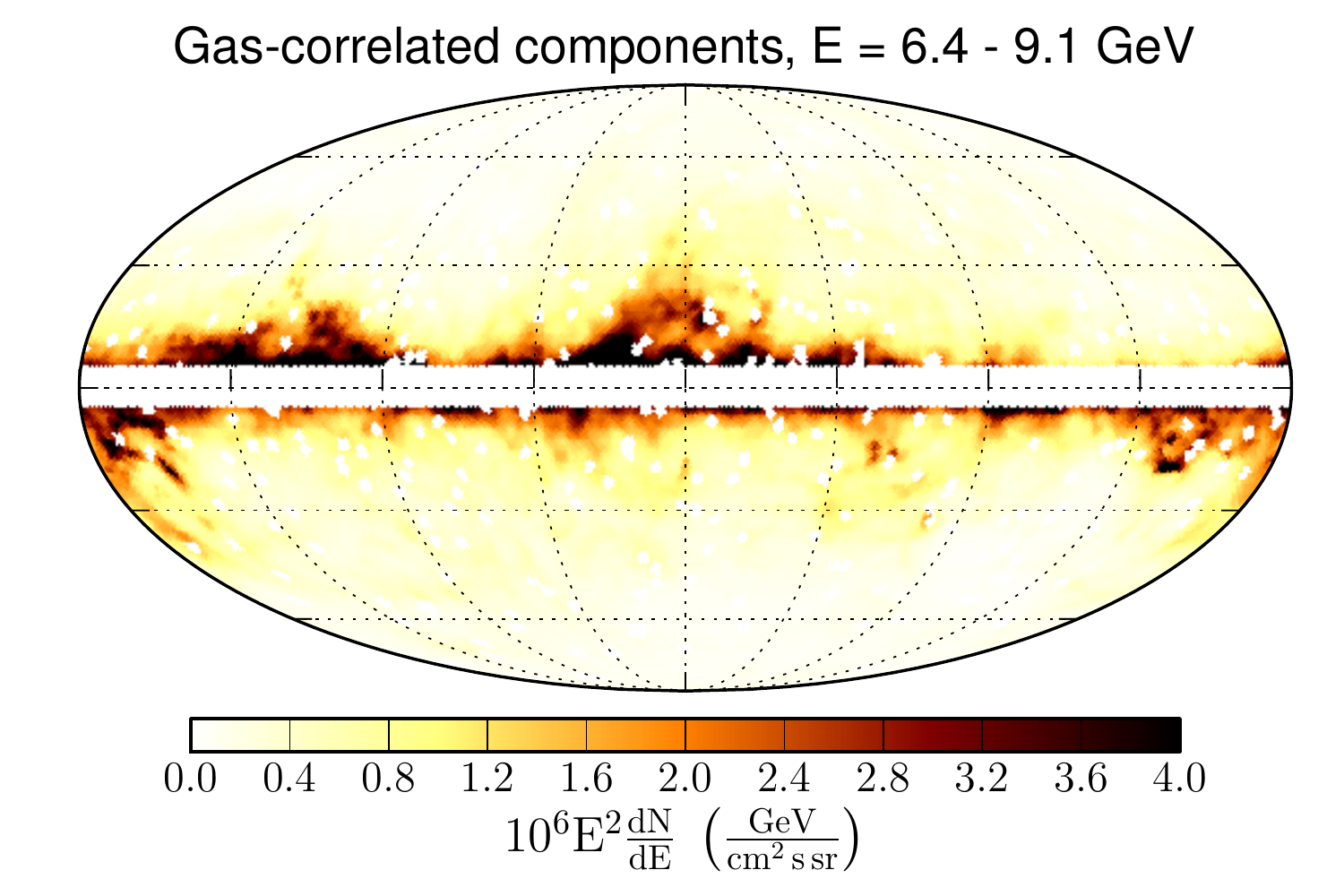, scale=\twopic}
\epsfig{figure = 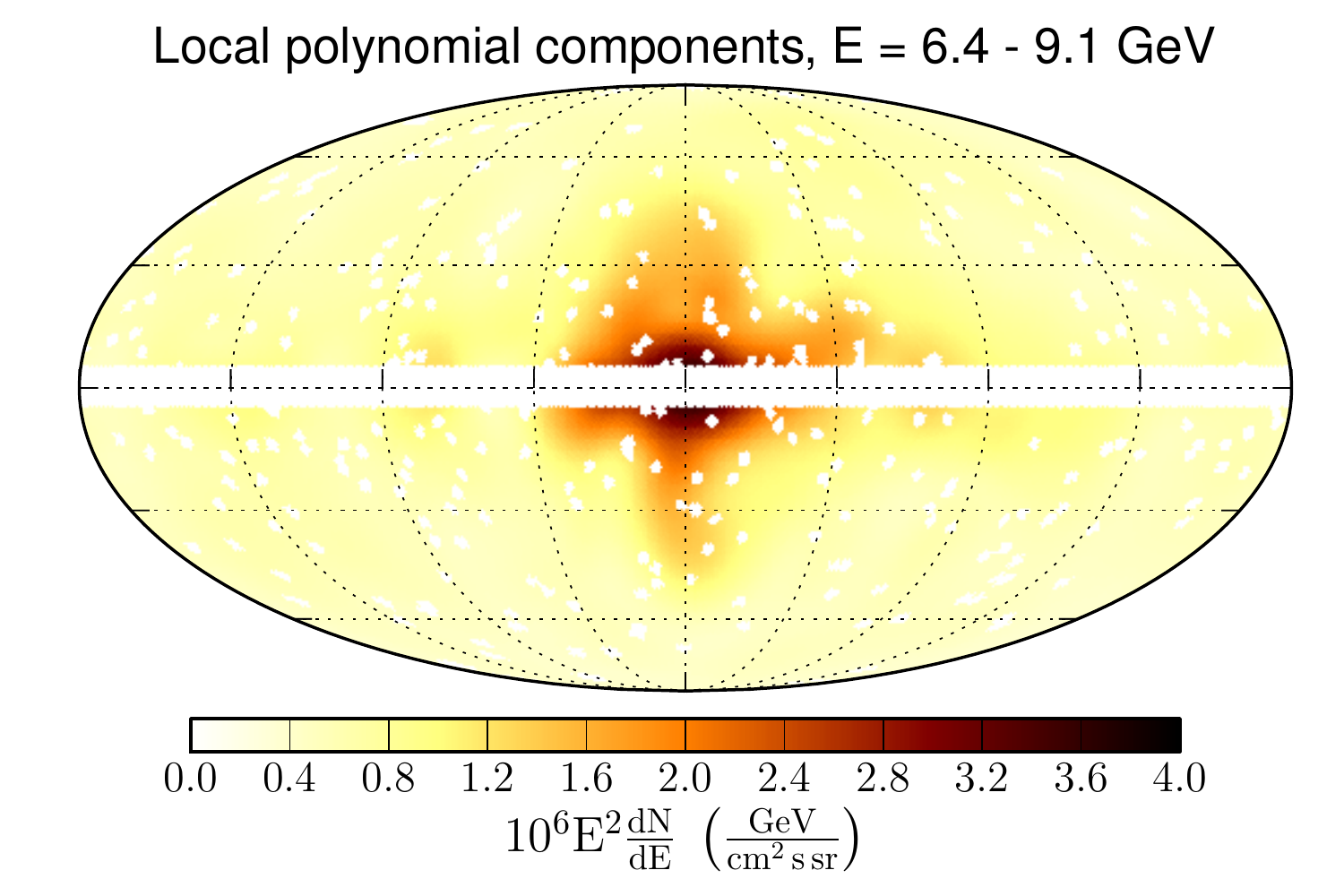, scale=\twopic}
\noindent
\caption{\small 
An example of fitting a combination of gas-correlated templates 
and a local polynomial model to the data.
Top left: gamma-ray intensity in the 6.4 - 9.1 GeV energy bin.
Bottom left: gas-correlated gamma-ray emission determined from the fit to the gamma-ray data.
Bottom right: emission components not correlated with the gas templates modeled
by a combination of local polynomials.
Top right: the residual map.
The details of the fitting procedure are described in Section \ref{sect:gas-cor}.
}
\label{fig:44fit_example}
\end{center}
\vspace{1mm}
\end{figure}

In each energy bin, all-sky models are obtained as a weighted average of the models in the patches
\be
\lb{eq:merge}
\mu_{ij} = \frac{\sum_\al w_j^\al \mu_{ij}^\al}{\sum_\al w_j^\al}.
\ee
An example of gamma-ray counts, template maps and residuals
is presented in Figure \ref{fig:44fit_example}.
Sharp features uncorrelated with the gas templates remain in the residual map in Figure \ref{fig:44fit_example}, 
e.g., the left edge of the southern bubble.
In the following, we use the weighted sum of the gas-correlated components as an all-sky template to determine 
the templates and the spectrum of the other components.

\subsection{IC and isotropic components}
\label{sect:IC_comp}

\begin{figure}[htbp] 
\vspace{-1mm}
\begin{center}
\epsfig{figure = 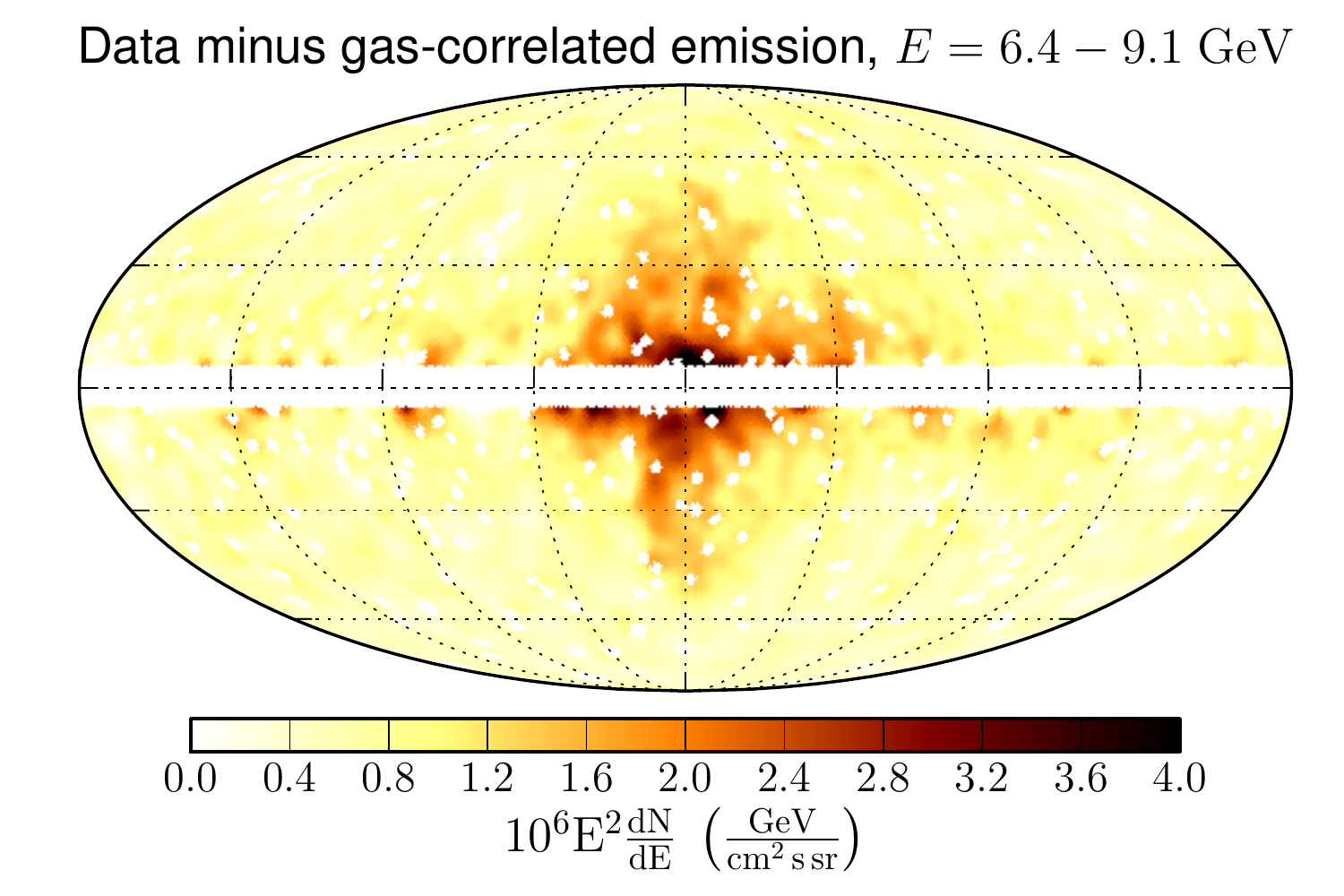, scale=\twopic}
\epsfig{figure = 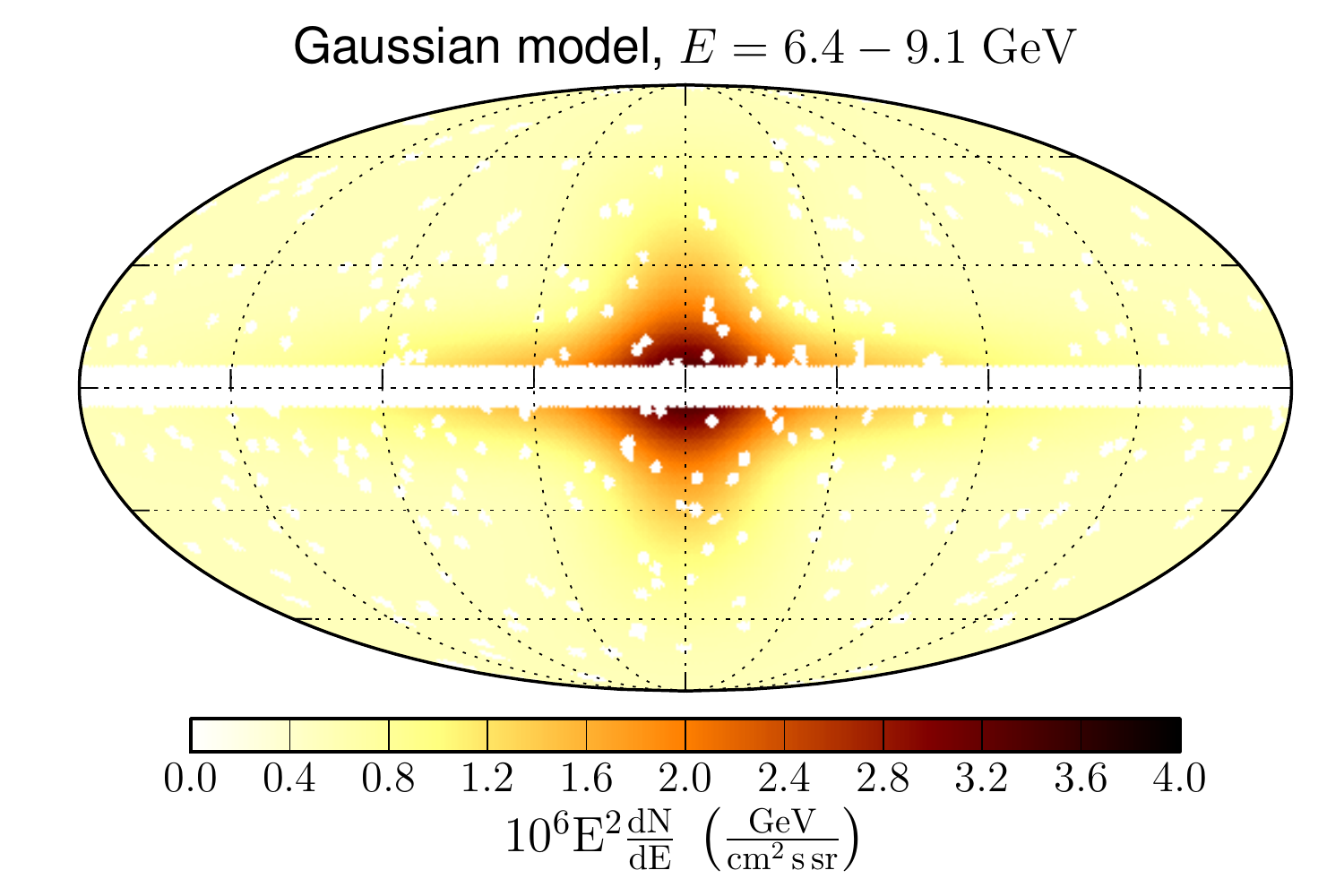, scale=\twopic}
\noindent
\caption{\small 
Left: data minus gas-correlated emission residuals in the energy bin $E = 6.4 - 9.1$ GeV
(smoothed with a $2^\circ$ Gaussian kernel).
Right: a model of the residual
with two Gaussians templates and an isotropic template.
The Gaussian along the Galactic plane models the IC emission.
The Gaussian that is more extended perpendicular to the plane
is a proxy template for Loop I and the bubbles.
}
\label{fig:45pi0resid}
\end{center}
\vspace{1mm}
\end{figure}

The next step is to model the IC and isotropic components.
First,
we subtract the point sources and the gas-correlated component found in the previous subsection from the data.
Examples of the polynomial models and the residuals after subtraction of the gas-correlated components are shown
in Figures \ref{fig:44fit_example} and \ref{fig:45pi0resid}.
One can notice the presence of two distinct components: 
a component along the Galactic disk (mostly IC) 
and a halo component (mostly Loop~I and the {\Fermi} bubbles). 

We model both the disk and the halo components by bivariate Gaussians with parameters
$\sm^{\rm disk}_{\rm b}$ and $\sm^{\rm disk}_{\rm \ell}$ and
$\sm^{\rm halo}_{\rm b}$ and $\sm^{\rm halo}_{\rm \ell}$ respectively.
The centers of the Gaussians are fixed at the GC.
We fit the two Gaussians together with the isotropic template to the residuals obtained by subtracting the gas-correlated
emission components and the point sources from the data.
The Gaussian for the halo is a proxy template for the bubbles and Loop~I, and is necessary to avoid a bias in the determination of the 
disk template.
The parameters of the Gaussians 
are fitted independently in each energy bin below 30 GeV. 
At higher energies, the parameters of the Gaussians are determined from a fit to the flux integrated above 30 GeV.
The Gaussian model in the energy bin $(6.4 - 9.1)$ GeV is shown in Figure \ref{fig:45pi0resid}.
In this section and below, we use the global $\chi^2$ fitting procedure described in Equation (\ref{eq:loglike})
without the additional weight factors introduced for the local templates analysis in Equation (\ref{eq:logLL}),
i.e., we perform an all-sky fit instead of the local fit in patches.

\subsection{Bubbles and Loop~I}
\label{subsec:localBubbles}

\begin{figure}[htbp] 
\vspace{-1mm}
\begin{center}
\epsfig{figure = 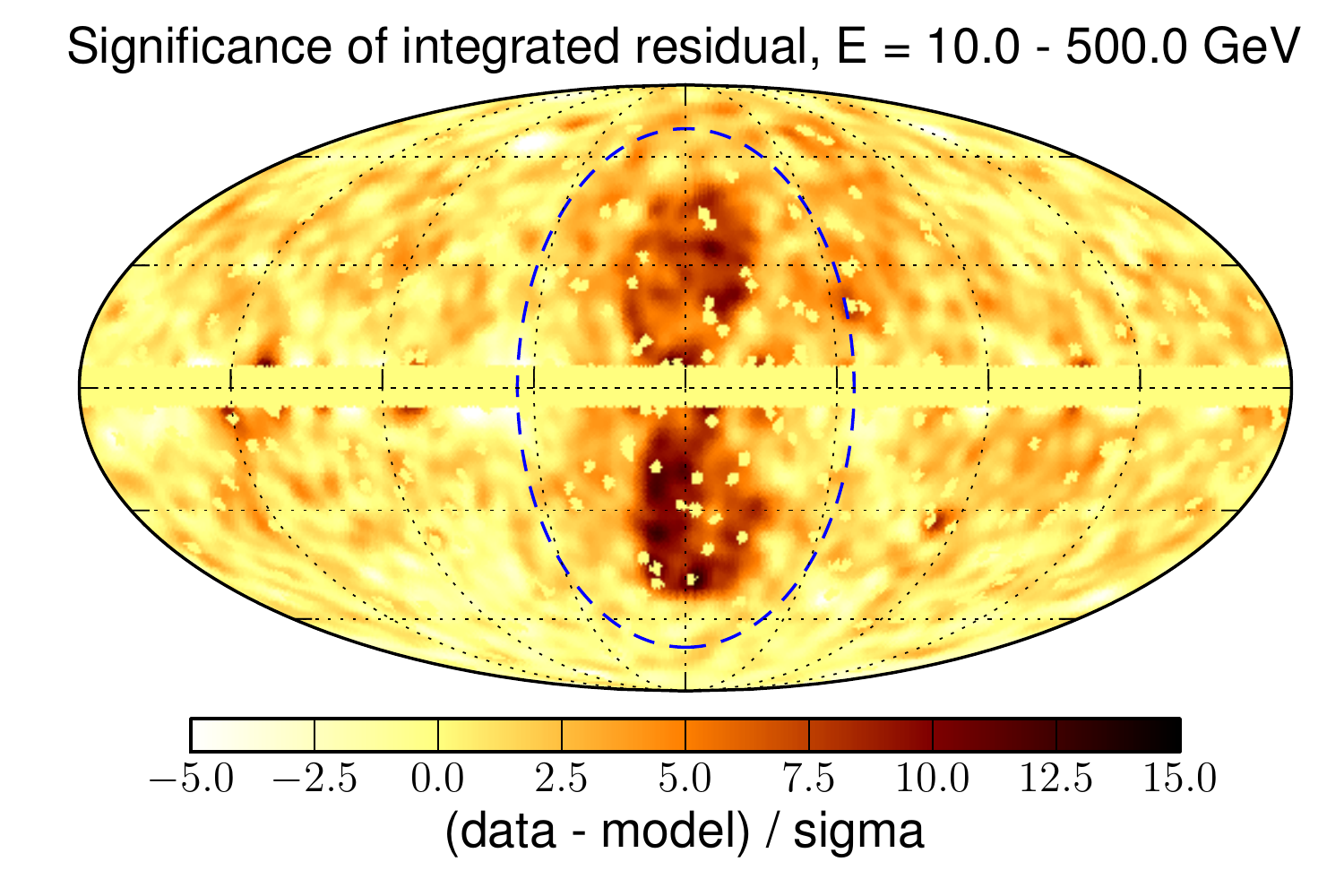, scale=\twopic}
\noindent
\vspace{-2mm}
\caption{\small 
Residuals after subtracting the gas-correlated, disk, and isotropic components.
The map shows the residuals integrated above 10 GeV in significance units 
(data minus model over the standard deviation of the data).
Dashed ellipse: the region that includes the bubbles.
}
\label{fig:47bbl_resid}
\end{center}
\vspace{1mm}
\end{figure}

\begin{figure}[htbp] 
\vspace{-1mm}
\begin{center}
\epsfig{figure = 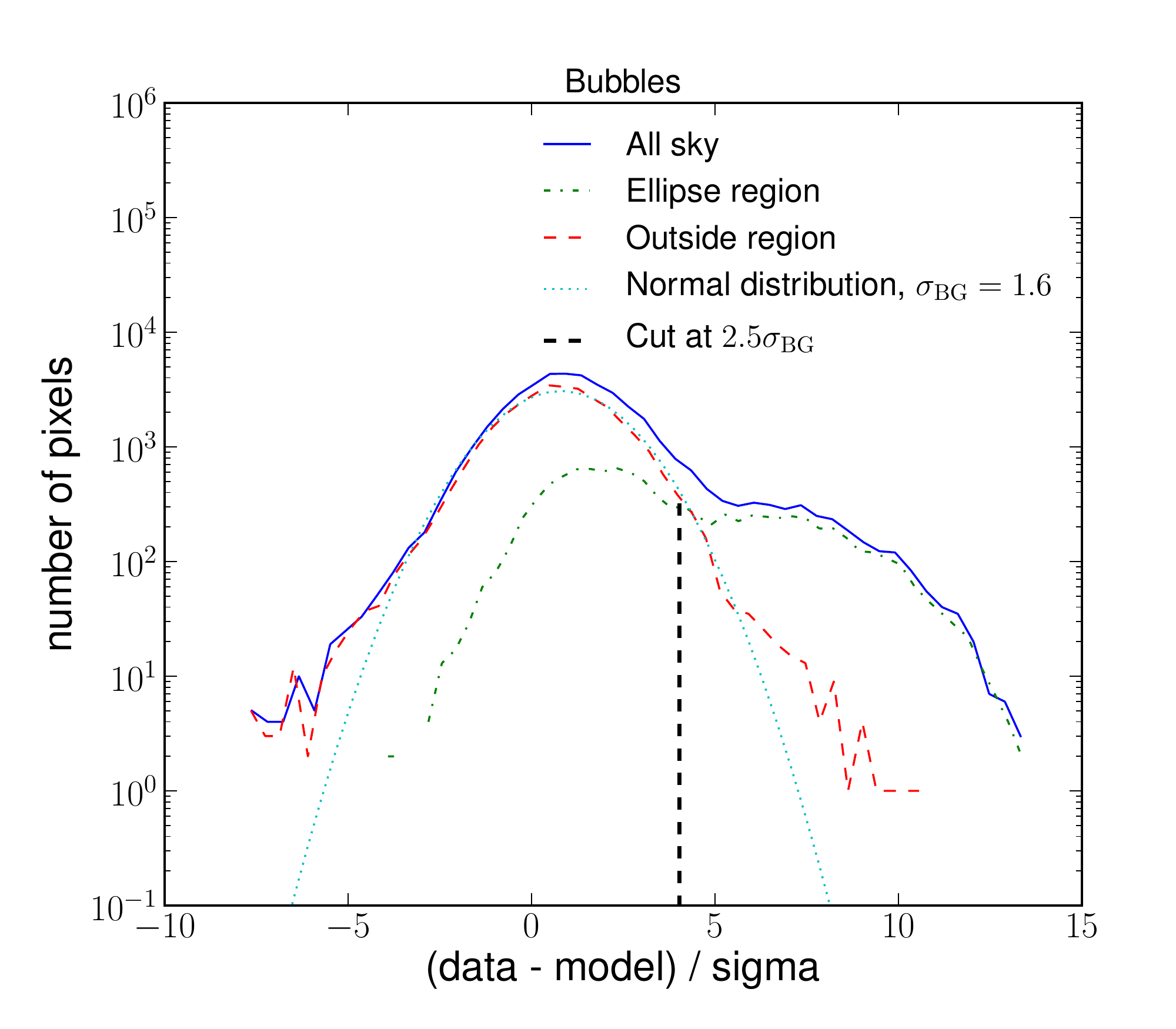, scale=\threepic}
\epsfig{figure = 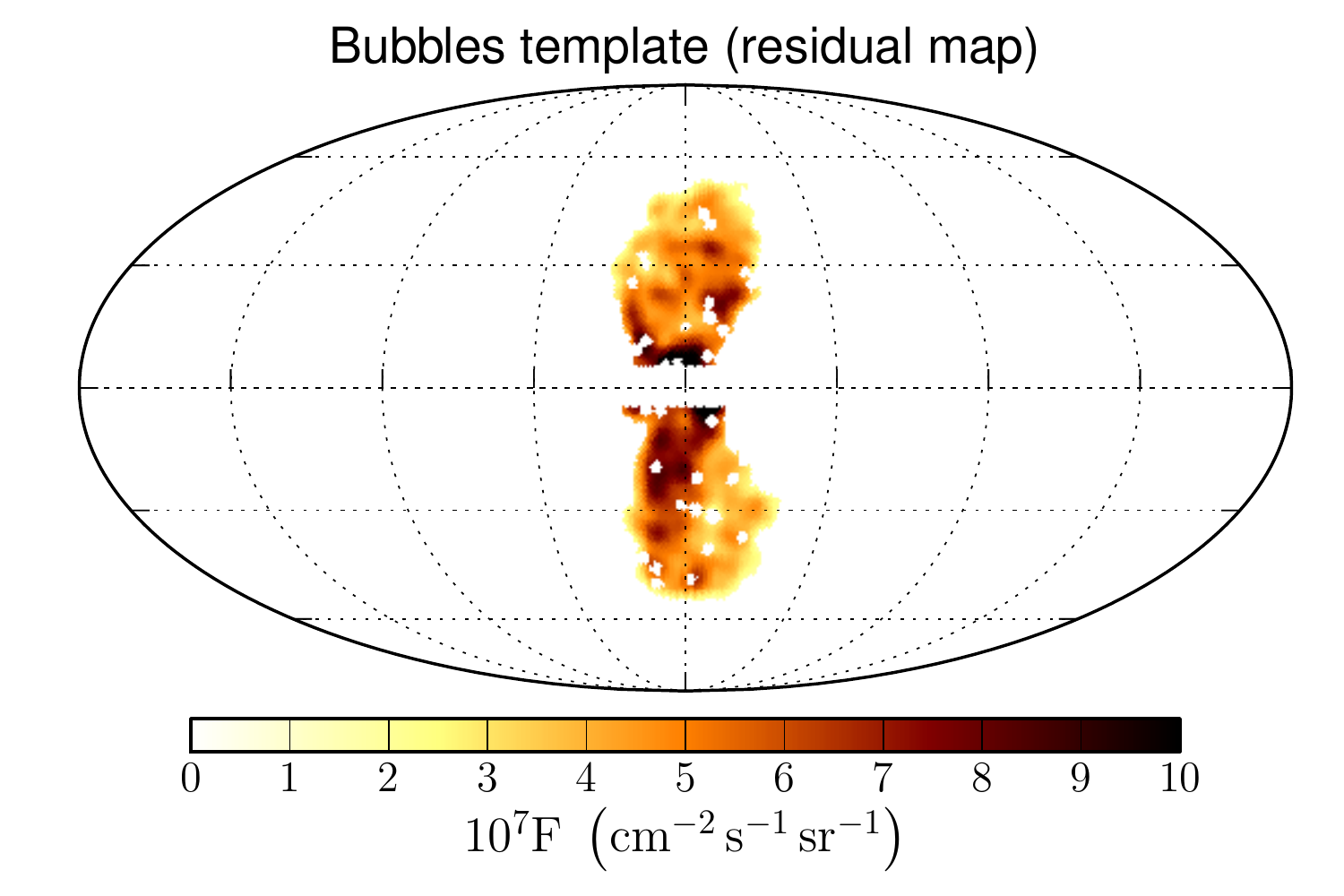, scale=\twopic} 
\noindent
\caption{\small 
Left: histogram of pixel counts.
Solid blue line: total counts for the residual map in Figure \ref{fig:47bbl_resid}.
Green dash-dotted line: counts inside the elliptical region including the bubbles.
Red dashed line: counts outside of the elliptical region.
Cyan dotted line: Gaussian fit to the background counts in the outside region.
The width of the distribution is $\sigma_{\rm BG} = 1.6$, 
which is larger than the Gaussian noise due to unresolved residual features on the map, such as 
faint point sources and structures in the Galactic diffuse emission not included in the model.
Black dashed vertical line: the threshold level of $2.5 \sigma_{\rm BG}$ used to determine the template of the bubbles.
Right: the template of the bubbles, determined by 
applying the threshold to the residual significance map.
}
\label{fig:48bbl_template}
\end{center}
\vspace{1mm}
\end{figure}

We define the template of the bubbles from the residual flux after
subtracting the gas-correlated, isotropic, and disk components from the data. 
We do not subtract the halo component, which only served as a proxy for bubbles and Loop~I in the previous step. 
The template for the bubbles is derived from the residual flux
integrated above 10 GeV (Figure \ref{fig:47bbl_resid}).
Compared to the derivation of the template of the bubbles in Section \ref{subsec:BubbleTemp},
here we use the energy range above 10 GeV to test the uncertainty related to the choice of the lower energy bound
(compared to 6.4 GeV in Section \ref{subsec:BubbleTemp}).
The histogram of pixel counts inside and outside of the bubbles' region and the template of the bubbles
are shown in Figure \ref{fig:48bbl_template}.
For the energy range above 10 GeV the pixel counts in the background region intersect the distribution of pixel counts in the ellipse region
around $2.5 \sigma_{\rm BG}$, which we use in the definition of the template of the bubbles.

In order to separate Loop~I from the {\Fermi} bubbles, we determine these templates from a correlation with 
the spectra of the two components between 0.7 GeV and 10 GeV,
where the contribution from both Loop~I and the bubbles is significant.
The energy range is chosen to be relatively small so that the spectra are well approximated by a simple power-law function.

The derivation of templates correlated with the known spectra
is similar to the derivation of the spectra for known templates.
If we represent the residuals after subtracting the gas-correlated, IC, 
and isotropic components in $k$ energy bins and in $N$ pixels as a $k \times N$ matrix $D$,
then, assuming that we can neglect the statistical uncertainty,
the problem of separating this residual into $m$ components
is equivalent to the following matrix separation problem \citep[e.g.,][]{2012arXiv1202.1034M}
\be
D = F  \cdot T,
\ee
where $F$ is a $k \times m$ matrix of the spectra and $T$ is an $m \times N$ matrix of templates.
If the spectra $F$ are known,
then the corresponding templates are determined as
\be
\lb{eq:sca}
T = (F^T \cdot F)^{-1} \cdot (F^T \cdot D).
\ee
This solution also works in the case of uniform statistical uncertainties. 
In the case of a non-uniform uncertainties, one has to minimize the $\chi^2$
\be
\chi^2 = \sum_i^{\rm E\, bins} \sum_j^{\rm pixels} \frac{(D_{ij} - \sum_m F_{im} T_{mj})^2}{{\sm_{ij}}^2}.
\ee
The only effect of this more general derivation is an inclusion of a factor $\frac{1}{{\sm_{ij}}^2}$ in all sums over 
energy bins and over pixels in Equation (\ref{eq:sca}).

\begin{figure}[htbp] 
\vspace{-1mm}
\begin{center}
\epsfig{figure = 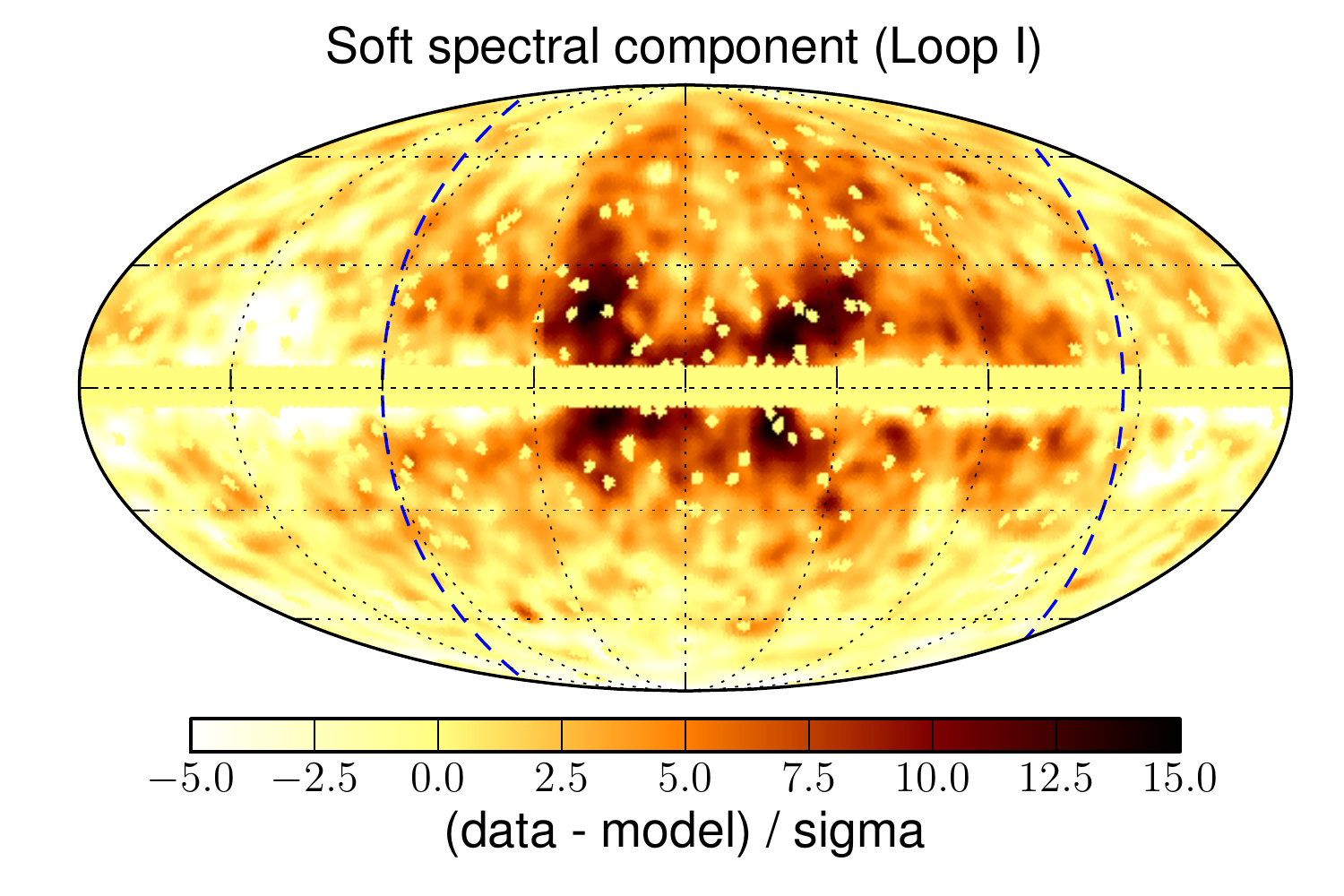, scale=\twopic}
\epsfig{figure = 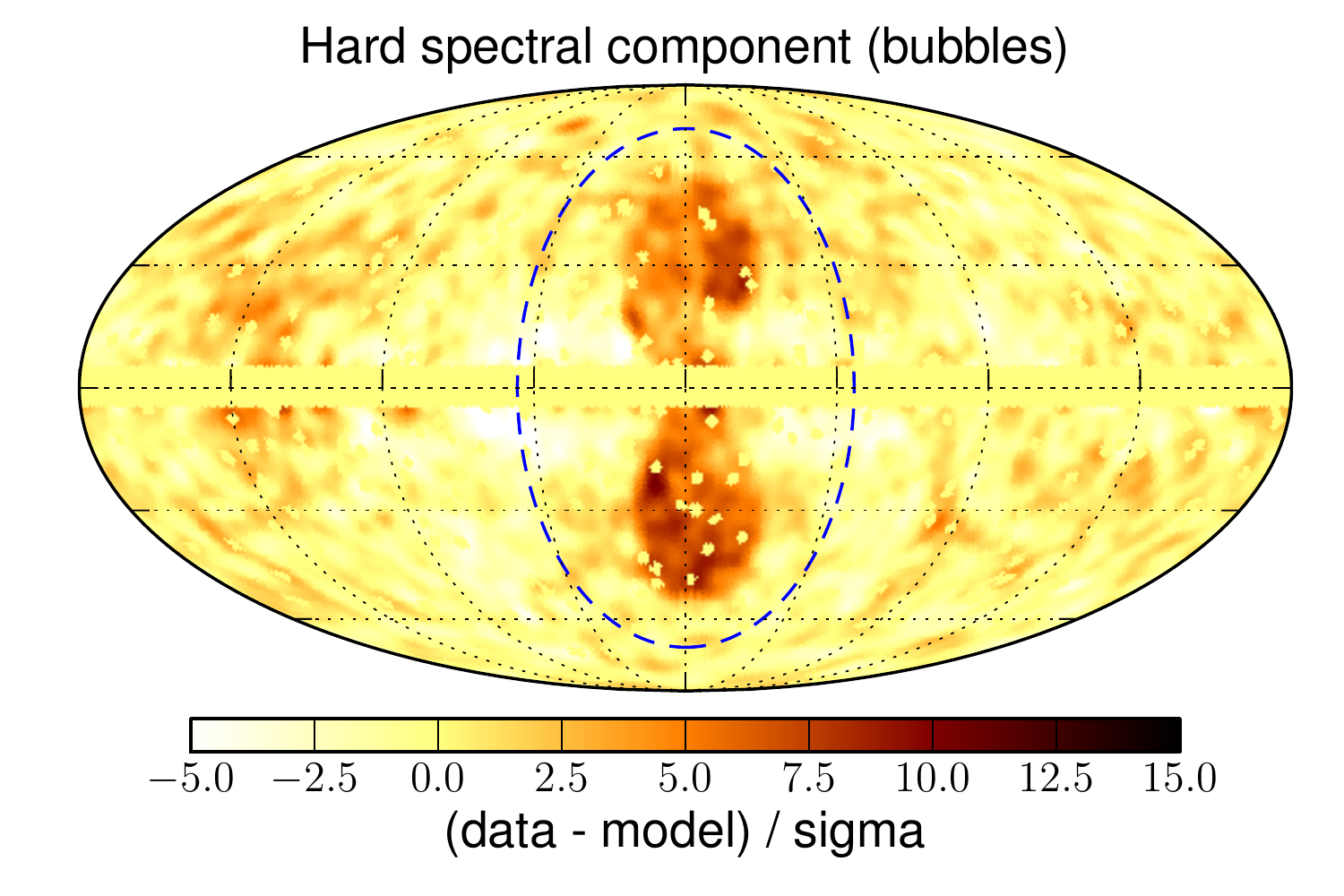, scale=\twopic}
\noindent
\caption{\small 
Soft and hard spectral components of the residuals between 700 MeV and 10 GeV.
Left: soft component  $\sim E^{-2.4}$.
Right: hard component $\sim E^{-1.9}$.
The maps are in significance units.
Dashed ellipses: regions that include Loop I and the bubbles.
}
\label{fig:49sca_maps}
\end{center}
\vspace{1mm}
\end{figure}

We assume that the residuals between 0.7 GeV and 10 GeV are dominated by two components:
hard and soft components corresponding to the bubbles and Loop~I respectively.
We estimate the spectrum of the soft component from residuals outside of the region of the {\Fermi} bubbles:
$45^\circ < b < 60^\circ$ and $285^\circ < \ell < 330^\circ$. 
A power-law fit in this region has a photon index $2.4$.
The spectrum of the hard component is determined from the residuals near the southern edge of the {\Fermi} bubbles:
$-55^\circ < b < -40^\circ$ and $-15^\circ < \ell < 15^\circ$.
A power-law fit in this region has a photon index $1.9$.
Thus, to determine the Loop~I template we use $F_{\rm soft} \propto E^{-2.4}$,
while for the template of the bubbles we use $F_{\rm hard} \propto E^{-1.9}$.
The corresponding hard and soft components are shown in Figure \ref{fig:49sca_maps}.
To check the systematic uncertainty, 
we vary the indices in the definitions of the templates in Section \ref{subsec:SysErrorLocal}.

\begin{figure}[htbp] 
\vspace{-1mm}
\begin{center}
\epsfig{figure = 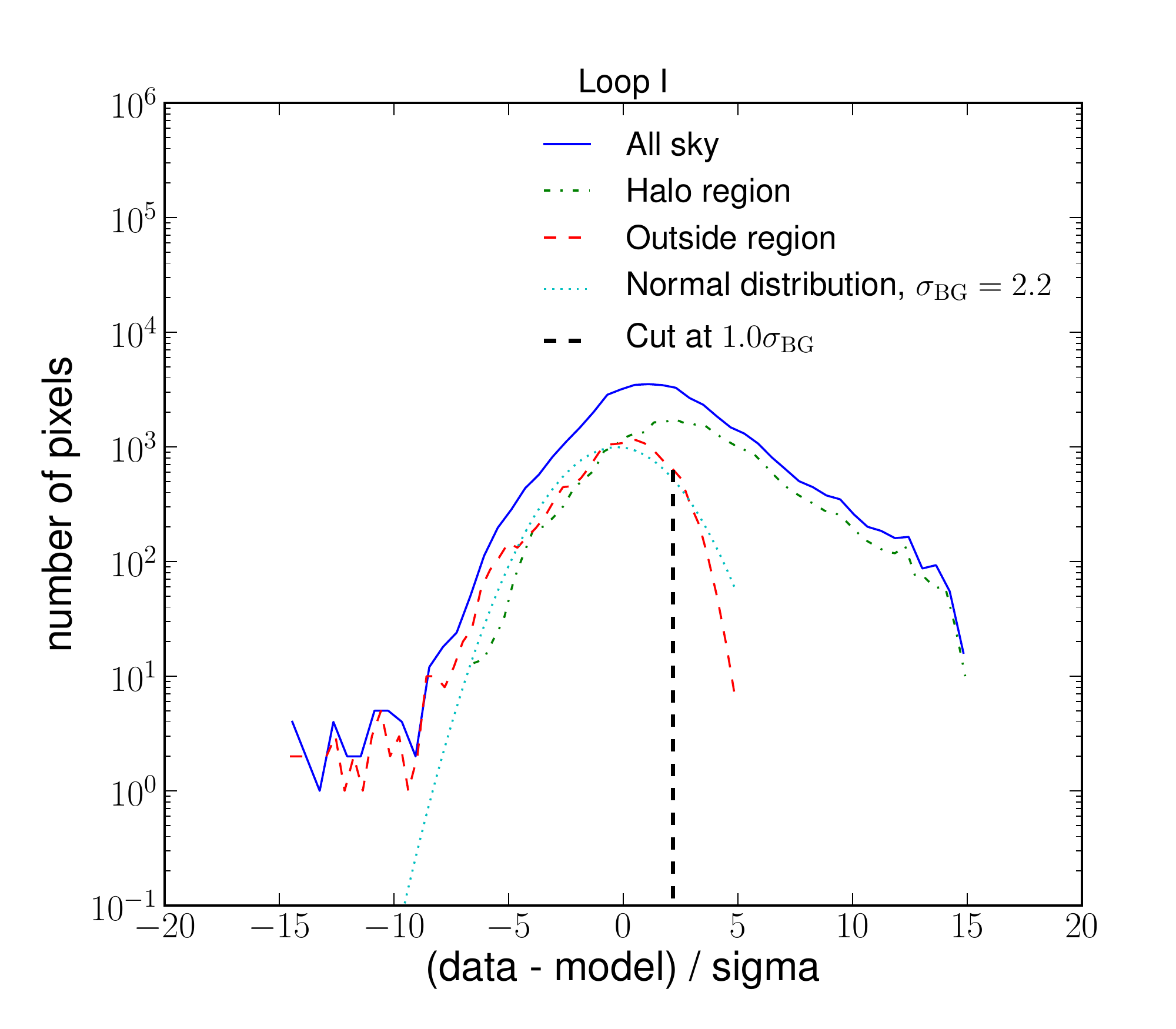, scale=\threepic}
\epsfig{figure = 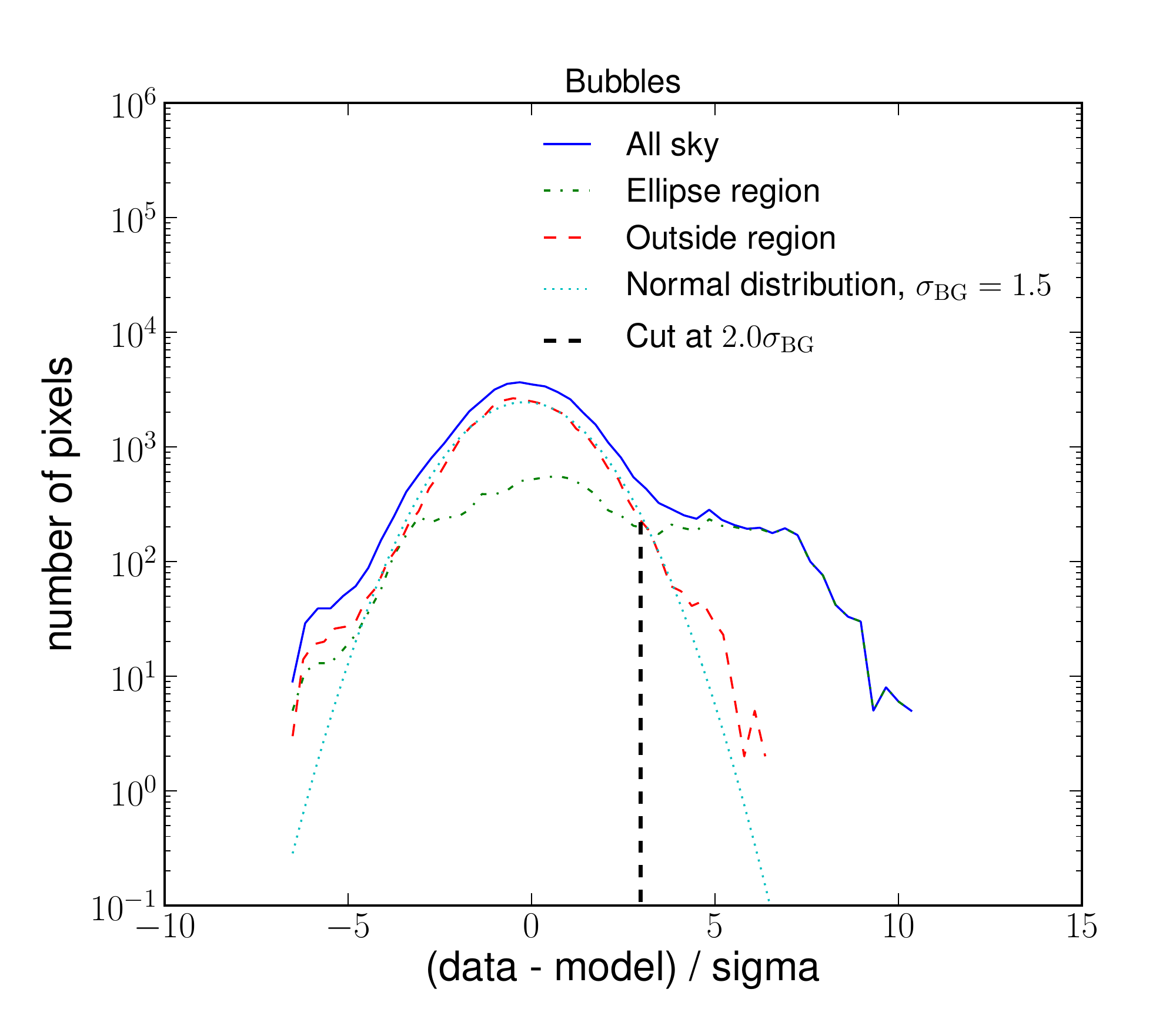, scale=\threepic} \\
\epsfig{figure = 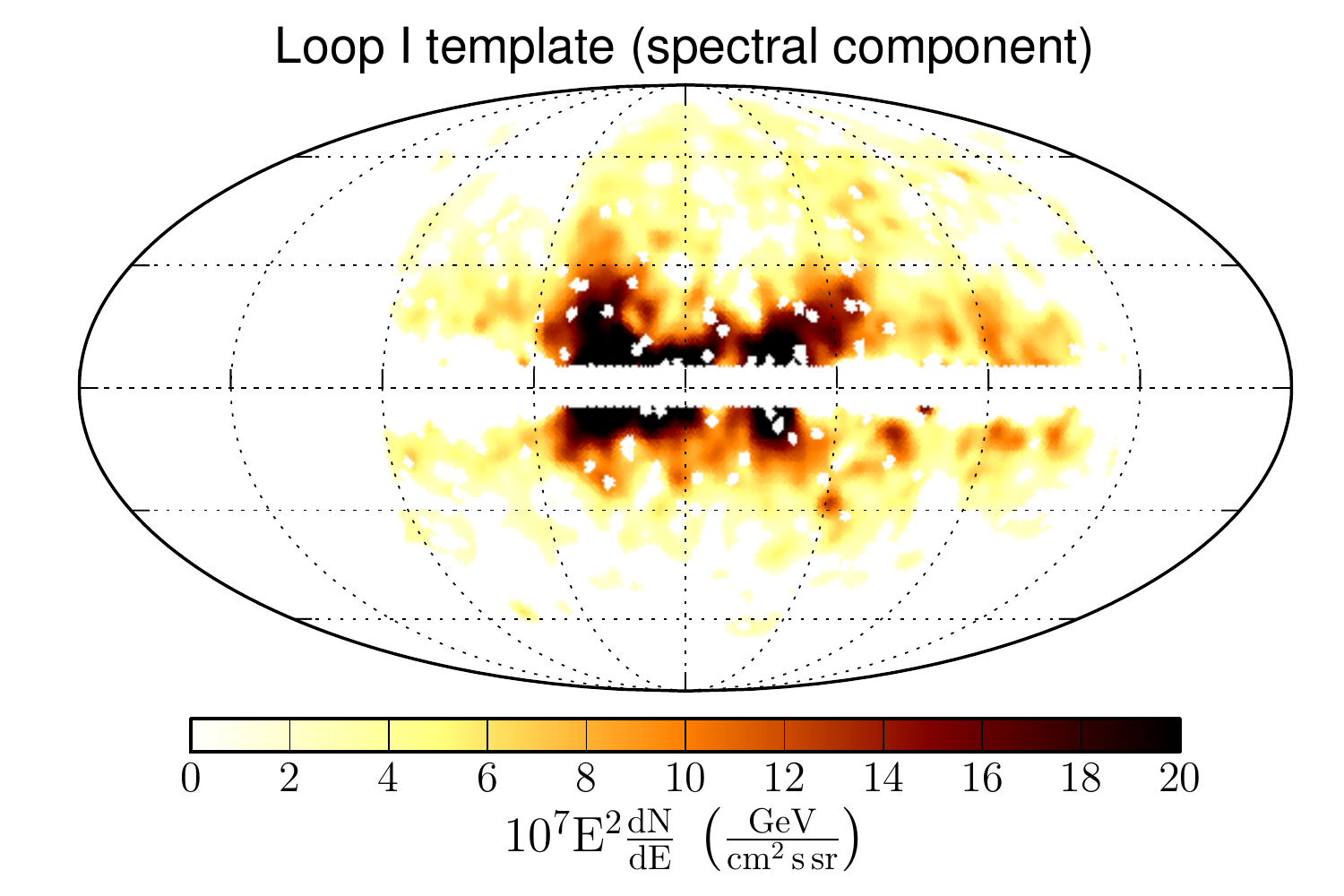 , scale=\twopic}
\epsfig{figure = 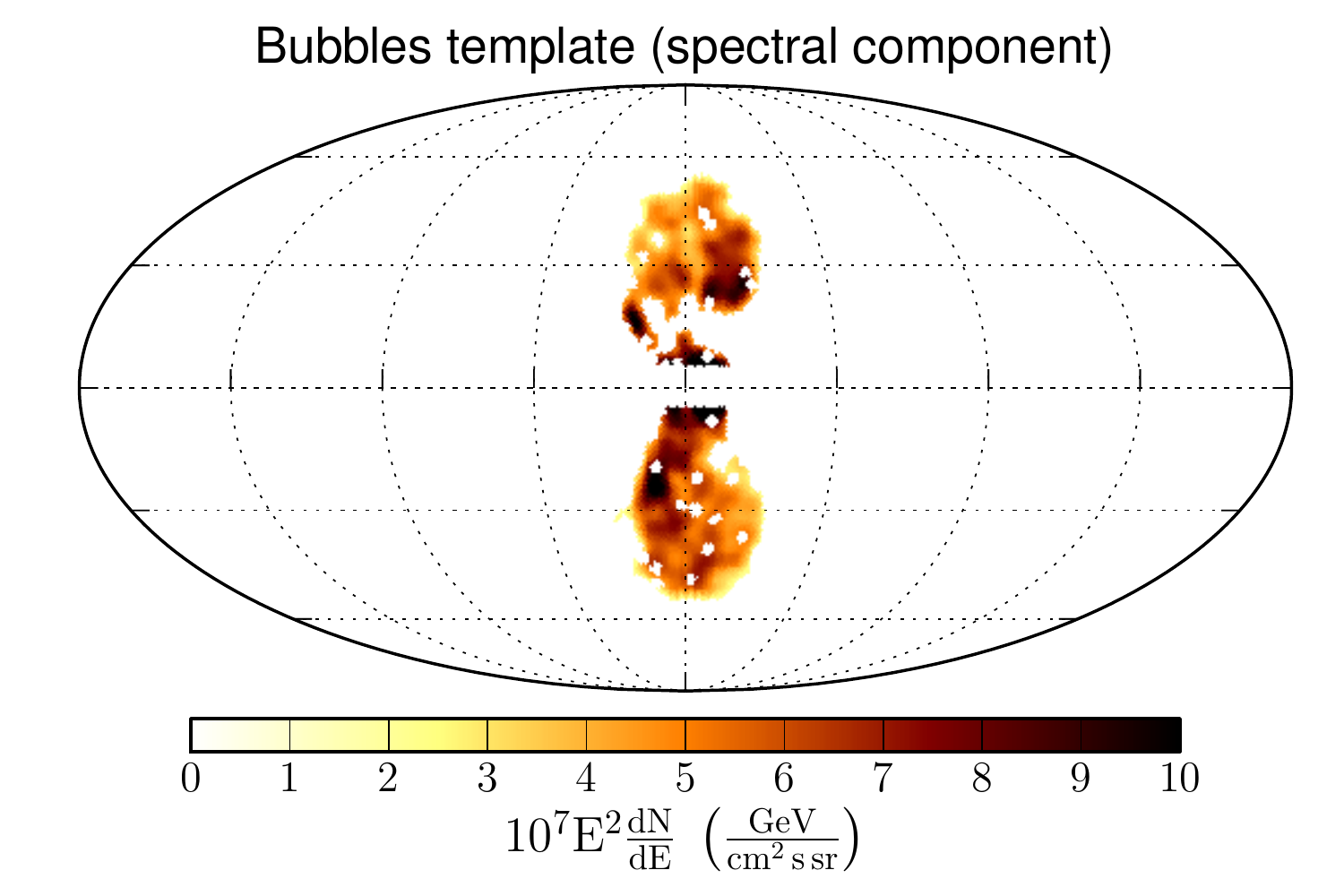, scale=\twopic}
\noindent
\caption{\small 
The histograms of the pixels counts and the templates for the Loop I and the bubbles derived 
from the maps in Figure \ref{fig:49sca_maps}
analogously to the derivation of the histogram and the template for the bubbles from the integrated residual flux
in Figures \ref{fig:47bbl_resid} and \ref{fig:48bbl_template}.
The threshold for the Loop I template is $1\sigma_{\rm BG}$. 
The threshold for the bubbles is $2\sigma_{\rm BG}$.
}
\label{fig:410sca_maps}
\end{center}
\vspace{1mm}
\end{figure}

The templates in Equation \ref{eq:sca} can be written as a linear combination of 
the residual maps
\be
\lb{eq:sca_alt}
T_{mj} = \sum_i k_{mi} D_{ij},
\ee
where $i$ labels the energy bins, $j$ labels the pixels, and $m$ labels the emission components (Loop~I and the bubbles).
$k_{mi}$ are linear decomposition coefficients.
The cuts are relative to the standard deviation outside of the regions containing the bubbles and Loop~I.
The standard deviation of the linear combinations of maps in Equation (\ref{eq:sca_alt})
is the root mean square of the standard deviations of the terms in the linear combination
\be
\lb{eq:sca_noise}
\sm_{mj} (T) = \sqrt{\sum_i k_{mi}^2 \sm_{ij}^2},
\ee
where $\sm_{ij}$ is the statistical uncertainty of the data in energy bin $i$ in pixel $j$
(derived from the square root of the observed photon counts).
The templates of the bubbles and Loop~I are derived by applying threshold cuts of $2\sigma_{\rm BG}$
and $\sigma_{\rm BG}$ to the spectral components maps, which are chosen from the comparison of the background
pixel counts to the pixel counts in the Loop~I and the bubbles regions
(Figure \ref{fig:410sca_maps}).
We note that both methods considered in this work give comparable results for the template of the {\Fermi} bubbles 
(Figures \ref{fig:structTemp}, \ref{fig:48bbl_template}, \ref{fig:410sca_maps}).

\begin{figure}[htbp] 
\vspace{-1mm}
\begin{center}
\epsfig{figure = 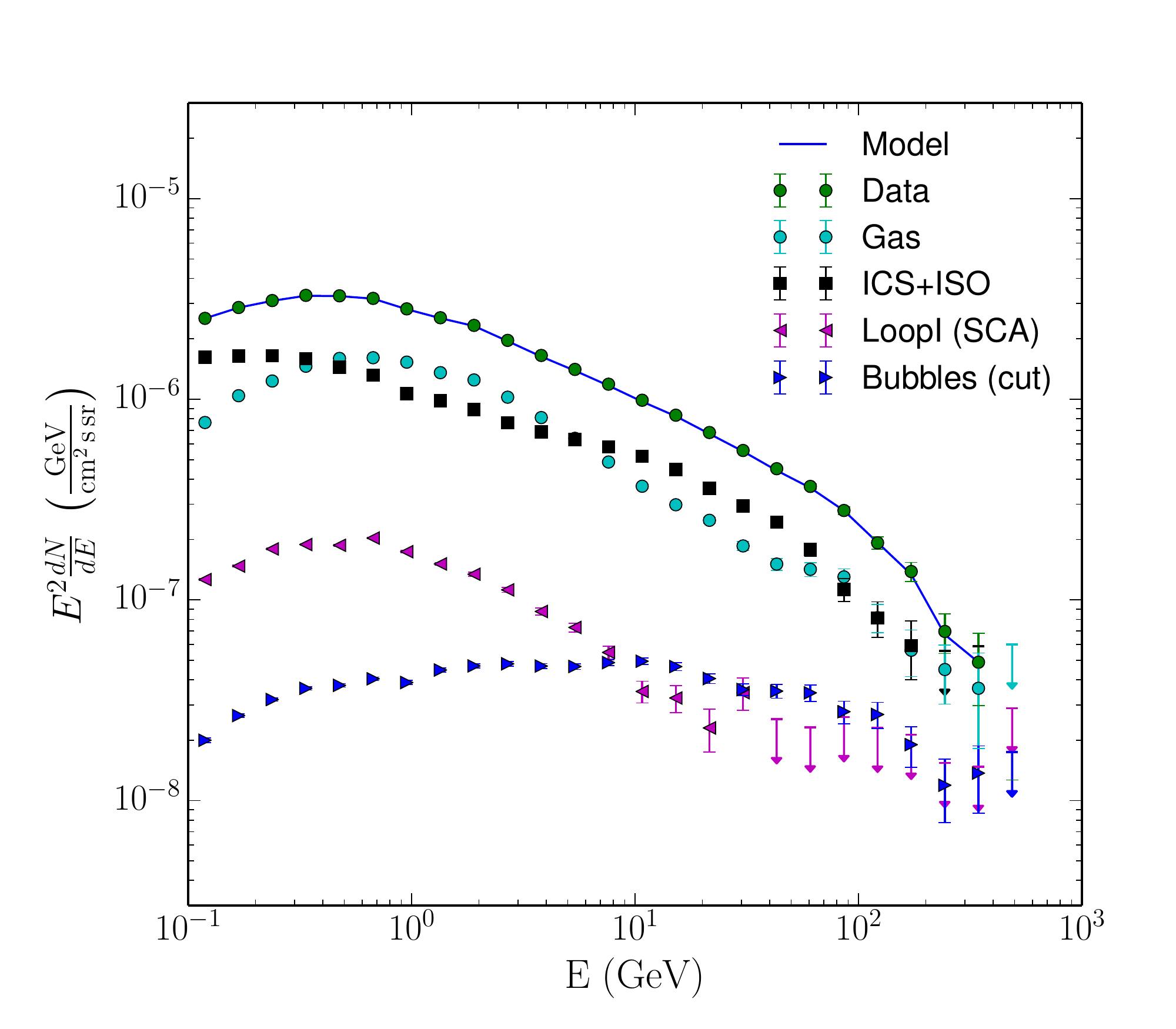, scale=\onepic}
\noindent
\caption{\small 
Spectra of the gas-correlated, IC, isotropic, Loop I, and bubbles components obtained by
fitting the corresponding templates to the data.
The template of the bubbles is derived from the residuals integrated above 10 GeV (Figure \ref{fig:48bbl_template}). 
The Loop I template is derived in the SCA, see Figure \ref{fig:410sca_maps}.
}
\label{fig:411tmpl_sp}
\end{center}
\vspace{1mm}
\end{figure}

\begin{figure}[htbp] 
\vspace{-1mm}
\begin{center}
\epsfig{figure = 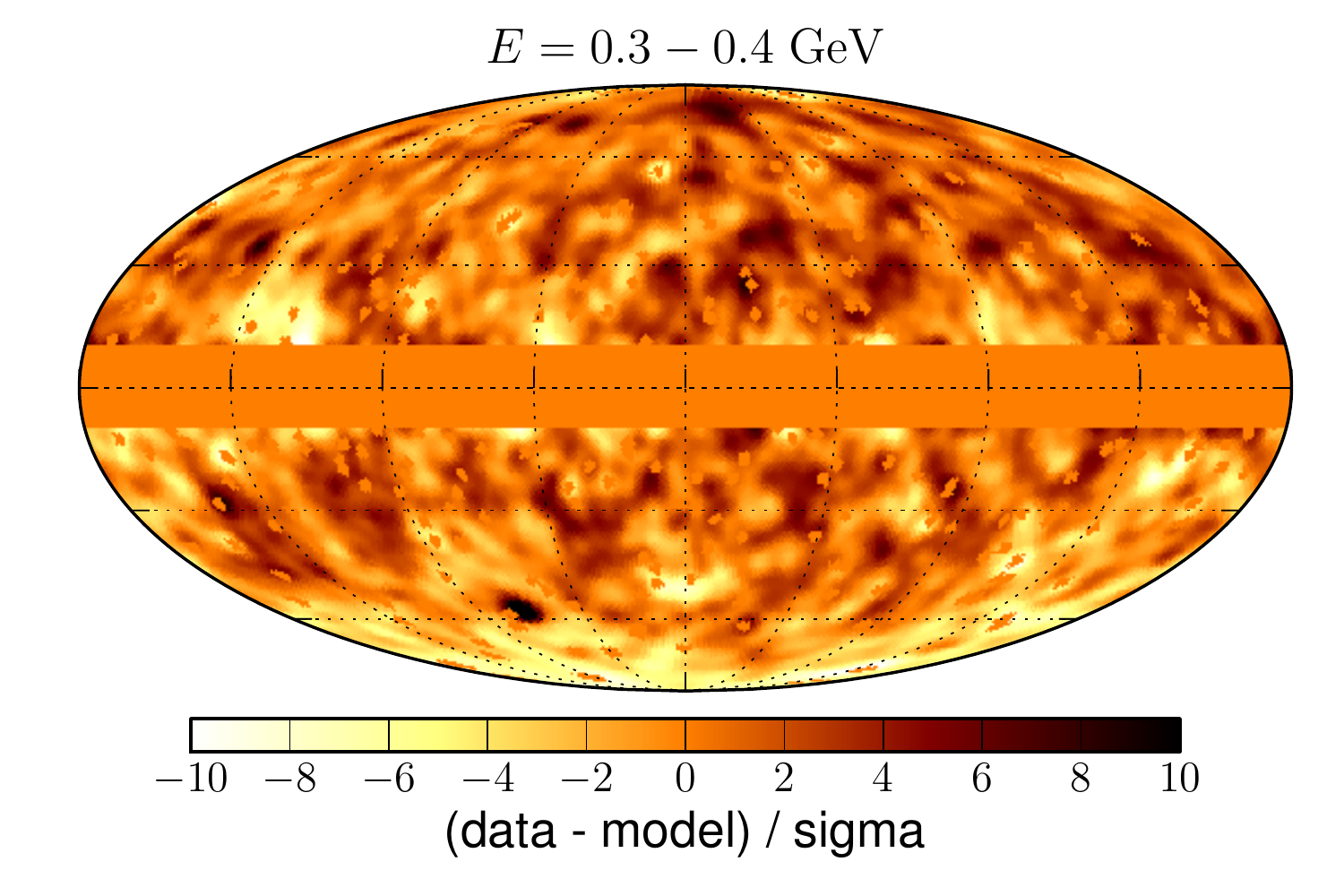, scale=\twopic}
\epsfig{figure = 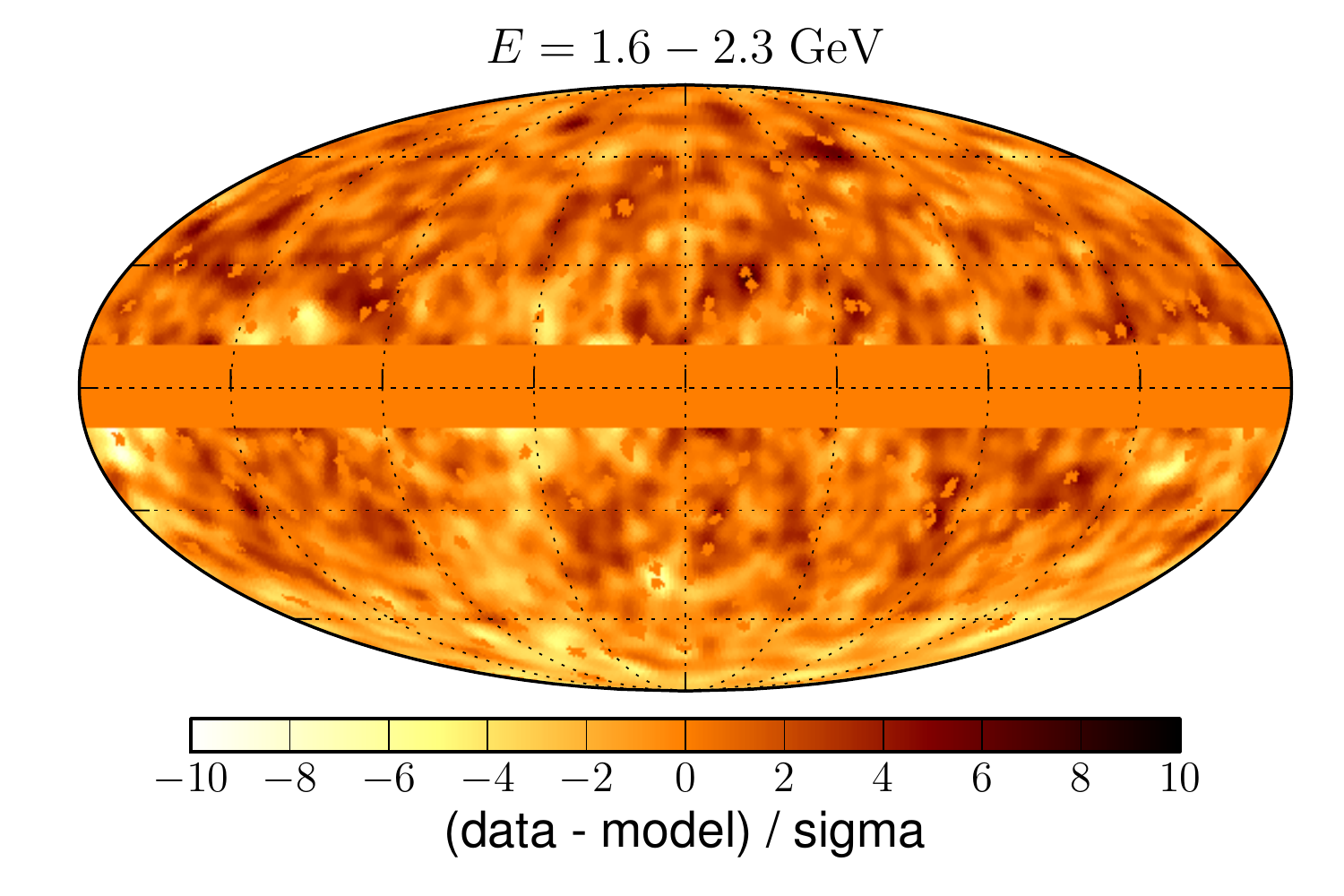, scale=\twopic} \\
\epsfig{figure = 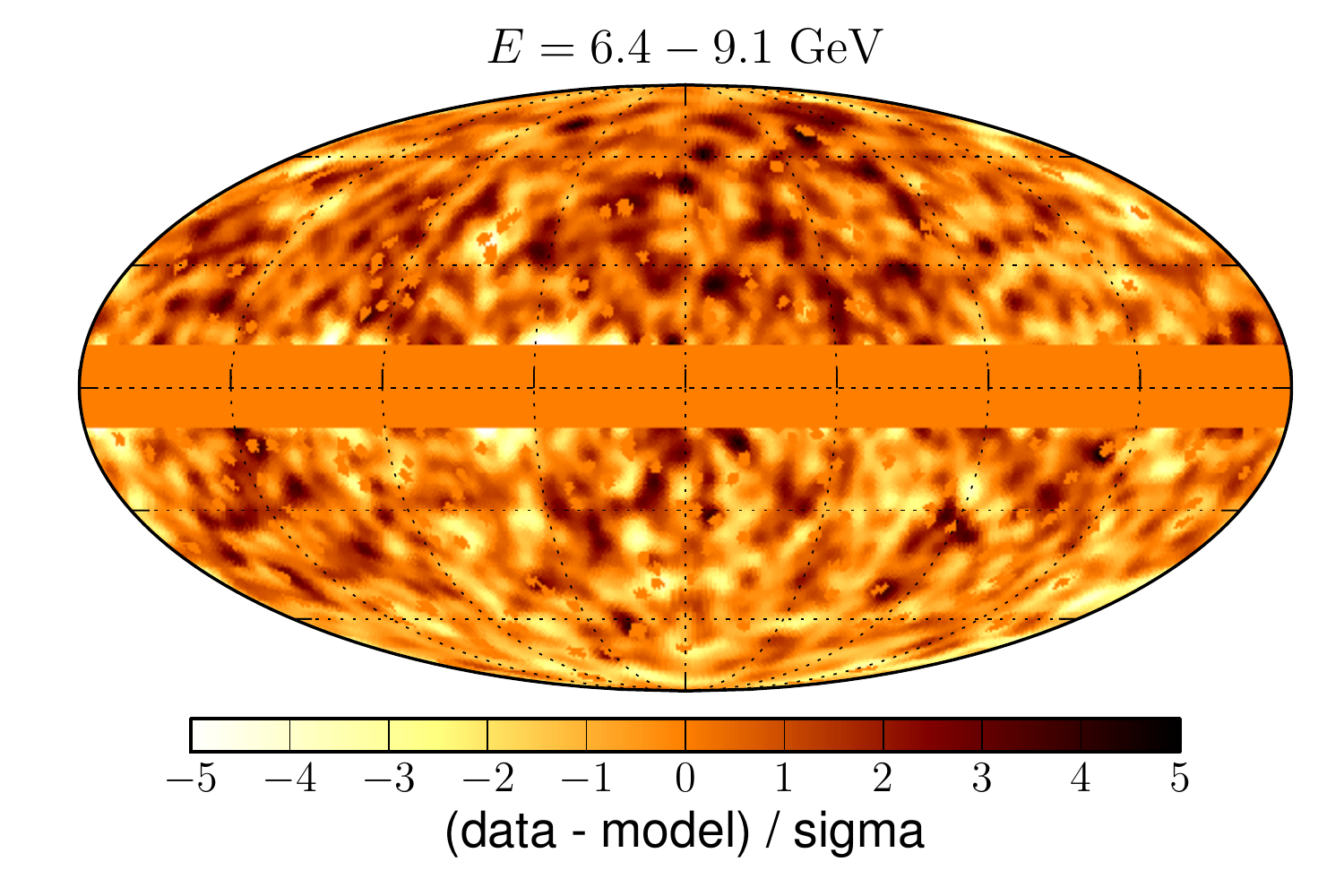, scale=\twopic} 
\epsfig{figure = 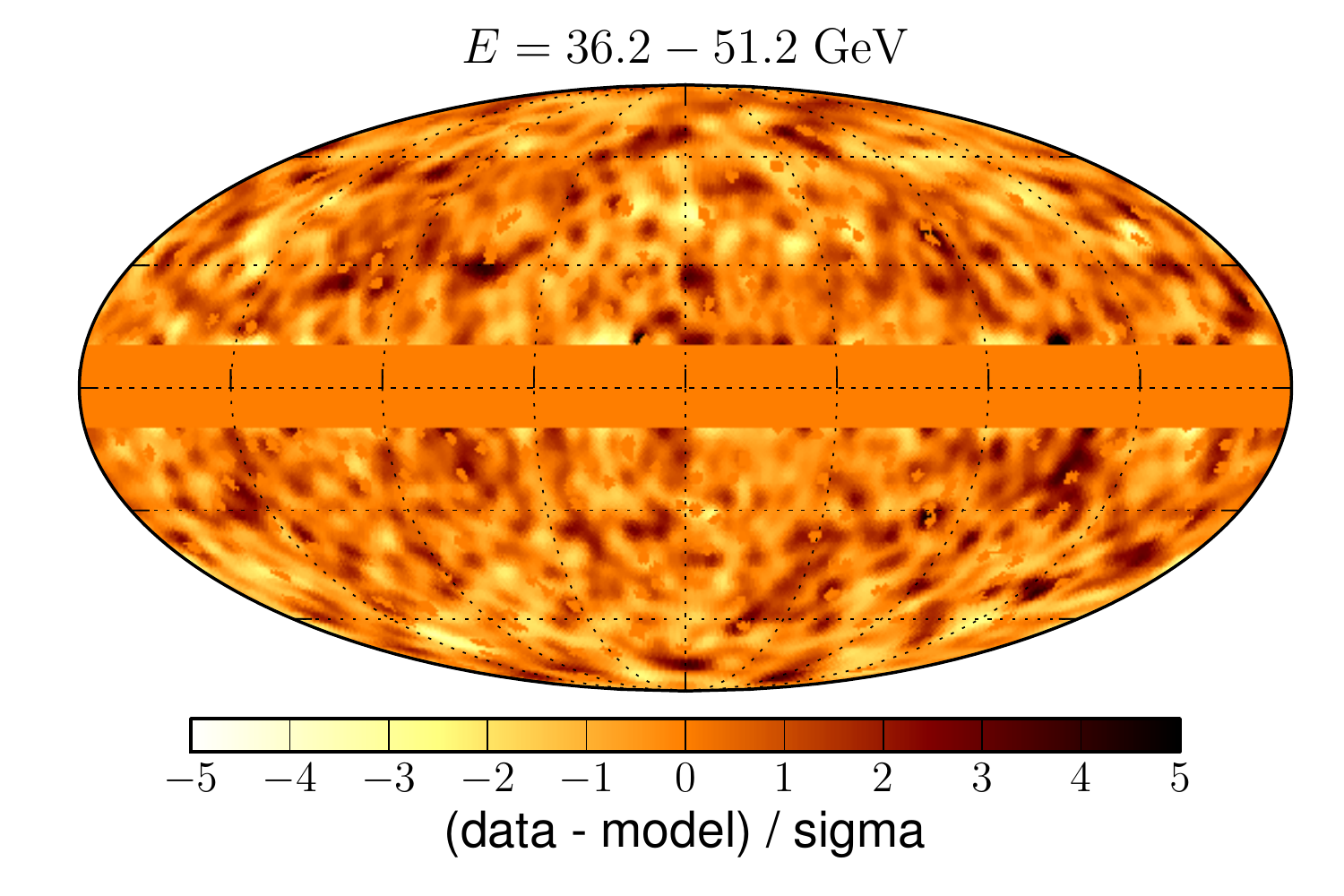, scale=\twopic}
\noindent
\caption{\small 
Residual significance maps for the analysis in Section \ref{subsec:localBubbles}.
The units are data minus model over the standard deviation of the data.
The maps are smoothed with a $2^\circ$ Gaussian kernel for display.
}
\label{fig:412global_resid}
\end{center}
\vspace{1mm}
\end{figure}

The spectra obtained by fitting the five templates (gas-correlated, isotropic, IC, bubbles, and Loop~I) 
to the gamma-ray data for $|b| > 10^\circ$ 
are shown in Figure \ref{fig:411tmpl_sp}.
Let us summarize the characteristics of this model as it will be used as a reference
model in the study of the systematic uncertainties in the next subsection.
We use the 2FGL spectra for point sources,
the local patches have a radius of $50^{\circ}$,
gas-correlated components are modeled by a combination of H I gas templates in two rings,
the H$_2$ template, and the SFD dust template,
the spectrum of the gas-correlated components at high energies is a power law,
the template of the bubbles is obtained from the residual above 10 GeV, while the Loop~I template is obtained with the 
spectral components analysis method. We use structured flux templates for the bubbles and Loop~I.
The significance of the residuals for this fit is presented in Figure \ref{fig:412global_resid}.

\subsection{Systematic uncertainties}
\label{subsec:SysErrorLocal}

\begin{figure}[htbp] 
\vspace{-1mm}
\begin{center}
\epsfig{figure = 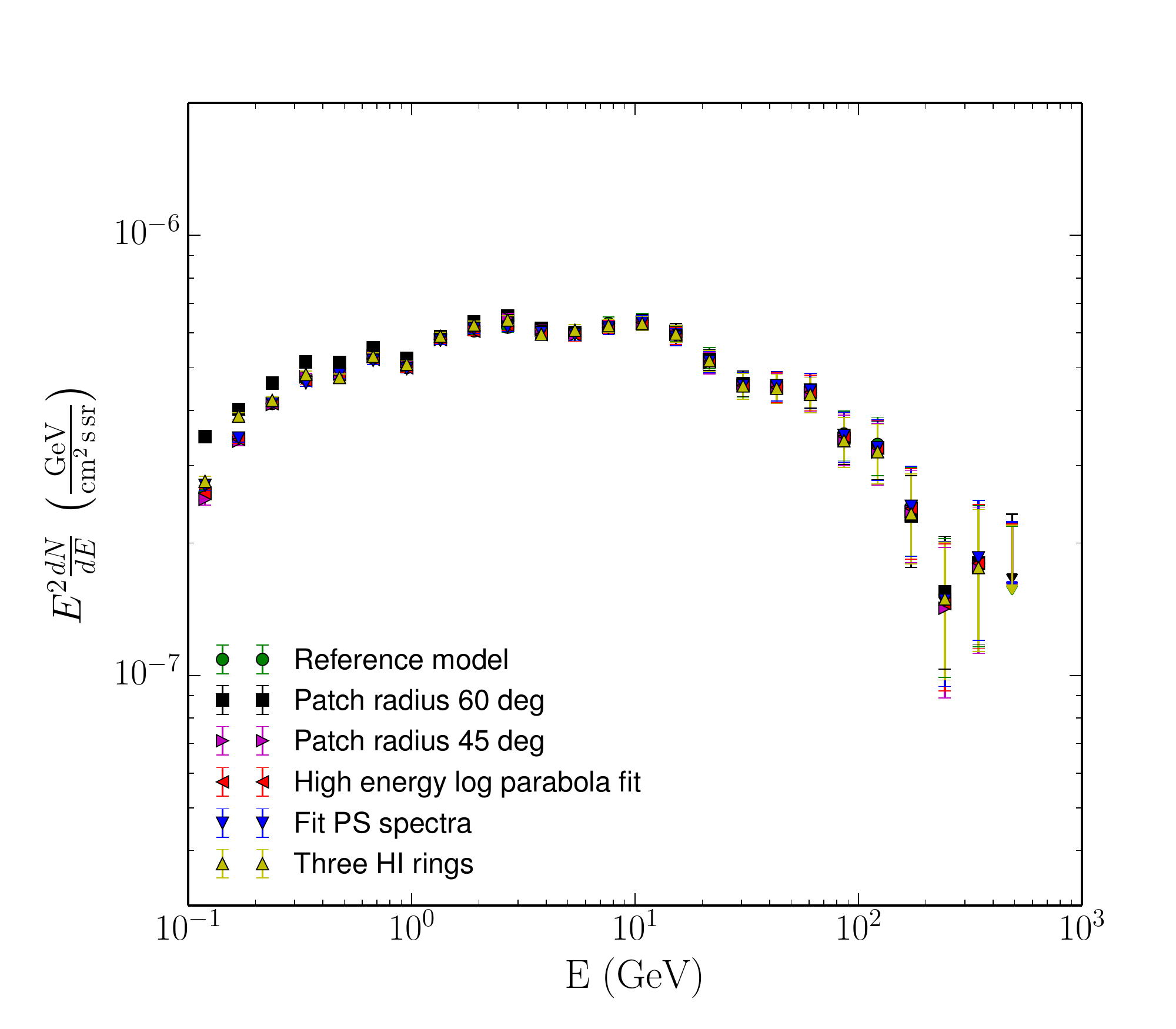, scale=\twopic}
\epsfig{figure = 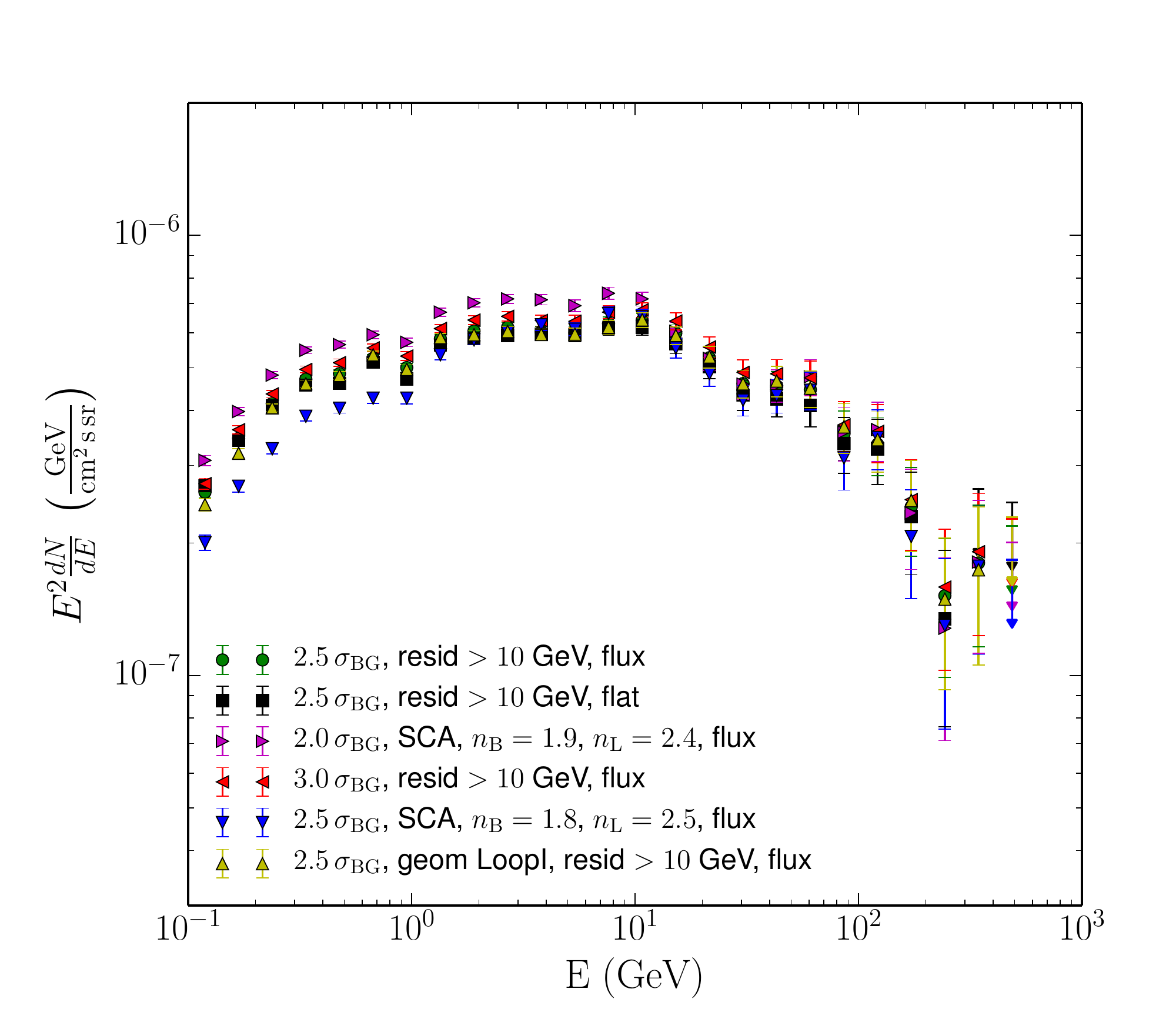, scale=\twopic}
\noindent
\caption{\small 
Determination of the spectrum of the bubbles for different choices of analysis parameters described in Section 
\ref{subsec:SysErrorLocal}.
Left: determination of the bubble spectrum for different choices
of point source subtraction method and
the local template fitting strategy.
Right: systematic uncertainty related to the definition of templates of Loop I and the {\Fermi} bubbles.
The template of the bubbles is determined either from the residuals integrated above 10 GeV,
or from the SCA,
where the bubbles and Loop~I spectra are described by power-law functions with indices
$n_{\rm B}$ and $n_{\rm L}$.
For Loop I, we use either the template determined from the SCA
or the geometric template described in Section \ref{sect:galprop}.
The ``reference model'' (Figure \ref{fig:411tmpl_sp}) 
is shown by green circles in both plots.
}
\label{fig:413local_syst}
\end{center}
\vspace{1mm}
\end{figure}

In order to estimate the systematic uncertainties in the local templates analysis,
we take the model presented in the previous subsection and vary
some aspects of the fitting procedure.
At first, we vary the parameters relevant to the derivation of the Galactic emission components
(Figure \ref{fig:413local_syst} on the left):
\ben
\item
We try different patch radii in the determination of the gas-correlated components: $45^{\circ}$ and $60^{\circ}$.
\item
To test the dependence on the assumption of a power-law spectrum for the gas-correlated 
emission at high energy, we use a log parabola function to model the gas-correlated spectra at high energies%
\footnote{This test is motivated by possible deviations from a simple power law in the gamma-ray spectrum
due to features in the hadronic CR spectra \citep{2011Sci...332...69A}. 
The log parabola function is the simplest generalization of a power law.}.
\item
We also try to use three H I rings (R $<$ 8 kpc, 8 kpc $<$ R $<$ 10 kpc, and 10 kpc $<$ R, see Table \ref{tab:templateList}) as opposed to 2 rings (R $<$ 8 kpc and 8 kpc $<$ R).
\item
To test the dependence on the model of the point sources, we refit bright point sources with ${\rm TS} > 200$ 
(472 sources).
We keep the positions given in the 2FGL catalog and refit the spectra of the point sources
assuming the same spectral function (e.g., power law, power law with a cutoff, or log parabola).
The fit is performed in a small patch around each point source: 
the radius is either $2^\circ$ or the 95\% containment angle, whichever is larger.
This choice of the radius is motivated by the requirement
that there are sufficiently many pixels to perform the fit,
but not too many pixels, so that a low order polynomial model of the background is appropriate:
the background is modeled by a combination of local polynomials of degree 4 
(the degree was found from Monte Carlo tests).
During a fit for a particular point source, all other 2FGL point sources with TS $< 200$
are subtracted from the data with the 2FGL fluxes.
\item
We also mask the cores of the point sources without subtracting them.
This has a relatively important effect at low energies, where the PSF is large.
But even in this case, the difference in the spectrum of the bubbles is not significant 
compared to the effect from modifying the definitions of the bubbles' and Loop~I templates considered below.
\een

In order to test the systematic uncertainty related to the definition of the {\Fermi} bubbles and the Loop~I templates,
we consider the following definitions of the templates
(Figure \ref{fig:413local_syst} on the right):
\ben
\item
Bubbles' template from residual maps above 10 GeV (we tested two threshold levels: 2.5$\sigma_{\rm BG}$ and 3$\sigma_{\rm BG}$).
\item
Bubbles' template from spectral components with indices 1.9, 1.8 (2$\sigma_{\rm BG}$ and 2.5$\sigma_{\rm BG}$).
\item
Loop~I template from spectral components with indices 2.4, 2.5 (1$\sigma_{\rm BG}$ and 1.2$\sigma_{\rm BG}$).
\item
Geometric Loop~I template (Section \ref{sect:galprop}).
\item
We use structured (proportional to flux) and flat (0 - 1) templates both for the bubbles and for Loop~I.
\een
The largest effect at low energies comes from the assumptions on the spectrum of the components in the spectral components analysis (SCA) derivation of the templates for the bubbles and for the Loop I.

%% file: 5comparison.tex
\section{The overall spectrum of the bubbles}
\lb{sect:comparison}

\begin{figure}[htbp] 
\begin{center}
\includegraphics[scale=\twopic]{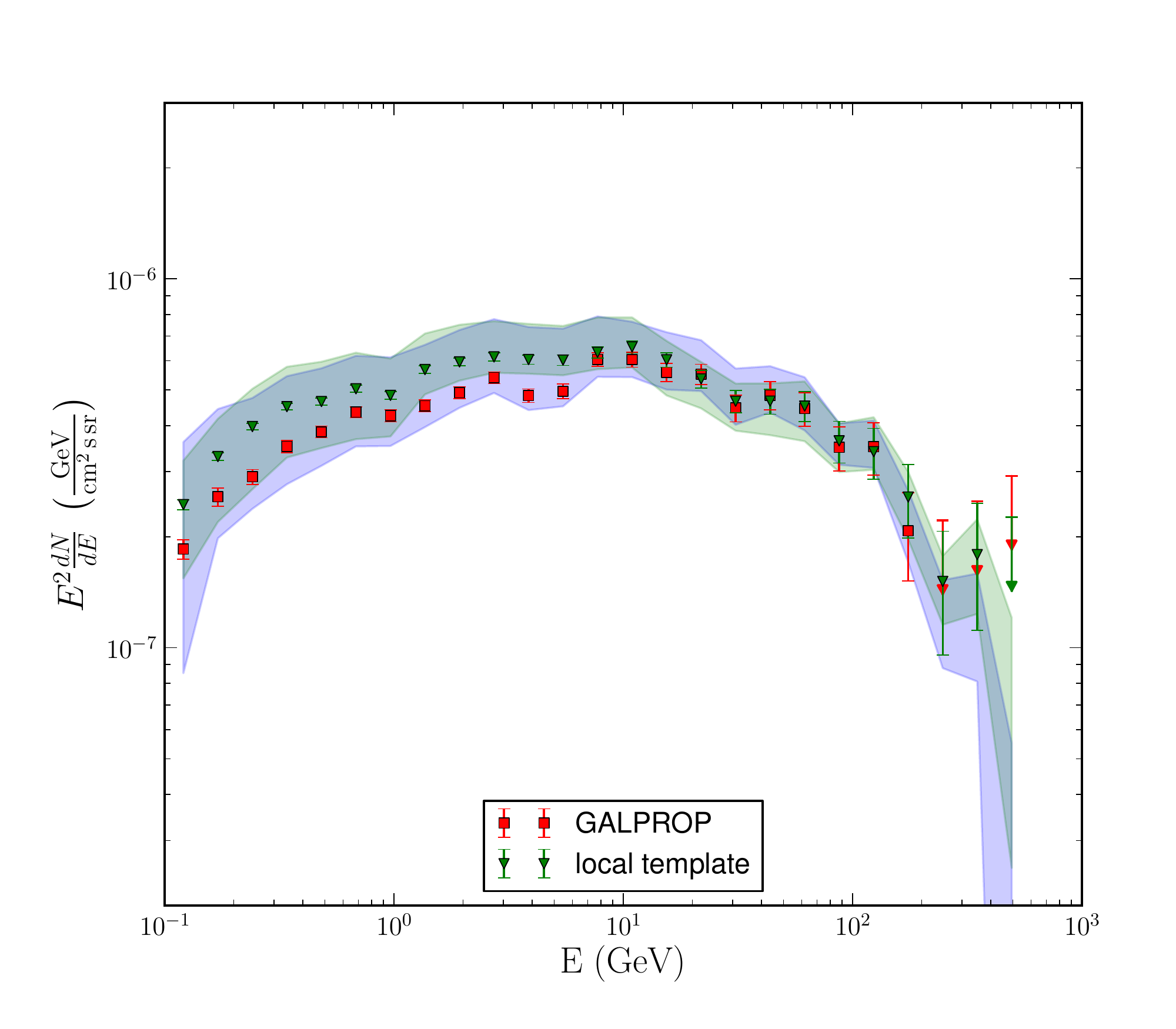}
\includegraphics[scale=\twopic]{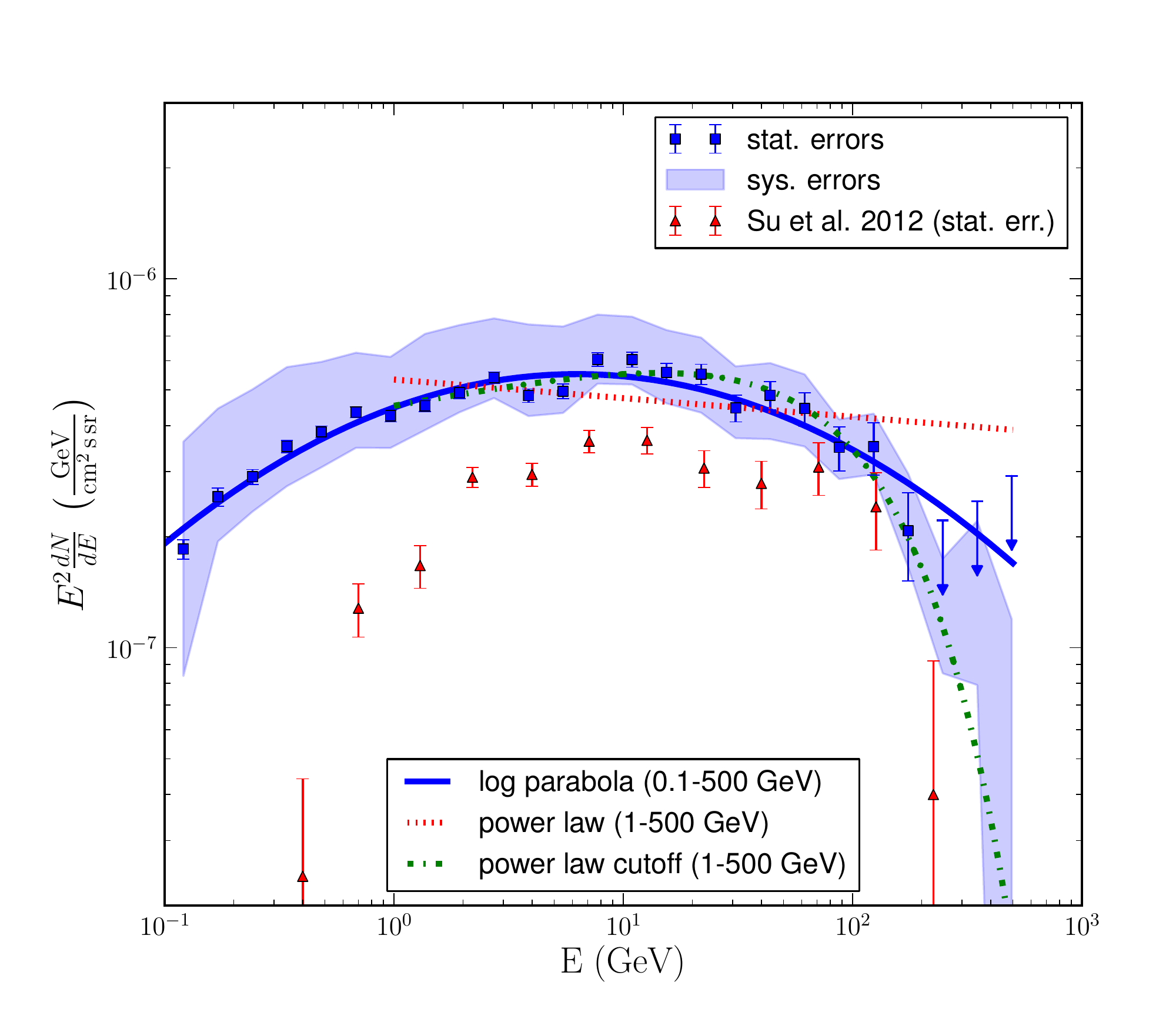}
\noindent
\caption{\small 
Left: SED of the bubbles for $|b|>10^{\circ}$ obtained using the GALPROP template analysis (red squares) and local template analysis (green triangles). 
The points with error bars represent the spectra obtained with the two methods 
(Figures \ref{fig:fitWithBubble} and \ref{fig:411tmpl_sp}).
The shaded bands are the systematic uncertainties due to the analysis procedure and Galactic foreground modeling as described in \ref{sec:gadgetSys} and \ref{subsec:SysErrorLocal}.
Right: combined bubble SED compared to the earlier result from~\cite{2012ApJ...753...61S} for $|b|>20^\circ$.
The baseline model is the same as the GALPROP curve in the left plot. 
The systematic uncertainties are the envelope of all possible spectra obtained from the two methods. 
In the combined spectrum we include the uncertainties in the LAT effective area~\citep{Ackermann:2012kna} 
by adding them in quadrature to the envelope of the other systematic uncertainties.
The curves show the functional forms fitted to the SED points.
Solid blue line: log parabola.
Dotted red line: simple power law. 
Dash-dotted green line: power law with an exponential cutoff. 
}
\label{fig:52compareSpectra}
\end{center}
\vspace{1mm}
\end{figure}

The spectra of the bubbles 
derived with the two methods presented in Sections~\ref{sect:galprop} and~\ref{sect:local}
are shown in Figure \ref{fig:52compareSpectra}, left.
In the following, we take the results from the GALPROP template analysis as a baseline for the spectral energy distribution (SED) and combine all the spectra
obtained with the two methods to get an envelope of the systematic uncertainties. The envelope includes uncertainties introduced by the diffuse modeling and uncertainties related to the analysis strategy, e.g., the threshold to define the template of the bubbles, or the size of the local patches. We add the systematic error of the LAT effective area~\citep{Ackermann:2012kna} in quadrature to the envelope obtained for different models. The systematic errors of the LAT PSF and the effect of energy dispersion are negligible given the spatial and energy binning chosen for this analysis. The uncertainties due to the modeling of Galactic foregrounds and analysis strategy 
(see Table~\ref{tab:fluxValues}) dominate the uncertainty compared to the effective area, which has a relative flux error of $\le10\%$.
The baseline model with its statistical and systematic uncertainties is presented in
Figure \ref{fig:52compareSpectra} on the right and in Table~\ref{tab:fluxValues}.
We also compare our results with the {\Fermi} bubbles' SED derived by
\cite{2012ApJ...753...61S}.
Our intensity is significantly higher than the spectrum of \cite{2012ApJ...753...61S},
especially at low energies.
The difference is due to a combination of several effects, namely a 
smaller Galactic plane mask 
($10^\circ$ in this work compared to $20^\circ$ in \cite{2012ApJ...753...61S}), a smaller area of the bubbles' template in this analysis
resulting in larger intensities, the inclusion of a separate template for the cocoon in  \cite{2012ApJ...753...61S},
and different modeling of the Galactic foregrounds. Our results agree with the spectrum in latitude strips at $|b|>20^\circ$ reported by~\cite{Hooper:2013rwa}.

We fit the baseline SED with a log parabola function, a power law with an exponential cutoff 
and a simple power law (Figure~\ref{fig:52compareSpectra}, right).
The log parabola and the power law with an exponential cutoff are defined, respectively, as
\bea
\label{eq:LogPara}
\frac{dN}{dE} &=& I \left( \frac{E}{10\rm{GeV}} \right)^{-\alpha - \beta \log(E/1\rm{GeV})} \textrm{;} \\
\label{eq:ExpCut}
\frac{dN}{dE} &=& I \left( \frac{E}{1\rm{GeV}} \right)^{-\gamma} e^{-E/E_{\rm{cut}}}.
\eea

\begin{figure}[htbp] 
\begin{center}
\includegraphics[scale=\twopic]{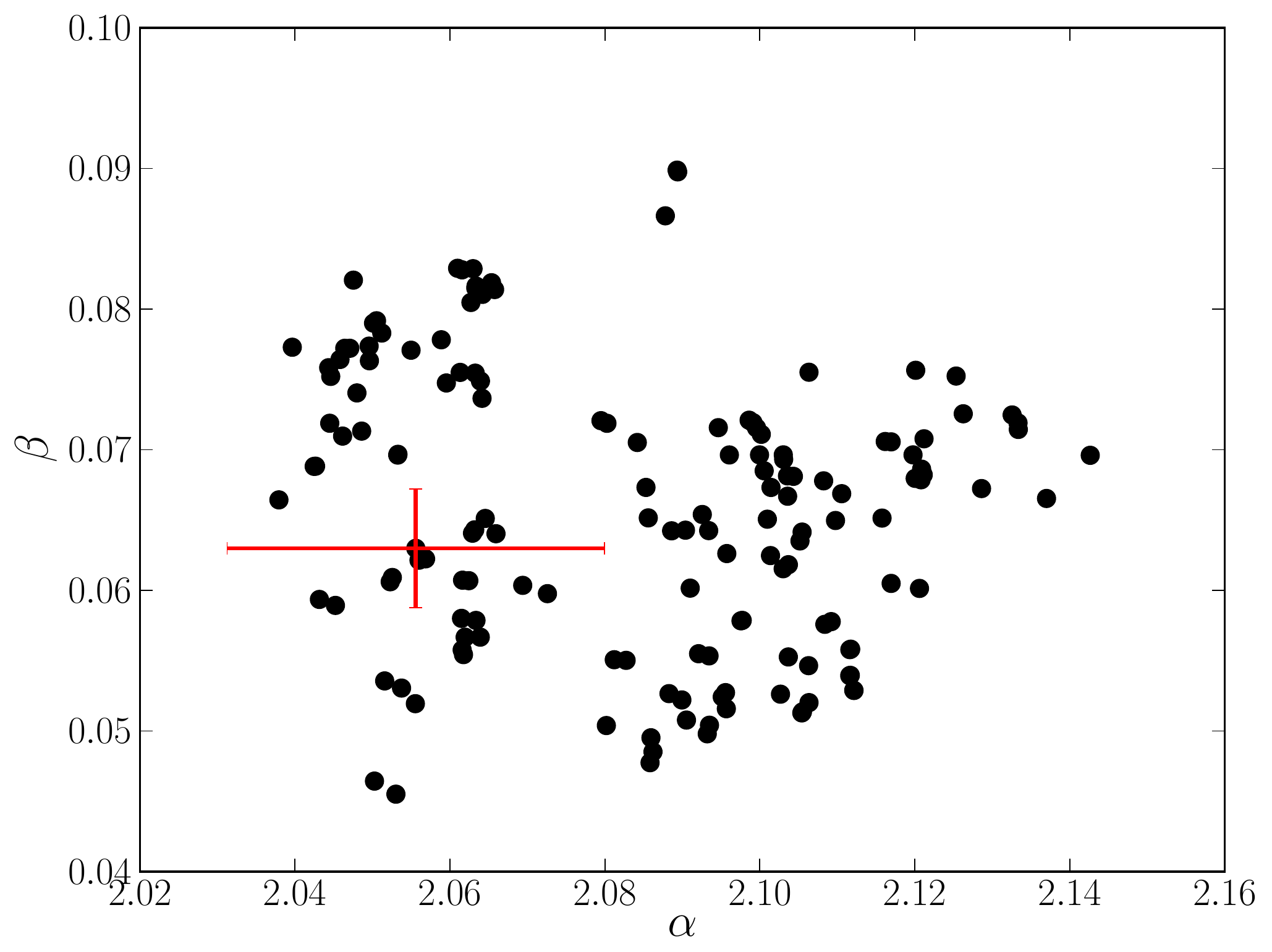}
\includegraphics[scale=\twopic]{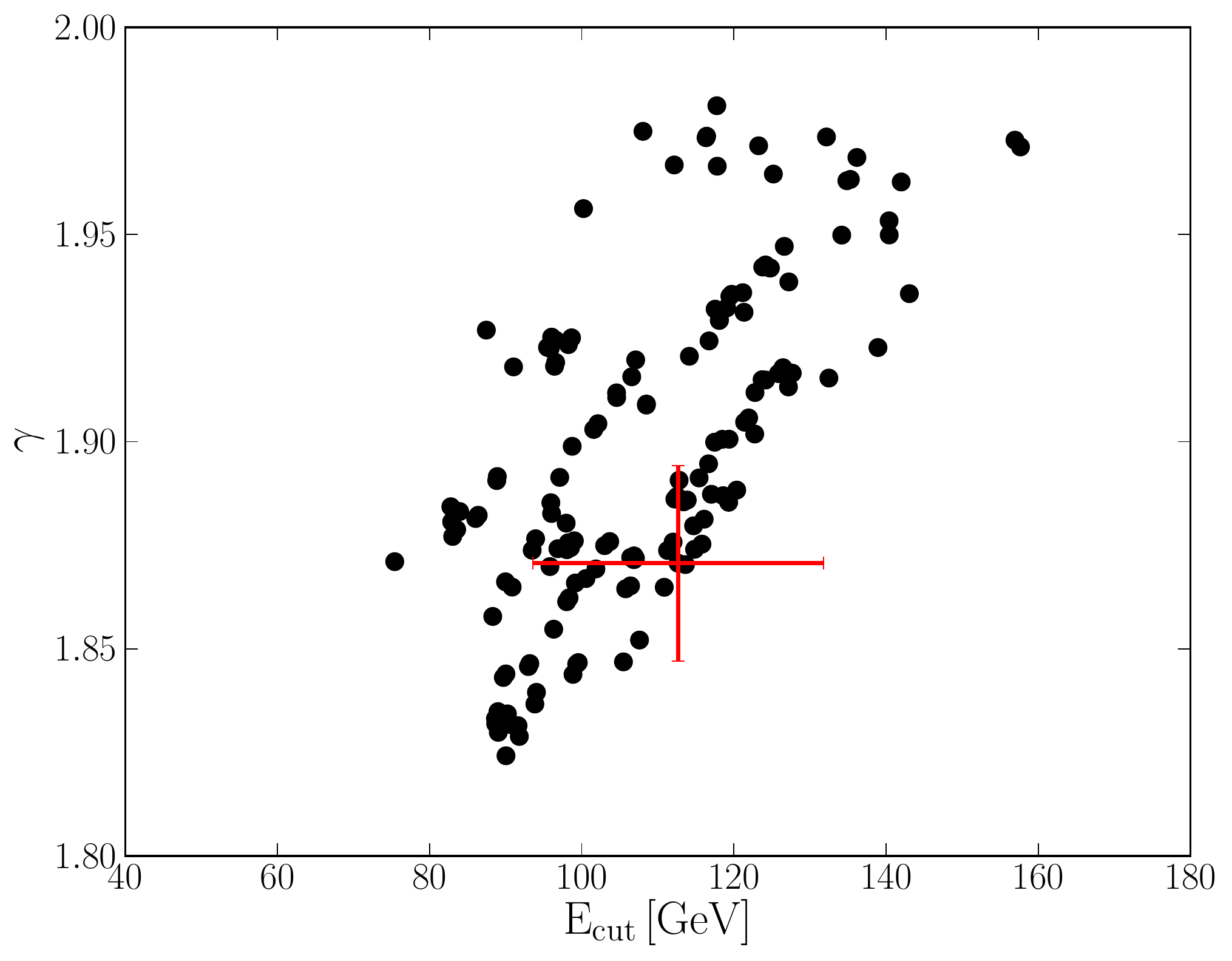}
\noindent
\caption{\small 
Left: distribution of log parabola fit parameters (energy range of the fit: 100 MeV to 500 GeV). 
Right: distribution of power law with exponential cutoff fit parameters (energy range of the fit: 1 GeV to 500 GeV) . Red crosses represent the baseline model values with their statistical uncertainties.
}
\label{fig:54fitParaDist2D}
\end{center}
\vspace{1mm}
\end{figure}

We repeat the fits for the bubbles' spectrum for different Galactic models and different definitions
of the bubbles and Loop I templates.
We obtain the following parameters for the
log-parabola function in the fit range 100 MeV to 500 GeV: 
$\alpha = 1.77\pm0.01\rm{[stat]}^{+0.10}_{-0.22}\rm{[syst]}$, 
$\beta = 0.063 \pm 0.004\rm{[stat]} ^{+0.047}_{-0.018}\rm{[syst]} $.
The values are given for the baseline model and the systematic uncertainties are estimated from the SEDs 
obtained for different Galactic foreground models and choices in the analysis strategy.
The systematic errors include the uncertainties of the LAT effective area~\cite{Ackermann:2012kna}. 
The distributions of the fit parameters $\alpha$ and $\beta$ for the log parabola fits 
are shown in Figure~\ref{fig:54fitParaDist2D} on the left.

\begin{figure}[htbp] 
\begin{center}
\includegraphics[scale=\twopic]{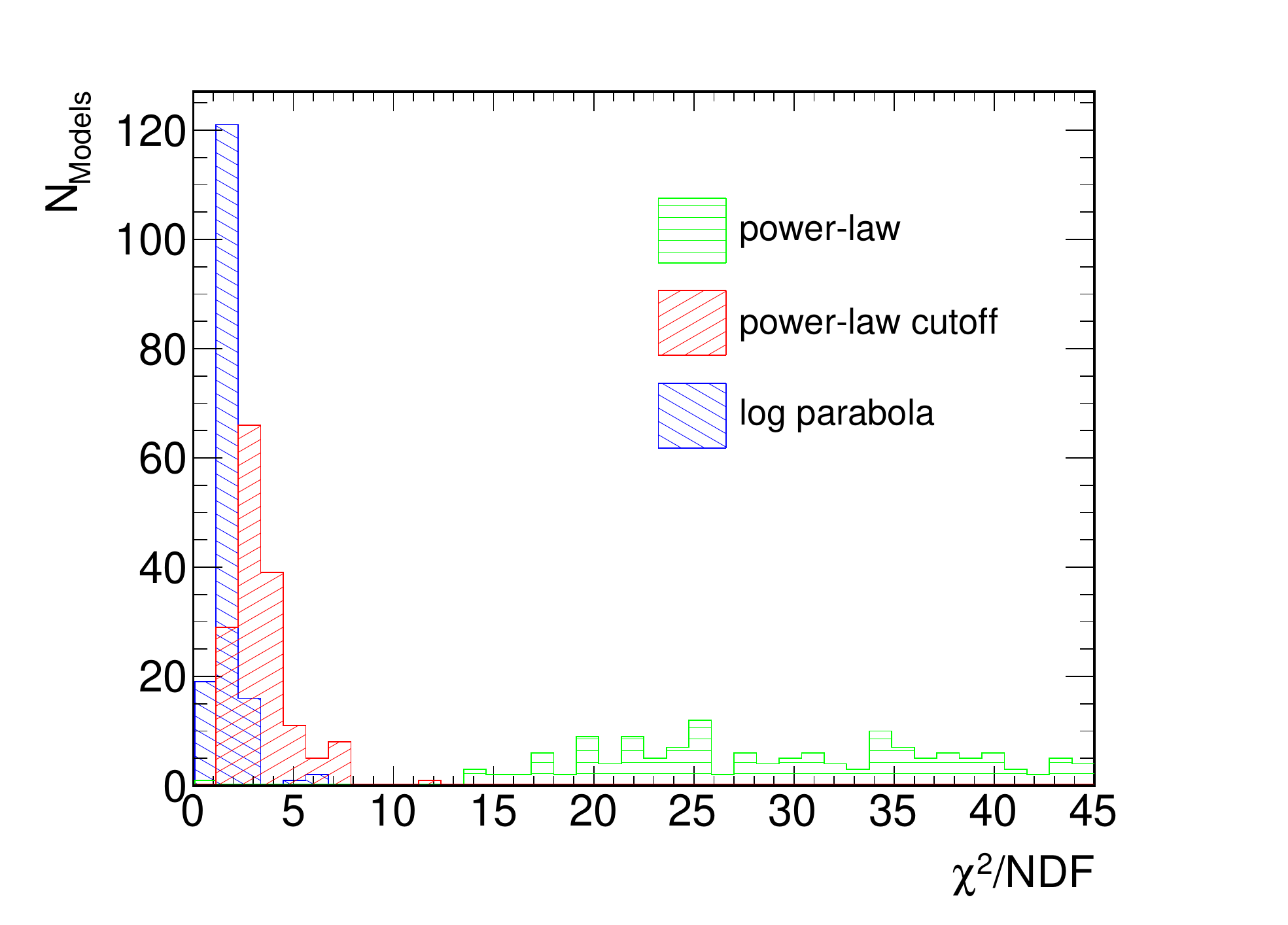}
\includegraphics[scale=\twopic]{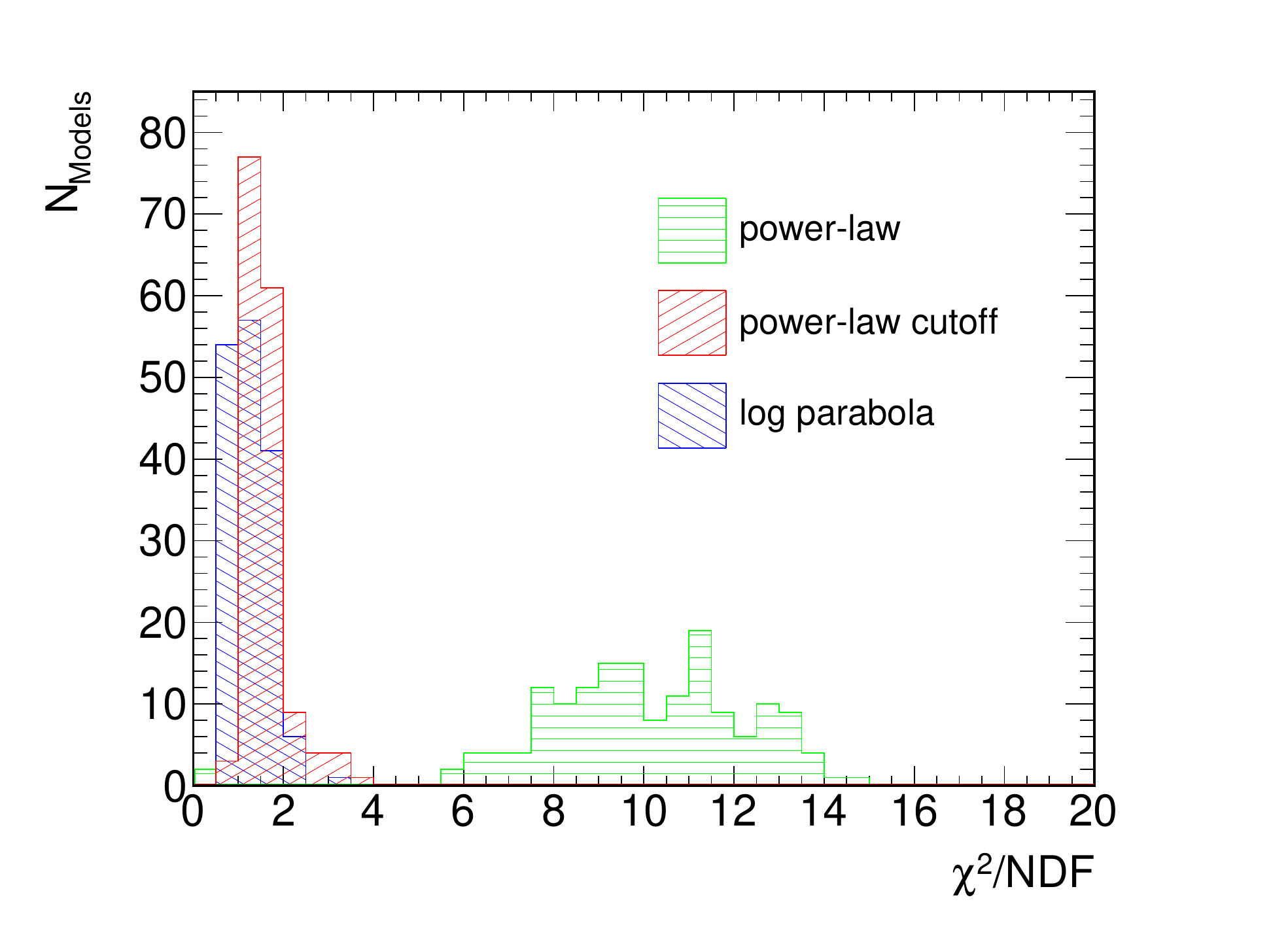}
\noindent
\caption{\small 
Distribution of $\chi^2$ per degree of freedom for all models 
fitted with a simple power law (green), a power law with exponential cutoff (red), and a log parabola function (blue).
All fits are performed in the energy range from 100 MeV to 500 GeV (left) and from 1 GeV to 500 GeV (right)
}
\label{fig:55chi2}
\end{center}
\vspace{1mm}
\end{figure}

The power law with a cutoff fit above 100 MeV is dominated by low and intermediate energies.
In order to find a value of the high-energy cutoff unbiased by low energies, we fit the power law with a cutoff in the range 1 GeV to 500 GeV.
We obtain 
$E_{\rm{cut}} = 113\pm 19 \rm{[stat]} ^{+45}_{-53}\rm{[syst]}$ GeV and
$\gamma=1.87\pm{0.02}\rm{[stat]}^{+0.14}_{-0.17}\rm{[syst]}$. 
The distribution of indices and cutoff energies of the power law with exponential cutoff fits
are shown in Figure~\ref{fig:54fitParaDist2D} on the right.
The corresponding distributions of $\chi^2$ per number of degrees of freedom (NDF) are presented in Figure~\ref{fig:55chi2}.
The log parabola gives a good description of the data over the whole energy range.
The simple power law does not describe the data well even above 1 GeV.
The power law with a cutoff is preferred over a power law with at least 7$\sigma$ significance.

\begin{figure}[htbp] 
\begin{center}
\includegraphics[scale=\onepic]{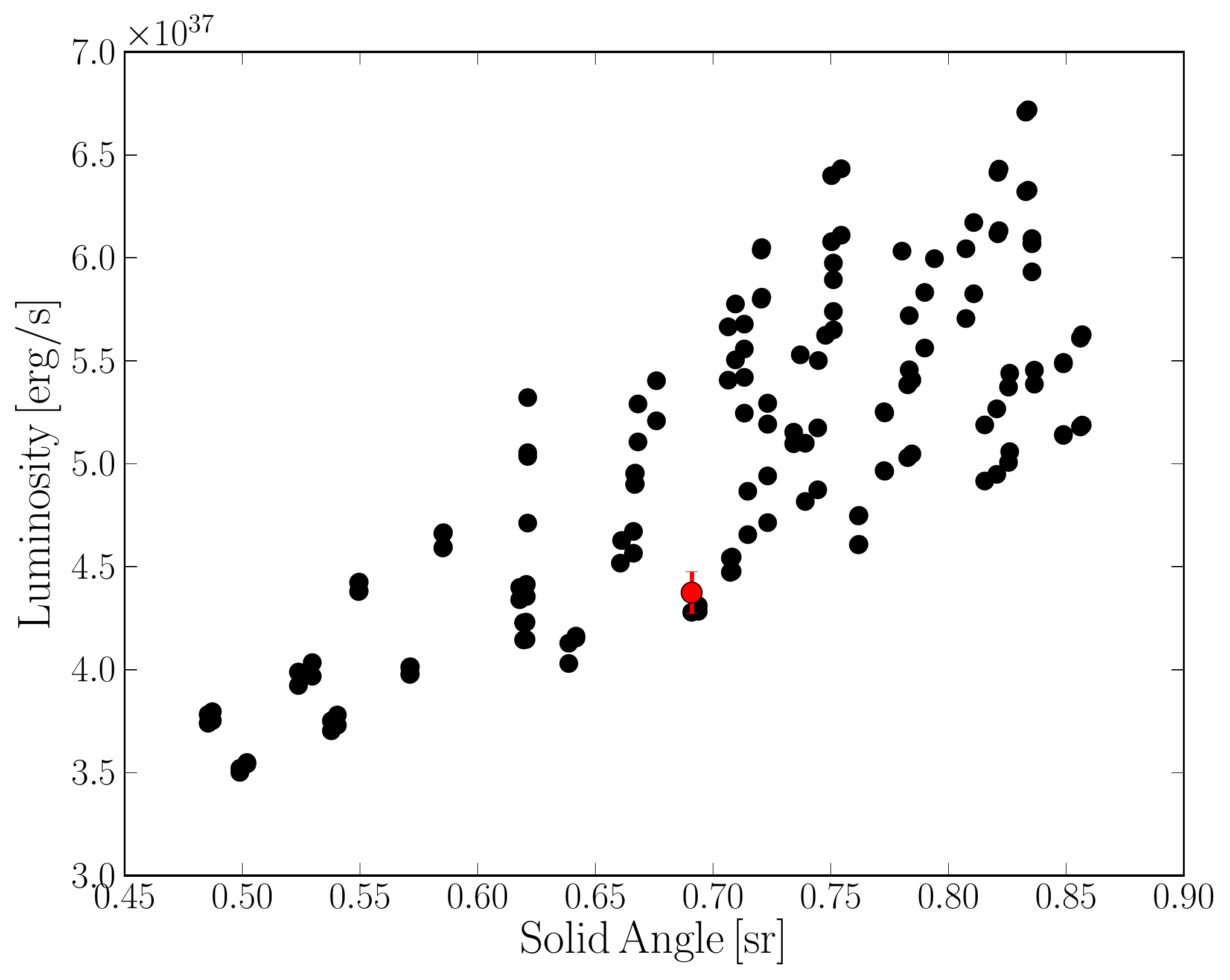}
\noindent
\caption{\small 
The 100 MeV - 500 GeV bubble luminosity vs. solid angle subtended by the bubbles at $|b| > 10^{\circ}$ for different models
of foreground emission and definitions of bubble templates.
}
\label{fig:57lum}
\end{center}
\vspace{1mm}
\end{figure}

We calculate the total luminosity of the bubbles for $|b| > 10^{\circ}$ for each determination of the spectrum in the energy range from 100 MeV to 500 GeV. The bubbles are found to have a luminosity of 
$(4.4\pm0.1 \rm{[stat]}  ^{+2.4}_{-0.9}\rm{[syst]} )\times 10^{37}$ erg s$^{-1}$. 
The distribution of the solid angle subtended by the bubbles, and the luminosity for the models considered are shown 
in Figure~\ref{fig:57lum}.

\begin{deluxetable}{ccccccc}
\tabletypesize{\scriptsize}
\tablecaption{
Differential energy spectrum per unit solid angle for the {\Fermi} bubbles.
$E_{\rm{min}}$ and $E_{\rm{max}}$ are the boundaries of the energy bins, and $E$ is the geometric mean of the bin.
$F_{\rm{min}}$ and $F_{\rm{max}}$ define the systematic error band, 
$\Delta F_{\rm{stat}}$ is the statistical error. The last entry is zero, which is the lowest value allowed in the fit (we do not allow negative values).\label{tab:fluxValues}}

\tablewidth{0pt}
\tablehead{
\colhead{$E$} & \colhead{$E_{\rm min}$} & \colhead{$E_{\rm max}$} &   \colhead{$E^2 F$}  & \colhead{$E^2  F_{\rm min} $}  & \colhead{$E^2  { F_{\rm max}}$} & \colhead{$E^2  \Delta F_{\rm stat}$}
}
\startdata
\multicolumn{3}{c}{[GeV] }           &  \multicolumn{4}{c}{$\rm 10^{-7}\, [ \frac{GeV}{cm^2\, s\, sr}$] }
\vspace{0.2cm} \\
\tableline
0.12 & 0.10 & 0.14 & 1.85 & 0.83 & 3.61 & 0.11\\ 
0.17 & 0.14 & 0.20 & 2.57 & 1.37 & 4.44 & 0.15\\ 
0.24 & 0.20 & 0.28 & 2.91 & 1.70 & 5.35 & 0.13\\ 
0.34 & 0.28 & 0.40 & 3.51 & 2.27 & 6.11 & 0.13\\ 
0.48 & 0.40 & 0.57 & 3.85 & 2.51 & 6.29 & 0.14\\ 
0.67 & 0.57 & 0.80 & 4.35 & 2.67 & 6.69 & 0.15\\ 
0.95 & 0.80 & 1.13 & 4.26 & 2.90 & 6.47 & 0.16\\ 
1.35 & 1.13 & 1.60 & 4.53 & 3.99 & 7.49 & 0.17\\ 
1.90 & 1.60 & 2.26 & 4.91 & 4.52 & 7.80 & 0.18\\ 
2.69 & 2.26 & 3.20 & 5.40 & 4.98 & 7.88 & 0.20\\ 
3.81 & 3.20 & 4.53 & 4.83 & 4.48 & 7.73 & 0.20\\ 
5.38 & 4.53 & 6.40 & 4.96 & 4.58 & 7.56 & 0.23\\ 
7.61 & 6.40 & 9.05 & 6.05 & 5.42 & 8.01 & 0.26\\ 
10.76 & 9.05 & 12.80 & 6.04 & 5.42 & 7.91 & 0.28\\ 
15.22 & 12.80 & 18.10 & 5.58 & 4.73 & 7.17 & 0.31\\ 
21.53 & 18.10 & 25.60 & 5.52 & 4.23 & 6.82 & 0.35\\ 
30.44 & 25.60 & 36.20 & 4.47 & 3.80 & 5.71 & 0.37\\ 
43.05 & 36.20 & 51.20 & 4.84 & 3.73 & 5.80 & 0.42\\ 
60.89 & 51.20 & 72.41 & 4.45 & 3.76 & 5.41 & 0.46\\ 
86.11 & 72.41 & 102.40 & 3.49 & 2.95 & 4.06 & 0.48\\ 
121.77 & 102.40 & 144.82 & 3.51 & 3.07 & 4.23 & 0.57\\ 
172.22 & 144.82 & 204.80 & 2.07 & 1.71 & 2.77 & 0.56\\ 
243.55 & 204.80 & 289.63 & 1.07 & 0.88 & 1.79 & 0.57\\ 
344.43 & 289.63 & 409.60 & 1.23 & 0.81 & 2.06 & 0.63\\ 
487.10 & 409.60 & 579.26 & 0.00 & 0.00 & 1.28 & 1.46\\ 
\enddata

\end{deluxetable}

%% file: 6morphology.tex
\section{Morphology and spectral variations}
\lb{sect:morphology}

The average spectrum of the bubbles is an important characteristic,
but it may be insufficient for distinguishing among the models of the bubbles' formation
and the mechanisms of the gamma-ray emission.
In this section, we calculate the spectrum of the bubbles in latitude strips,
and estimate the significance and the spectrum of the
enhanced gamma-ray emission in the south-eastern part of the bubbles,
called the ``cocoon''~\citep{2012ApJ...753...61S}.
We search for a jet inside the bubbles
and determine the location and the width of the boundary of the bubbles.

\subsection{Longitude Profiles}
\label{sect:profile}

\begin{figure}[htbp] 
\begin{center}
\epsfig{figure = 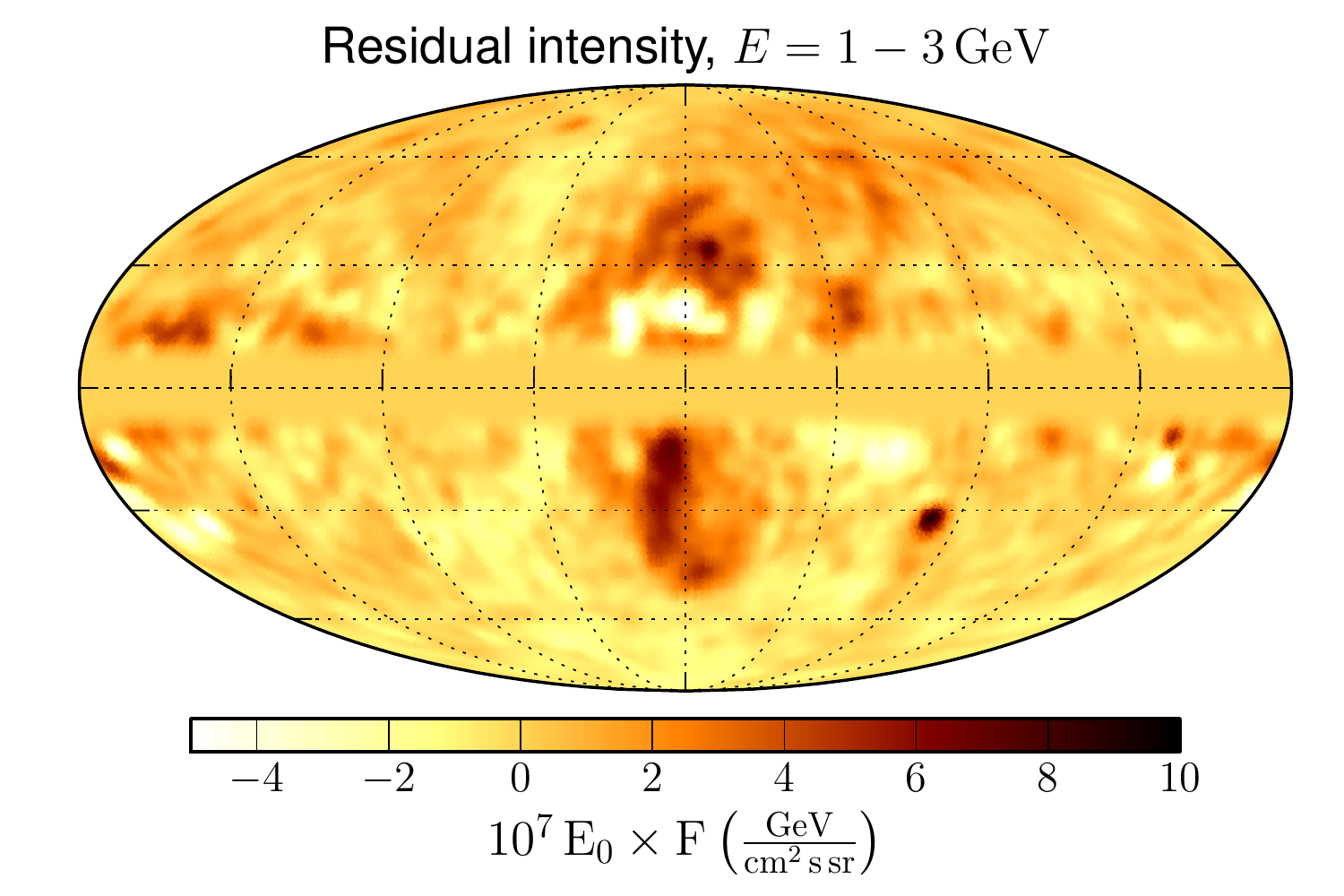, scale=\threepic}
\epsfig{figure = 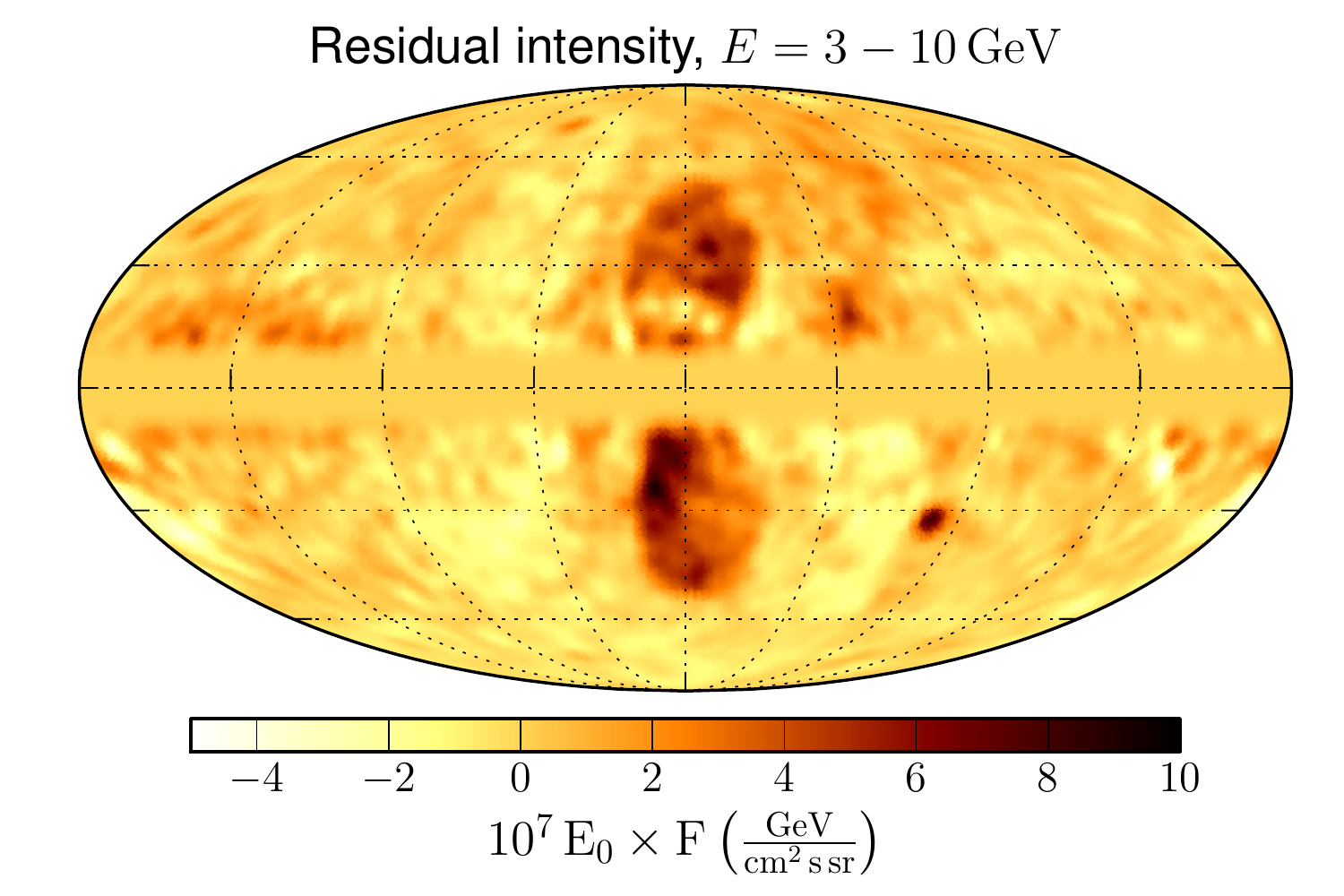, scale=\threepic}
\epsfig{figure = 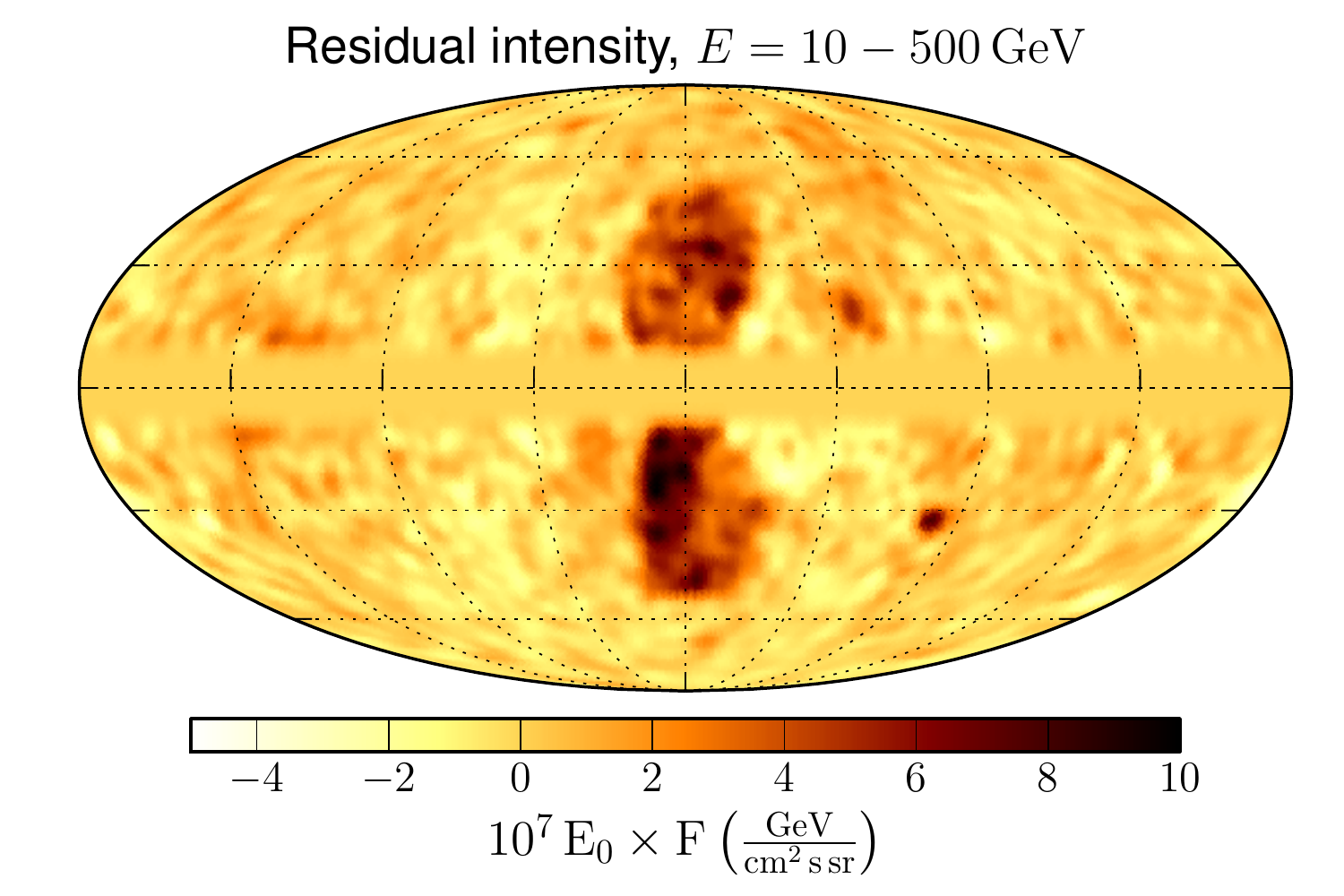, scale=\threepic} \\
\epsfig{figure = 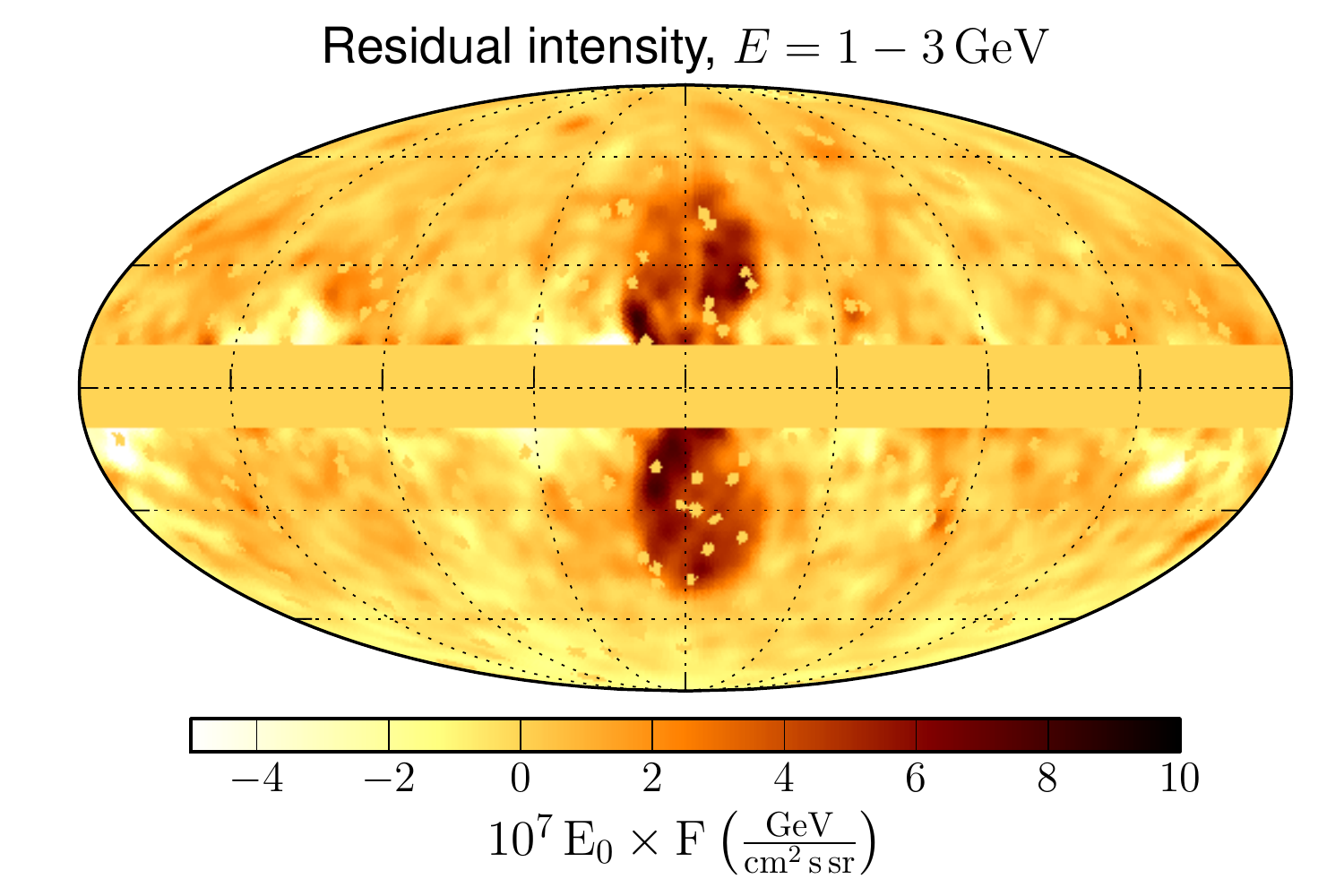, scale=\threepic}
\epsfig{figure = 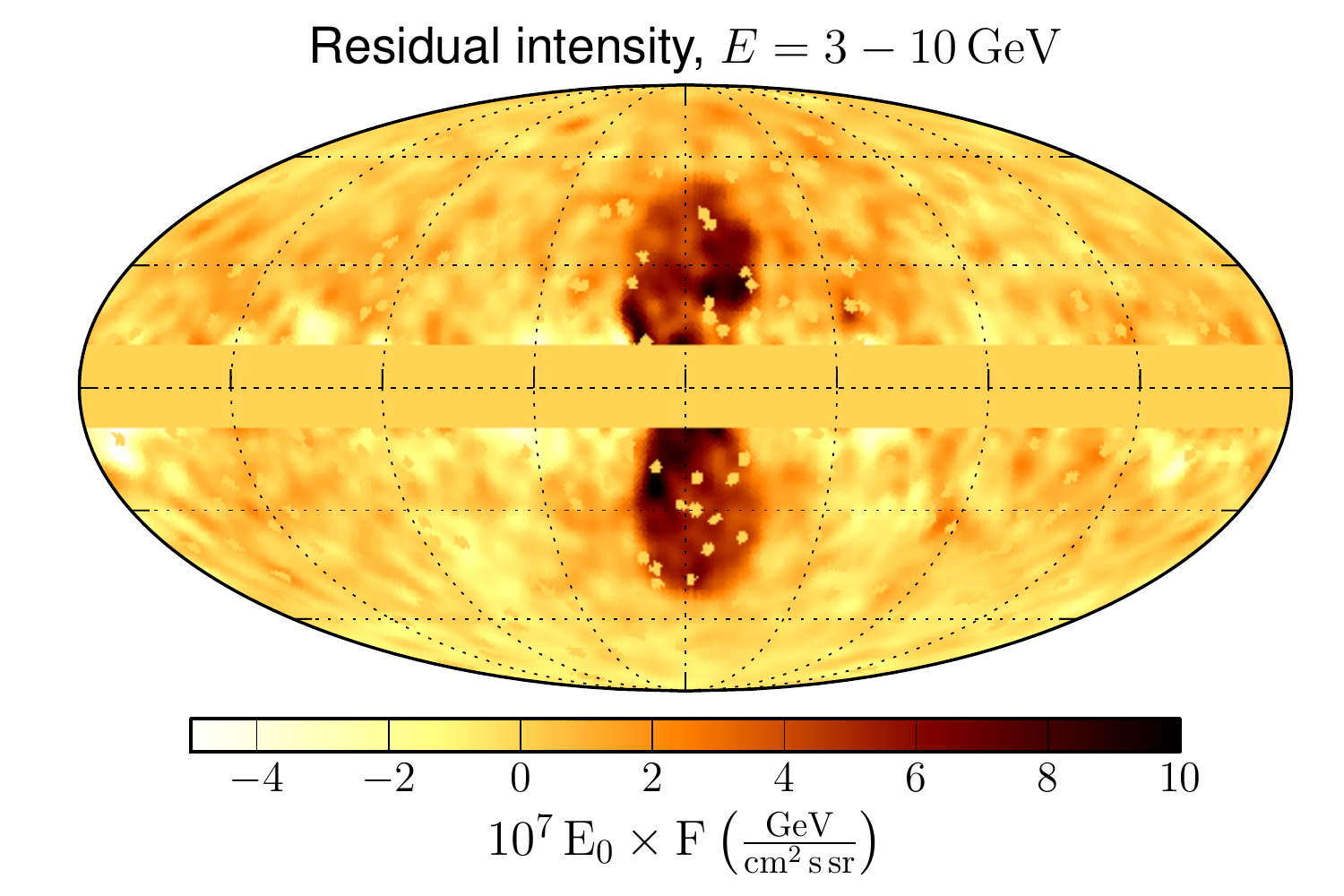, scale=\threepic}
\epsfig{figure = 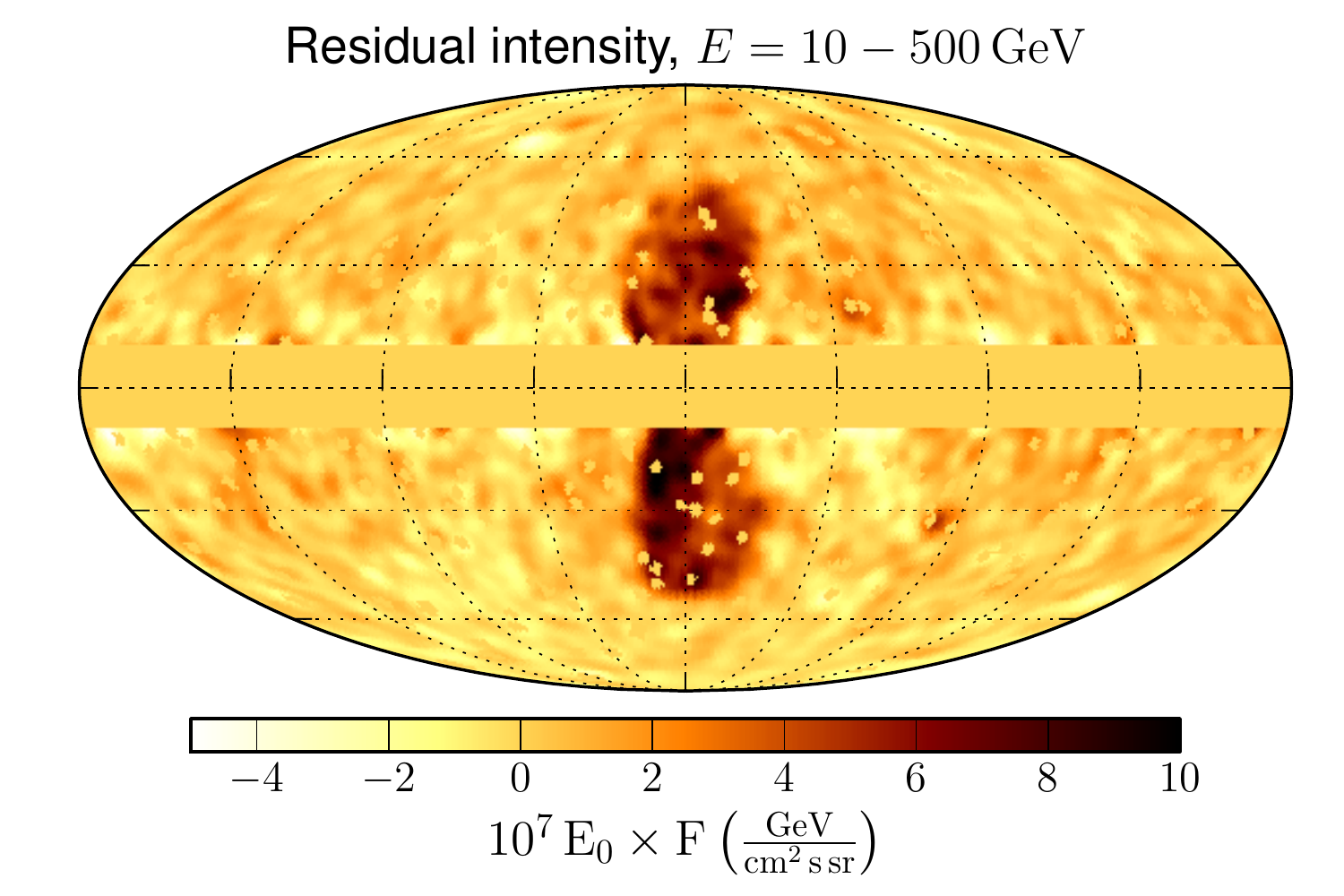, scale=\threepic}
\noindent
\caption{\small 
Residual intensity integrated in different energy bands for the baseline model derived with GALPROP templates in Section \ref{subsec:BubbleTemp} (top)
and for the example model derived with the local templates analysis in Section \ref{subsec:localBubbles} (bottom).
}
\label{fig:51resid10GeV}
\end{center}
\vspace{1mm}
\end{figure}

\begin{figure}[htbp] 
\begin{center}
\epsfig{figure = 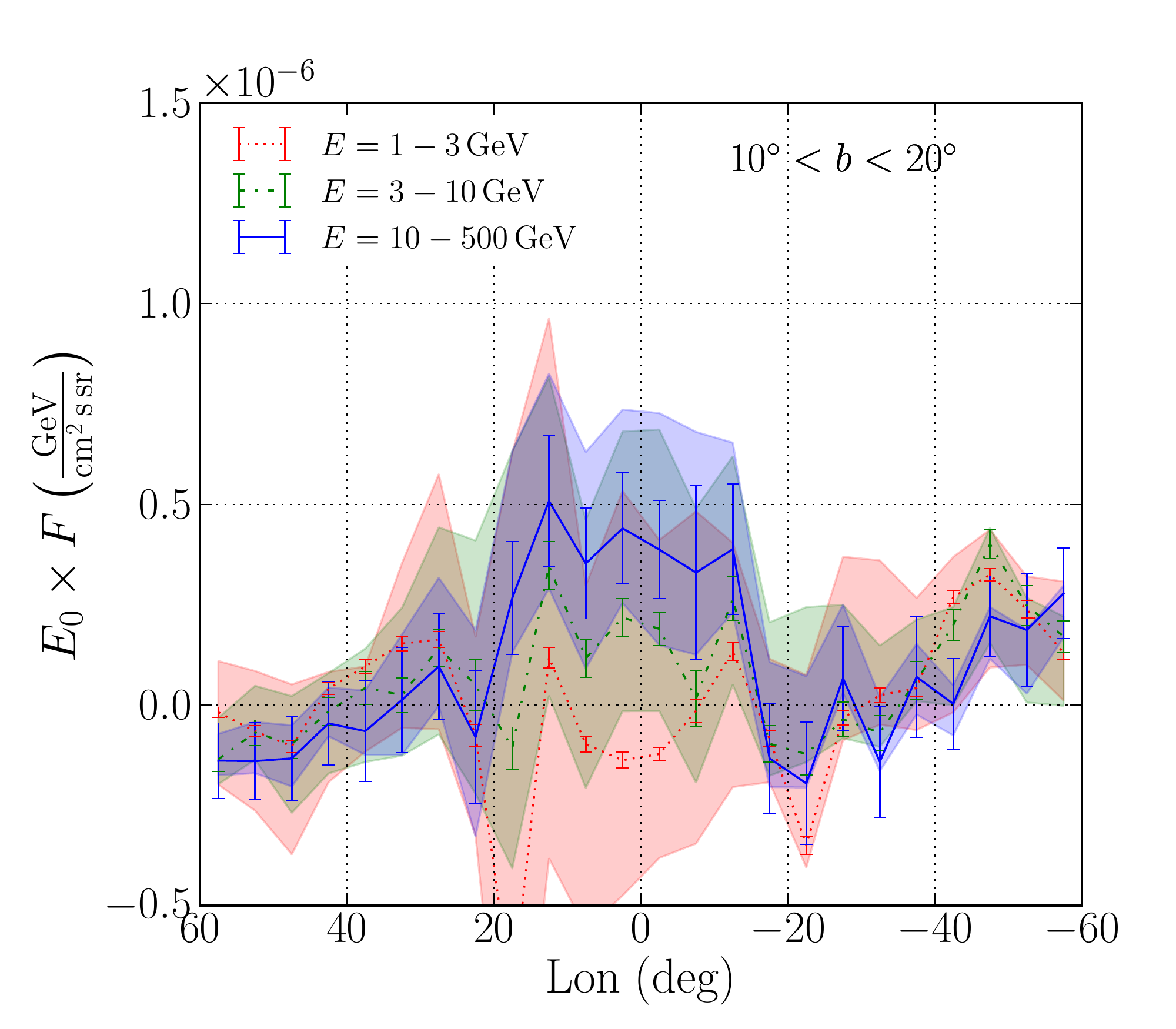, scale=\fourpic}
\epsfig{figure = 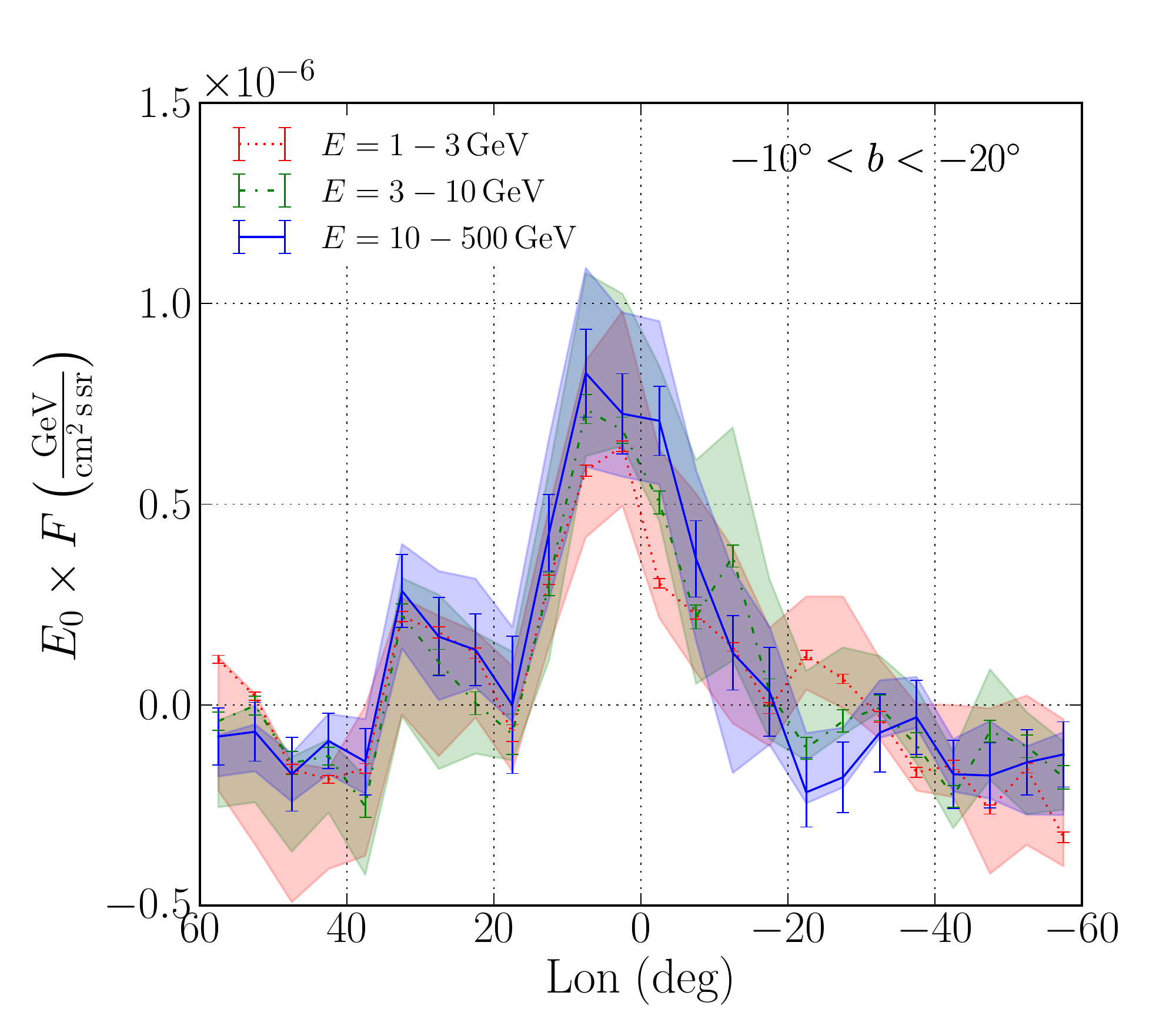, scale=\fourpic} \\
\epsfig{figure = 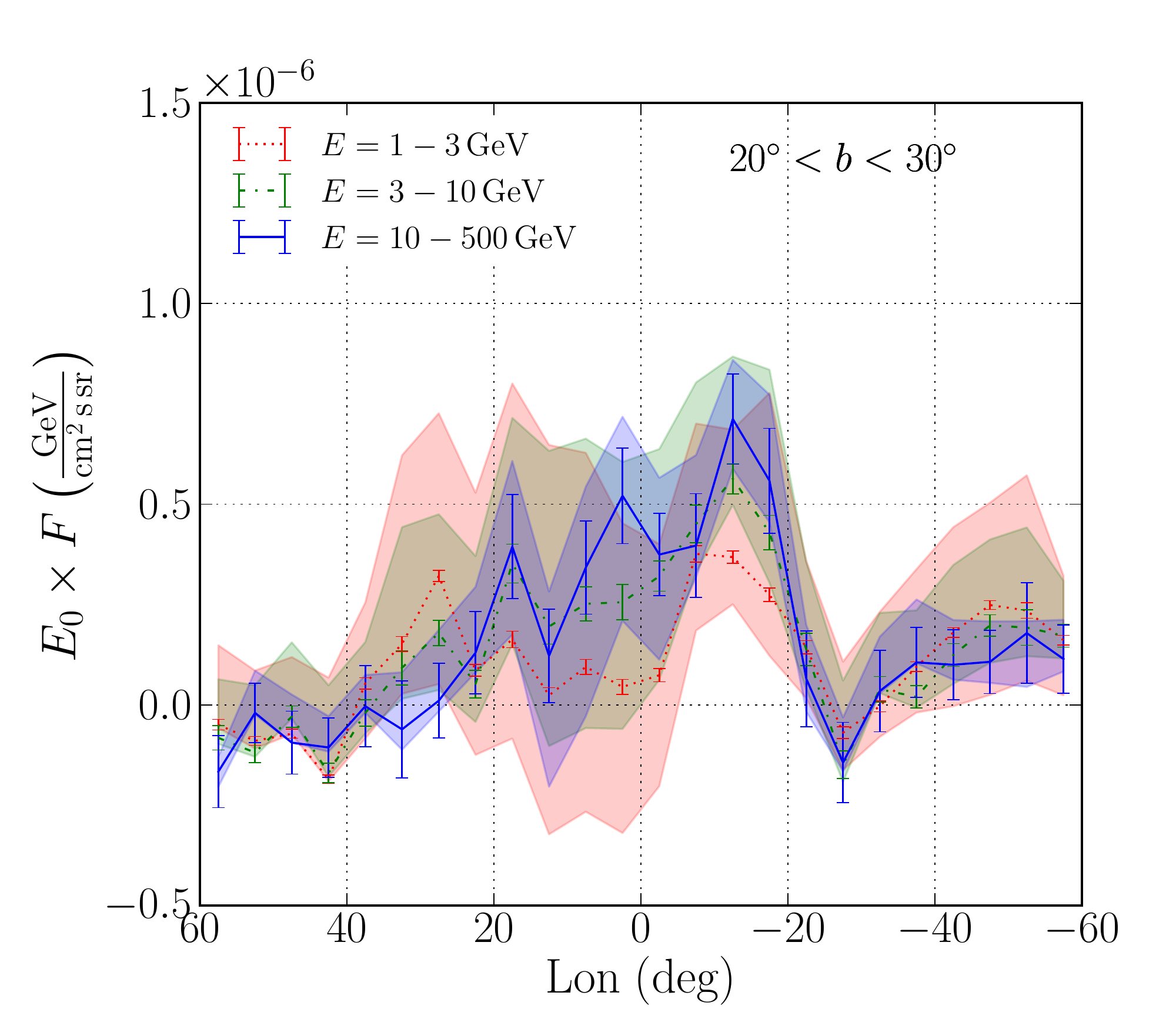, scale=\fourpic}  
\epsfig{figure = 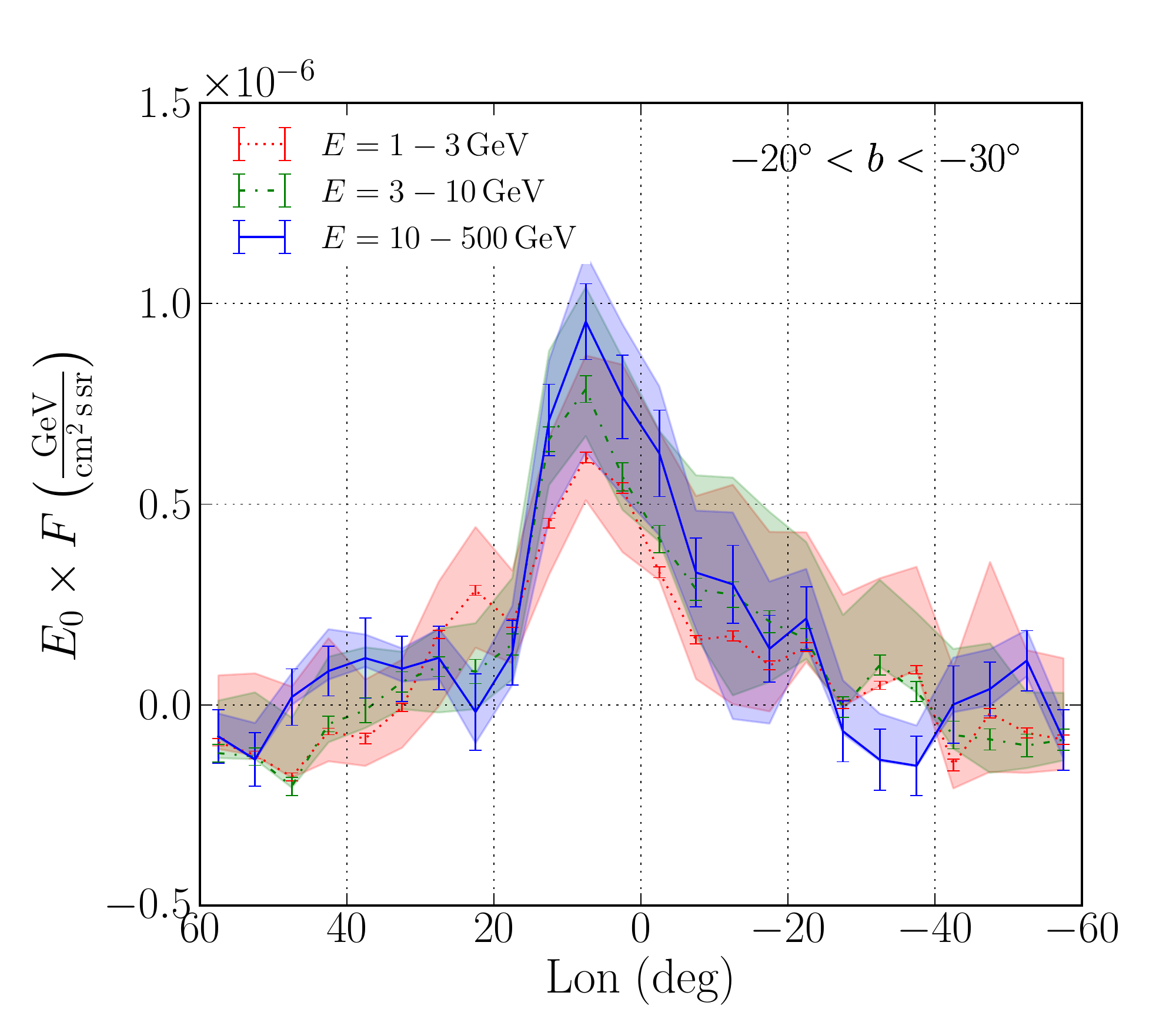, scale=\fourpic} \\
\epsfig{figure = 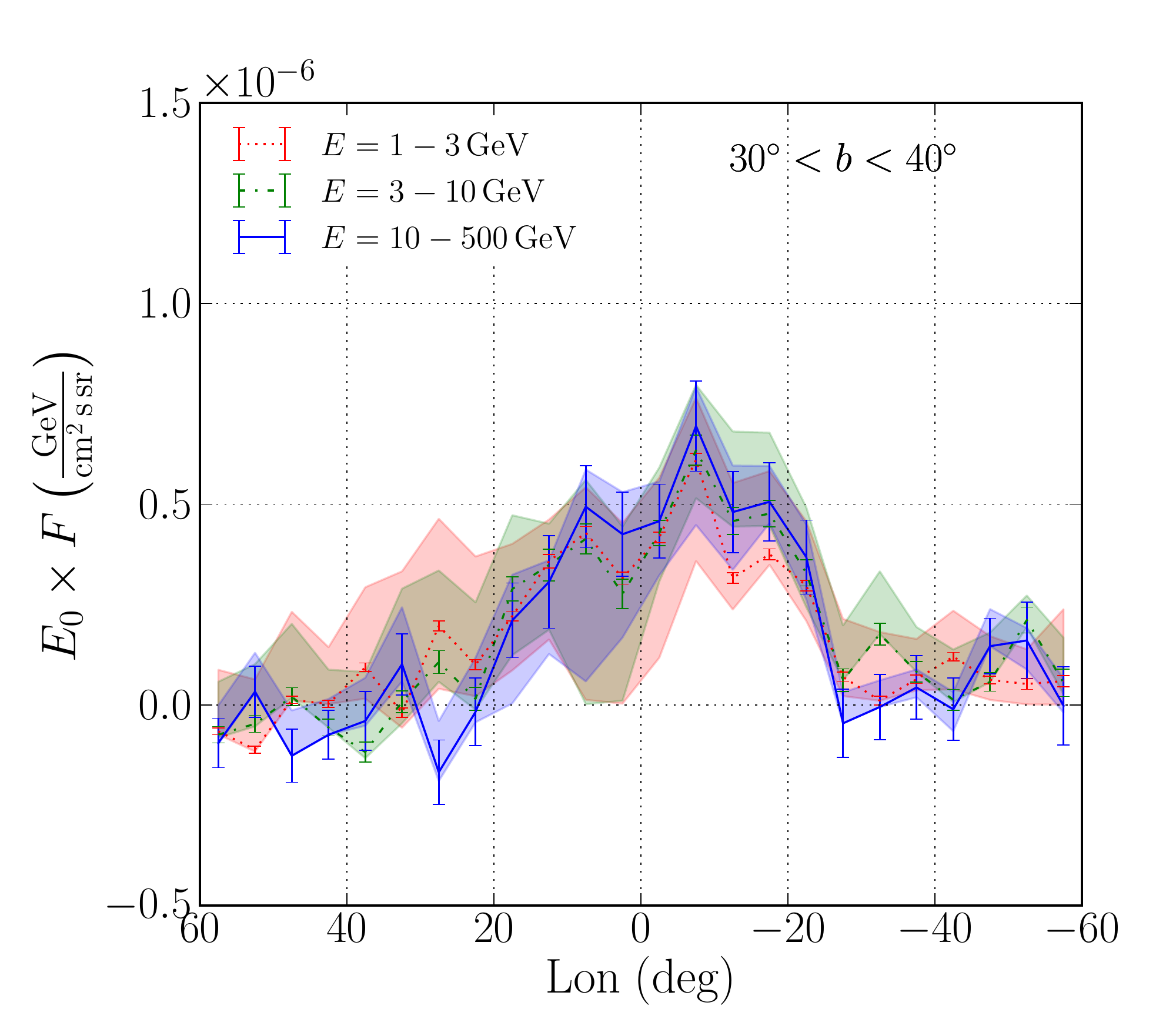, scale=\fourpic}
\epsfig{figure = 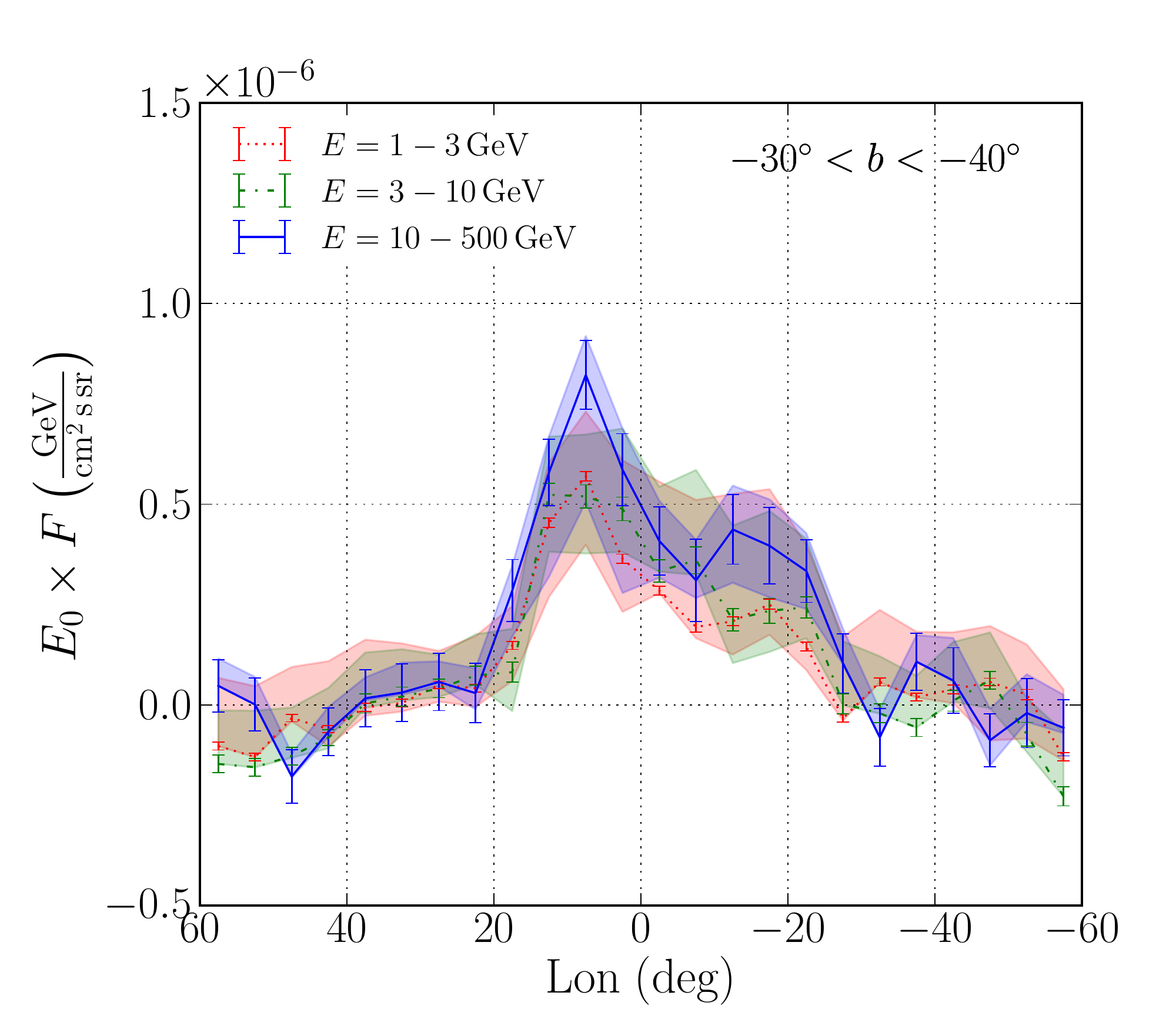, scale=\fourpic} \\
\epsfig{figure = 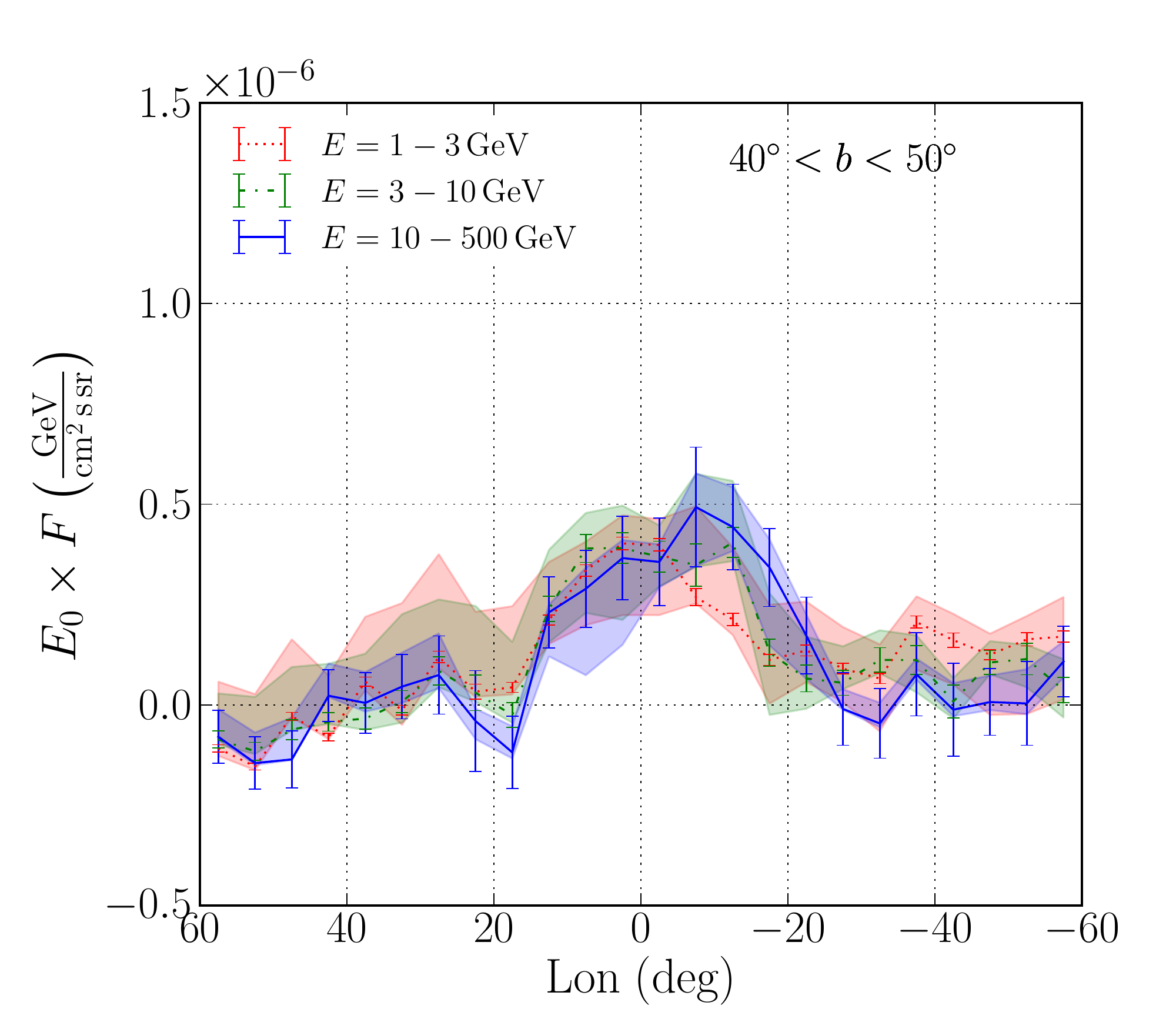, scale=\fourpic}
\epsfig{figure = 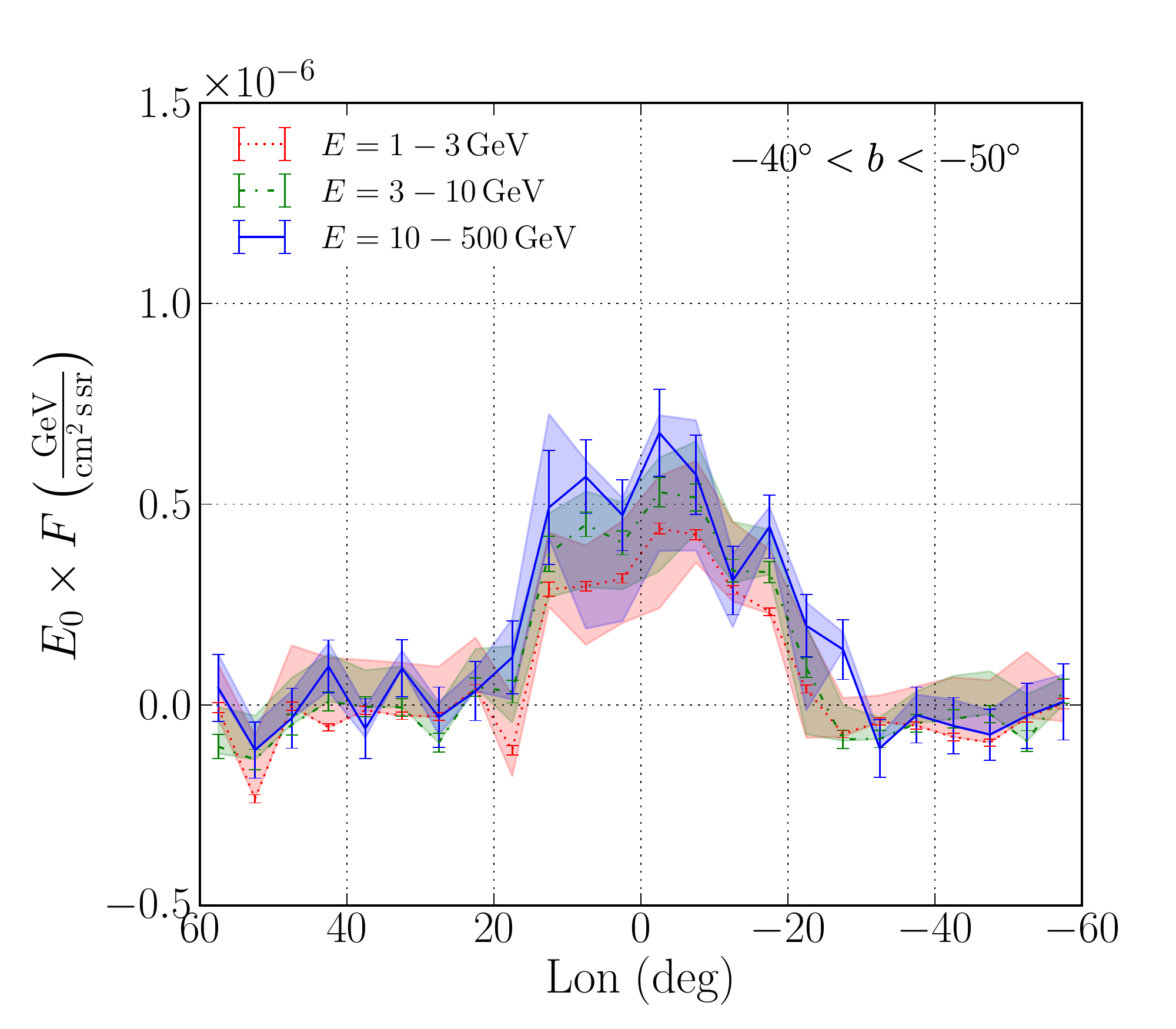, scale=\fourpic}
\noindent
\caption{\small 
Longitude profiles of the residual intensity including the bubbles
integrated over different energy bands. The profile plots are obtained by dividing the sky into $10^\circ$ strips in latitude. 
Points correspond to the GALPROP residuals (baseline model defined in Section \ref{sect:galprop}) in Figure \ref{fig:51resid10GeV}.
Shaded bands are computed as an envelope of the residuals for different
models of the foregrounds and different definitions of the templates for the bubbles and Loop~I (Section~\ref{sec:gadgetSys} and~\ref{subsec:SysErrorLocal}).
The width of the longitude bins is $5^\circ$.
}
\label{fig:52profiles}
\end{center}
\vspace{1mm}
\end{figure}

To give a general idea about the morphology of the bubbles,
we present the profile plots of the residual intensity corresponding to the {\Fermi} bubbles
at different latitudes integrated in three energy bands: 1 - 3 GeV, 3 - 10 GeV, 10 - 500 GeV.
The residual intensity is shown in Figure \ref{fig:51resid10GeV}. There is an L-shaped over-subtraction at low energies in the GALPROP residuals in the low latitude part of the northern bubble. This residual is spatially correlated with the star forming region $\rho$ Ophiuchi, which might have a different CR spectrum compared to the average. Notice that this feature is not present in the residuals obtained from the local template analysis, which allows the adjustment of the normalization of the CR density in local patches. 
The profile plots in $10^\circ$ latitude strips are shown in Figure \ref{fig:52profiles}.

An excess of emission in the southern bubble for latitudes 
$-40^\circ < b < -20^\circ $ and longitudes $0^\circ < \ell < 15^\circ$
corresponds to the cocoon proposed by \cite{2012ApJ...753...61S}.
There is also a slight excess of emission for  
$20^\circ < b < 40^\circ$ around $\ell = 10^\circ$.
At some latitudes, the width of the boundary of the bubbles is approximately or smaller than $5^\circ$.
We study the width of the edge in more detail in Section \ref{sect:edges}.

\subsection{Substructures}
\lb{sect:cocoon}

In this section, we present an analysis of substructures within the bubbles. 
In the residual maps we find an enhanced gamma-ray emission mostly in the south east side of the {\Fermi} bubbles.
Following \cite{2012ApJ...753...61S}, we will denote the region of enhanced gamma-ray
emission as the ``cocoon", although the physical origin of this emission
is not known.
In order to study the significance and the spectrum of the cocoon, 
we separate the cocoon template from the bubbles and fit both the cocoon and the bubbles 
templates together with the other diffuse foreground templates to the data.

\begin{figure}[htbp] 
\begin{center}
\epsfig{figure = 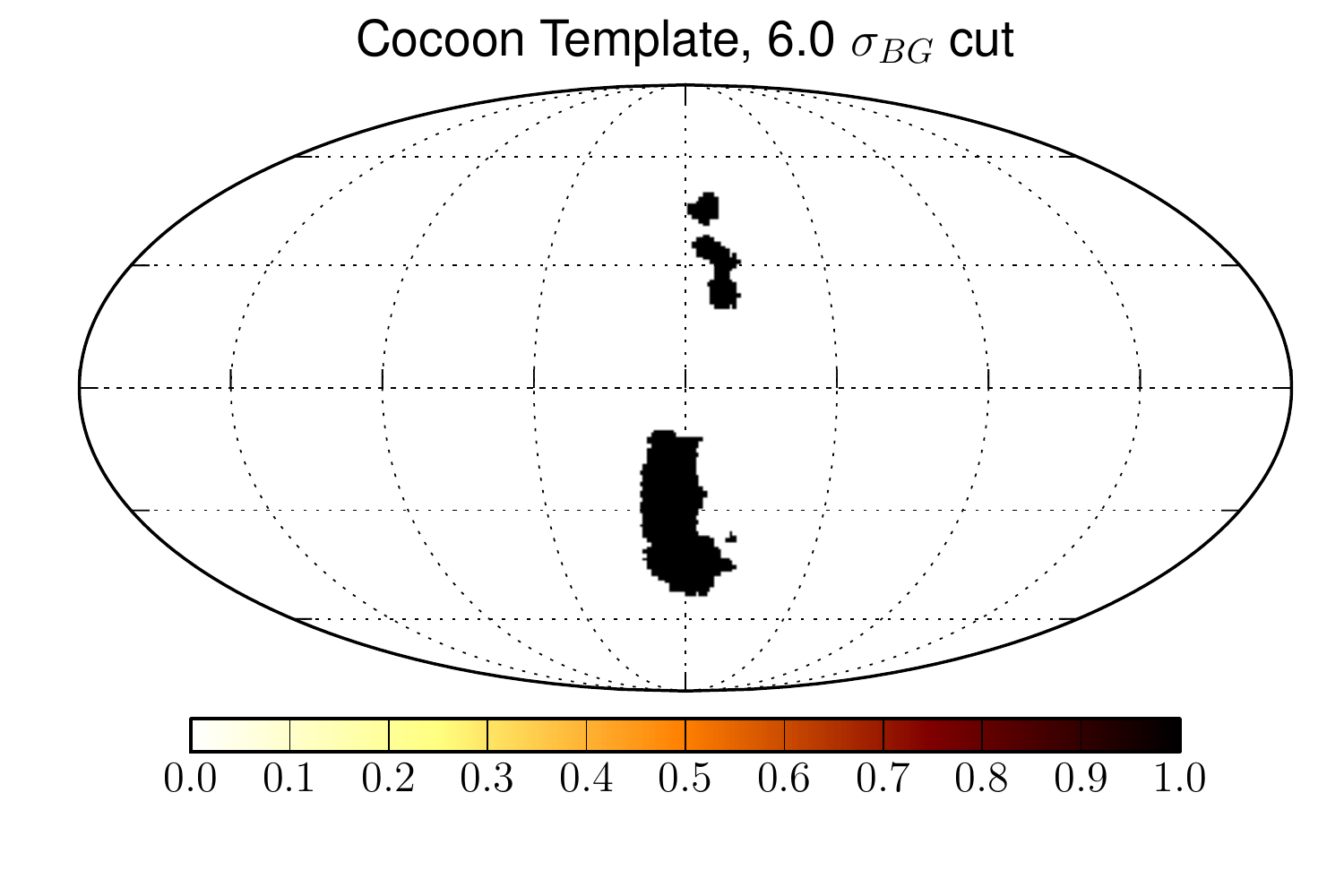, scale=\twopic}
\epsfig{figure = 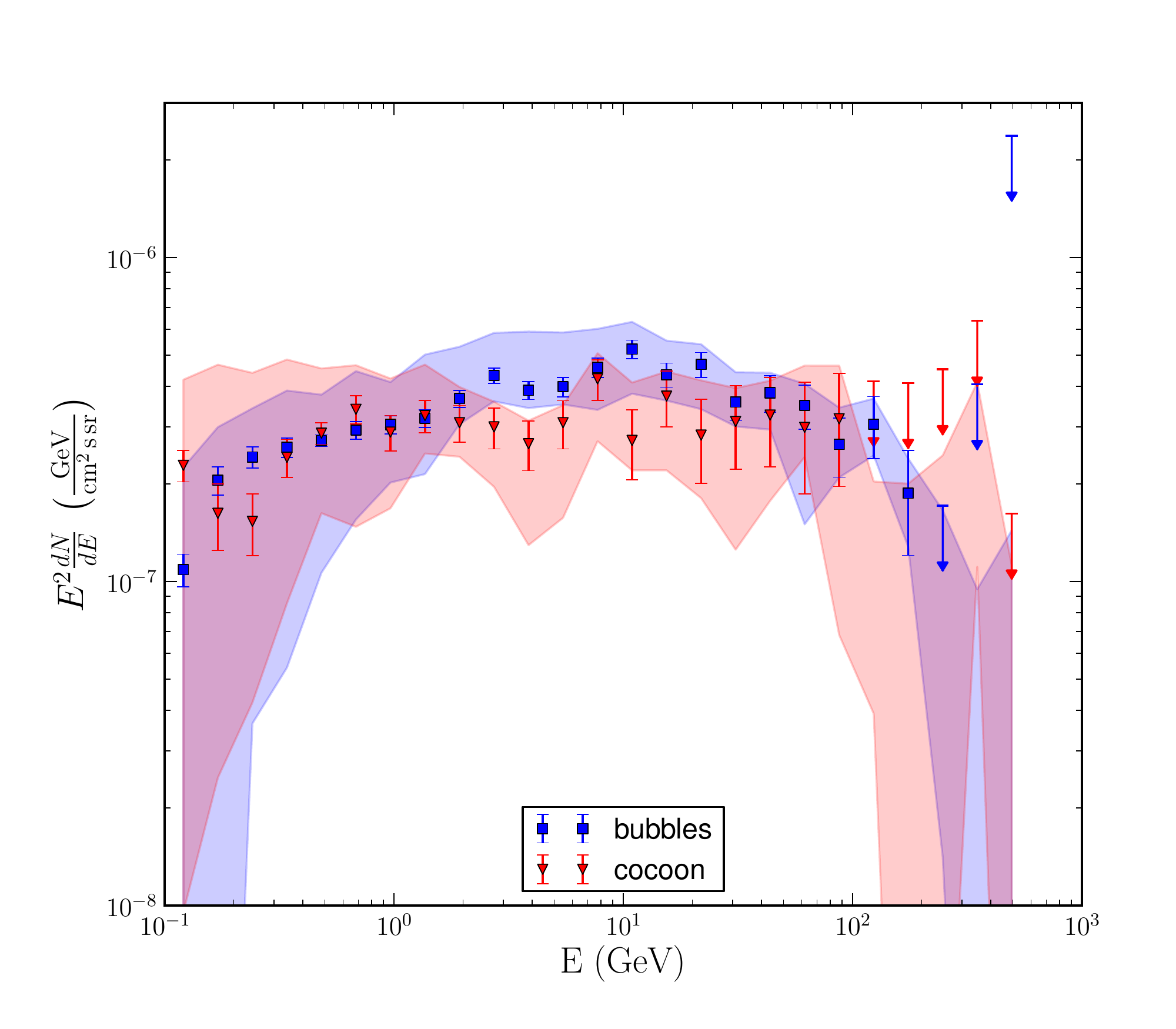, scale=\twopic}
\noindent
\caption{\small 
Left: Cocoon template derived analogously to the template of the bubbles
in Section \ref{subsec:BubbleTemp}. 
Right: The spectrum of the cocoon compared to the spectrum of the bubbles.
The points correspond to the baseline model from Section \ref{sect:galprop}.
The bands are the envelope resulting from different derivations of foreground
models and different definitions of the bubbles' and Loop~I templates.
}
\label{fig:54cocoon_template}
\end{center}
\vspace{1mm}
\end{figure}

\begin{figure}[htbp] 
\begin{center}
\epsfig{figure = 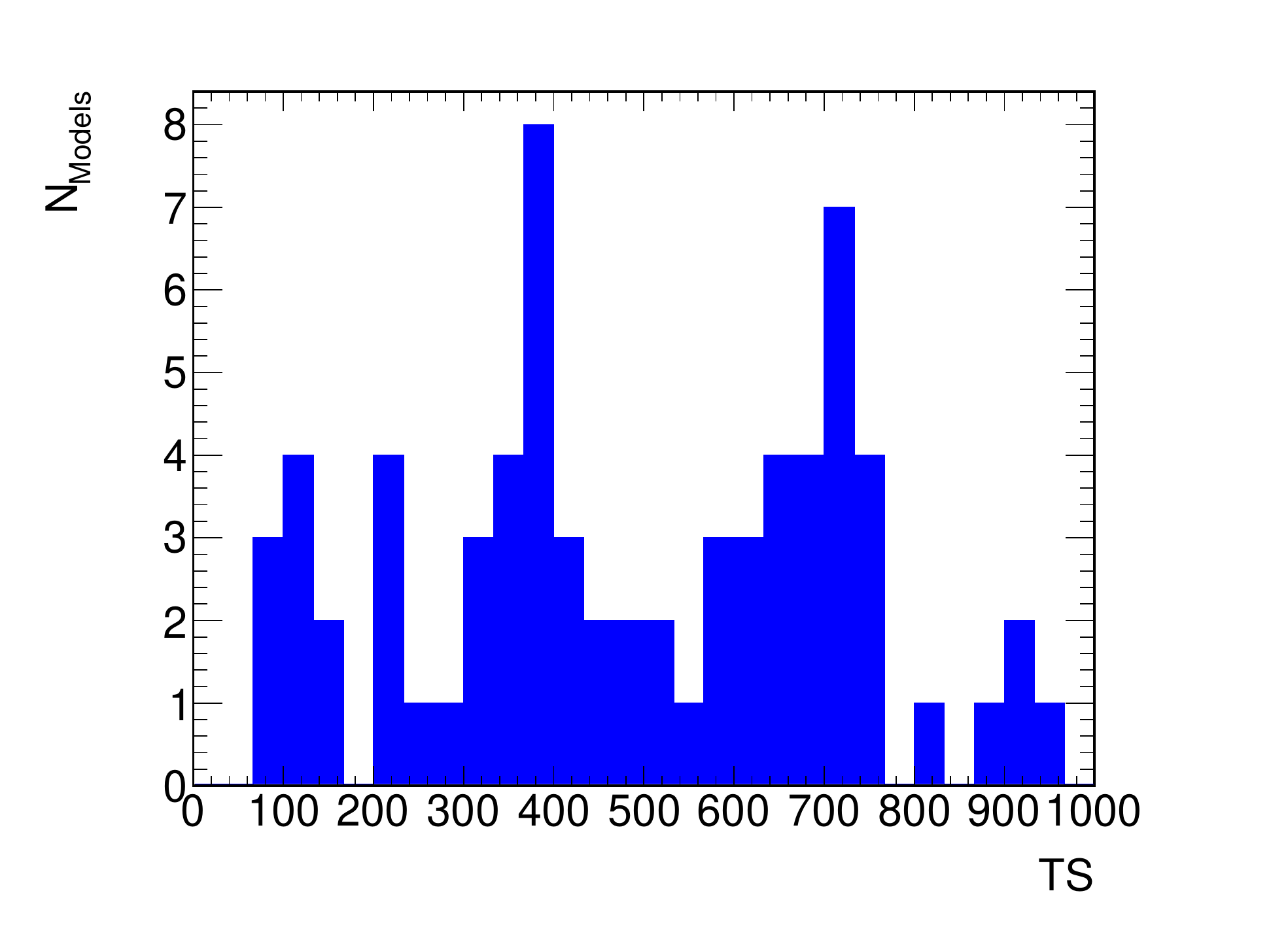, scale=\twopic}
\noindent
\caption{\small 
$\rm{TS}$ of the cocoon template for different derivations of the foreground emission
models and different definitions of the templates of the bubbles and Loop~I.
}
\label{fig:55cocoon_spectrum}
\end{center}
\vspace{1mm}
\end{figure}

We derive the cocoon template from the same residual maps that we use for the 
derivation of the {\Fermi} bubbles by applying a higher cut in significance. We take
6$\sigma_{\rm BG}$ in the GALPROP templates analysis and
5$\sigma_{\rm BG}$ in the local templates analysis.
The difference is due to the difference in the energy ranges used to define the templates in the two methods (see Section \ref{subsec:localBubbles}).
The cocoon template for the baseline model (defined in Section \ref{sect:galprop}) is shown in 
Figure~\ref{fig:54cocoon_template} (left). Notice that in contrast to the cocoon template in~\cite{2012ApJ...753...61S} this template is not restricted to the southern hemisphere and includes also excess emission in the North.
In the fits we use the flat template for the bubbles; otherwise the structures in the template of the bubbles 
can absorb a significant part of the cocoon emission.
We also use a flat cocoon template to get a conservative estimate of the significance of the 
cocoon.
Note that the cocoon template is inside the template of the bubbles, i.e., the cocoon emission is on top of the emission
from the bubbles modeled by a flat template.
Using both the cocoon and the bubbles' templates improves the likelihood of the fit
relative to the flat template for the bubbles alone.
We find that TS is between 95 and 975, 
depending on the foreground emission model, while the number of additional
free parameters in the fit is 25 (one for each energy bin). 
Figure~\ref{fig:55cocoon_spectrum} shows the distribution of TS
for different foreground emission models.
The probability that the cocoon is a statistical fluctuation is $<10^{-9}$. 
This probability does not include the trials factor.

The intensity spectrum of the cocoon is compared to the spectrum of the bubbles in Figure~\ref{fig:54cocoon_template} (right).
We find that the spectrum of the cocoon is consistent with the spectrum of the rest of the bubbles. 
However, due to large statistical uncertainties in the cocoon spectrum we cannot rule out a simple power-law spectrum of the cocoon emission.
The absolute value of intensity of the gamma-ray emission from the cocoon detected on top of the flat bubbles template
is, by coincidence, very similar to the gamma-ray intensity of the bubbles.
In other words, the intensity in the cocoon region is about two times larger than the intensity inside the bubbles 
but outside of the cocoon.

\begin{figure}[htbp] 
\begin{center}
\epsfig{figure = 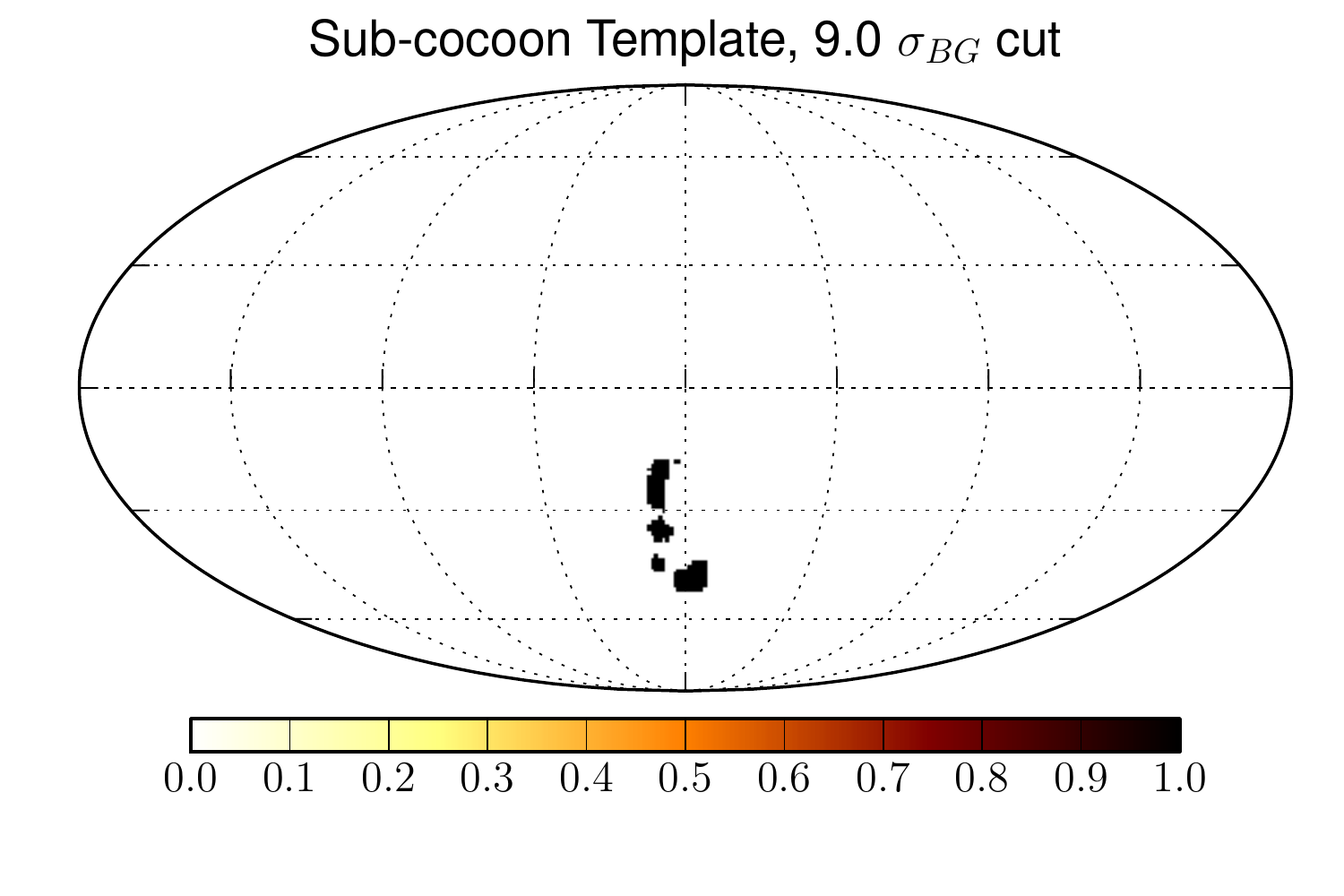, scale=\twopic}
\epsfig{figure = 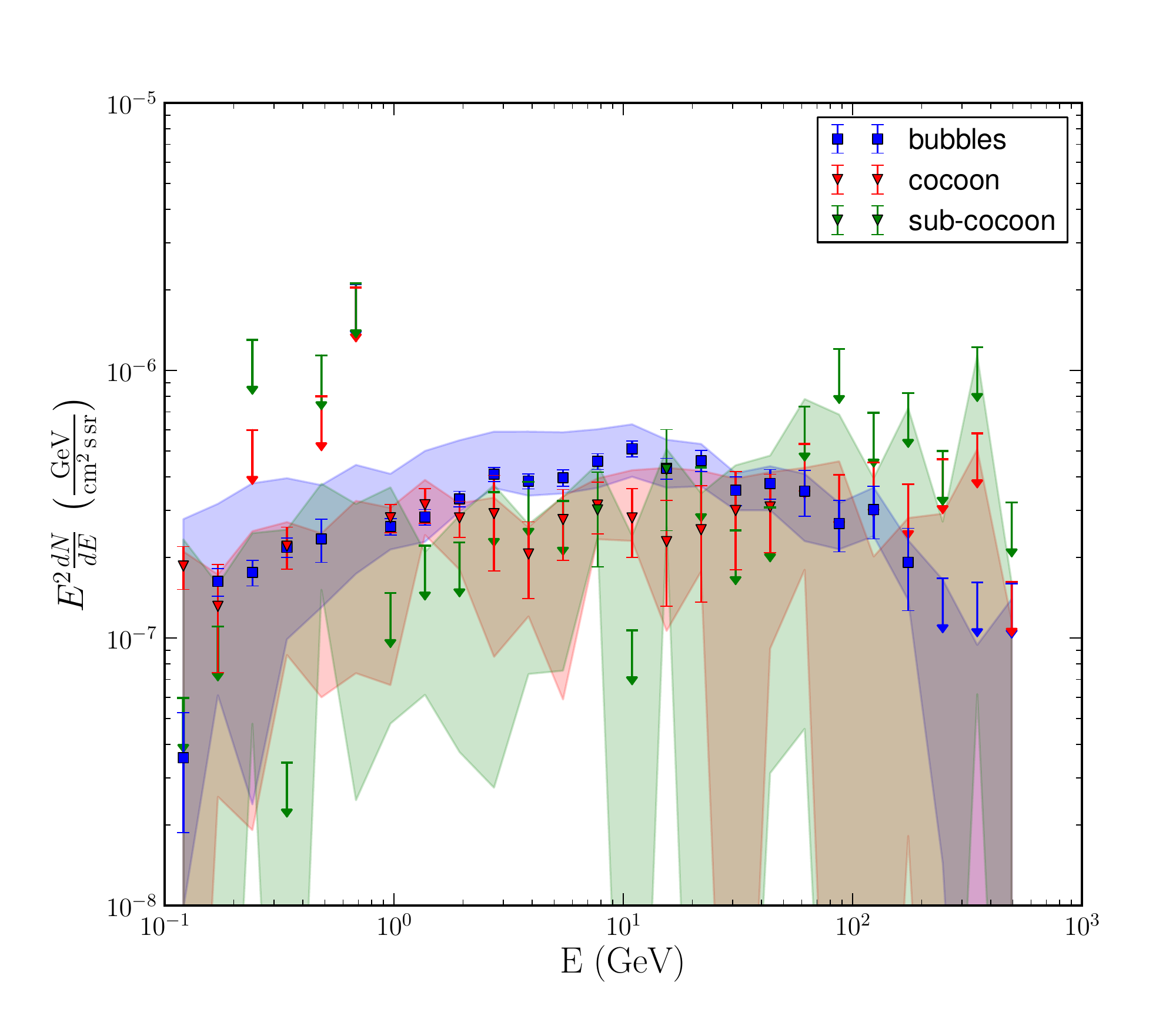, scale=\twopic}
\noindent
\caption{\small 
Left: A template for a high significance residual, the sub-cocoon,
created from the residual map with threshold 9$\sigma_{BG}$.
Right: The spectrum of the sub-cocoon compared to the spectrum of the cocoon and the bubbles. }
\label{fig:55ajet_temp}
\end{center}
\vspace{1mm}
\end{figure}

\begin{figure}[htbp] 
\begin{center}
\epsfig{figure = 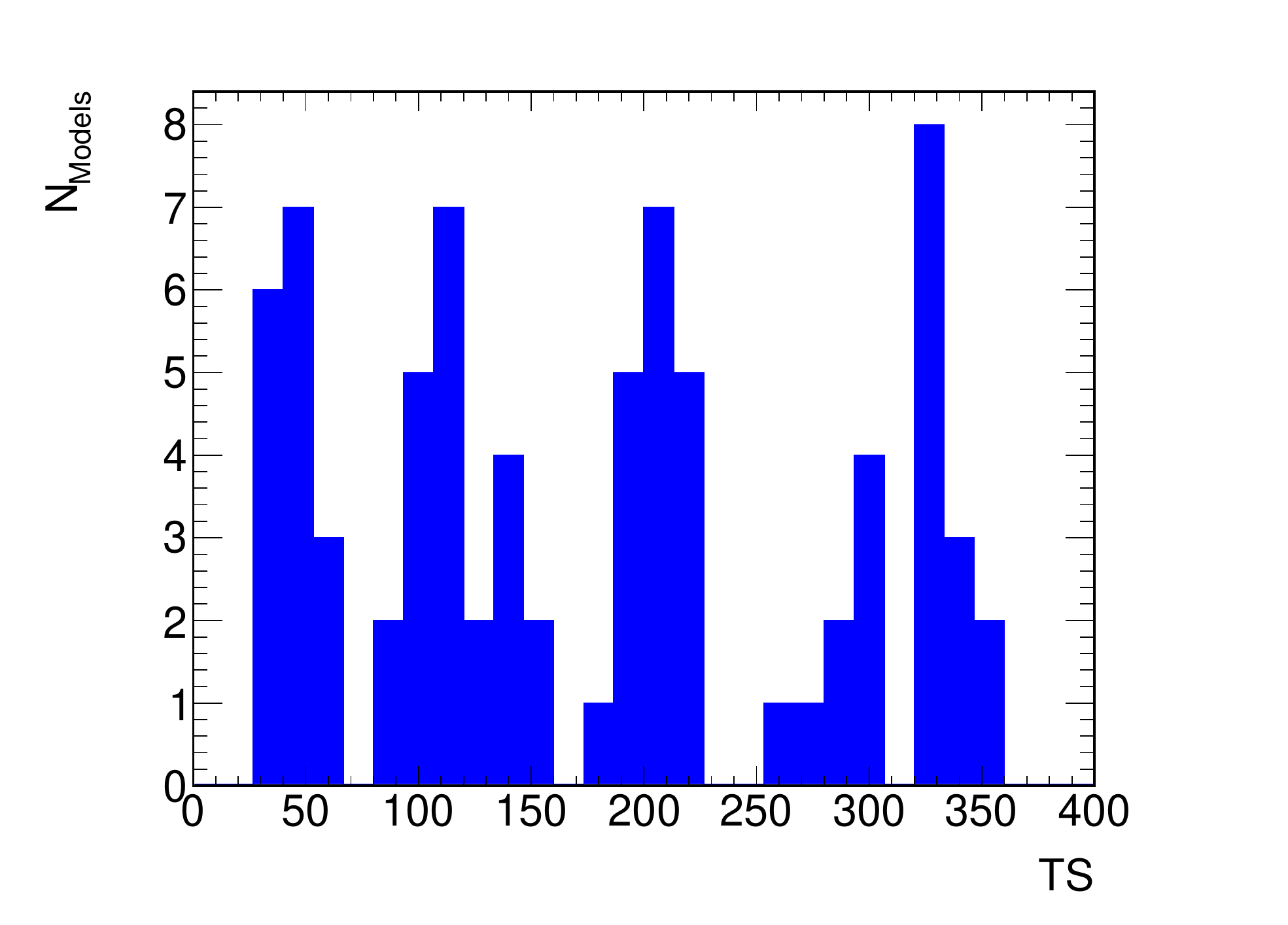, scale=\twopic}
\noindent
\caption{\small 
$\rm{TS}$ of the sub-cocoon template for different foreground models.}
\label{fig:55bjet_temp}
\end{center}
\vspace{1mm}
\end{figure}

In a second step we apply an even higher significance threshold of 9$\sigma_{\rm{BG}}$ to the residual map. We call the resulting structure the ``sub-cocoon''.
The sub-cocoon template is displayed in Figure~\ref{fig:55ajet_temp} (left). We include this template in the all-sky 
fit together with a flat bubbles and a flat cocoon template. The resulting spectra are shown in Figure~\ref{fig:55ajet_temp} (right). The improvement of the fit obtained by including the additional template is displayed in Figure~\ref{fig:55bjet_temp}. We find that TS for different foreground models is between 30 and 360 (for 25 additional free parameters).
The probability that the sub-cocoon is a statistical fluctuation is about $20\%$ considering the smallest TS value among all models.

In the following we investigate the existence of jet-like emission as proposed by \cite{2012ApJ...753...61S}, who
tentatively observed a pair of jets along the cocoon's axis of symmetry 
with a harder spectral index compared to the spectrum of the bubbles.
The existence of a jet within the bubbles would be an important indication of 
an AGN-like activity of the black hole in the Galactic center.

\begin{figure}[htbp] 
\begin{center}
\epsfig{figure = 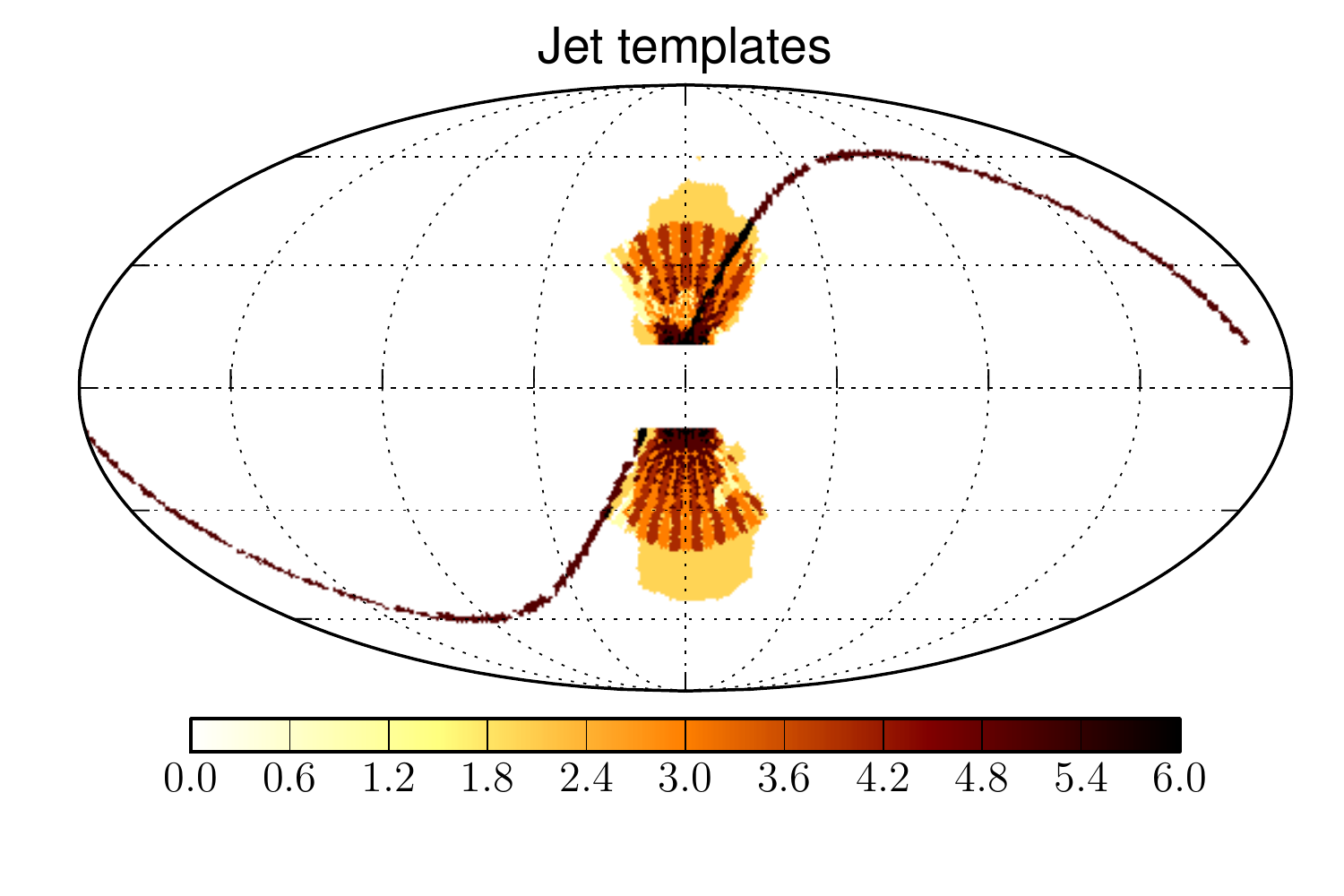, scale=\twopic}
\epsfig{figure = 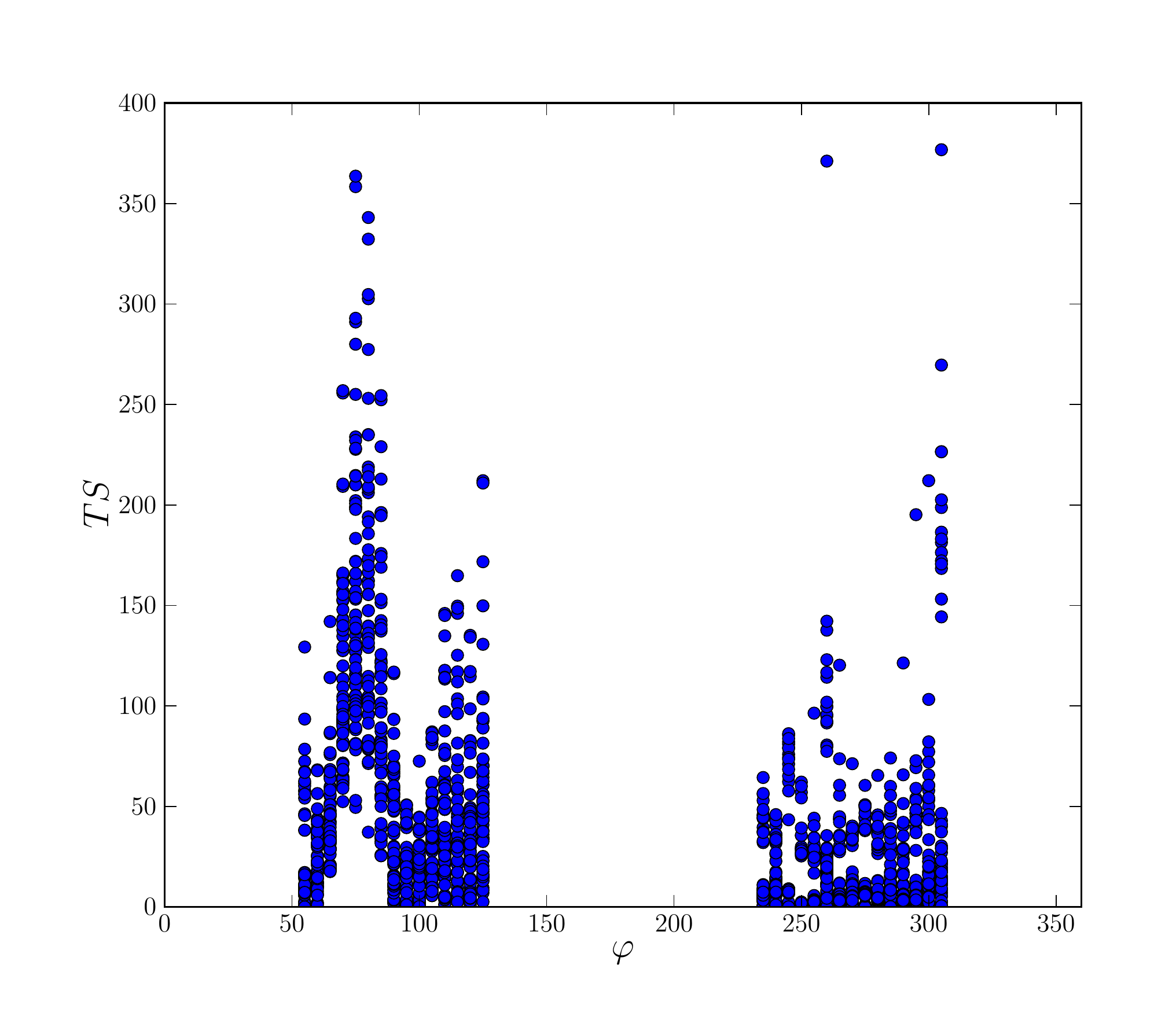, scale=\twopic}
\noindent
\caption{\small 
Left: jet templates and solar emission template. Each template has a width of $2.5^{\circ}$ and a length of $40^{\circ}$. Every second template is multiplied by 2 for better visibility. We only include one template at a time in the all-sky fit. The yellow shaded region corresponds to the flat bubble template.
Right: $\rm{TS}$ of the jet template as a function of the counterclockwise jet angle $\varphi$ defined with respect to the 
positive longitude direction, so that the $90^{\circ}$ jet points to the South. 
Different points at the same angle correspond to different foreground models and analysis strategy.
}
\label{fig:55jet_sig}
\end{center}
\vspace{1mm}
\end{figure}

To test this hypothesis we use a generic jet template, which is a strip of width $2.5^{\circ}$ and length $40^{\circ}$ originating in the Galactic center. We rotate this template in $5^{\circ}$ steps. 
Note, that the physical distance corresponding to $5^{\circ}$ at $40^{\circ}$ is $5^{\circ} \sin(40^{\circ}) \approx 3.2^\circ$,
i.e., the jets cover practically all pixels within $40^{\circ}$ from the GC.
Figure~\ref{fig:55jet_sig} (left) shows an illustration of all jet templates. Note that we include only one jet template at a time in each fit. Because the emission of the sun could mimic a jet as the Sun moves along the ecliptic, which passes near the Galactic center, we also add a template for the gamma-ray emission from cosmic-ray interactions in the outer atmosphere and radiation field of the Sun\footnote{A description of the solar template can be found at the \Fermi Science Support Center:
\hbox{\url{http://fermi.gsfc.nasa.gov/ssc/data/analysis/scitools/solar_template.html}}}
\citep{2013arXiv1307.0197J}.
For each position of the jet we calculate the improvement of the fit (compared to a fit with only flat bubbles and flat cocoon template). Figure~\ref{fig:55jet_sig} (right) shows the distribution of $\rm{TS}$ for the jet template in different orientations for different diffuse models. 
There is a broad enhancement towards several directions centered at $\varphi\cong 75^\circ$, which cover the cocoon. In those cases, the jet templates account for some remaining excess emission on top of the cocoon, that is not modeled by the flat cocoon template.   
In summary, 
we do not find significant residuals aligned along a specific direction that could 
be interpreted as a jet.

\subsection{Width of the boundary of the bubbles}
\lb{sect:edges}

The sharpness of the edge is one of the main arguments for a transient process of bubble formation
\citep{2010ApJ...724.1044S}.
The main difficulties in estimating the width of the edge 
are the statistical error and the systematic uncertainty
due to modeling of the diffuse foreground gamma-ray emission.
In order to get maximal information from the available data,
we have chosen to perform a parametric fit of a smoothed step function (modeled by a hyperbolic tangent)
across the edge of the bubbles.
The algorithm has the following steps:
\ben
\vspace{-3mm}
\item
Choose a reference point approximately on the bubble edge (by visual inspection of residuals above 10 GeV).
\item
Find the gradient of the residual by fitting a plane to the residual intensity 
in pixels within $10^{\circ}$ from the reference point. 
The gradient is along the direction of maximal change in the
residual flux.
We use it to define the direction perpendicular to the edge.
\item
In order to find the position of the edge and its width, we choose a strip along the gradient and project the data onto the gradient direction. This allows us to reduce the fitting to a one-dimensional fitting problem. 
We take the length of the strip to be $40^{\circ}$ ($\pm 20^{\circ}$ from the reference point), 
and the width to be $20^{\circ}$ ($\pm 10^{\circ}$ from the reference point). 
The size of the bins in the projection is $1.3^{\circ}$, which is larger than the pixel size but small enough that resolving the
sharp transitions is possible.
\item
Fit the data along the strip with a smoothed step function modeled by the hyperbolic tangent plus a constant:
\be
\lb{eq:edge}
f(\vp) = \rm A \tanh \ld(\vp - \vp_0) + C.
\ee
In total we have the four parameters A, $\ld$, $\vp_0$, and C. Two of them are non-linear ($\ld$ and $\vp_0$).
The best-fit position of the edge is determined by $\vp_0$, the width of the edge is defined as $\Delta\vp = 2 / \ld$,
and the constant $C$ models the residual background emission.
During the fit to the data, the model is convolved with the PSF for each energy range.
\item
Test the systematic uncertainty by changing the length and the width of the strip and the size of the bins. The lengths are $40^{\circ}$, $50^{\circ}$, $60^{\circ}$, the widths are $10^{\circ}$, $20^{\circ}$, $30^{\circ}$, and the bins are $1.3^{\circ}$ and $1.5^{\circ}$ wide.
\item
Test the systematic uncertainty related to the derivation of the foreground models and the templates of the bubbles and Loop~I.
\een

\begin{figure}[htbp] 
\begin{center}
\epsfig{figure = 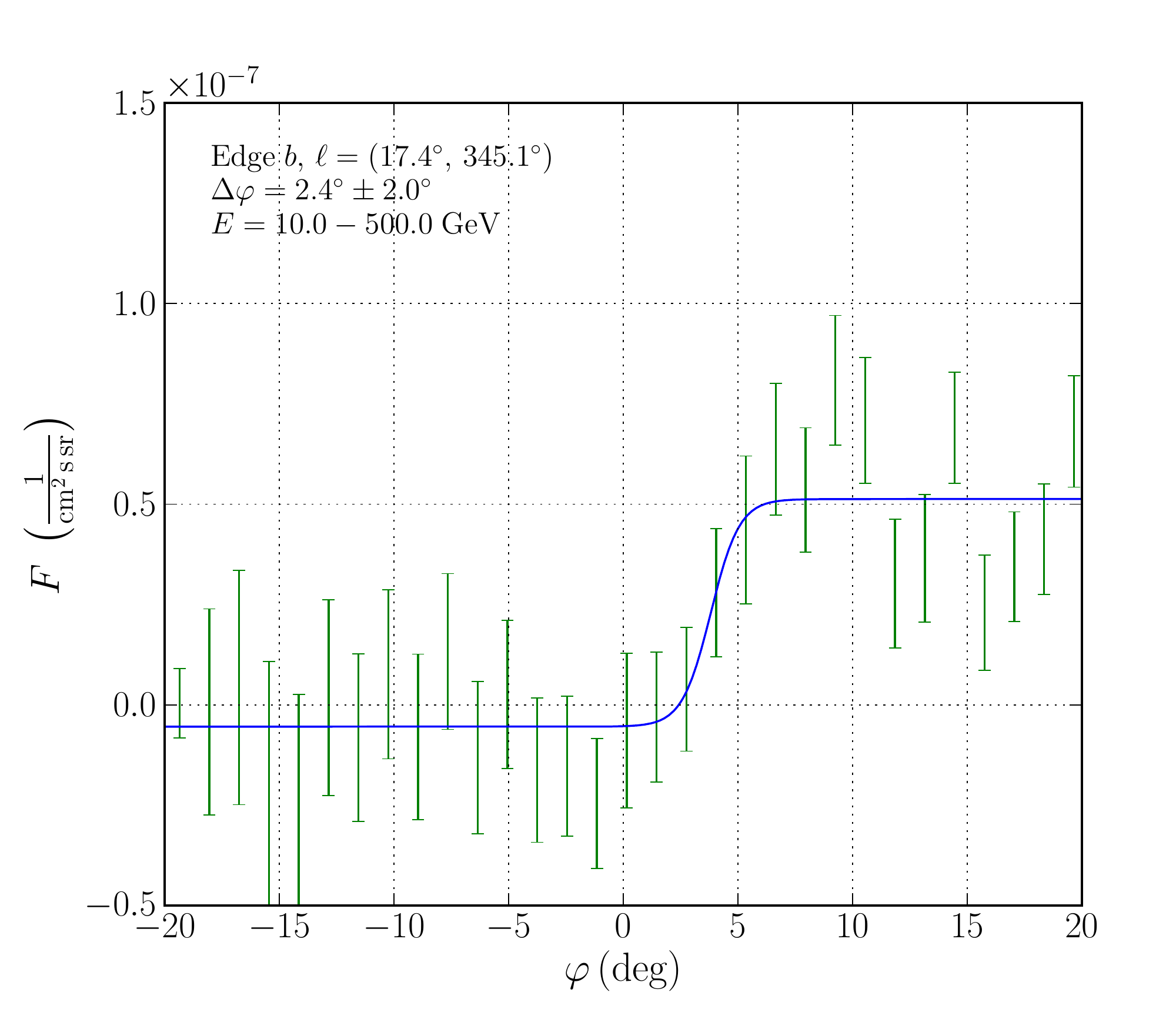, scale=\twopic}
\epsfig{figure = 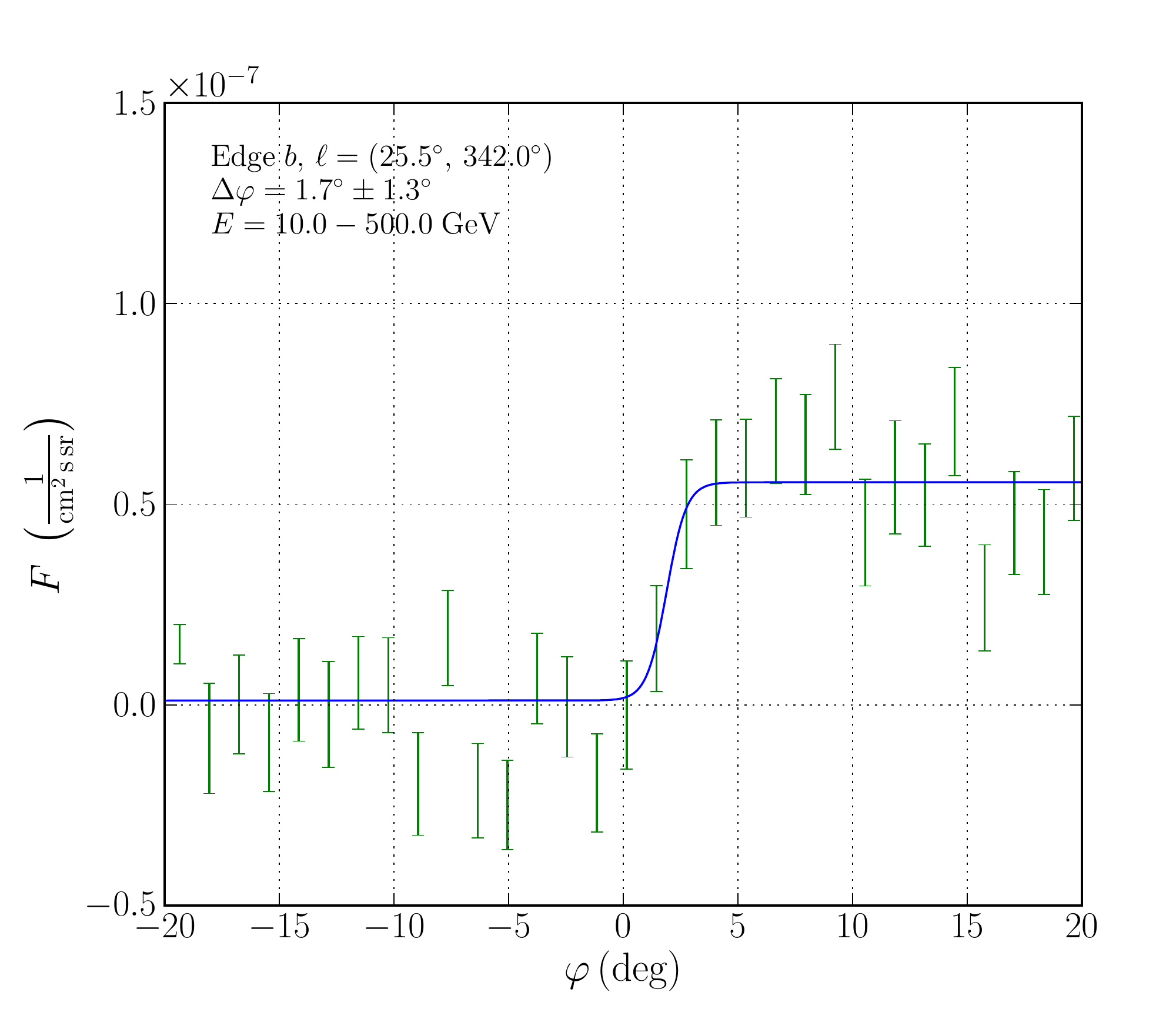, scale=\twopic}
\epsfig{figure = 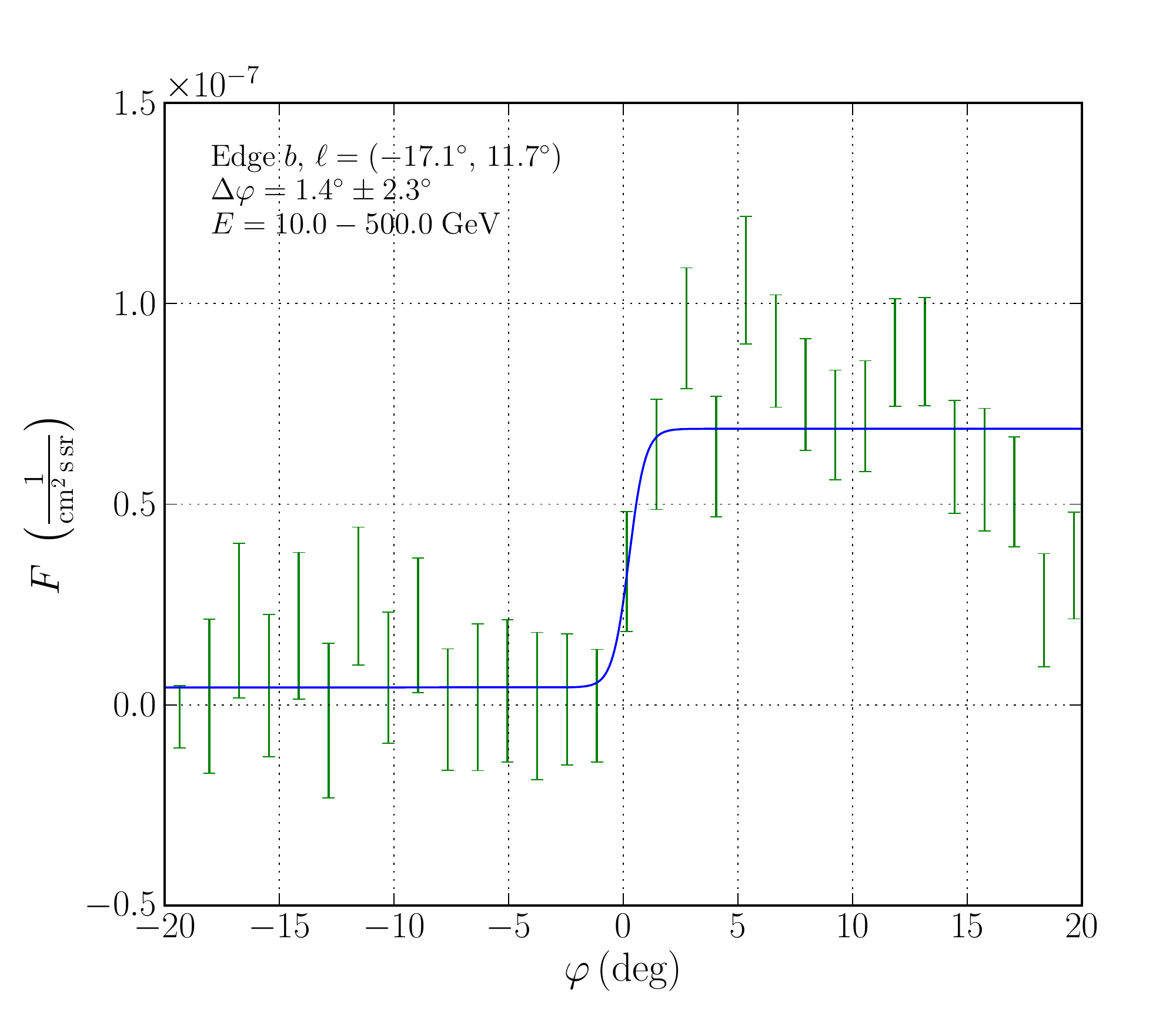, scale=\twopic}
\epsfig{figure = 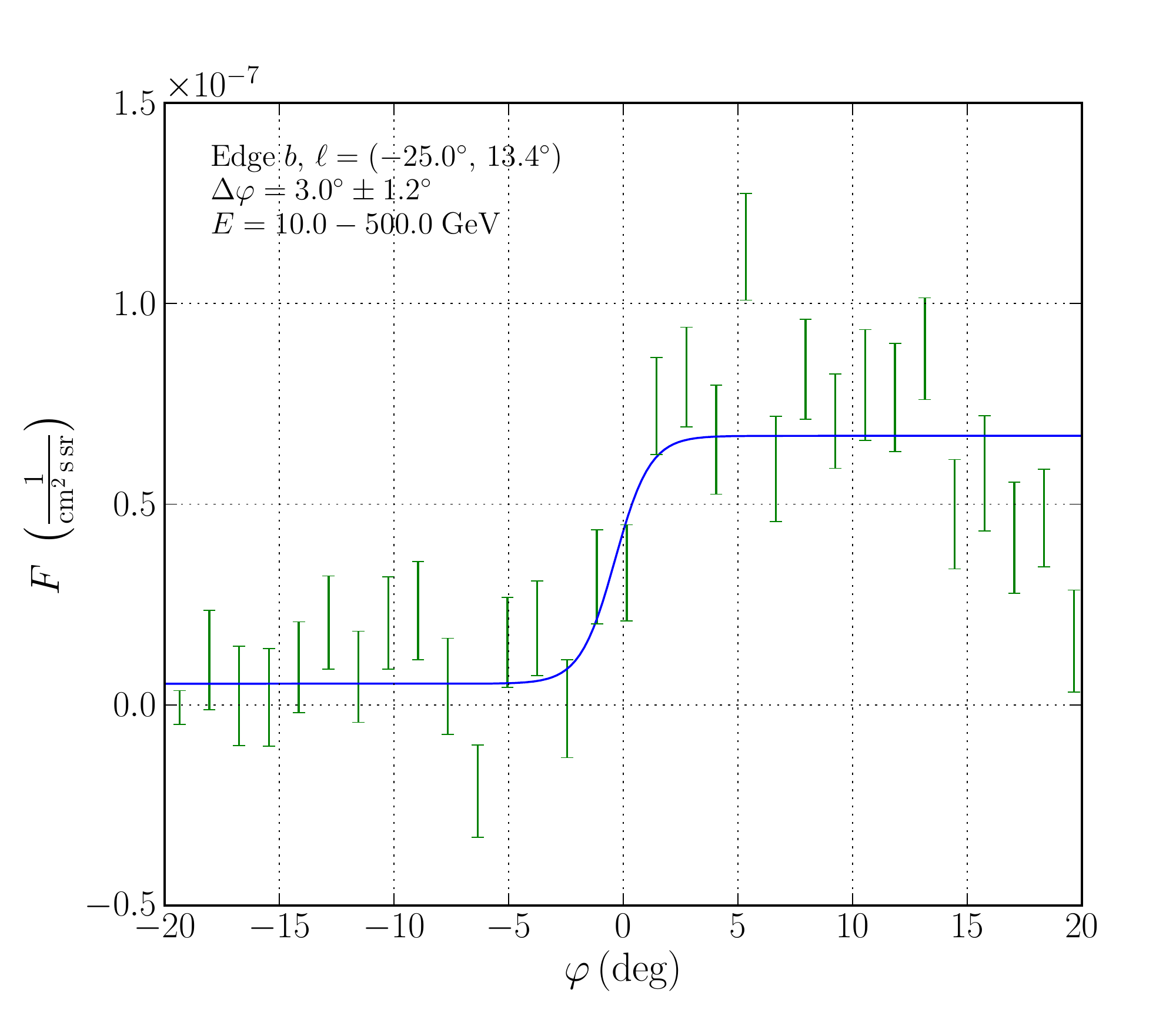, scale=\twopic}
\noindent
\caption{\small 
Examples of fitting the edges of the bubbles in the GALPROP 
residual map, integrated above 10 GeV (Figure \ref{fig:51resid10GeV}).
}
\label{fig:57edge_examples}
\end{center}
\vspace{1mm}
\end{figure}

\begin{figure}[htbp] 
\begin{center}
\epsfig{figure = 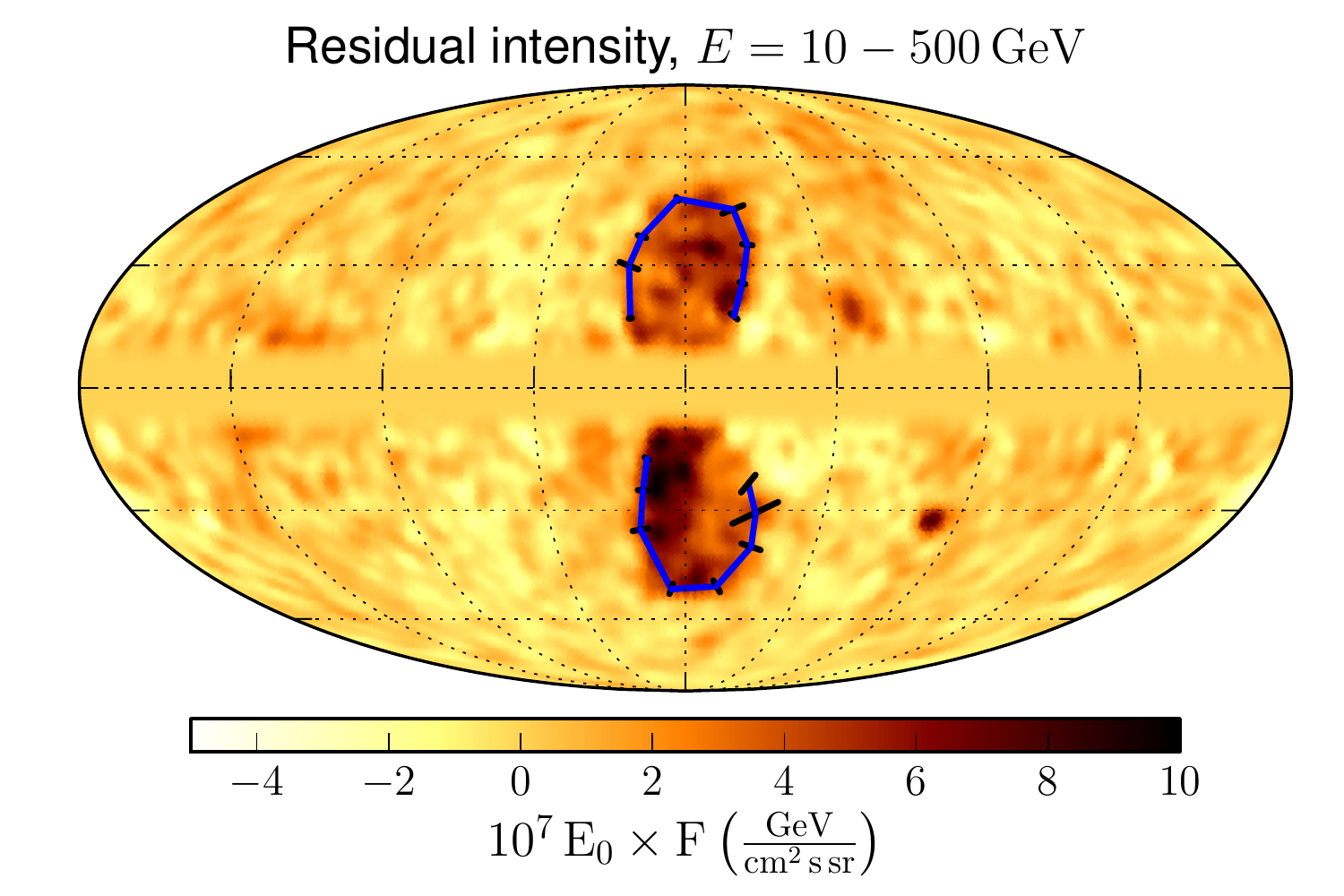, scale=0.8}
\noindent
\caption{\small 
Residual map (Figure \ref{fig:51resid10GeV}, top right) overplotted with the edge of the bubbles.
The direction of the bars perpendicular to the edge corresponds to the local gradient in the residual map;
the length of the bars represents the width of the edge.
The location of the curve along the edge corresponds to the locus of the best fit values of $\varphi_0$ (Equation~\ref{eq:edge}).
}
\label{fig:58edge_map}
\end{center}
\vspace{1mm}
\end{figure}

\begin{figure}[htbp] 
\begin{center}
\epsfig{figure = 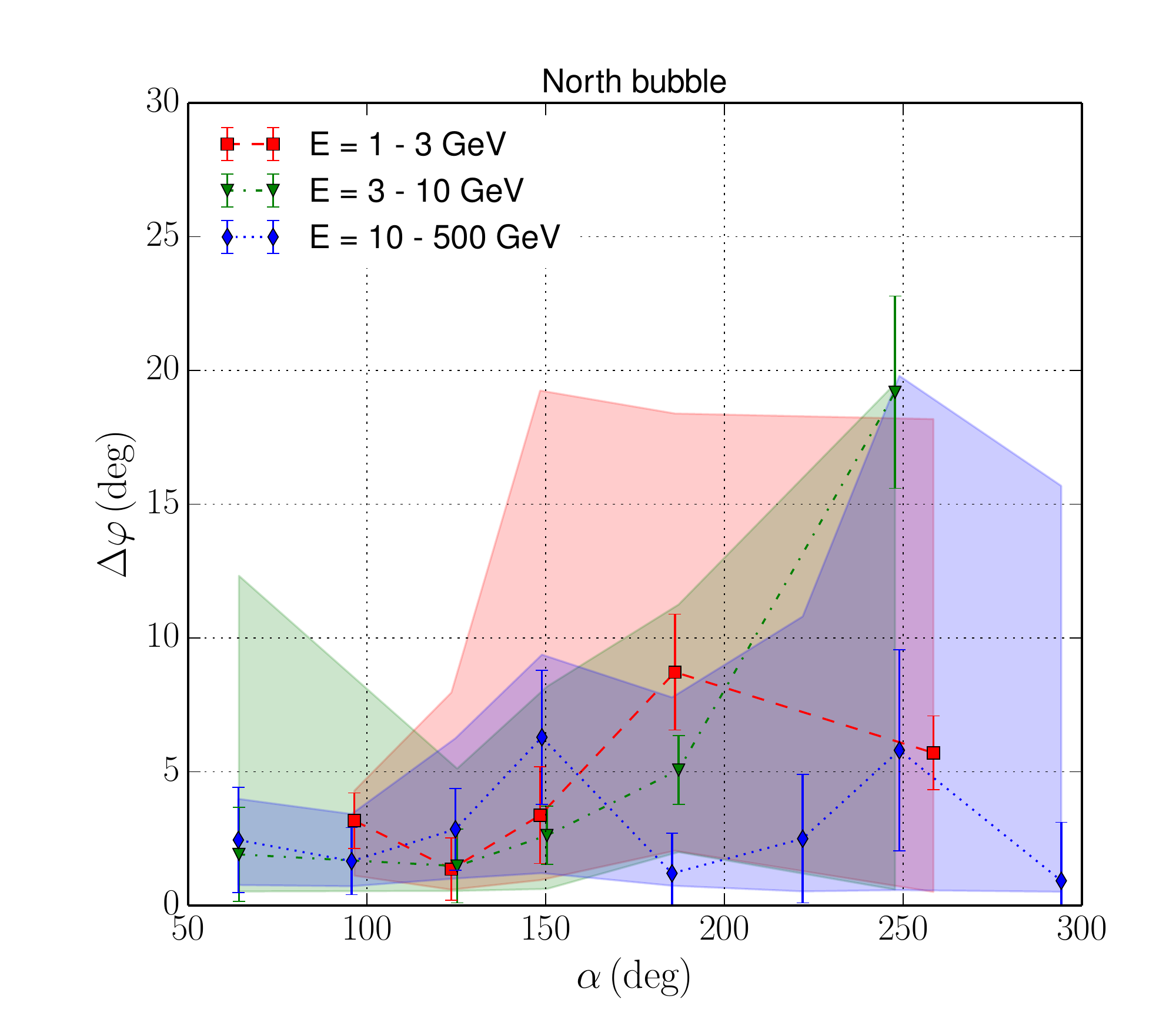, scale=\twopic}
\epsfig{figure = 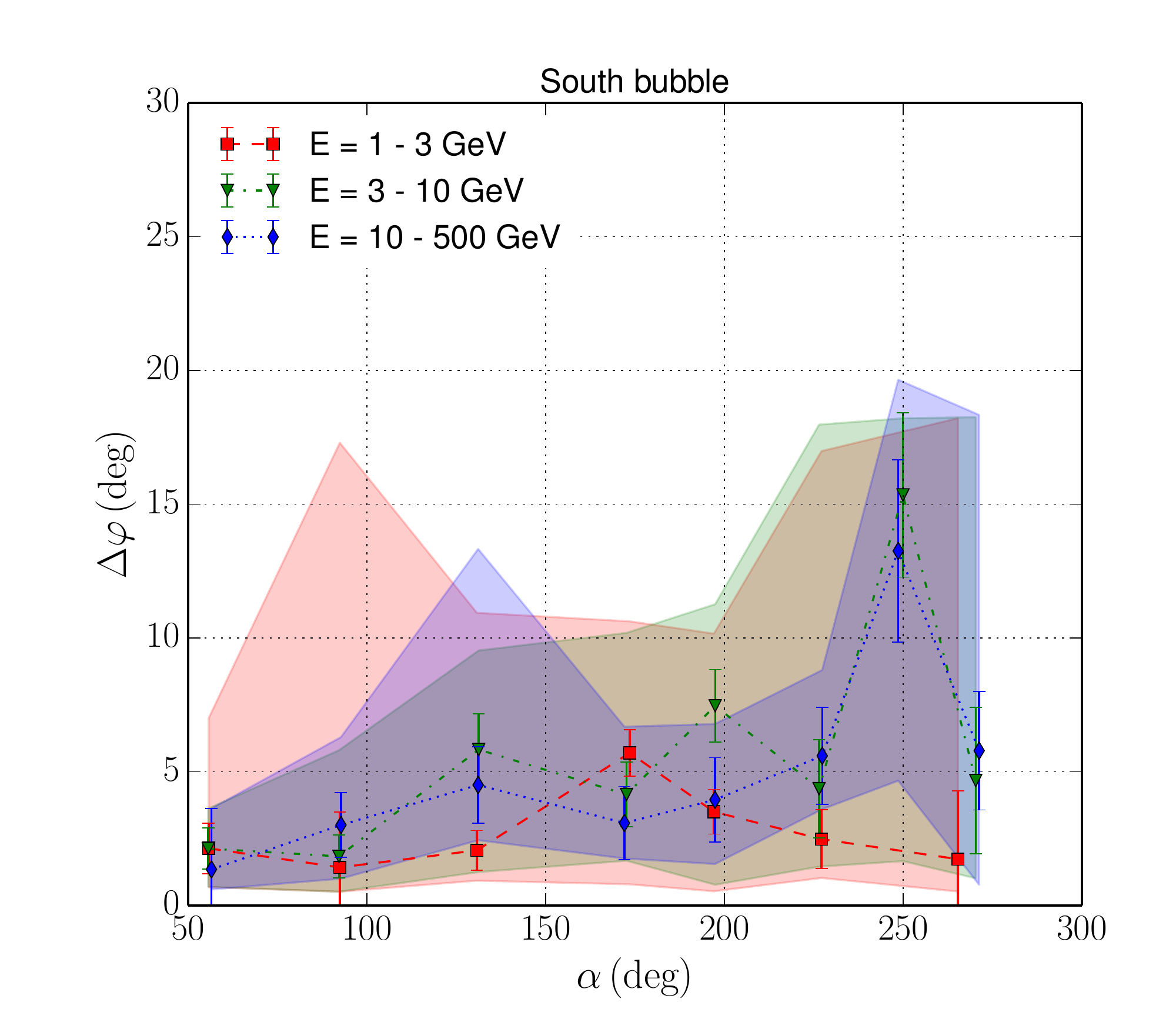, scale=\twopic}
\noindent
\caption{\small 
The width of the edge for residual maps integrated in three energy ranges
for the baseline model. 
The angle on the x-axis is determined relative to the centers of the bubbles, which we choose to be at
$b = \pm 25^\circ$, $\ell = 0^\circ$.
The angle is in the clockwise (counterclockwise) direction starting from the Galactic center for the 
northern (southern) bubble.
The points with the error bars correspond to the baseline model and its statistical uncertainties derived in Section \ref{sect:galprop}.
The shaded areas give the systematic uncertainty due to different binning of the data perpendicular to the edge
and different derivations of the foregrounds.
}
\label{fig:59edge_width}
\end{center}
\vspace{1mm}
\end{figure}

We derive the width of the bubbles' edge in three energy ranges:
1 - 3 GeV, 3 - 10 GeV, 10 - 500 GeV.
The average PSF is dominated by the events at the lower boundaries of the energy ranges.
The 68\% PSF containment angles at 1 GeV, 3 GeV, and 10 GeV are $0.5^\circ$, $0.25^\circ$, and 
$0.12^\circ$ respectively.
Examples of fitting the hyperbolic tangent function across the bubble edge for the residual 
maps above 10 GeV are shown in Figure \ref{fig:57edge_examples}.
In Figure \ref{fig:58edge_map} we show the location of the edge and the width overplotted on the residual map.
A summary of the edge widths, including statistical and systematic uncertainties, is presented in 
Figure \ref{fig:59edge_width}.
The values are reported in Table \ref{tab:north_south_edge}.
Sometimes the fit of the width does not converge, either due to oversubtractions in the foreground modeling
or due to lack of statistical significance.
In this table, we do not report values of the width larger than $20^{\circ}$, which is comparable to the size of the bubbles themselves,
or less than $0.5^{\circ}$, which is smaller than the resolution of the pixelation.
We find that in most locations along the bubbles edge, the width of the boundary varies between 
$1^\circ$ and $6^\circ$. The value of $13.3^\circ$ in one location in the southern bubble is because the edge lies on top of a local
excess (see Figure \ref{fig:58edge_map}). This high value is likely due to poor convergence of the fit rather than the actual
width of the boundary.

We do not detect a significant difference in the width of the bubbles' boundary for the three energy ranges.
The median width of the bubbles' boundary
among all locations along the boundary and among all models of foreground
emission and templates of the bubbles is 
$\Delta\vp = 3.4 \pm 2.0 \rm{[stat]}^{+3.1}_{-1.7}\rm{[syst]}$ deg.
The systematic uncertainty boundaries are estimated as values which enclose
$\pm 34$\% of the values above and below the median value.
We take the median instead of the mean in order to avoid bias due to outliers with large values of the width
either due to oversubtractions in the foreground modeling or poor convergence of the width estimation.

\begin{deluxetable}{cccccc}
\tabletypesize{\scriptsize}
\tablecaption{
Position and width of the bubbles' boundary for the baseline model residuals above 10 GeV
(Figures \ref{fig:58edge_map} and \ref{fig:59edge_width}).
The position and the width are determined from Equation (\ref{eq:edge}).
All values are in degrees.
The {\it Lat} and {\it Lon} columns give the positions at the center of the boundary in Galactic coordinates.
The {\it Stat} column is the statistical error.
{\it Min} and {\it Max Widths} correspond to the envelope of the systematic uncertainties.
Values larger than $20^\circ$ or smaller than $0.5^\circ$ are not reported (see text for explanation).
\label{tab:north_south_edge}}
\tablewidth{0pt}
\tablehead{\colhead{Lat} & \colhead{Lon}  & \colhead{Width}  & \colhead{Stat} & \colhead{Min Width} & \colhead{Max Width}}
\startdata
\multicolumn{6}{c}{North}\\
\tableline
17.4 & 345.1 & 2.4 & 2.0 & 0.9 & 4.0 \\ 
 25.5 & 342.0 & 1.7 & 1.3 & 0.7 & 3.4 \\ 
 35.3 & 339.1 & 2.9 & 1.5 & 2.0 & 6.3 \\ 
 44.8 & 342.5 & 6.3 & 2.5 & 1.2 & 9.4 \\ 
 47.7 & 3.1 & 1.2 & 1.5 & 0.7 & 5.3 \\ 
 37.5 & 14.9 & 2.5 & 2.4 & 0.5 & 10.6 \\ 
 30.0 & 18.3 & 5.8 & 3.8 & 0.6 & 19.8 \\ 
 16.8 & 16.8 & 0.9 & 2.2 & 0.5 & 15.7 \\ 
\tableline
\multicolumn{6}{c}{South}\\
\tableline
-17.1 & 11.7 & 1.4 & 2.3 & 1.0 & 3.6 \\ 
 -25.0 & 13.4 & 3.0 & 1.2 & 1.0 & 6.3 \\ 
 -35.0 & 15.1 & 4.5 & 1.4 & 2.4 & 13.3 \\ 
 -51.1 & 5.6 & 3.1 & 1.4 & 1.8 & 6.7 \\ 
 -50.3 & 347.8 & 4.0 & 1.6 & 1.6 & 6.8 \\ 
 -39.5 & 337.1 & 5.6 & 1.8 & 3.6 & 8.8 \\ 
 -30.9 & 337.1 & 13.3 & 3.4 & 4.7 & 19.7 \\ 
 -23.3 & 340.3 & 5.8 & 2.2 & 0.8 & 18.3 \\
 \tableline
\enddata
\end{deluxetable}

\subsection{Spectrum in latitude strips}

\begin{figure}[htbp] 
\begin{center}
\epsfig{figure = 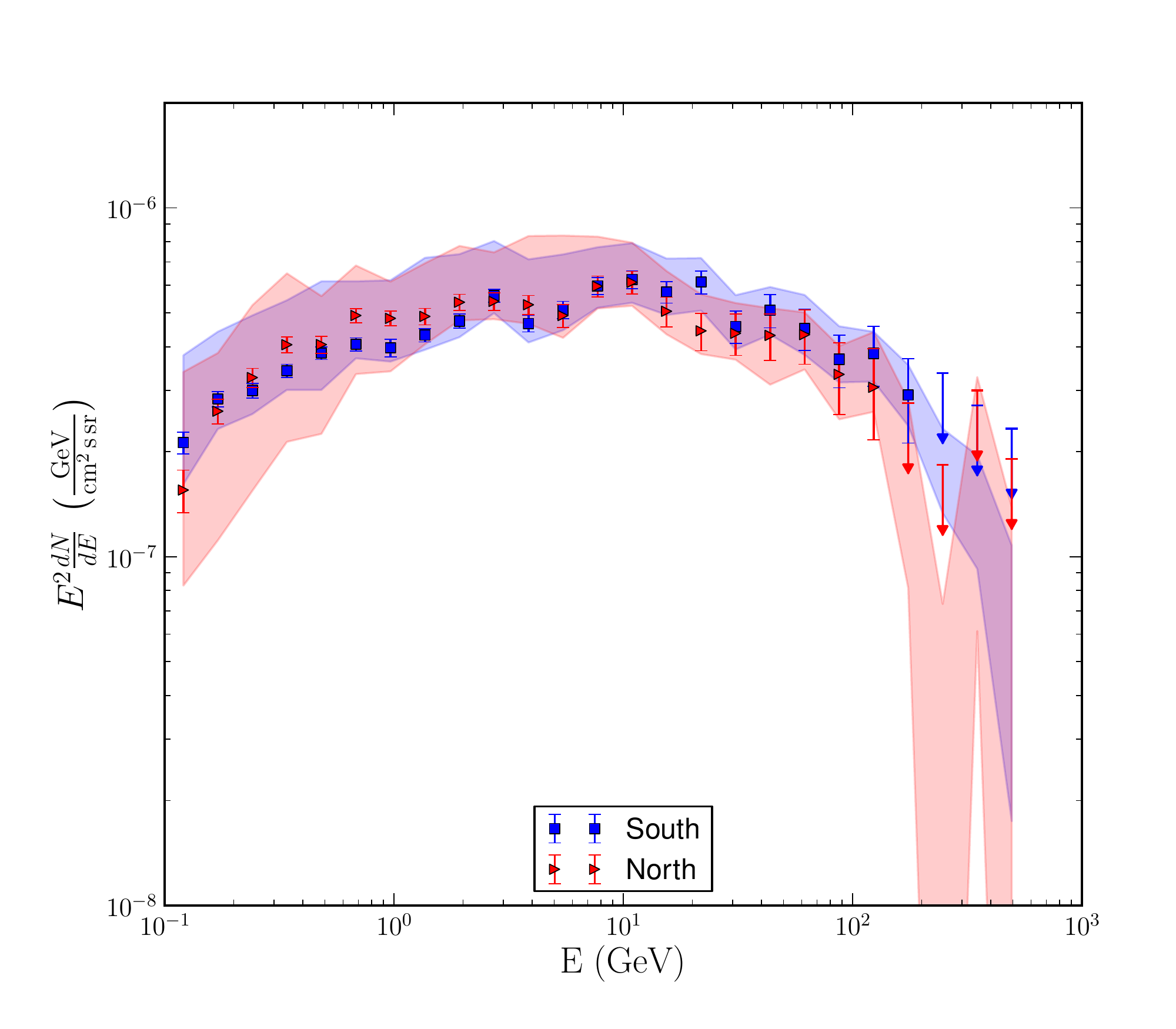, scale=\onepic}
\noindent
\caption{\small 
SED for the northern and southern bubbles. 
The points with statistical error bars correspond to the baseline SED.
The bands represent an envelope of the SEDs for different
derivations of the Galactic foreground emission and the definitions of the template of the bubbles.
The uncertainty of the effective area is added in quadrature to the other systematic uncertainties.
}
\label{fig:53NS}
\end{center}
\vspace{1mm}
\end{figure}

The spectra for northern and southern bubbles are shown in Figure~\ref{fig:53NS}.
These spectra are derived similarly to the overall spectrum of the bubbles,
but instead of one template of the bubbles, we fit two independent templates: for the northern and southern bubbles.
We find that the spectra in the North and in the South agree with each other within the uncertainties.
The southern bubbles has a region of enhanced emission, the cocoon, while the brightness in the northern bubbles is more
uniform. The overall intensities of the two bubbles are consistent with each other.

\begin{figure}[htbp] 
\begin{center}
\epsfig{figure = 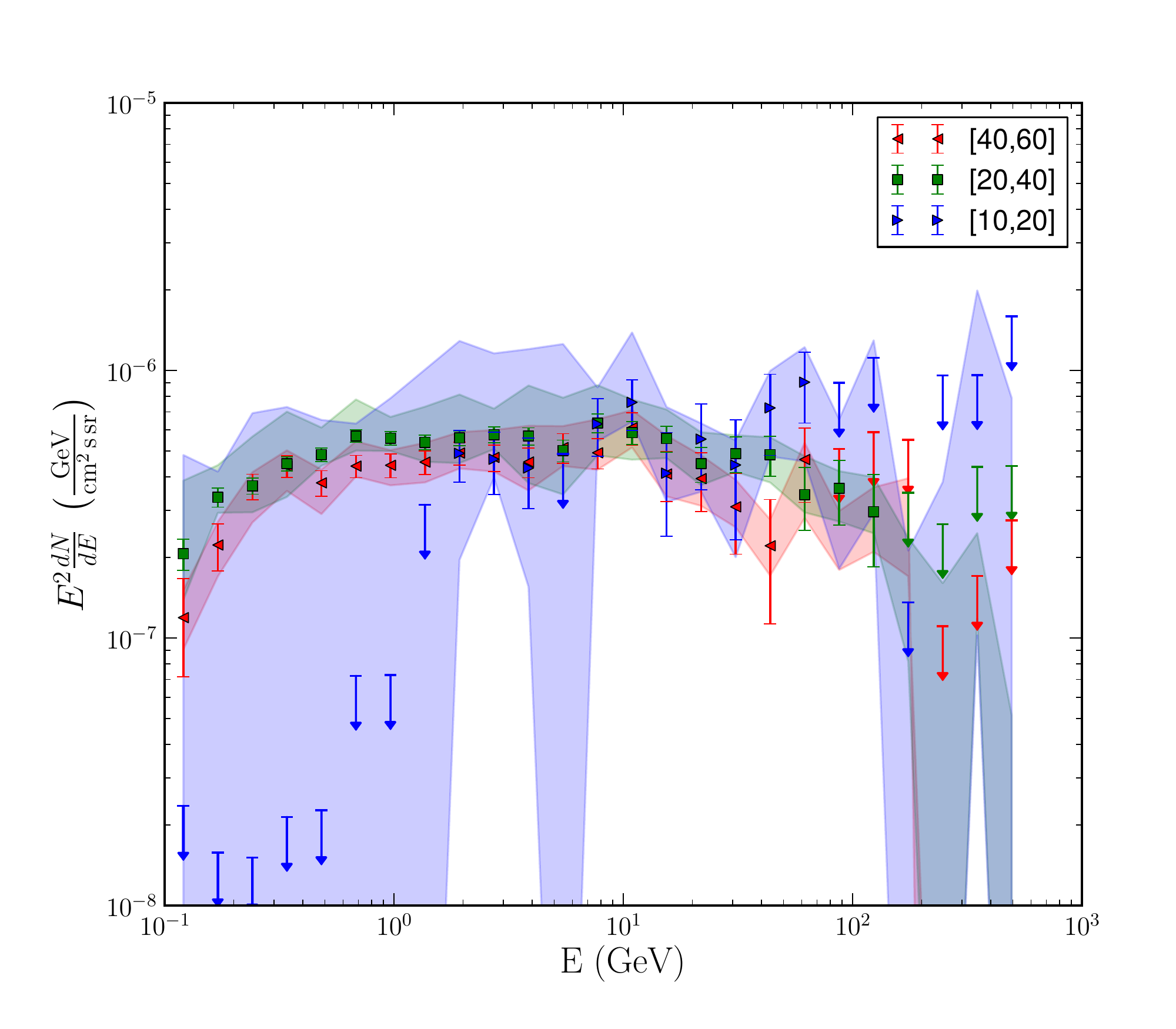, scale=\twopic}
\epsfig{figure = 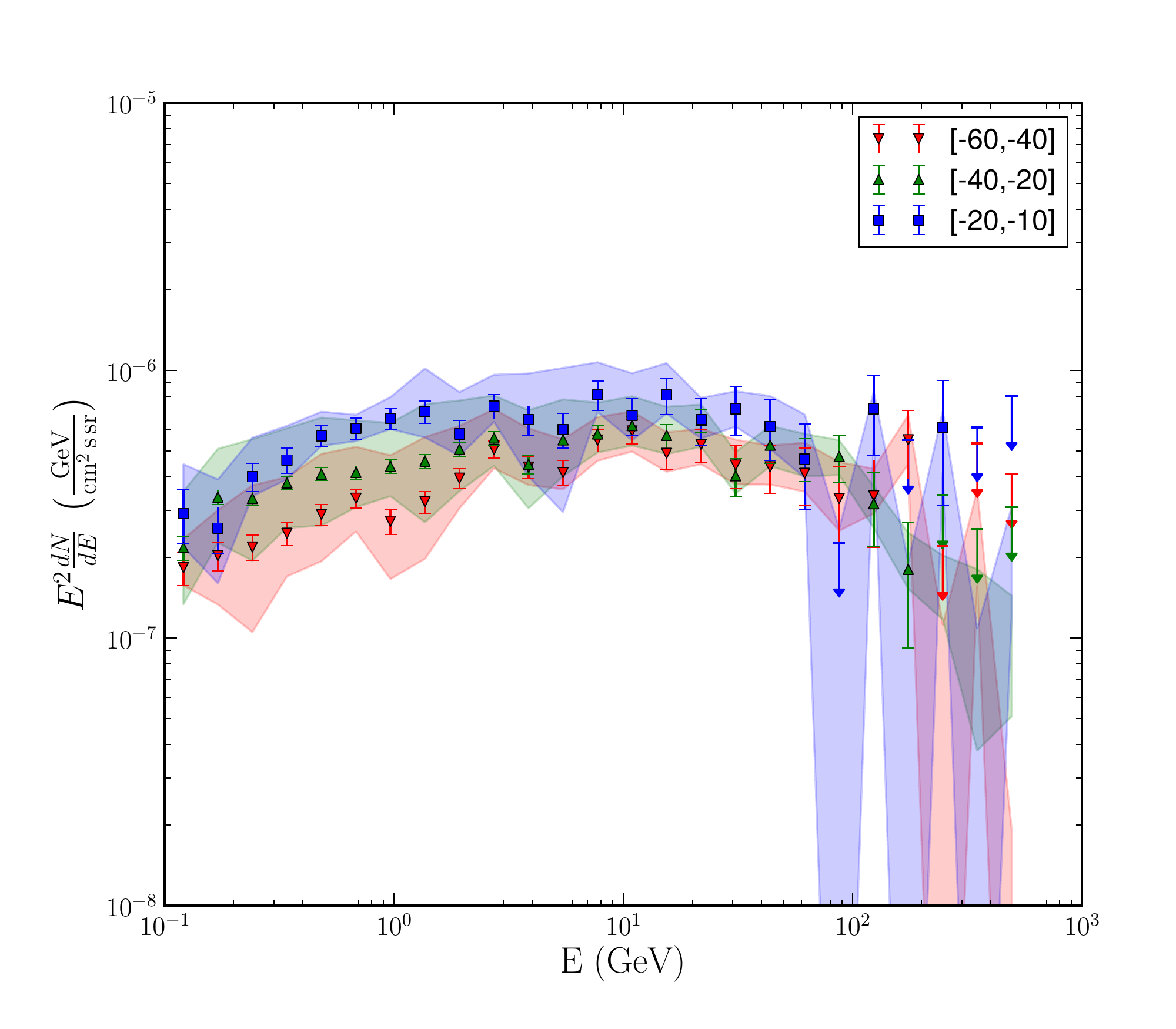, scale=\twopic}
\noindent
\caption{\small 
SED of the {\Fermi} bubbles in latitude strips. 
{Left}: northern bubble. {Right}: southern bubble.
For description of the points and bands, see caption of Figure \ref{fig:53NS}.
}
\label{fig:53strips}
\end{center}
\vspace{1mm}
\end{figure}

The spectra in latitude strips are shown in Figure~\ref{fig:53strips}. 
For the derivation of the spectra in strips we separated the template of the bubbles into 6 independent templates according to latitude. The latitude boundaries of the stripes are $-60^\circ$ to $-40^\circ$, $-40^\circ$ to $-20^\circ$ and $-20^\circ$ to $-10^\circ$ in the South and $10^\circ$ to $20^\circ$, $20^\circ$ to $40^\circ$ and $40^\circ$ to $60^\circ$ in the North.  
With the current level of statistical and systematic uncertainties, 
we cannot detect a variation of the spectrum with latitude.
Our results agree with \cite{Hooper:2013rwa} at latitudes $|b| > 20^\circ$, 
but we do not find a significant 
variation of the spectrum of the bubbles for $10^\circ < |b| < 20^\circ$ compared to higher latitudes.
There is a large systematic uncertainty in the energy spectrum at latitudes $10^\circ < b < 20^\circ$
mostly due to uncertainties of the model of the foreground gamma-ray emission from the interactions of CR with interstellar gas.
Manifestation of this uncertainty can be seen in the residual maps in Figure \ref{fig:51resid10GeV}
and in the profile plots in Figure \ref{fig:52profiles}.

%% file: 7interpretation.tex
\section{IC and hadronic models of the bubbles}
\lb{sect:interp}

In this section we fit the spectrum of the bubbles with IC and hadronic models
of gamma-ray production.
In addition,
we calculate the synchrotron emission from the population of electrons
in the IC model and from the secondary electrons and positrons in the hadronic
model.
The details of the calculations are presented in Appendix \ref{sec:IC_h_models}.

\subsection{IC model of the bubbles}
\lb{sect:ICmodel}

The IC scattering is calculated with the cross sections presented by~\cite{1970RvMP...42..237B}.
The interstellar radiation field (ISRF) is taken from the GALPROP v54 distribution
\citep{2005ICRC....4...77P, 2006ApJ...640L.155M}.
Since no significant variation of the gamma-ray spectrum across the bubbles has been found,
we will use the spectrum averaged over the area of the bubbles 
(Figure \ref{fig:52compareSpectra}, right,
and Table \ref{tab:fluxValues}).

As a benchmark model for the spectrum of electrons we take the 
spectrum derived using the ISRF
at 5 kpc above the GC.
We also compare it with the electron spectrum obtained for CMB photons only.

Since the gamma-ray spectrum of the {\Fermi} bubbles has a significant cutoff at high energies,
we model the electron spectrum by
a power law with an exponential cutoff
$\propto E^{-n} e^{-E/E_{\rm cut}}$.
The best fit parameters are 
$n = 2.17\pm 0.05\rm{[stat]}^{+0.33}_{-0.89}\rm{[syst]}$
and 
$E_{\rm cut} = 1.25\pm 0.13\rm{[stat]}^{+1.73}_{-0.68}\rm{[syst]}$ TeV.
The corresponding IC spectra are shown in Figure \ref{fig:71IC_spectra} on the left.
The details of the calculation can be found in Appendix \ref{sect:IC_model}.
The indices and the cutoff values for different foreground models and 
definitions of the templates of Loop I and the bubbles
are shown in Figure \ref{fig:71IC_spectra} on the right.
The bremsstrahlung emission is at least two orders of magnitude smaller than the IC emission
for a characteristic gas density $n_{\rm H} \lesssim 0.01\, {\rm cm}^{-3}$ at a few kpc from the Galactic plane~\citep{1997ApJ...485..125S},
and can be neglected.

\begin{figure}[htbp] 
\begin{center}
\epsfig{figure = 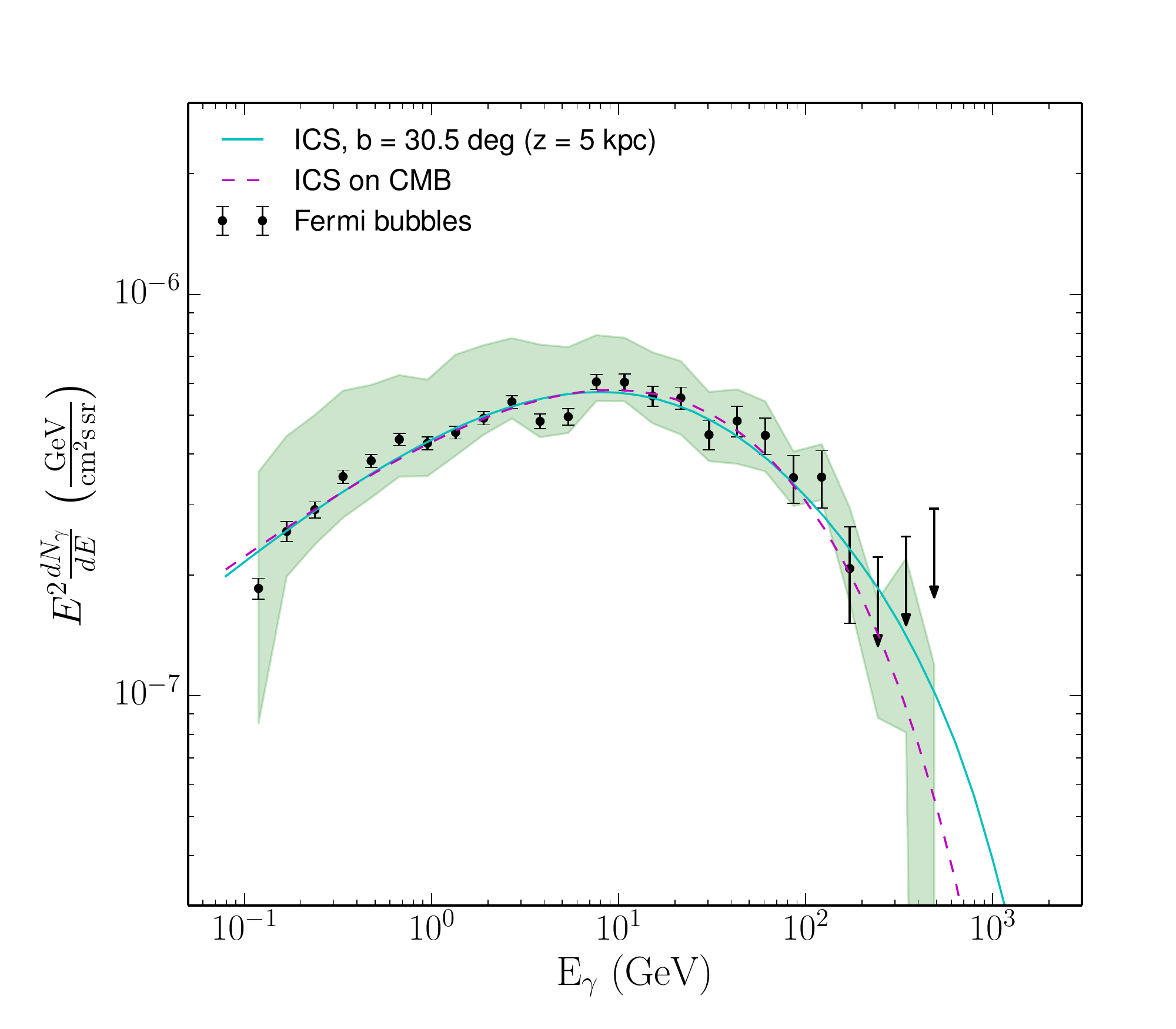, scale=\twopic}
\epsfig{figure = 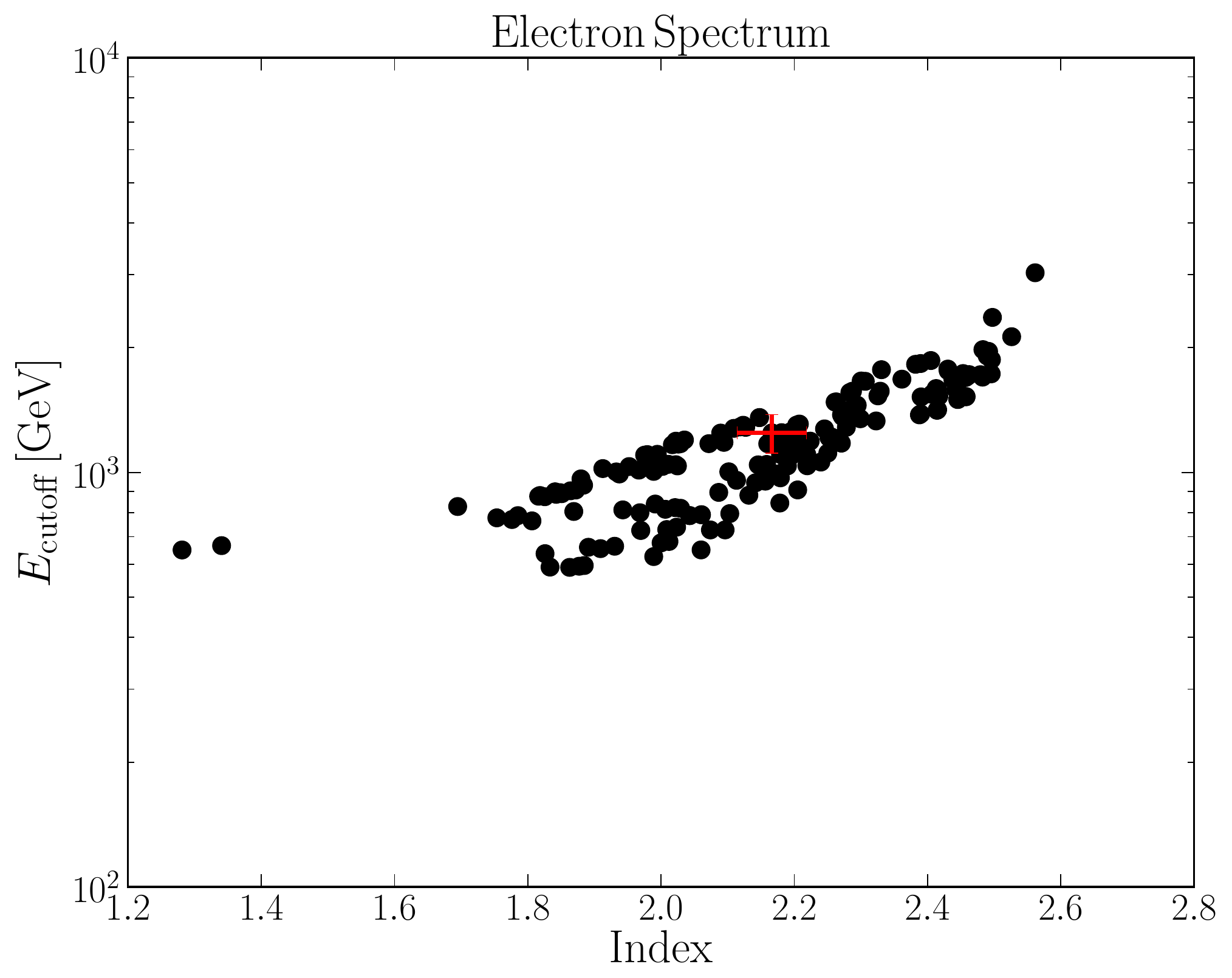, scale=\twopic}
\noindent
\caption{\small 
Left: IC model fit to the baseline gamma-ray spectrum of the {\Fermi} bubbles 
(Section \ref{sect:comparison}).
The spectrum of electrons for the ISRF at 5 kpc is a power law with an index 2.2
and an exponential cutoff at 1.25 TeV (Section \ref{sect:ICmodel}).
If we take into account only IC scattering on CMB photons,
then the electron spectrum has an index 2.3 and a cutoff at 2.0 TeV.
Right: index and cutoff energy for electron spectra determined for different derivations of gamma-ray foregrounds
and different definitions of the {\Fermi} bubbles templates (for the ISRF at 5 kpc above the Galactic center). 
The red cross corresponds to the baseline model values with the statistical errors. 
}
\label{fig:71IC_spectra}
\end{center}
\vspace{1mm}
\end{figure}

We will assume that the center of the bubbles is at $b = 25^\circ$,
i.e., the distance to the center of the bubbles is $R = R_\odot / \cos b = 9.4$ kpc,
where $R_\odot = 8.5$ kpc is the distance to the GC.
The total energy contained in the electron population inside the bubbles above 1 GeV is
$(1.0\pm 0.2\rm{[stat]}^{+6.0}_{-1.0}\rm{[syst]} )\times 10^{52}$ erg, 
where the value corresponds to the baseline model;
the statistical uncertainty is calculated by marginalizing over the index and cutoff of the electron spectrum.
The systematic uncertainty is estimated by calculating the electron spectrum for 
different models of the foreground emission and definitions of the templates of the bubbles and Loop~I.

\begin{figure}[htbp] 
\begin{center}
\epsfig{figure = 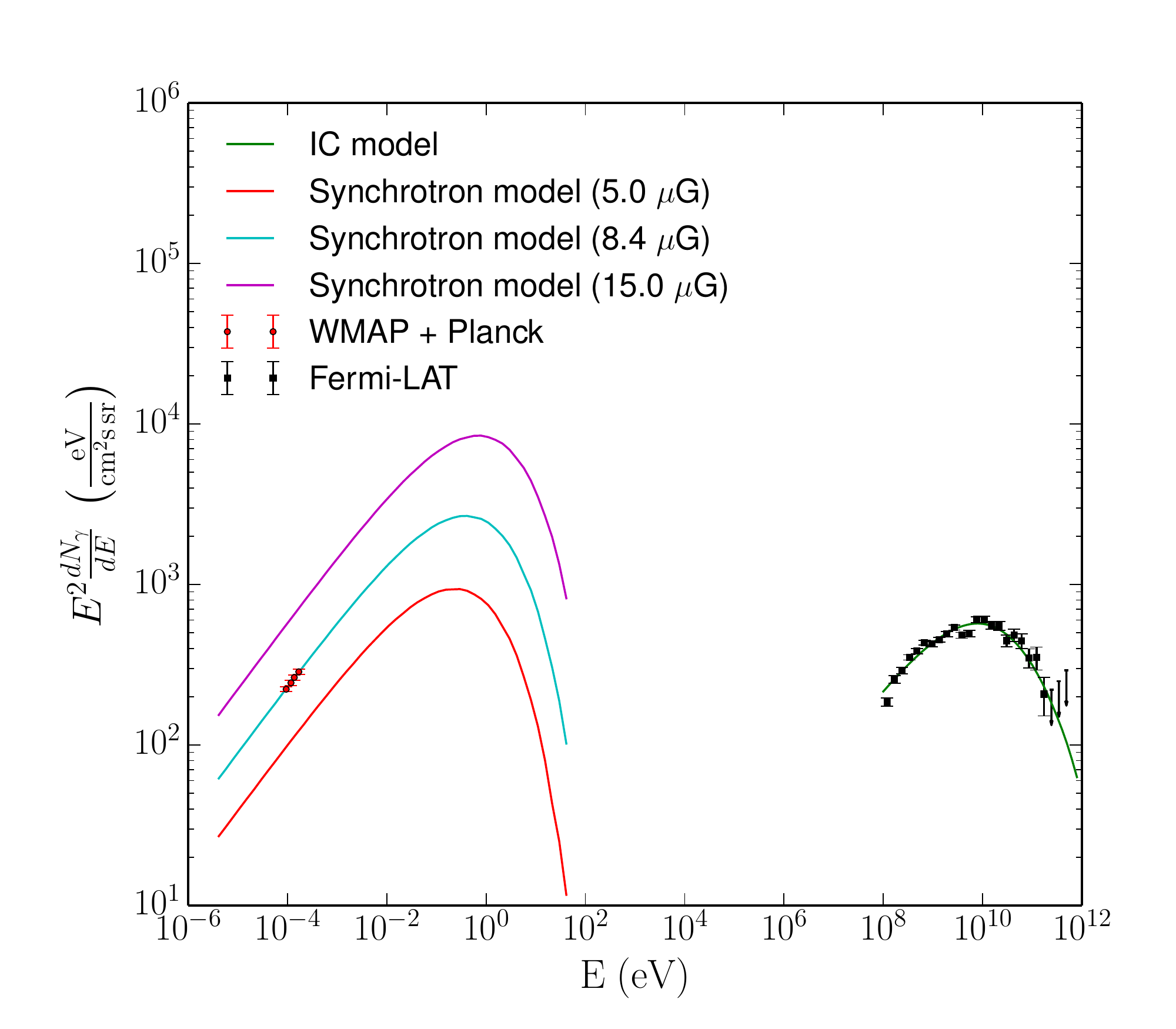, scale=\twopic}
\epsfig{figure = 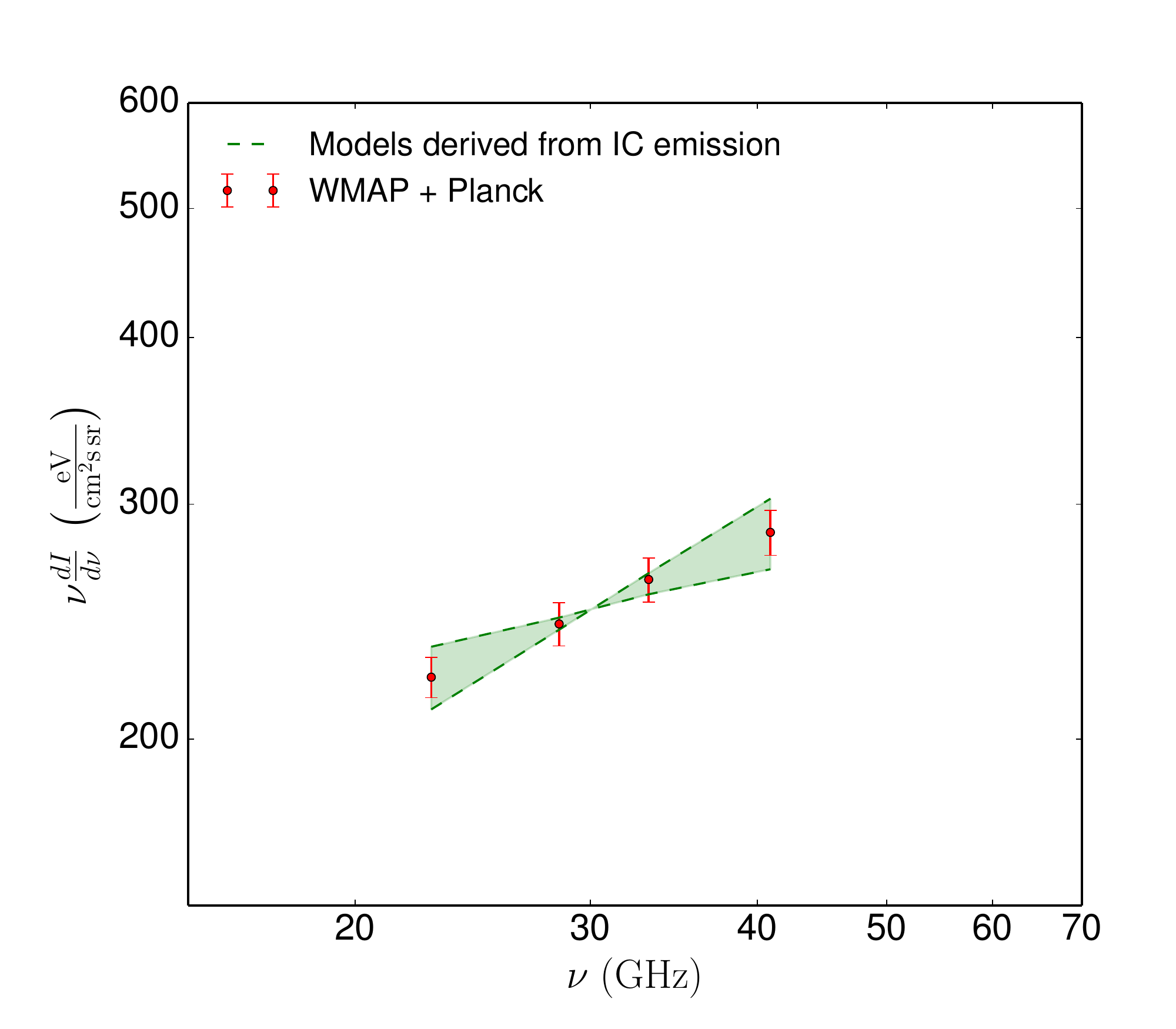, scale=\twopic}
\noindent
\caption{\small 
Left: IC and synchrotron emission from the same benchmark population of electrons.
The electron energy density is derived from fitting the IC model to the gamma-ray data.
We use the synchrotron emission from the same population of electrons to fit the {\it Planck} microwave haze data
\citep{2013A&A...554A.139P}
by optimizing the value of the magnetic field.
The best-fit magnetic field is about $8.4\, \upmu{\rm G}$.
Right: microwave haze spectrum compared to the synchrotron emission from the electrons in the IC model of the {\Fermi} bubbles. 
The green band shows the systematic uncertainties introduced by the systematic uncertainty in the gamma-ray spectrum of the bubbles.}
\label{fig:72synch}
\end{center}
\vspace{1mm}
\end{figure}

The synchrotron emission from the benchmark population of electrons for different values
of the magnetic field is shown in Figure \ref{fig:72synch} together with the IC signal.
On the same plot, we also include the {\it Planck} and {\it WMAP} microwave haze spectrum
\citep{Pietrobon:2011hh, 2013A&A...554A.139P}.
The index of the microwave haze emission is
harder than the synchrotron emission for a stationary population of electrons in the Galaxy.
The microwave haze spatially overlaps with the gamma-ray bubbles at $|b|<35^{\circ}$.
We confirm that the population of electrons that produces the gamma-ray emission of the {\Fermi} bubbles 
via IC scattering can also produce the microwave haze
\citep{2010ApJ...717..825D, 2010ApJ...724.1044S, 2012ApJ...753...61S,
2012ApJ...750...17D, 2013A&A...554A.139P}.

The range of spectra of the synchrotron emission corresponding to the systematic uncertainty 
in the electron spectrum (Figure \ref{fig:71IC_spectra})
is shown in Figure \ref{fig:72synch} on the right.
For each electron spectrum, we find the magnetic field that gives the best fit to the microwave data.
We find
$B = 8.4\pm 0.2\rm{[stat]}^{+11.2}_{-3.5}\rm{[syst]}\; \upmu\rm{G}$,
where the value is for the baseline model, the statistical uncertainty is calculated
using the statistical errors of the \WMAP and \Planck haze spectra,
the systematic uncertainty is due to modeling of the gamma-ray foregrounds
and the definition of the template of the bubbles.
The allowed magnetic fields are approximately in the range from 5 to 20 $\upmu$G.
A larger index (softer spectrum) corresponds to a greater number density of electrons at lower energies;
in this case the magnetic field is $\sim 5 \upmu$G.
A harder index requires a magnetic field of $\sim 20 \upmu$G.
The uncertainties of the index and magnetic field are due to large uncertainties in the 
distribution of electrons around $10 - 30$ GeV.
We note that the spectrum of the microwave haze was obtained for latitudes
$-35^\circ < b < -10^\circ$ \citep{Pietrobon:2011hh}, i.e., the derived magnetic field
corresponds to the region of the bubbles encompassed by these latitudes. 
At higher latitudes the microwave haze emission has smaller intensity, which can be explained
if the magnetic field decreases with height above the Galactic plane.

The main contribution to the IC signal comes from electrons at energies $>100$ GeV.
We show in Appendix \ref{sec:IC_h_models} that
the main contribution to the \WMAP and \Planck
frequencies, where the microwave haze is detected, 
comes from electrons between 10 GeV and 30 GeV.
Thus, although the gamma-ray bubbles and the microwave haze can be
produced by the same population of electrons, the presence of two populations
of electrons cannot be excluded: one population producing the gamma-ray signal and the other
producing the microwave signal.
In this scenario, the magnetic field can have a lower value.
As a result, the electron cooling time can be longer than in the case of a single population of electrons.

The cooling time for 1 TeV electrons in a 5 $\upmu$G magnetic field and in the ISRF at 5 kpc
is only 500 kyr,  while taking into account only the IC losses gives a cooling time of $\sim1$ Myr
(Appendix \ref{sect:mw_haze}).
If the bubbles were formed by a jet or an outflow from the Galactic center,
where most of the acceleration happened during the initial stages of the expansion,
then the expansion velocity should be greater than 10,000 km/s so that the bubbles formation time
is smaller than the cooling time of the 1 TeV electrons.
The lower bound on the expansion velocity becomes 20,000 km/s, if the magnetic field is 5 $\upmu$G.
In scenarios with electron reacceleration inside the volume of the bubbles \citep{2011PhRvL.107i1101M},
the characteristic acceleration time for 1 TeV electrons should be 
shorter than 1 Myr for IC losses only, or 500 kyr for IC and synchrotron losses in
a 5 $\upmu$G magnetic field.
Since the synchrotron losses at these energies are about the same as the IC losses,
in the case of a steady injection, the electron injection rate should be about two times larger than the gamma-ray luminosity
of the bubbles, i.e. around $10^{38} \;{\rm erg\,s^{-1}}$.

\subsection{Hadronic model of the bubbles}
\lb{sect:hadrmodel}

\begin{figure}[htbp] 
\begin{center}
\epsfig{figure = 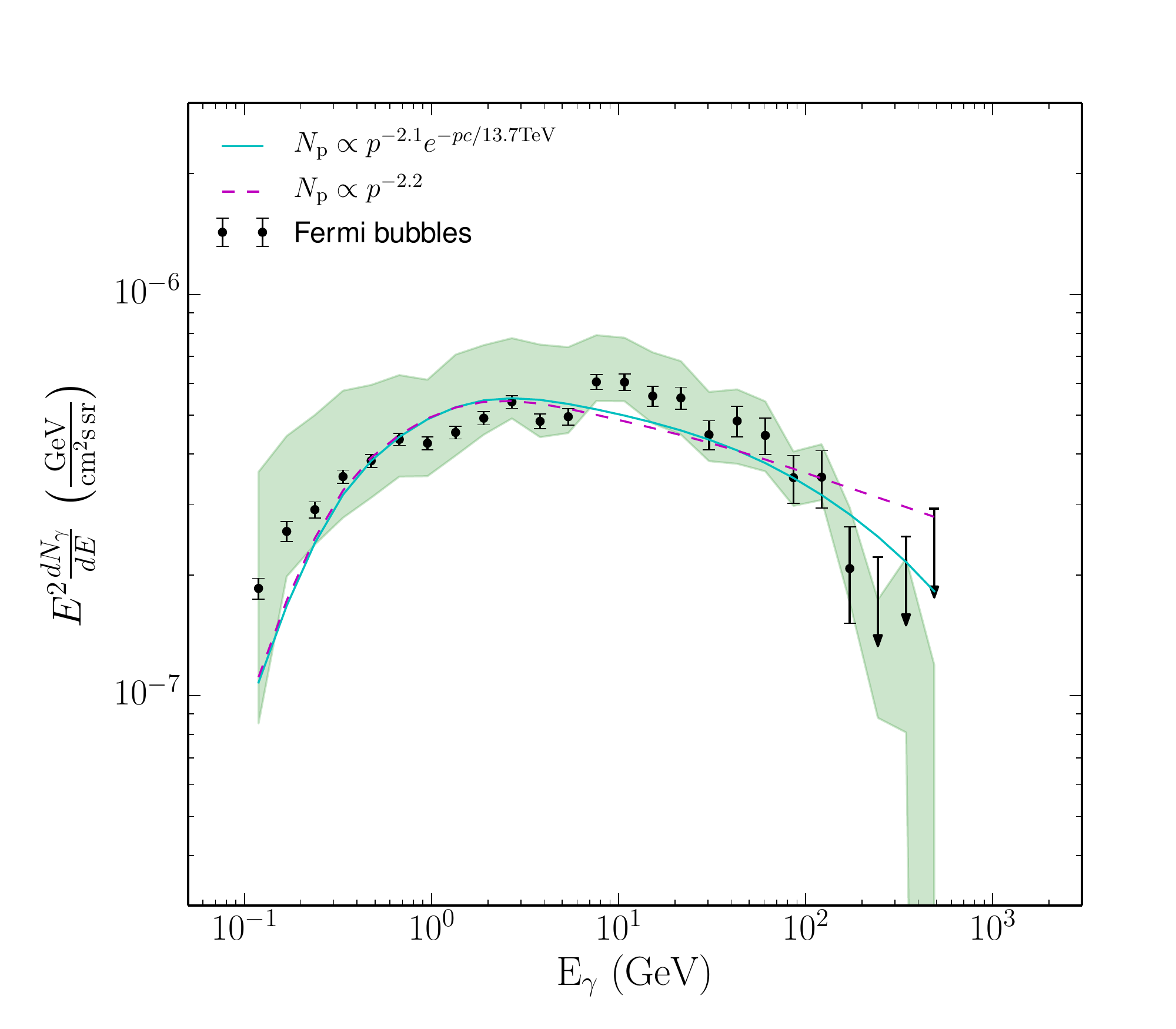, scale=\twopic}
\noindent
\caption{\small 
Primary gamma-ray emission of hadronic model of the {\Fermi} bubbles spectrum using a simple power law or a power law with an exponential cutoff
for the spectrum of protons.
}
\label{fig:73pi0_spectra}
\end{center}
\vspace{1mm}
\end{figure}

For the calculation of the spectrum of gamma rays produced in hadronic interactions, 
we use the cross sections described in 
\cite{2006ApJ...647..692K}  and \cite{2008ApJ...674..278K}, which are
implemented in the cparamlib package%
\footnote{\url{https://github.com/niklask/cparamlib}}.
In this analysis,
we consider only proton cosmic rays in the hadronic model of the gamma-ray emission in the bubbles.
We parameterize the spectrum of the protons as a function of momentum.
A power law with an exponential cutoff spectrum of the CR protons 
\be
\frac{dn (p)}{dp} \propto p^{-n} e^{- pc / E_{\rm cut}}
\ee
gives a better fit
at high energies than a simple power-law spectrum (Figure \ref{fig:73pi0_spectra}).
The parameters of the power law with a cutoff function are
$n = 2.13\pm 0.01\rm{[stat]}^{+0.15}_{-0.52}\rm{[syst]}$
and 
$E_{\rm cut} = 14\pm 7\rm{[stat]}^{+6}_{-13}\rm{[syst]}\, TeV$.
In order to estimate the amount of energy in hadronic cosmic rays required to produce the
gamma-ray signal, one needs to know the density of gas inside the bubbles.
We take into account only ionized hydrogen and we use $n_{\rm H} = 0.01\; {\rm cm^{-3}}$
as a reference value for the density\footnote{Note, that the gamma-ray emissivity integrated along the line of sight, which is relevant for the template fitting, is dominated by H~I from the local ring, while the emissivity a few kpc above the Galactic plane is dominated by ionized hydrogen.}.
It is of the same order of magnitude
as the plasma density $n_{\rm H} \sim 0.0035\; {\rm cm^{-3}}$ at 2 kpc above the Galactic center \citep{1997ApJ...485..125S}.

The total energy in hadronic CRs above 1 GeV
required to produce the gamma rays is 
$(3.5\pm 0.1\rm{[stat]}^{+4.7}_{-3.0}\rm{[syst]} )\times 10^{55}
\left(\frac{0.01\; {\rm{cm}^{-3}}}{\textit n_{\rm{H}}}\right) \; {\rm erg} $.
Including heavier nuclei may change this estimate.
However, 
the evaluation of this effect depends on the uncertain composition of CRs and gas in the bubbles, 
hence it is beyond the scope of the modeling in this paper.
In the relevant energy range of the proton kinetic energy $E_p \sim$ 0.1 - 10 TeV,
the center of mass energy is
$E_{\rm CM} \sim$ 10 - 100 GeV and
the inelastic cross section is $\sigma_{\rm pp} \approx 30\; {\rm mb}$.
The average time for a collision is 
$t_p = {(n_{\rm H}\, c\, \sigma_{\rm pp})^{-1}} \approx 3.5 \times 10^9
\left(\frac{0.01\; {\rm cm^{-3}}}{n_{\rm H}}\right)\; {\rm yr}$.
In a steady state, the minimal injection rate of cosmic-ray protons is
$L_p \sim {W_p}/{t_p} \approx 3.1 \times 10^{38}\; {\rm erg\,s^{-1}}$.
This calculation assumes that the main energy loss process is inelastic proton-proton collisions.
The injection rate actually required may be an order of magnitude higher due to, e.g., adiabatic losses.

The proton spectrum at high energies inside the bubbles 
must be much harder than the spectrum of CR protons in the Galactic plane~\citep[e.g.,][]{2011Sci...332...69A,FermiLAT:2012aa, 2013arXiv1303.6482D}.
If we assume that the proton spectrum injected in the ISM by the supernova explosions
is the same everywhere in the Galaxy and is $\propto E^{-2.0 - 2.2}$,
then the softening of the spectrum in the Galactic plane can be explained by energy-dependent
escape,
while the hard proton spectrum inside the bubbles can be explained if the escape time
from the bubbles is longer than the interaction time \citep{2011PhRvL.106j1102C}.
In other words, protons escape from the Galactic plane before they interact,
but the protons inside the bubbles should interact before they can escape, which means that they have to remain inside the bubbles for several Gyr.

\begin{figure}[htbp] 
\begin{center}
\epsfig{figure = 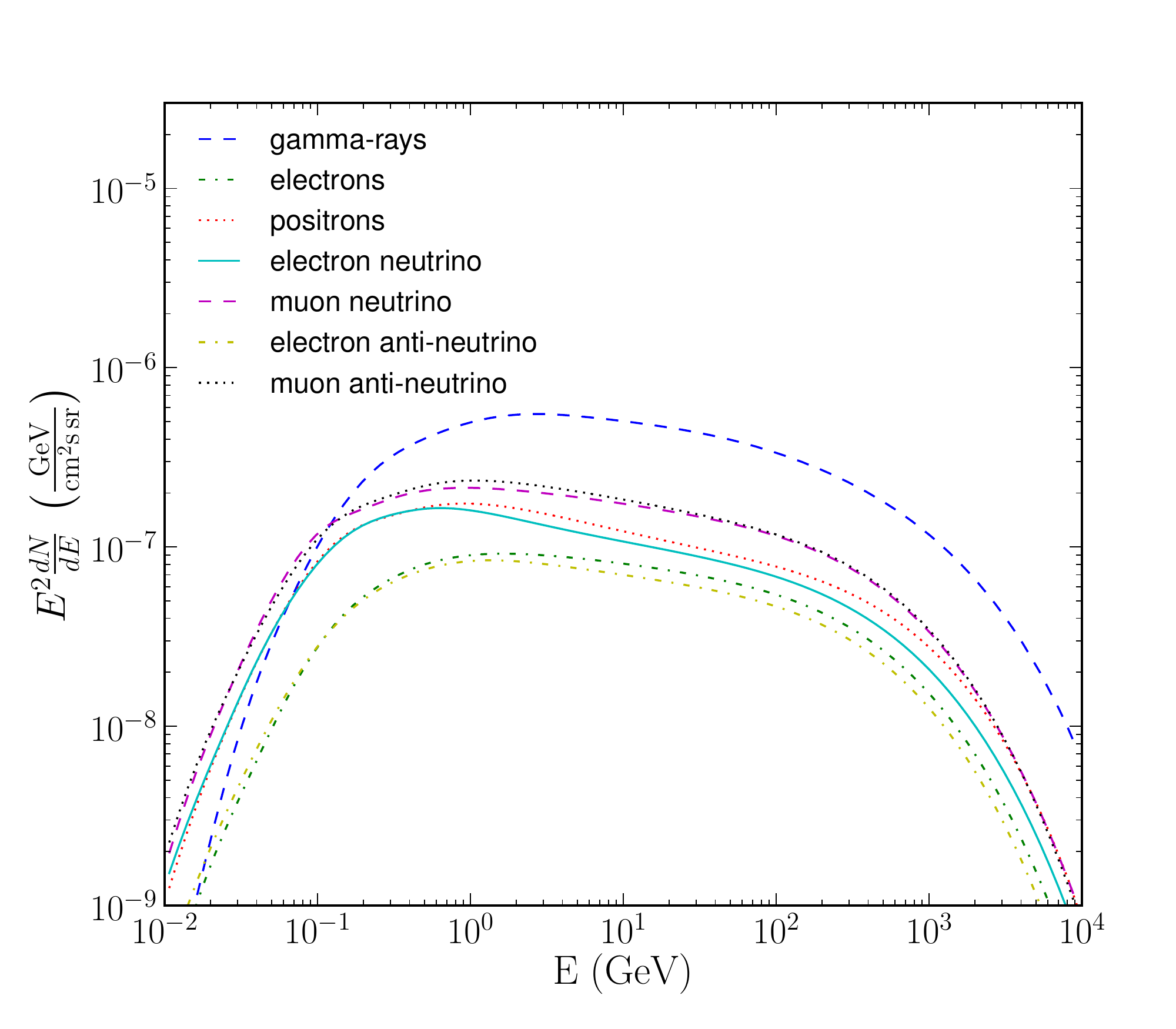, scale=\twopic}
\epsfig{figure = 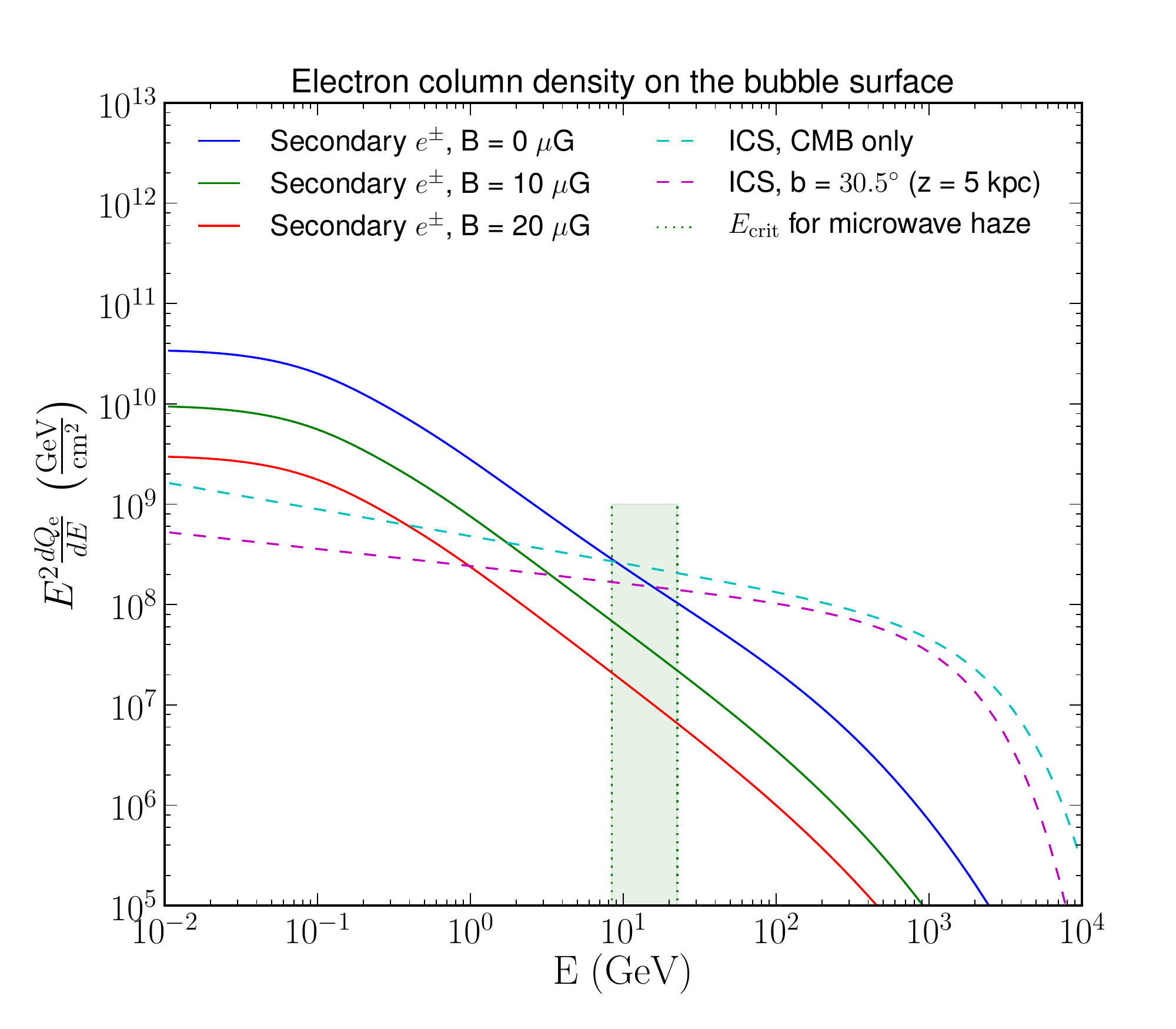, scale=\twopic}
\noindent
\caption{\small 
Left: flux of neutrinos and gamma rays produced in the hadronic model of the gamma-ray emission in the bubbles.
We also formally include the flux of secondary electrons and positrons as it would be observed if charged
particles could propagate in straight lines and were not trapped inside the bubbles.
Right: stationary energy density of secondary electrons and positrons (taking into account energy losses)
produced in hadronic interactions compared to the energy density of electrons necessary for 
the IC production of the {\Fermi} bubbles gamma-ray flux.
The shaded region is an approximate range of energies relevant for production of synchrotron emission
at the \WMAP and \Planck microwave haze frequencies.
}
\label{fig:74nu_flux}
\end{center}
\end{figure}

In addition to producing gamma rays, interactions of high-energy protons also produce electrons, positrons,
and several species of neutrinos. 
The flux of neutrinos from the hadronic interactions in the {\Fermi} bubbles has been previously 
considered by \cite{2012PhRvL.108v1102L, 2013arXiv1308.5260A, 2013arXiv1309.4077A, 2013arXiv1311.7188L}.
We present our calculation of the fluxes of all particles produced in the hadronic interactions in 
Figure \ref{fig:74nu_flux} on the left.
As above, in this calculation the cross sections are taken from the cparamlib package.
The electrons and positrons are included in this plot only formally to show their relative production cross sections 
compared to the cross sections of neutrinos and gamma rays.
In reality, the secondary leptons are assumed to be trapped inside the bubbles together with the protons.
Note that we calculate the neutrino spectrum based on the average gamma-ray spectrum of the bubbles. However, we cannot rule out that the cocoon spectrum follows a simple power law, which might produce neutrinos at higher energies.

We now calculate the synchrotron emission from the secondary $e^\pm$
produced in the hadronic collisions inside the bubbles.
The secondary electrons and positrons undergo cooling due to IC and synchrotron losses.
We denote the energy loss function as \hbox{$\dot E = - b(E)$}, where $b(E) \propto E^2$ for the synchrotron and
the IC energy loss in the Thomson approximation (Appendix \ref{sec:IC_h_models}).
Assuming that the high-energy leptons lose their energy inside the bubbles,
the stationary energy density of secondary electrons and positrons is
\be
\lb{eq:sec_lept}
\frac{dQ_{e^\pm}(E)}{dE} = \frac{1}{b(E)} \int_E^\infty \frac{dJ_{e^\pm} (\td E)}{d\td E} d\td E,
\ee
where $dJ_{e^\pm} / dE$ is the spectrum of electrons or positrons produced in interactions of hadronic CR 
with interstellar gas.
We compare the energy density of secondary $e^\pm$ to the energy density of electrons necessary
to produce the gamma-ray flux of the bubbles by IC scattering in Figure \ref{fig:74nu_flux} on the right.

In the range of energies relevant for the \WMAP and \Planck haze, the density of the secondary electrons
and positrons is comparable to the electrons producing the IC.
Above the pion cutoff at $\sim 100$ MeV,
the injection spectrum for the secondary leptons is
proportional to the proton spectrum $\propto E^{-2}$.
There is an additional softening by one power of $E$ due to cooling.
As a result, the spectrum of the secondary leptons above 100 MeV is proportional to $E^{-3}$.

Below the pion cutoff, the spectrum of secondary leptons is harder than
$E^{-1}$ and the integral in Equation \ref{eq:sec_lept} is insensitive to the lower energy bound.
As a result, the stationary spectrum of the secondary leptons below 100 MeV is simply proportional
to ${b(E)}^{-1} \propto {E^{-2}}$.
For the benchmark gas density  $n_{\rm H} \sim 0.01\, {\rm cm}^{-3}$,
the bremsstrahlung emission from the secondary leptons is at least an order of magnitude smaller 
than the gamma-ray signal for all energies above 100 MeV and we neglect it in our analysis.

\begin{figure}[htbp] 
\begin{center}
\epsfig{figure = 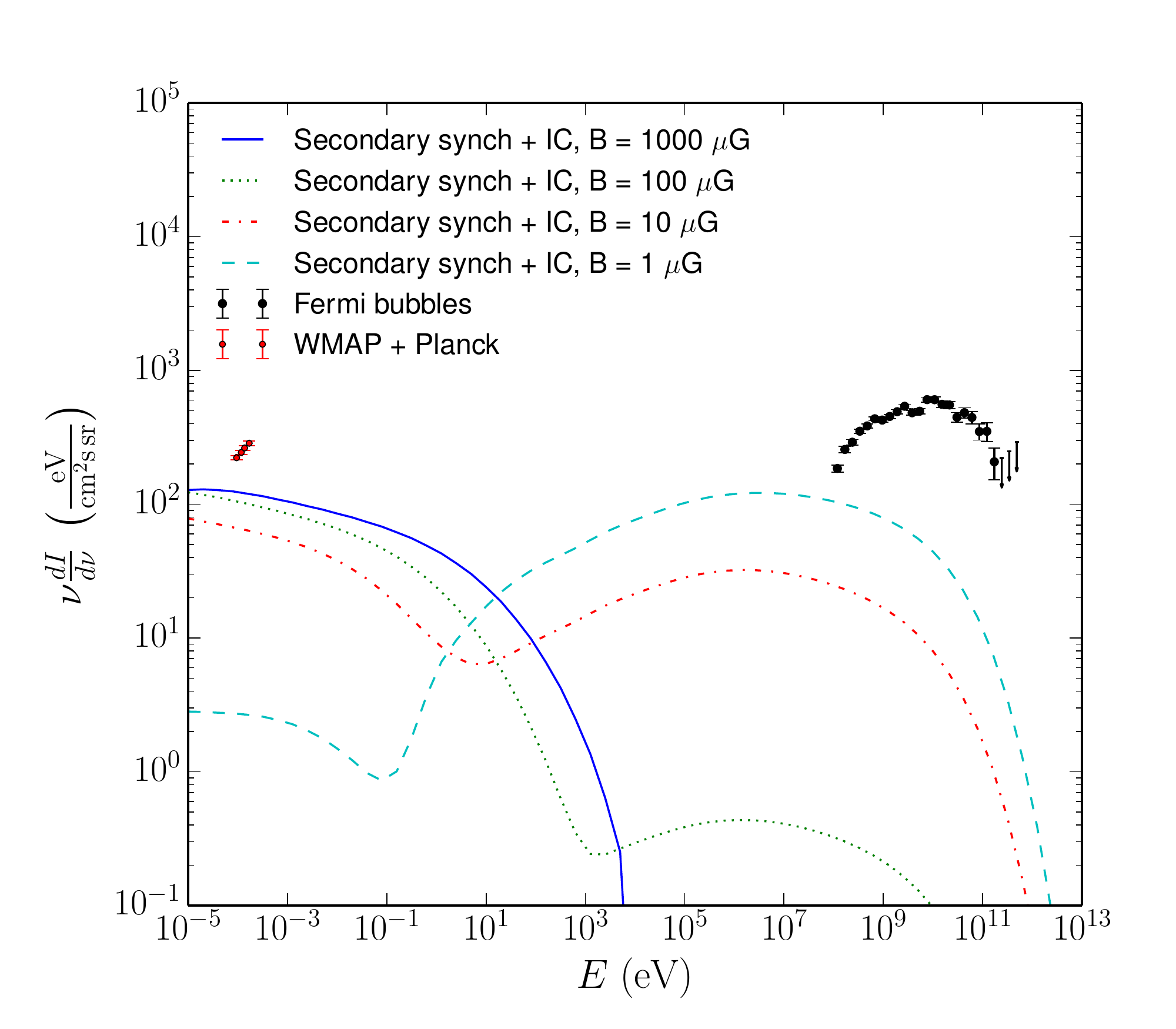, scale=\twopic}
\epsfig{figure = 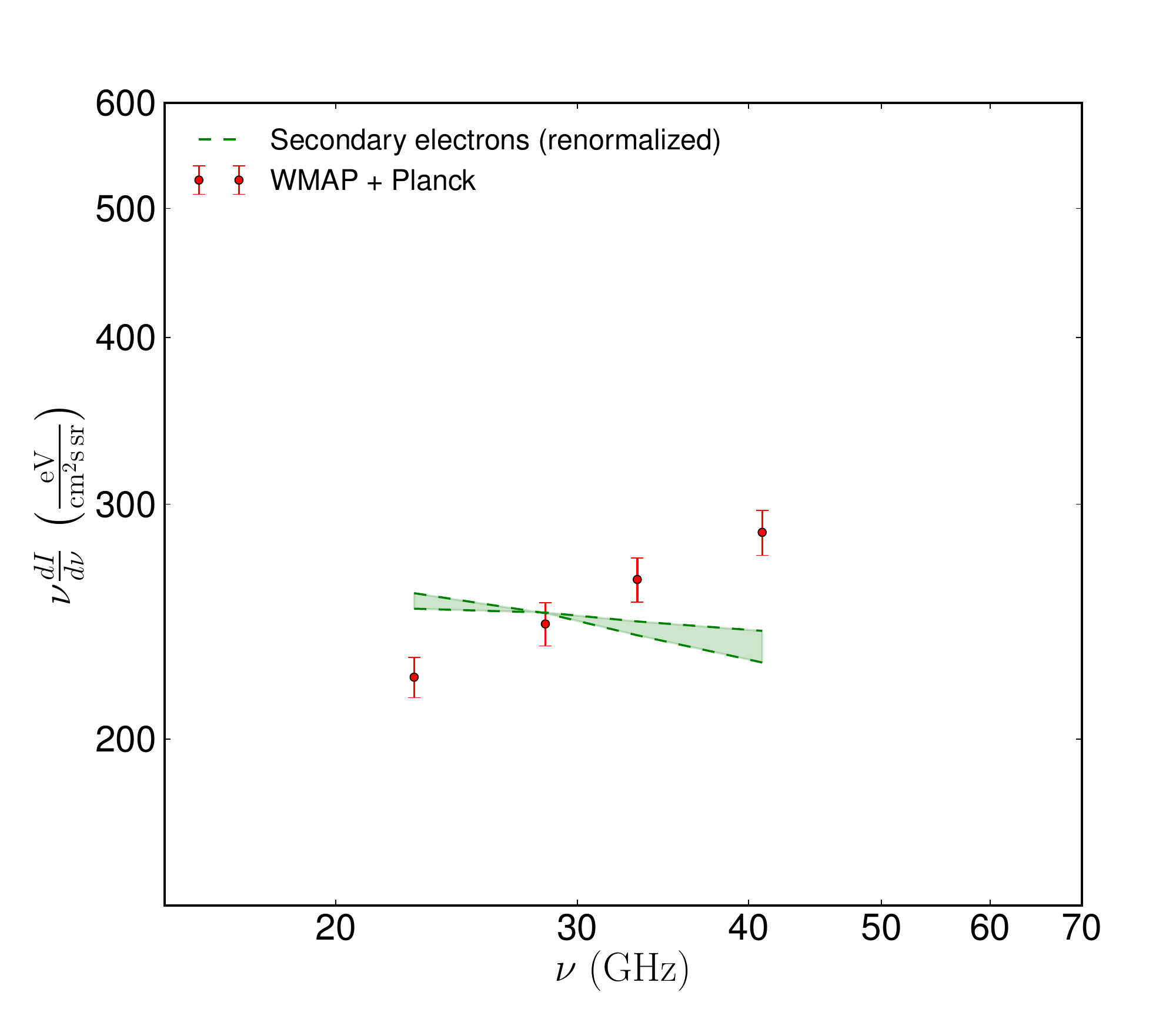, scale=\twopic}
\noindent
\caption{\small 
Left: synchrotron radiation produced by the secondary leptons in the hadronic model of the bubbles emission
in comparison with the microwave haze \citep{2013A&A...554A.139P}.
Right: the range of spectra for the synchrotron radiation from secondary leptons that correspond to different
models of the foreground gamma-ray emission and different definitions of the bubbles template.
}
\label{fig:75ssynch}
\end{center}
\end{figure}

The IC and synchrotron spectra from the secondary leptons produced in the hadronic model
for different magnetic fields are shown in Figure \ref{fig:75ssynch} on the left.
We note that the synchrotron intensity is a factor of 3 - 4 lower than the \WMAP and \Planck points
and one cannot correct for this offset by tuning the magnetic field.
The reason is that for electrons with a spectrum $\propto E^{-3}$  the synchrotron radiation scales as $B^2$,
but for large magnetic field strengthes the synchrotron energy losses dominate and the
normalization of the secondary electrons is proportional to 
${\dot E}^{-1} \propto {B^{-2}}$.
Consequently the dependence on the magnetic field cancels out.
This is the reason 
that the synchrotron intensity is saturated
for $B \gtrsim 10\, \upmu G$.
The IC flux is proportional to the stationary energy density of the secondary leptons and it decreases with increasing magnetic field.

The synchrotron radiation produced by $E^{-3}$ electrons is $I_\nu \propto \nu^{-1}$ which is significantly softer 
than the microwave haze spectrum $I_\nu \propto \nu^{-0.55}$ \citep{2013A&A...554A.139P}.
The distribution of indices for the synchrotron radiation from the secondary leptons is shown in 
Figure \ref{fig:75ssynch} on the right.
The lines on this plot are obtained by calculating the synchrotron radiation from the secondary
leptons in a 10 $\upmu$G magnetic field with a subsequent rescaling (for illustration)
by an overall normalization factor to fit the \WMAP and \Planck points.

\begin{figure}[htbp] 
\begin{center}
\epsfig{figure = 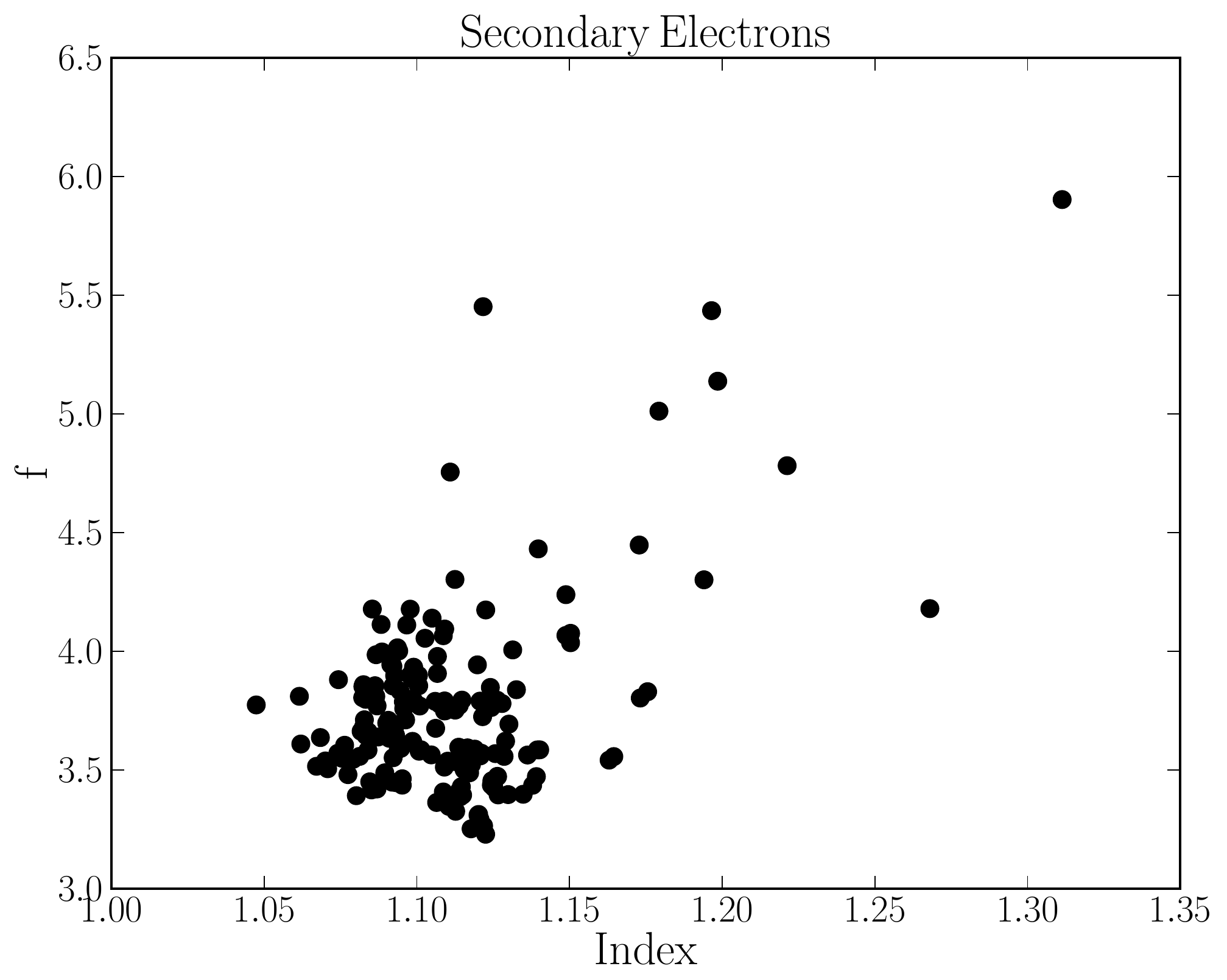, scale=\onepic}
\noindent
\caption{\small 
Distribution of synchrotron indices and rescaling factor, $f$,  for synchrotron emission from the secondary leptons in hadronic models in a $B = 10\, \upmu$G field 
necessary to fit the microwave haze intensities.
}
\label{fig:76synch_inds}
\end{center}
\end{figure}

The range of indices and the renormalization factors that we use to rescale the synchrotron
radiation from the secondary leptons to fit the microwave haze are shown in Figure \ref{fig:76synch_inds}.
Both the index and the rescaling factor have a relatively small range of 
values because both the leptons around 10 GeV, which are responsible for the synchrotron radiation,
and the gamma rays around 10 GeV, where the {\Fermi} bubbles spectrum has small statistical and systematic errors, are produced by the same protons with energies around 100 GeV.
As a result, the synchrotron radiation at \WMAP and \Planck haze frequencies is directly linked 
to the gamma-ray radiation around 10 GeV in this scenario.
The conclusion is that there should be either an additional source of primary electrons or a
reacceleration of the secondary leptons
inside the bubbles to increase the overall normalization by a factor of 3 and to produce
the spectrum $\propto E^{-2}$ of electrons and positrons around 10 GeV required to fit the microwave haze data.
The timescale of this reacceleration should be smaller than the cooling time of electrons
around 10 GeV, which is about 10 Myr for a 10 $\upmu$G magnetic field (Appendix \ref{sect:mw_haze}).
Another possibility in the hadronic scenario of the \Fermi bubbles is transporting the electrons from the Galactic plane
in the wind from supernovae together with the hadronic CR \citep{2011PhRvL.106j1102C}.
The timescale of 10 Myr is sufficient to bring the electrons to 10 kpc above the Galactic plane with a wind of about 1000 km\,s$^{-1}$.
These electrons can produce the microwave haze by synchrotron radiation.
They may also contribute to the gamma-ray spectrum at energies $\lesssim 10$ GeV by IC scattering off starlight.

\begin{figure}[htbp] 
\begin{center}
\epsfig{figure = 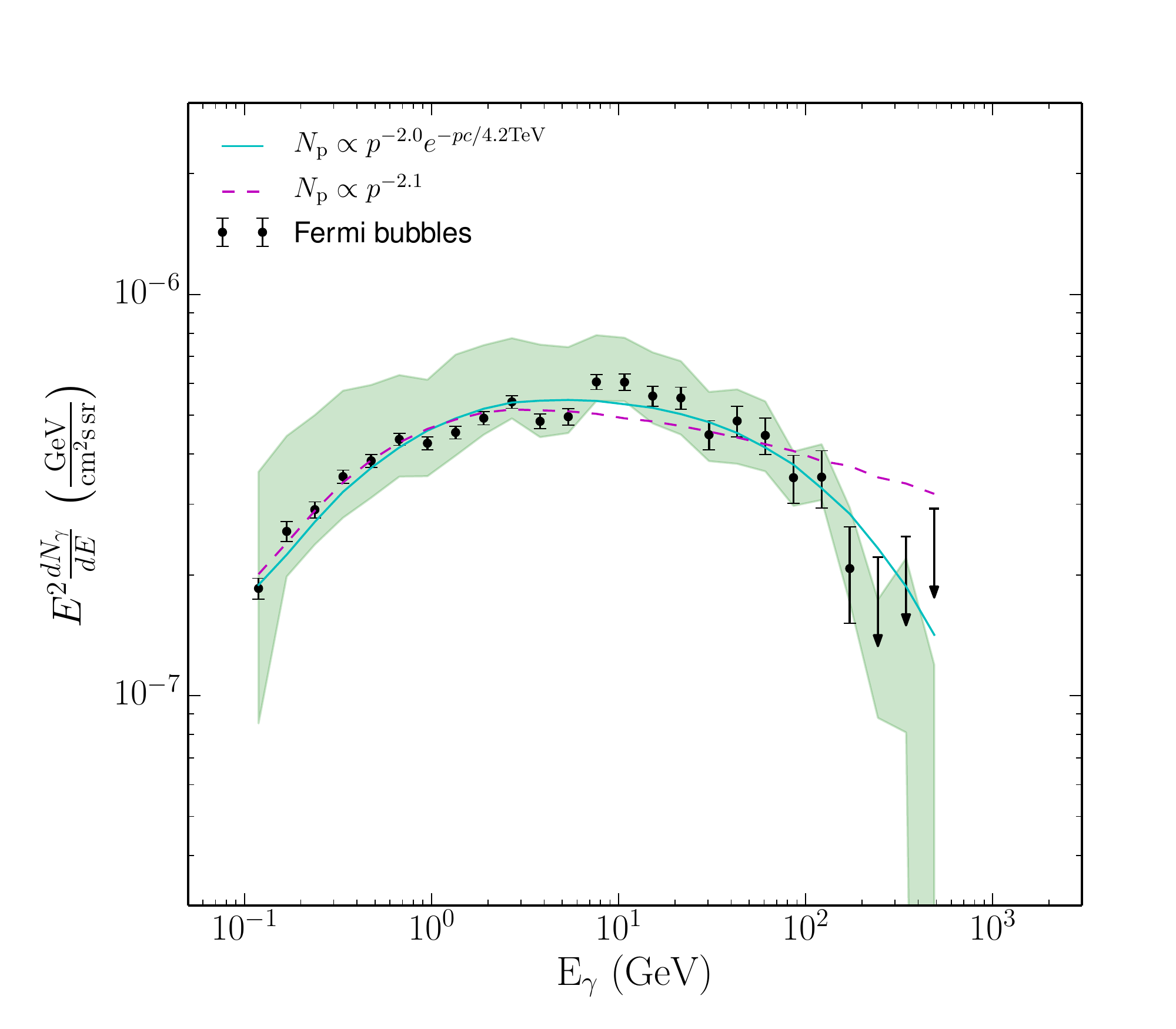, scale=\twopic}
\epsfig{figure = 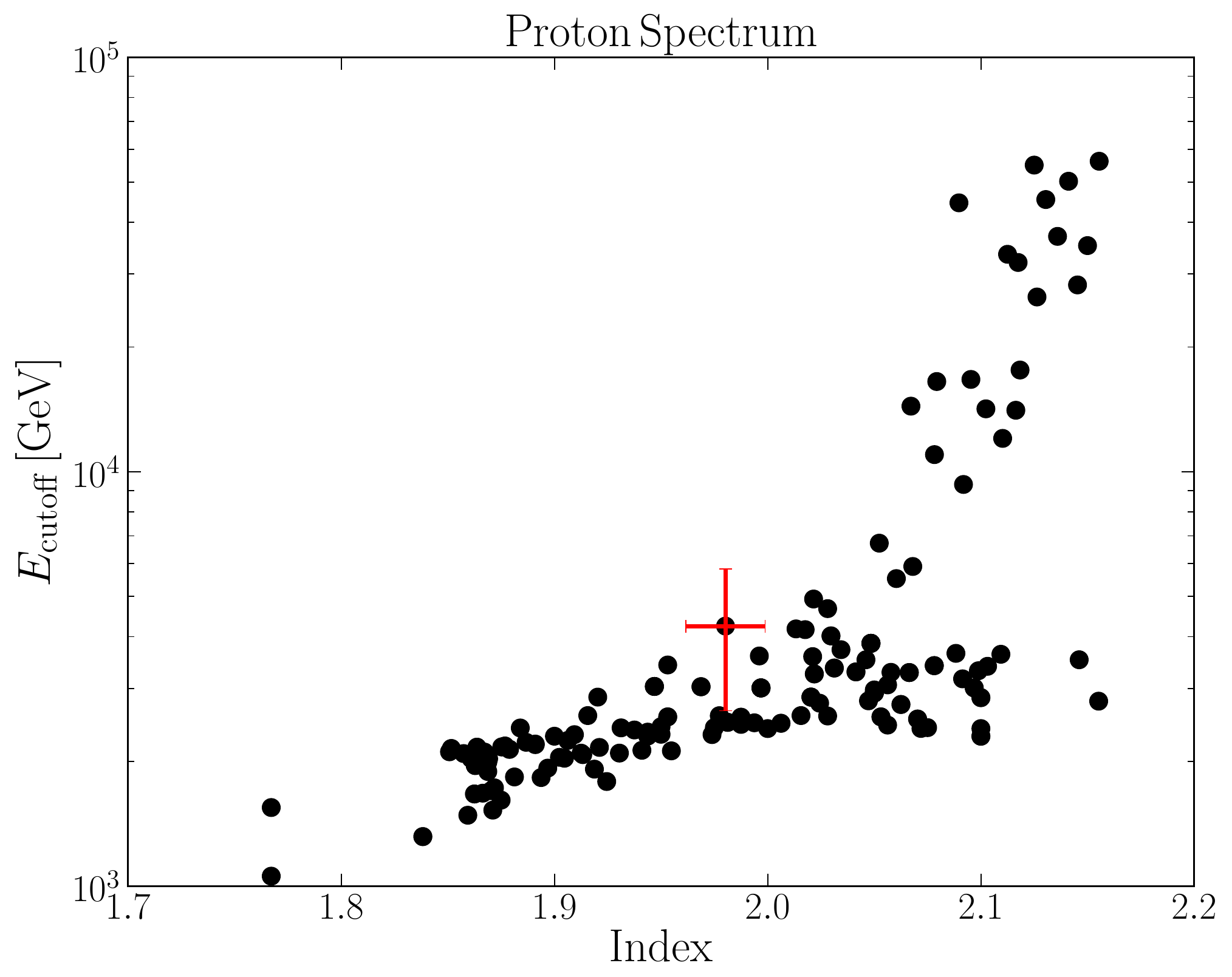, scale=\twopic}
\noindent
\caption{\small 
Left: same as Figure~\ref{fig:73pi0_spectra} (left), but including IC gamma-ray emission from secondary leptons assuming a $2\upmu$G magnetic field.  
Right: index and cutoff energy for proton spectra determined for different derivations of gamma-ray foregrounds
and different definitions of the {\Fermi} bubbles templates. The red cross corresponds to the baseline model values with the statistical errors. 
}
\label{fig:78chi2}
\end{center}
\end{figure}

Including emission from secondary leptons (Equation~\ref{eq:sec_lept}) in the hadronic models can improve the fit. The secondary gamma-ray spectrum depends on the value of the magnetic field: for magnetic fields larger than approximately 5 $\upmu$G, most of the power injected in secondary leptons is dissipated into synchrotron emission and the gamma-ray spectrum is similar to the primary only emission. The total primary and secondary gamma-ray spectrum for a 2 $\upmu$G magnetic field is shown in Figure~\ref{fig:78chi2} on the left. A sample of proton spectral indices and cutoffs for the energy spectra of the {\Fermi} bubbles derived
for different foreground models and definitions of the bubbles and Loop I templates
is shown in Figure \ref{fig:78chi2} on the right.

\begin{figure}[htbp] 
\begin{center}
\epsfig{figure = 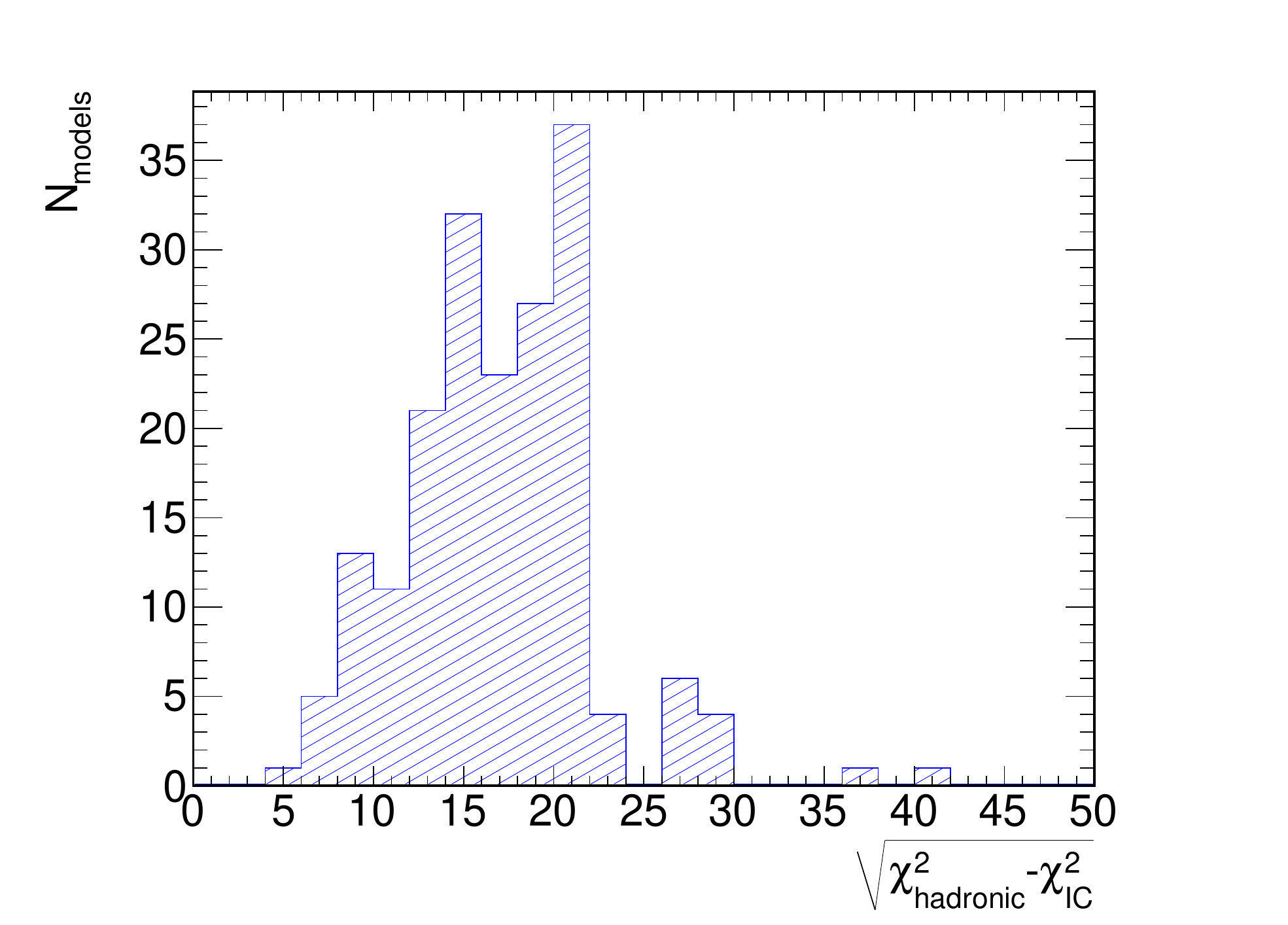, scale=\twopic}
\epsfig{figure = 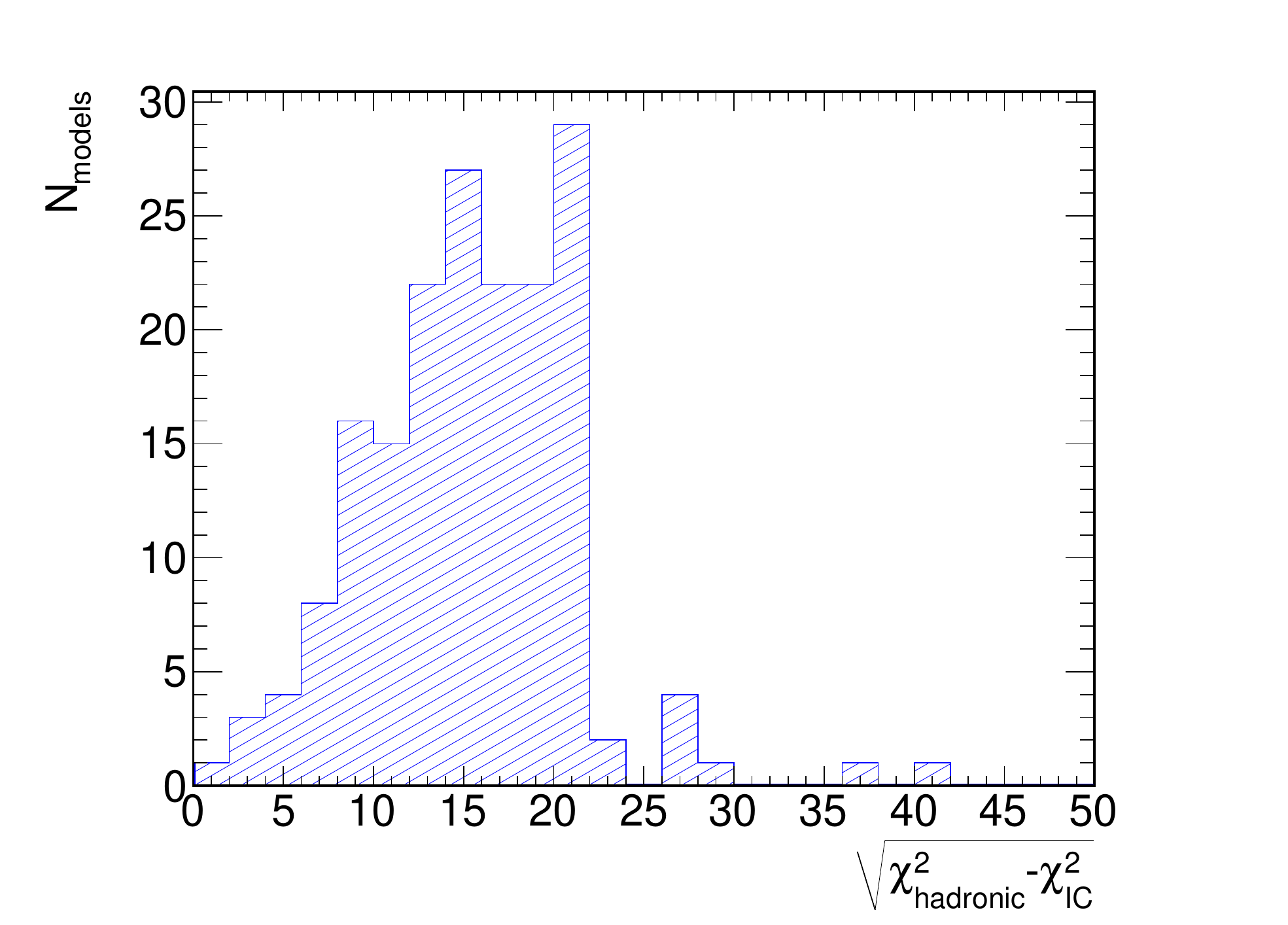, scale=\twopic}
\noindent
\caption{\small 
Difference of $\chi^2$ values obtained from fits with IC models and primary emission of hadronic models (left) 
and including the secondary emission in the hadronic scenario (right).
}
\label{fig:77chi2}
\end{center}
\end{figure}

In Figure \ref{fig:77chi2} we compare the reduced $\chi^2$ for IC and hadronic models of the {\Fermi} bubbles.
In general, we find that the IC models fit the spectrum of the bubbles better than the primary gamma-ray spectrum in hadronic models
with a significance of at least $4.9 \sigma$ (Figure~\ref{fig:77chi2} left). 
The distribution of the difference between the $\chi^2$ in leptonic and hadronic models including secondary emission is presented in Figure~\ref{fig:77chi2} on the right. 
For some cases, the hadronic model including secondary gamma-ray emission has a $\chi^2$ comparable to the IC model. 
Therefore, we cannot discriminate between hadronic and IC models based on the current gamma-ray data alone.
This calculation does not include uncertainties in the nuclear production models, which could affect the gamma-ray 
spectrum at low energies up to $30\%$~\citep{2012PhRvL.109i1101D}%
\footnote{Note that the microwave haze is produced by electrons at energies above 10 GeV, where the uncertainties in the nuclear production cross sections
are lower $\sim 10\%$~\citep{2012PhRvL.109i1101D}, hence, they are subordinate to the systematic uncertainties
related to modeling of the foreground diffuse emission.}.


%% file: 8conclusion.tex
\section{Summary and conclusions}
\lb{sect:conclusions}

In this paper, we analyze 50 months of {\Fermi}-LAT data in order to
determine the morphology and the spectrum of the {\Fermi} bubbles.
The main challenge 
is the spatial overlap with the other components of diffuse gamma-ray emission
due to collisions of hadronic CRs with the interstellar
gas, IC production of gamma rays, bremsstrahlung, extragalactic gamma-ray emission,
and contamination from CRs.

We model the foreground gamma-ray emission with two independent methods.
In the first method we use maps generated by the GALPROP CR propagation and interaction code as templates
of the emission components. Fitting these templates to the data makes this method less sensitive 
to the distribution of CR sources and to the propagation model.
In the second method we develop a more general data-driven model
of diffuse gamma-ray foregrounds.
We use gas maps derived from the 21-cm, CO, and dust surveys as tracers of the gamma rays produced by
interactions of hadronic CRs and by bremsstrahlung.
We fit the gas-correlated templates on small patches of the sky 
together with a combination of 2D polynomial functions that model components of emission not correlated
with the distribution of the gas.
In this method no assumptions are made on CR sources or propagation, except for an assumption 
that the CR density inside each patch can be approximated by a constant.
After subtracting the gas-correlated component from the data,
we model the IC component as a bivariate Gaussian along the Galactic plane.

One of the largest systematic uncertainties is the definition of the template for Loop I, 
which is a large structure in the sky, overlapping the {\Fermi} bubbles mostly in the northern hemisphere.
In the paper we use three different definitions of the Loop I template: a template based
on the radio data at 408 MHz \citep{Haslam:1982zz}, 
a geometrical model based on the 1.4 GHz polarization data \citep{Wolleben:2007pq}, 
and a template derived from the gamma-ray residuals based on a correlation with the gamma-ray
spectrum in the residuals outside the {\Fermi} bubbles.
We determine templates for the bubbles from the residuals obtained after subtracting
the foreground emission components.

We find that the gamma-ray spectra of the {\Fermi} bubbles derived with the two methods agree with each other.
At energies above approximately 10 GeV the uncertainty in the spectrum is dominated by statistics while below 10 GeV systematic uncertainties are dominant.
The total luminosity of the bubbles between 100 MeV and 500 GeV is
$(4.4\pm0.1 \rm{[stat]}  ^{+2.4}_{-0.9}\rm{[syst]} )\times 10^{37}$ erg s$^{-1}$ (assuming that they are located at the distance of the Galactic center).
The spectrum of the bubbles is well described either by a log parabola function
or by a power law with an exponential cutoff.
The fit of the photon spectrum above 1 GeV by a power law with an exponential cutoff has an index 
$\gamma=(1.87\pm{0.02}\rm{[stat]}^{+0.14}_{-0.17}\rm{[syst]})$ 
and a cutoff 
$E_{\rm{cut}} = (113\pm 19 \rm{[stat]} ^{+45}_{-53}\rm{[syst]})$ GeV.
A simple power-law spectrum is excluded with very high confidence (Figure \ref{fig:55chi2}).
Both the log parabola and a power law with an exponential cutoff fit the spectrum of the 
{\Fermi} bubbles with $\Delta \chi^2 / {\rm NDF} \approx 1$ for some models of gamma-ray
diffuse foregrounds and the definition of the bubbles and Loop I templates.
In the energy range between 100 MeV and 500 GeV, a log parabola gives a slightly better fit
than a power law with a cutoff.
The reported spectrum has a higher normalization and a softer spectrum at energies below 1 GeV compared to previous results~\citep{2012ApJ...753...61S}. For the first time we show that a significant cutoff exists at high energies.

The brightness of the emission is not uniform across the bubbles: 
we confirm the previously reported~\citep{2012ApJ...753...61S} excess of emission on the south-eastern side of the bubbles 
(referred to as the cocoon).
We define a template for the cocoon similarly to the bubbles template by applying a higher
cut in significance to the residual maps.
We find that fitting a flat cocoon template together with a flat bubbles template improves the fit
compared to the flat bubbles template case by $2\Delta \log\mathcal{L} / {\rm NDF} > 3.7$ for NDF = 25
(Figure \ref{fig:55cocoon_spectrum}).
The probability that the cocoon is a statistical fluctuation is $<10^{-9}$ 
for some models of foreground emission
that we have tested and for all definitions of the bubbles and Loop I templates.
The energy spectrum of the cocoon is consistent with the spectrum of the {\Fermi} bubbles.

We search for a 
jet-like structure inside the bubbles using two different techniques.
First, we apply a higher cut in the residual significance map than in the definition of the cocoon
(Figure \ref{fig:55ajet_temp}).
This structure is not significant, with a probability of statistical fluctuation of $~20\%$ 
for all models of foreground emission.
We also searched for a narrow linear feature emanating from the Galactic center (Figure \ref{fig:55jet_sig}).
We do not find any significant linear jet ~\citep[in contrast to tentative claims by][]{2012ApJ...753...61S} within our systematic uncertainty,
although the scatter in significance for the different foreground models
is very high.

The spectra of the northern bubble and the southern bubble are consistent with each other
(Figure \ref{fig:53NS}).
We find no variation of the spectrum with latitude within the statistical and 
systematic uncertainties in contrast to claims by~\cite{Hooper:2013rwa} and~\cite{2014arXiv1402.0403Y}.
The systematic uncertainties are especially high for the strips in the North 
due to strong contamination from gamma rays produced in hadronic interactions
and a significant overlap with Loop I.

We estimate the position and the width of the boundary of the bubbles by 
fitting a hyperbolic tangent function across the edge of the bubbles.
The width is calculated for residuals in three energy ranges:
1 - 3 GeV, 3 - 10 GeV, and 10 - 500 GeV.
We find that, on average, the width is smaller than $\sim 6^\circ$.
We do not find a significant variation of the width with energy
but the width changes with position along the boundary of the bubbles
(Figure \ref{fig:58edge_map}).

We fit the spatially integrated spectrum of the bubbles with IC and hadronic models of gamma-ray production.
The spectrum of the electrons in the IC model is well described by a power law with an exponential cutoff
with
$n = 2.17\pm 0.05\rm{[stat]}^{+0.33}_{-0.89}\rm{[syst]}$ and
$E_{\rm cut} = 1.25\pm 0.13\rm{[stat]}^{+1.73}_{-0.68}\rm{[syst]}\,TeV$
and the total energy in electrons above 1 GeV is
$(1.0\pm 0.2\rm{[stat]}^{+6.0}_{-1.0}\rm{[syst]} )\times 10^{52}$ erg.
The spectrum of the protons in the hadronic scenario is expressed as a function of the 
proton momentum $dn / dp \propto p^{-n} e^{-pc/E_{\rm cut}}$ with 
$n = 2.13\pm 0.01\rm{[stat]}^{+0.15}_{-0.52}\rm{[syst]}$ and
$E_{\rm cut} = 14\pm 7\rm{[stat]}^{+6}_{-13}\rm{[syst]}\, TeV$.
The total energy in proton CRs above 1 GeV is
$(3.5\pm 0.1\rm{[stat]}^{+4.7}_{-3.0}\rm{[syst]} )\times 10^{55}
\left(\frac{0.01\; {\rm cm^{-3}}}{n_{\rm H}}\right) \; {\rm erg} $.
The IC model fits the energy spectrum of the {\Fermi} bubbles better than the primary gamma-ray emission in the hadronic
model with a significance of at least $4.9 \sigma$ without taking into account the uncertainties in the nuclear cross sections of~\cite{2006ApJ...647..692K}. 
If we include gamma-ray emission from secondary leptons and assume
a magnetic field weaker than approximately 5 $\upmu$G, then
the quality of the fits of the hadronic models to the gamma-ray data becomes comparable to the IC models (Figure \ref{fig:78chi2}).
The derived spectra of CRs in leptonic and hadronic models of the {\Fermi} bubbles agree with the previous results
\citep[e.g.,][]{2011PhRvL.106j1102C, 2011PhRvL.107i1101M}.

For IC models of the bubbles,
the homogeneous energy spectrum within the area of the bubbles and the rather sharp edges
may favor a transient bubble formation scenario with either fast transport of leptonic CRs to high latitudes
or continuous reacceleration of leptons within the volume of the bubbles.
The cutoff in the energy spectrum can be explained by cooling of electrons
due to IC and synchrotron energy losses.
The presence of a cutoff in the proton spectrum for the hadronic model of gamma-ray production 
\citep{2011PhRvL.106j1102C} would indicate that either the mechanism of proton acceleration becomes less
efficient around a few TeV or that the protons with higher energies can escape from the bubbles.

We find that the electrons in the IC scenario can also explain the \WMAP and \Planck microwave haze
data for a magnetic field in the range between 5 $\upmu$G and 20 $\upmu$G
(Figure \ref{fig:72synch}).
The secondary electrons and positrons in the hadronic scenario produce synchrotron radiation
with a spectrum that is too soft compared to the microwave haze spectrum while
the overall normalization of the synchrotron radiation from the secondary particles is at least a factor
3 - 4 smaller than the microwave haze level 
(Figures \ref{fig:75ssynch} and \ref{fig:76synch_inds}).
A simultaneous explanation of the gamma-ray and the microwave data in a hadronic model requires an additional 
source of primary electrons or a reacceleration of the secondary leptons.

%% file: acknowledgement.tex
\section{Acknowledgements}

The {\Fermi}-LAT Collaboration acknowledges generous ongoing support
from a number of agencies and institutes that have supported both the
development and the operation of the LAT as well as scientific data analysis.
These include the National Aeronautics and Space Administration and the
Department of Energy in the United States, the Commissariat \`a l'Energie Atomique
and the Centre National de la Recherche Scientifique / Institut National de Physique
Nucl\'eaire et de Physique des Particules in France, the Agenzia Spaziale Italiana
and the Istituto Nazionale di Fisica Nucleare in Italy, the Ministry of Education,
Culture, Sports, Science and Technology (MEXT), High Energy Accelerator Research
Organization (KEK) and Japan Aerospace Exploration Agency (JAXA) in Japan, and
the K.~A.~Wallenberg Foundation, the Swedish Research Council and the
Swedish National Space Board in Sweden.

Additional support for science analysis during the operations phase is gratefully
acknowledged from the Istituto Nazionale di Astrofisica in Italy and the Centre National d'\'Etudes Spatiales in France.
Some parts of the data analysis have been
done with the HEALPix package \citep{Gorski:2004by}.

%% file: 9bias.tex
\section{$\chi^2$ approximation to the likelihood}
\lb{sect:bias}

In Section \ref{sect:local}, we use the quadratic approximation to the log likelihood 
\be
 \log \mathcal{L} \approx  \sum_i^{\rm E\, bins} \sum_j^{\rm pixels}  d_{ij} \log \frac{\mu_{ij}}{d_{ij}} - (\mu_{ij} - d_{ij}) \approx
- \frac{1}{2} \sum_i^{\rm E\, bins} \sum_j^{\rm pixels} \frac{(d_{ij} - \mu_{ij})^2}{\sm^2_{ij}},
\ee
where $d_{ij}$ is the gamma-ray data and $\mu_{ij}$ is the model
in an energy bin $i$ and a pixel $j$.
For a large number of counts the statistical uncertainty in the denominator is
$\sm^2_{ij} \approx d_{ij} \approx \mu_{ij}$.
For both choices $\sm^2_{ij} = d_{ij}$ and $\sm^2_{ij} = \mu_{ij}$
the fit has a statistically significant bias.
For example, 
consider an energy bin $i$
(in the following argument we omit the energy index $i$ for brevity).
If the underlying distribution of photons is isotropic (for example, in a high latitude patch),
then the best estimate of the model $\mu_j = \mu$ is the arithmetic mean 
$\mu_* = \bar{d} = \frac{1}{N_{\rm pix}}\sum_{i = 1}^{N_{\rm pix}} d_j$. 
If, however, one takes $\sm^2_j = d_j$, then the $\chi^2$ minimization gives
the harmonic mean
\be
\mu_* = \frac{N_{\rm pix}}{\sum_{i = 1}^{N_{\rm pix}} \frac{1}{d_j}} < \bar{d}.
\ee
The reason for this bias is that the fluctuations above the true model $d_j > \mu$ are underweighted
in the $\chi^2$, while the fluctuations below the true model $d_j < \mu$ are overweighted.

In the second case, if we take $\sm^2_j = \mu$ and minimize $\chi^2$ with respect to $\mu$,
then we get the quadratic mean
\be
\mu_* = \sqrt{\frac{1}{N_{\rm pix}} \sum_{i = 1}^{N_{\rm pix}}d_j^2} > \bar{d}.
\ee
The reason for this bias is that models with larger $\mu$ have smaller $\chi^2$ due to the presence of $\mu$ in the
denominator.

In order to get an unbiased $\mu_*$ from a $\chi^2$ minimization, 
one can take $\sm^2_j = \bar{d}$.
In this case both the best-fit parameter $\mu_*$ and the standard deviation found from the $\chi^2$ minimization
coincide with the best estimators in the true Poisson distribution.
Indeed, the estimators of the mean and the standard deviation of the mean 
in a Poisson distribution are
\be
\mu = \bar{d},\qquad \sm_\mu^2 = \frac{\bar{d}}{N}.
\ee
The same values are obtained by minimizing
\be
\chi^2(\mu)  = \sum_j \frac{(d_j - \mu)^2}{\bar{d}}
\ee
with respect to $\mu$.
In particular, the standard deviation of the mean is
\be
\frac{1}{\sm_\mu^2} = \frac{1}{2}\frac{d^2 \chi^2}{d\mu^2} = \frac{N}{\bar{d}}.
\ee

The observed gamma-ray flux is not uniform, i.e., we cannot average over the whole sky to
get an estimate of $\sm^2$.
The gamma-ray flux can be approximated as a constant only locally.
We find the value of the local average by smoothing the data with a Gaussian kernel.
At low energies the radius of the kernel is $2^\circ$, which corresponds to averaging over 
approximately 15 pixels in the pixelation scheme chosen for the paper.
At energies where on average there are fewer than 100 photons within a $2^\circ$ radius
we take the smoothing radius $r$ such that there are on average 100 photons within $r$.
The radius stays at $2^\circ$ up to 5 GeV, then it grows to about $20^\circ$ at 100 GeV,
and $100^\circ$ at 500 GeV.
Thus the estimate of the $\sm^2_{ij}$ that we use in the $\chi^2$ definition
is given by the smoothed counts map $\td{d}_{ij}$, where the smoothing radius depends on energy bin $i$.
There is still a small bias due to using the smoothed data counts instead 
of the true but unknown model. 
This bias manifests itself in a small underestimation of the total gamma-ray flux. The difference is less than 1\% below 2 GeV,
it increases up to 2 - 3\% around 10 GeV, and above 10 GeV it is smaller than the statistical uncertainty.
This bias is much smaller than the other sources of the systematic uncertainty.

%% file: 9models.tex
\section{IC and hadronic models of the bubbles}
\lb{sec:IC_h_models}

Here we present details 
about the calculation of IC and hadronic gamma-ray emission to model the spectrum of the {\Fermi} bubbles.
As in Section \ref{sect:interp},
we use the baseline model in Figure \ref{fig:52compareSpectra}
as an example of the gamma-ray spectrum of the bubbles,
while the other determinations of the {\Fermi} bubbles spectra 
are used to estimate the systematic uncertainties.

\subsection{IC model of the {\Fermi} bubbles}
\lb{sect:IC_model}

\begin{figure}[htbp] 
\vspace{-1mm}
\begin{center}
\epsfig{figure = 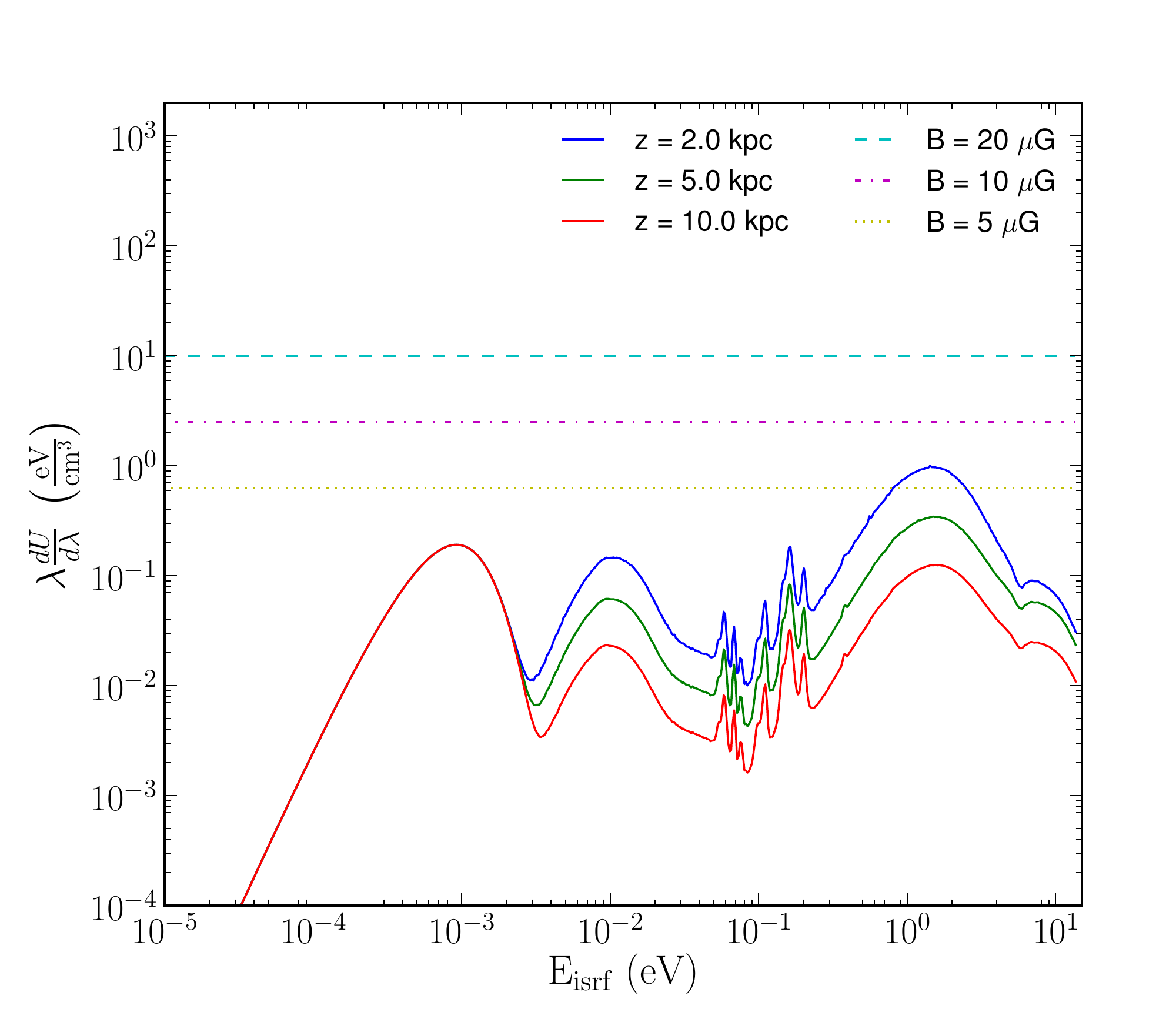, scale=\twopic}
\epsfig{figure = 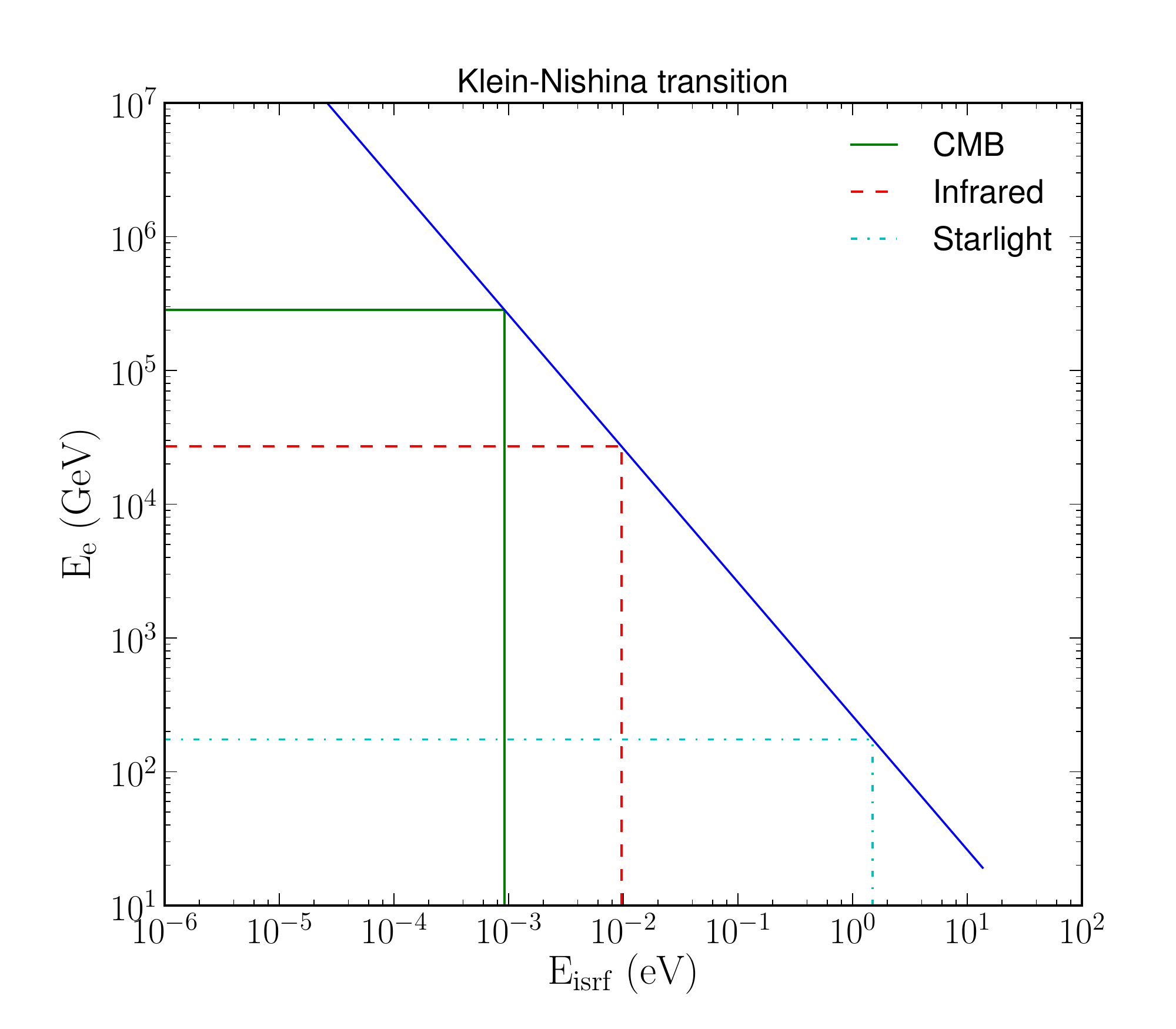, scale=\twopic}
\noindent
\caption{\small 
Left: energy density of the ISRF (including CMB) at different heights above the Galactic center $z$ = 2, 5, 10 kpc
\citep{2005ICRC....4...77P, 2006ApJ...640L.155M}.
For comparison we also plot the energy density of 5, 10, and 20 $\upmu$G magnetic fields.
Right: Klein-Nishina transition energies $E_{\rm e} = m_{\rm e}^2 c^2 / E_{\rm ISRF}$
for the peaks of the ISRF energy density fields at $z$ = 5 kpc.
The peaks of the ISRF at 5 kpc are
$E_{\rm CMB} = 9.2 \times 10^{-4}\,{\rm eV}$, 
$E_{\rm IR} = 9.6 \times 10^{-3}\,{\rm eV}$ and
$E_{\rm SL} = 1.5 \,{\rm eV}$.
}
\label{fig:91isrf}
\end{center}
\end{figure}

The IC emission is produced by the scattering of high energy electrons on the ISRF photons.
The density of the ISRF photons \citep{2005ICRC....4...77P, 2006ApJ...640L.155M}
that we use to calculate the IC emission is presented in
Figure \ref{fig:91isrf} on the left.
The three large bumps (going from left to right) correspond to CMB, IR, and starlight (SL) photons.
In this analysis we use the ISRF distribution at $z = 5$ kpc.

The energy spectrum of the IC photons is found from the convolution of the electron
and the ISRF densities
\be
\frac{dQ_\g}{dE_\g} = c \int \frac{d \sigma_{\rm IC}(E_\g, E_{\rm e}, E_{\rm ph})}{dE_\g}  \;
\frac{d n_{\rm e}}{dE_{\rm e}} dE_{\rm e} \; \frac{d n_{\rm ph}}{dE_{\rm ph}} dE_{\rm ph},
\ee
where $c$ is the speed of light, ${dQ_{\g}}/{dE_\g}$ is in units of $\rm ({GeV^{-1}\, cm^{-3}\, s^{-1}})$, and
${d n_{\rm e}}/{dE_{\rm e}}$ and ${d n_{\rm ph}}/{dE_{\rm ph}}$ are in units of
$(\rm {GeV^{-1}\, cm^{-3}})$.
For completeness we present the scattering cross section \citep{1970RvMP...42..237B}
\be
\frac{d \sigma_{\rm IC}}{dE_\g}  = \frac{3 \sigma_T}{E_{\rm e}} \frac{1}{\G_\epsilon}
\left(2 q \ln q + (1 + 2 q) (1 - q) + \frac{1}{2} \frac{(\G_\epsilon q)^2}{1 + \G_\epsilon q}(1 - q) \right),
\ee
where $\sigma_T$ is the Thomson cross section and
\be
\G_\epsilon = \frac{4 E_{\rm ph} E_{\rm e}}{m_{\rm e}^2 c^4},
\qquad
q = \frac{E_\g}{\G (E_e - E_\g)}.
\ee
The transition between the non-relativistic (Thomson) and relativistic (Klein-Nishina) scattering happens
when the energy of the ISRF photon in the center of mass is comparable to the mass of the electron
$h\nu \g \sim m_{\rm e} c^2$.
The maxima of the CMB, IR and starlight components of the ISRF at 5 kpc are
$E_{\rm CMB} = 9.2 \times 10^{-4}\,{\rm eV}$, 
$E_{\rm IR} = 9.6 \times 10^{-3}\,{\rm eV}$ and
$E_{\rm SL} = 1.5 \,{\rm eV}$.
The characteristic electron energies for the maxima 
of the ISRF components are shown in Figure \ref{fig:91isrf}
on the left.
For starlight photons the Klein-Nishina transition is around 100 GeV.

The gamma-ray flux is determined from the source energy density as a line of sight integral
\be
\lb{eq:source2flux}
\frac{dN_\g}{dE_\g} = \frac{1}{4\pi} \int \frac{dQ_{\g}}{dE_\g} dR.
\ee
If we assume a spatially uniform ISRF distribution, then the flux of gamma rays can be expressed in terms of the column
density of the electrons 
\be
\lb{eq:edens}
f_{\rm e} = \frac{1}{4\pi} \int \frac{d n_{\rm e}}{dE_{\rm e}} dR.
\ee
Then
\be
\lb{eq:gamma_flux}
\frac{dN_\g}{dE_\g}
= c \int \frac{d \sigma_{\rm IC}}{dE_\g} 
f_{\rm e}(E_{\rm e}) dE_{\rm e} \frac{d n_{\rm ph}}{dE_{\rm ph}} dE_{\rm ph} .
\ee
The best-fit electron spectrum is 
$f_{\rm e}(E_{\rm e}) = 3.6\times 10^8 \cdot E_{\rm e}^{-2.2} e^{-E_{\rm e}/1.3\, {\rm TeV}}$ in units of 
$\rm ({GeV^{-1}\, cm^{-2}\, sr^{-1}})$.
The total energy in electrons above 1 GeV is
\be
W_{\rm e} = \Om 4\pi R^2 \int_{1\rm\,GeV}^\infty
E_{\rm e} f_{\rm e}(E_{\rm e}) dE_{\rm e} \approx 1.0 \times 10^{52}\, {\rm erg}
\ee
where $\Om \approx 0.66\, {\rm sr}$ is the surface area of the bubbles (for $|b|>10^\circ$)
and $R \approx 9.4 $ kpc is the distance to the center of the bubbles at $|b| = 25^\circ$.

\begin{figure}[htbp] 
\vspace{-5mm}
\begin{center}
\epsfig{figure = 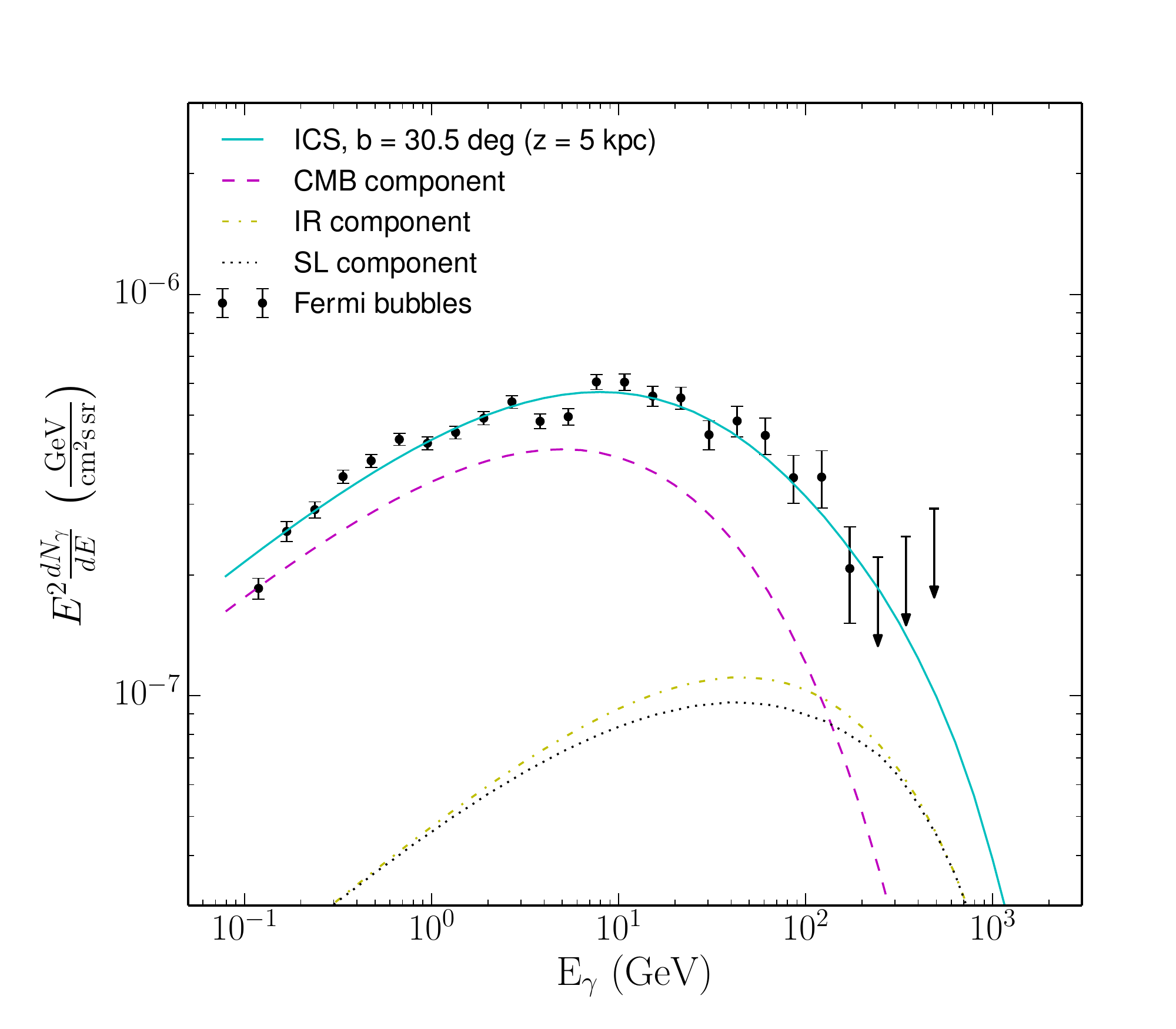, scale=\twopic}
\epsfig{figure = 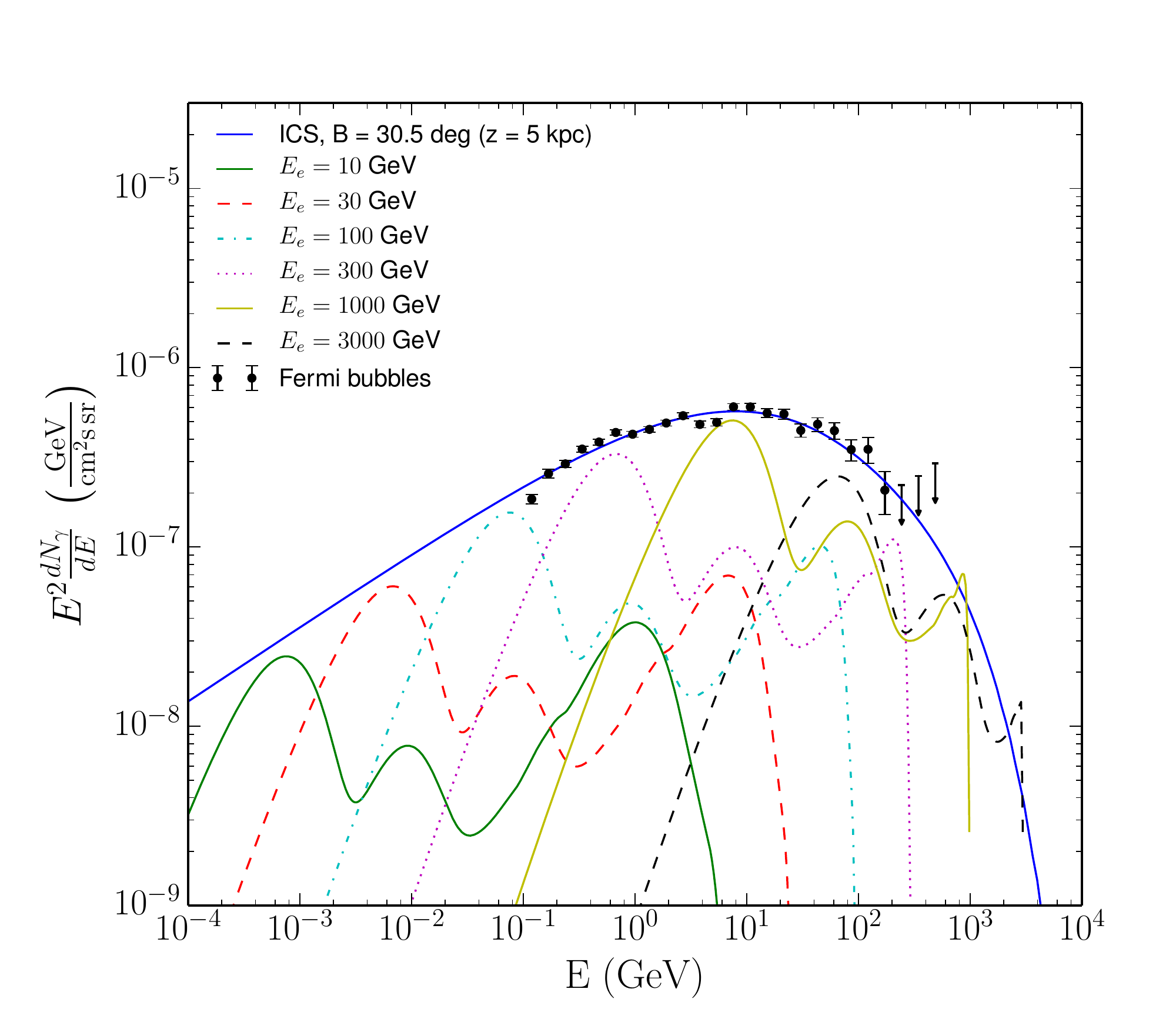, scale=\twopic}
\noindent
\caption{\small 
Left: contribution to the IC model of the {\Fermi} bubbles from different components of the ISRF.
Right: contribution to the IC model of the {\Fermi} bubbles from electrons of different energies.
}
\label{fig:93ICcases}
\end{center}
\vspace{1mm}
\end{figure}

The contribution of different ISRF fields and the contribution
of electrons of different energies to the gamma-ray flux is presented in Figure \ref{fig:93ICcases}.
Most of the contribution below 100 GeV comes from the CMB, which is the most abundant source of photons 
in terms of the number density.
Above 100 GeV the IC signal is dominated by starlight and IR photons.
In this calculation, we assume isotropic IC scattering cross section. The anisotropy of the starlight and IR
photon flux at high latitudes may introduce a correction to the calculations~\citep{2000ApJ...528..357M}
at energies above 100 GeV where the IR and starlight contribution is significant.
The magnitude of the change is not expected to be large as one can see from Figure \ref{fig:71IC_spectra},
where we compare the full ISRF model with CMB only IC emission.

\subsection{Microwave haze}
\lb{sect:mw_haze}

In this subsection, we calculate the synchrotron emission from the
same population of electrons derived in the previous subsection.
We find that this population of electrons can also explain the \WMAP and \Planck microwave haze data
\citep{2004ApJ...614..186F, 2013A&A...554A.139P}.

The power emitted by an electron with an energy $E = \g mc^2$ in a magnetic field $B$ with
an angle $\alpha$ between the electron velocity and the magnetic field is
\citep{1970RvMP...42..237B}
\be
P_{\rm emitted} (\nu, \al, E, B) = \frac{\sqrt{3} e^3 B \sin \alpha}{m c^2}
\frac{\nu}{\nu_c} \int_{\nu/\nu_c} d\xi K_{5/3}(\xi),
\ee
where $K_{5/3}(\xi)$ is the modified Bessel function of the second kind and $\nu_c$ is the critical frequency
\be
\nu_c = \frac{3 e B \g^2}{4\pi m c} \sin \alpha.
\ee
The electron distribution can be expressed as a product of a distribution related to pitch angle $\al$, $N(\al)$, and
the energy spectrum ${dn_{\rm e}}/{dE}$
\be
\frac{dN}{dE d\Om_\al dV} = \frac{N(\alpha)}{4\pi} \frac{dn_{\rm e}}{dE}.
\ee
The power emitted from a volume element is
\be
\frac{dW}{d\nu dt} = \int dE \int d\Om_\al \frac{N(\alpha)}{4\pi}
\frac{dn_{\rm e}}{dE} P_{\rm emitted} (\nu, \al, E, B).
\ee
The intensity of microwave flux is derived analogously to Equations (\ref{eq:source2flux}) 
and (\ref{eq:gamma_flux})
\be
\lb{eq:mw_flux}
\frac{dI}{d\nu} = \int dE \int d\Om_\al \frac{N(\alpha)}{4\pi}
f_{\rm e}(E) P_{\rm emitted} (\nu, \al, E, B),
\ee
where $f_{\rm e}(E)$ is the same distribution of electrons as in Equation (\ref{eq:edens}).
We assume that there is no dependence on the pitch angle, 
i.e., $N(\al) = 1$.

\begin{figure}[htbp] 
\begin{center}
\epsfig{figure = 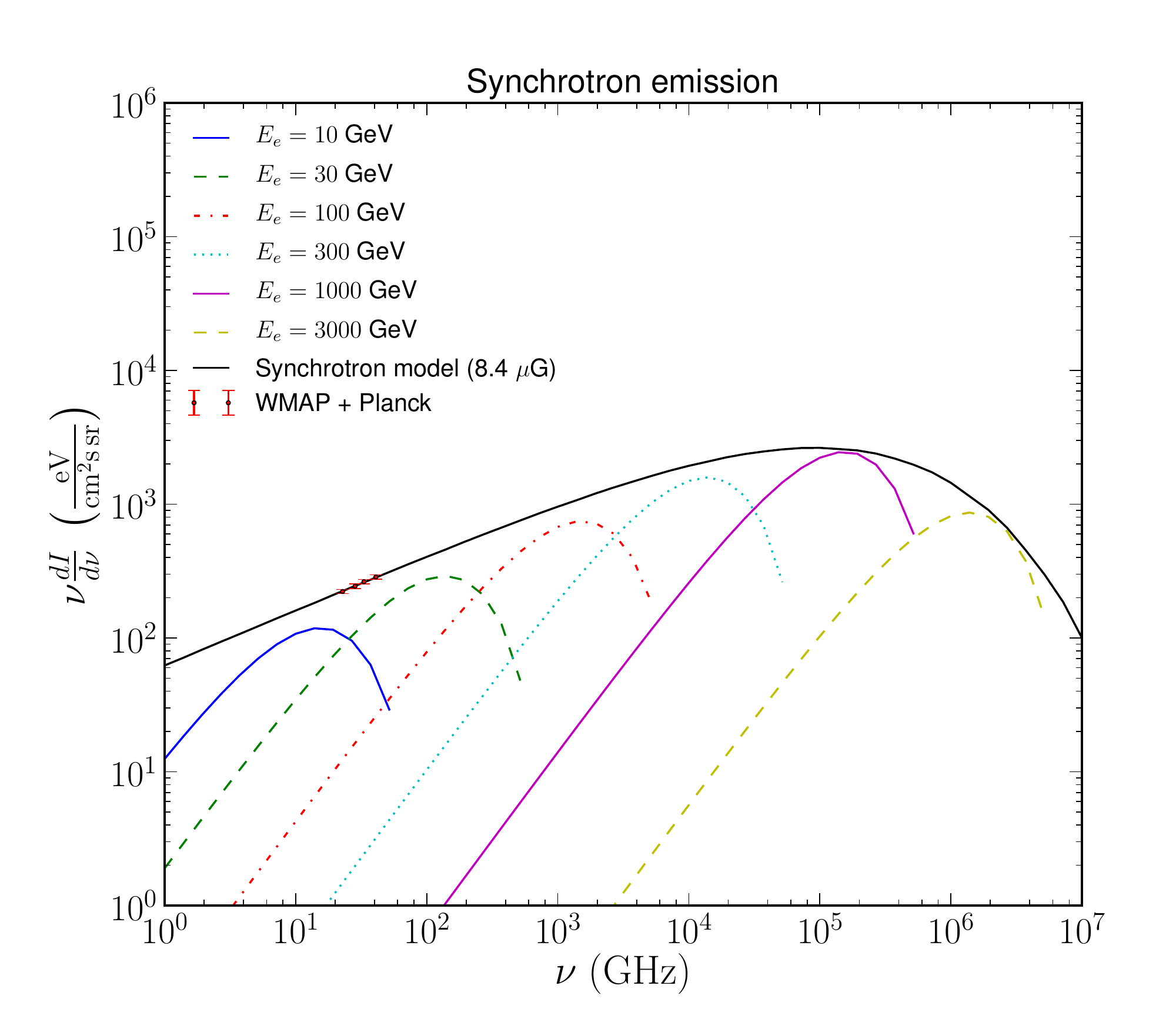, scale=\twopic}
\epsfig{figure = 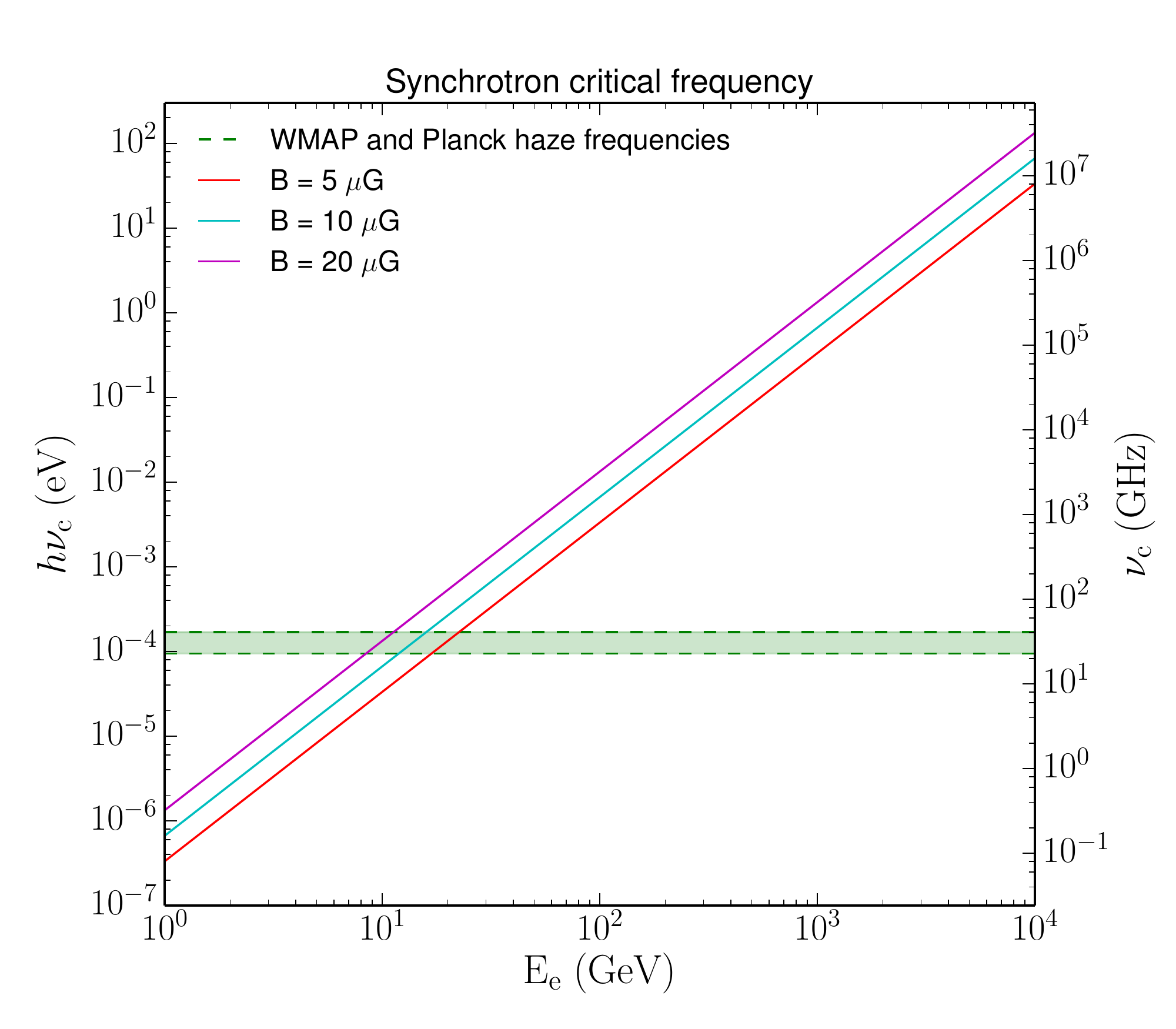, scale=\twopic}
\noindent
\caption{\small 
Left: synchrotron emission from electrons of different energies.
The points correspond to the {\WMAP} and {\Planck} microwave haze intensities.
Right: synchrotron critical frequency as a function of electron energy for different magnetic fields
at $\al = 90^\circ$.
The band corresponds to the \WMAP and \Planck haze frequencies 
\citep{2013A&A...554A.139P}.
}
\label{fig:95synch_energy}
\end{center}
\vspace{1mm}
\end{figure}

In Figure (\ref{fig:95synch_energy}) on the left we show the contribution of electrons at different energies
to the total synchrotron spectrum.
The curves are derived from Equation (\ref{eq:mw_flux}) by integrating only over the pitch angle $\al$.
For a given electron energy $E$, most of the emitted power is concentrated around the critical frequency.
In Figure (\ref{fig:95synch_energy}) on the right we show the critical frequency for a range of magnetic fields
relevant to the problem (we assume $\sin \al = 1$ on this plot).
The electrons at energies between 5 and 30 GeV contribute
most of the power in the synchrotron emission at the WMAP and Planck frequencies.
From Figure \ref{fig:93ICcases} we find that most of the contribution to the
gamma-ray emission of the bubbles comes from electrons with energies above 100 GeV,
i.e. the gamma-ray spectrum of the {\Fermi} bubbles and the microwave haze signal in
{\WMAP} and {\Planck} data are produced by electrons in different energy ranges.

\begin{figure}[htbp] 
\begin{center}
\epsfig{figure = 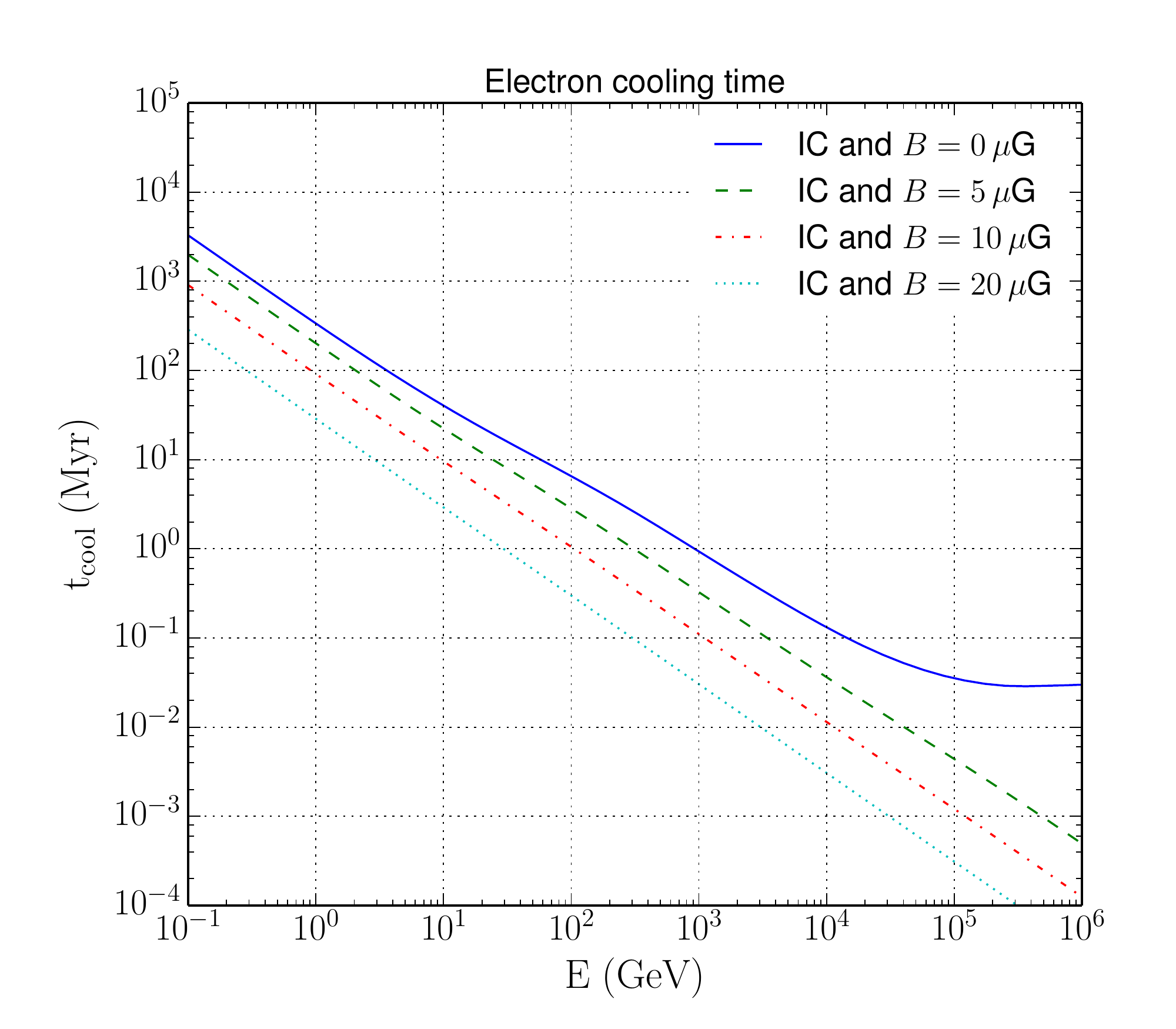, scale=\twopic}
\epsfig{figure = 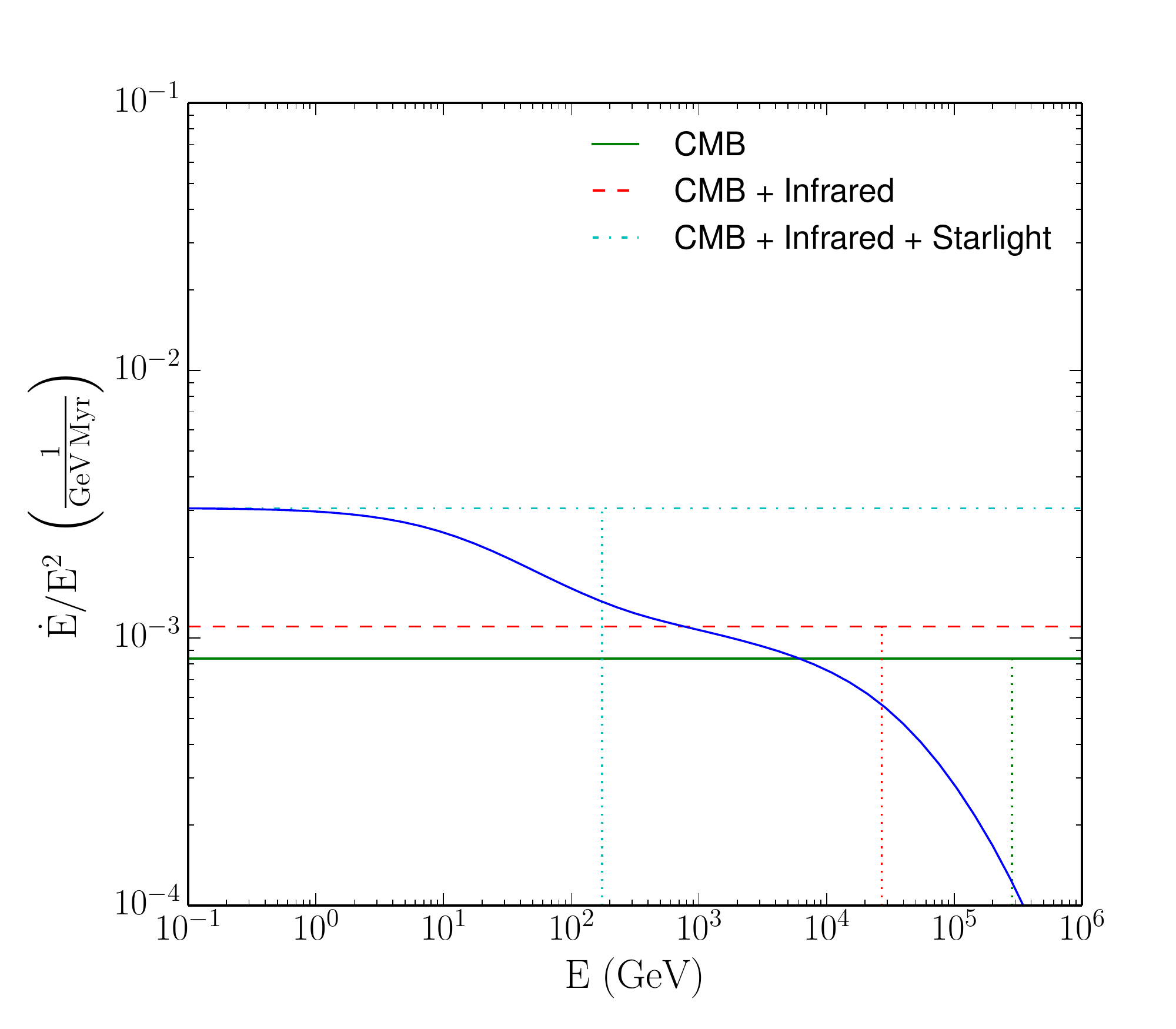, scale=\twopic}
\noindent
\caption{\small 
Left: IC and synchrotron characteristic cooling time for CR electrons,
which is defined as $t_{\rm cool} = - E / \dot{E}$.
Right: the IC energy loss rate for different ISRF fields.
The solid line represents the loss rate including the Klein-Nishina transition.
Horizontal lines correspond to the Thomson approximation of the energy loss
for different densities of the ISRF fields (CMB only, CMB+IR, CMB+IR+starlight).
Vertical lines correspond to the Klein-Nishina transition energy for starlight,
IR, and CMB (left to right).
The characteristic transition energies are the same as in Figure \ref{fig:91isrf}.
}
\label{fig:96tcool}
\end{center}
\vspace{1mm}
\end{figure}

In Figure \ref{fig:96tcool} we show the cooling time for electrons in different magnetic fields.
In the case of zero magnetic field, all energy losses in this figure are due to IC scattering.
The starlight photons do not contribute significantly to the energy loss above 100 GeV.
Above 100 TeV the IR and the CMB photons must be considered in the Klein-Nishina regime as well.
As a result, the IC cooling time above 100 TeV is leveling out (top curve on the left plot).

\subsection{Hadronic model of the {\Fermi} bubbles}

The gamma rays can be produced as a result of an interaction of high energy CR protons
with interstellar gas.
The rate of gamma-ray production per unit volume and unit energy is
\be
\frac{dQ}{dE_\g dt} = \int \frac{d \sigma(T_{\rm p}, E_\g)}{dE_\g} n_{\rm H} v_{\rm p} 
\frac{d n_{\rm p}}{d T_{\rm p}} dT_p ,
\ee
where $n_{\rm p}$ and $n_{\rm H}$ are the densities of CR protons and hydrogen atoms per unit volume,
$v_{\rm p}$ is the velocity of the CR,
and ${d \sigma(T_p, E_\g)}/{dE_\g}$ is the differential cross section to produce
a gamma ray with energy $E_\g$ in an interaction of a CR proton with kinetic energy 
$T_{\rm p} = E_{\rm p} - m_{\rm p}$
and the nucleus of a hydrogen atom at rest.

\begin{figure}[htbp] 
\begin{center}
\epsfig{figure = 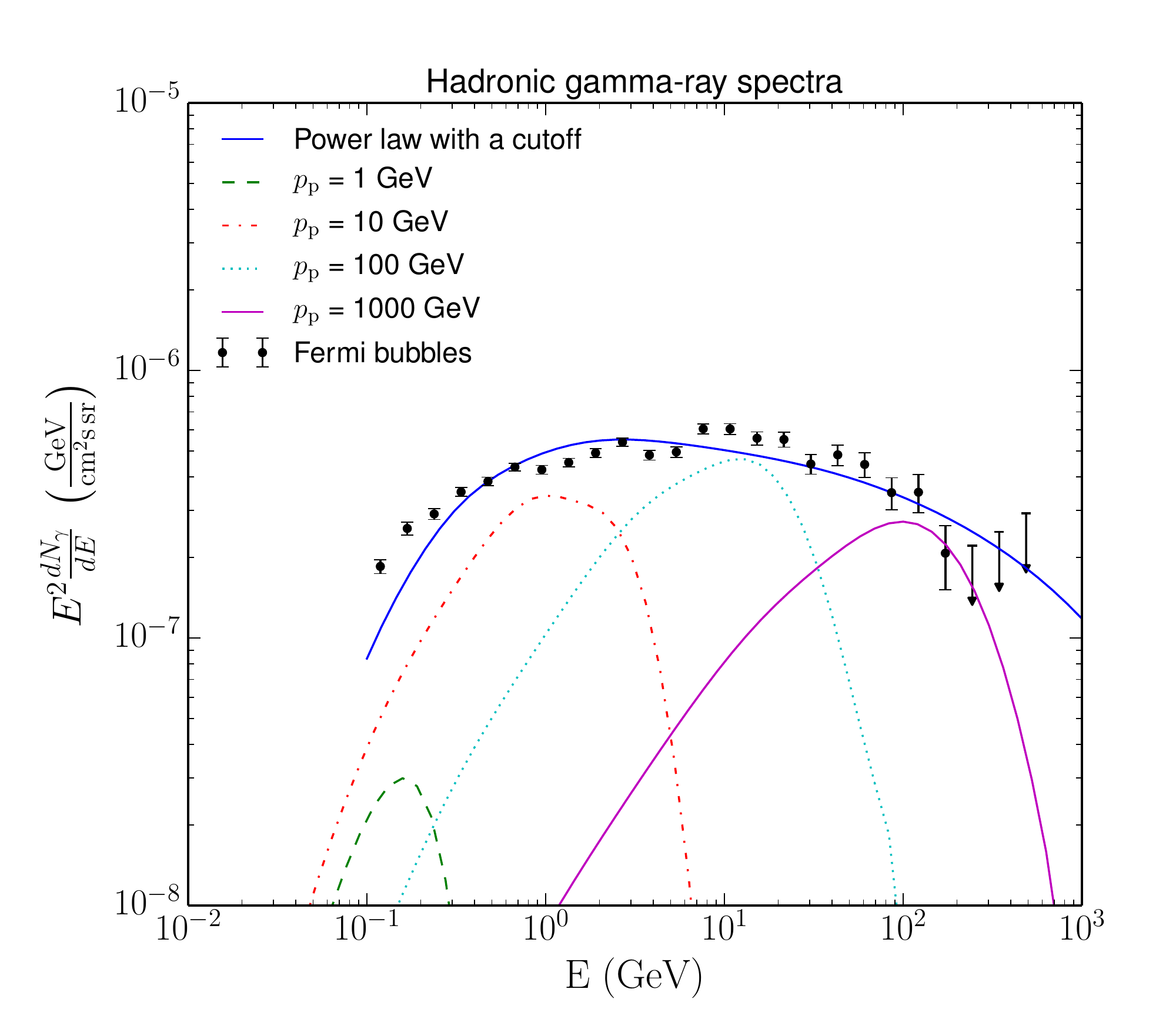, scale=\onepic}
\noindent
\caption{\small 
Contributions to the gamma-ray spectrum from protons at different momenta.
The overall spectrum of CR protons is derived from fitting
to the {\Fermi} bubbles spectrum in Section \ref{sect:hadrmodel}.
}
\label{fig:97pi0}
\end{center}
\vspace{1mm}
\end{figure}

We assume that the hadronic CR spectrum can be described as a power law in momentum at low energies.
The derivative with respect to $T_{\rm p}$ is transformed to the derivative with respect to the momentum $p_{\rm p}$  through the expression
\be
v_{\rm p} \frac{d n_{\rm p}}{d T_{\rm p}} = c \frac{d n_{\rm p}}{d p_{\rm p}}.
\ee
The gamma-ray flux at the in the vicinity of  Earth is
\be
\lb{eq:hadr_gamma}
\frac{dN_\g}{dE_\g} = \int \frac{d \sigma(T_{\rm p}, E_\g)}{dE_\g} n_{\rm H} c f_{\rm p}(p_{\rm p}) dT_p ,
\ee
where we assume a constant gas density $n_{\rm H}$ throughout the bubbles and denote
the line of sight integral of the CR density as
\be
\lb{eq:CRp_dens}
 f_{\rm p}(p_{\rm p}) = \frac{1}{4\pi} \int  \frac{d n_{\rm p}}{d p_{\rm p}} dR.
\ee
In Figure \ref{fig:97pi0} we show examples of fitting the gamma-ray data assuming a power law or a power law with a
cutoff form of $ f_{\rm p}$.
We also show the contribution from different proton momenta in the power law with an exponential cutoff case.
The parameters are $f_{\rm p} \propto p^{-2.1} e^{-p c/ 13.7\,{\rm TeV}}$.
In order to find the normalization and the total energy in protons, we need to make an assumption
on the number density of hydrogen atoms inside the bubbles.
We take $n_{\rm H} = 0.01\, {\rm cm^{-3}}$ as a reference value.

The source function for the secondary particles is
\be
\frac{dQ_s}{dE_s} = \int \frac{d \sigma_{\rm pp\ra s}(T_{\rm p}, E_s)}{dE_s} n_{\rm H} c 
\frac{d n_{\rm p}}{d p_{\rm p}} dT_p,
\ee
where $s$ denotes the particle species.
The flux of neutrinos can be calculated similarly to the flux of gamma rays in Equation (\ref{eq:hadr_gamma})
\be
\frac{dN_\nu}{dE_\nu} = \int \frac{d \sigma(T_{\rm p}, E_\nu)}{dE_\nu} n_{\rm H} c f_{\rm p}(p_{\rm p}) dT_p.
\ee
The flux of neutrinos from the bubbles was presented in Figure \ref{fig:74nu_flux} on the left.

\begin{figure}[htbp] 
\begin{center}
\epsfig{figure = 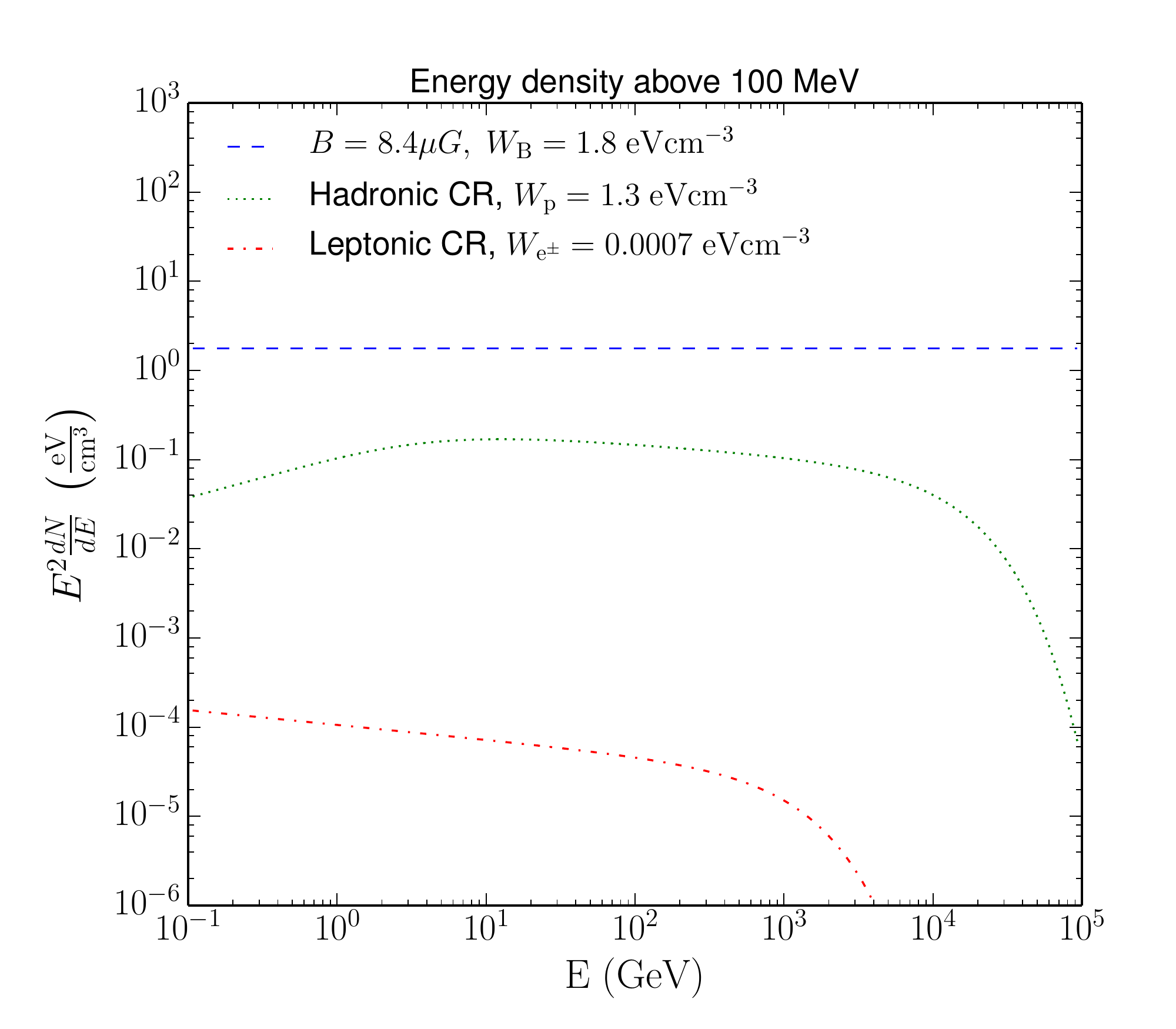, scale=\onepic}
\noindent
\caption{\small 
Comparison of the energy density of CRs in leptonic and hadronic models of the {\Fermi} bubbles 
and the energy density of an 8.4 $\upmu$G magnetic field.
The CR energy densities are obtained from Equations \ref{eq:edens} and \ref{eq:CRp_dens}
assuming that the distance to the center of the bubbles is 9.4 kpc.
}
\label{fig:98B_field_E}
\end{center}
\vspace{1mm}
\end{figure}

We compare the CR energy density in leptonic and hadronic models to the energy density of a 8.4 $\upmu$G magnetic field
in Figure \ref{fig:98B_field_E}.
We find that the energy density of CR in hadronic model of the {\Fermi} bubbles is approximately in equipartition with the 
magnetic field.
In IC model of the {\Fermi} bubbles the energy density in leptonic CR is much lower than the magnetic field energy density.

\begin{figure}[htbp] 
\begin{center}
\epsfig{figure = 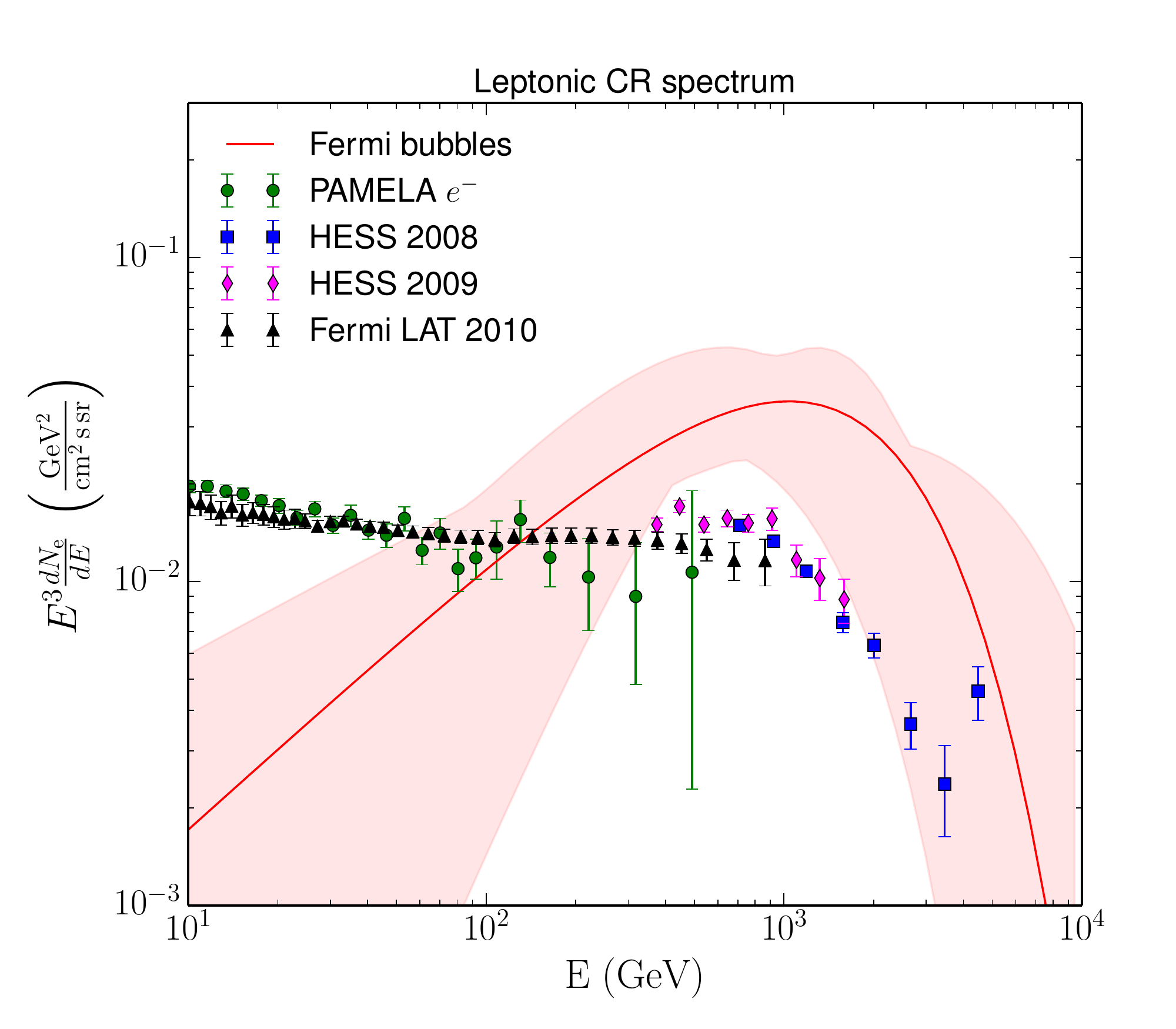, scale=\twopic}
\epsfig{figure = 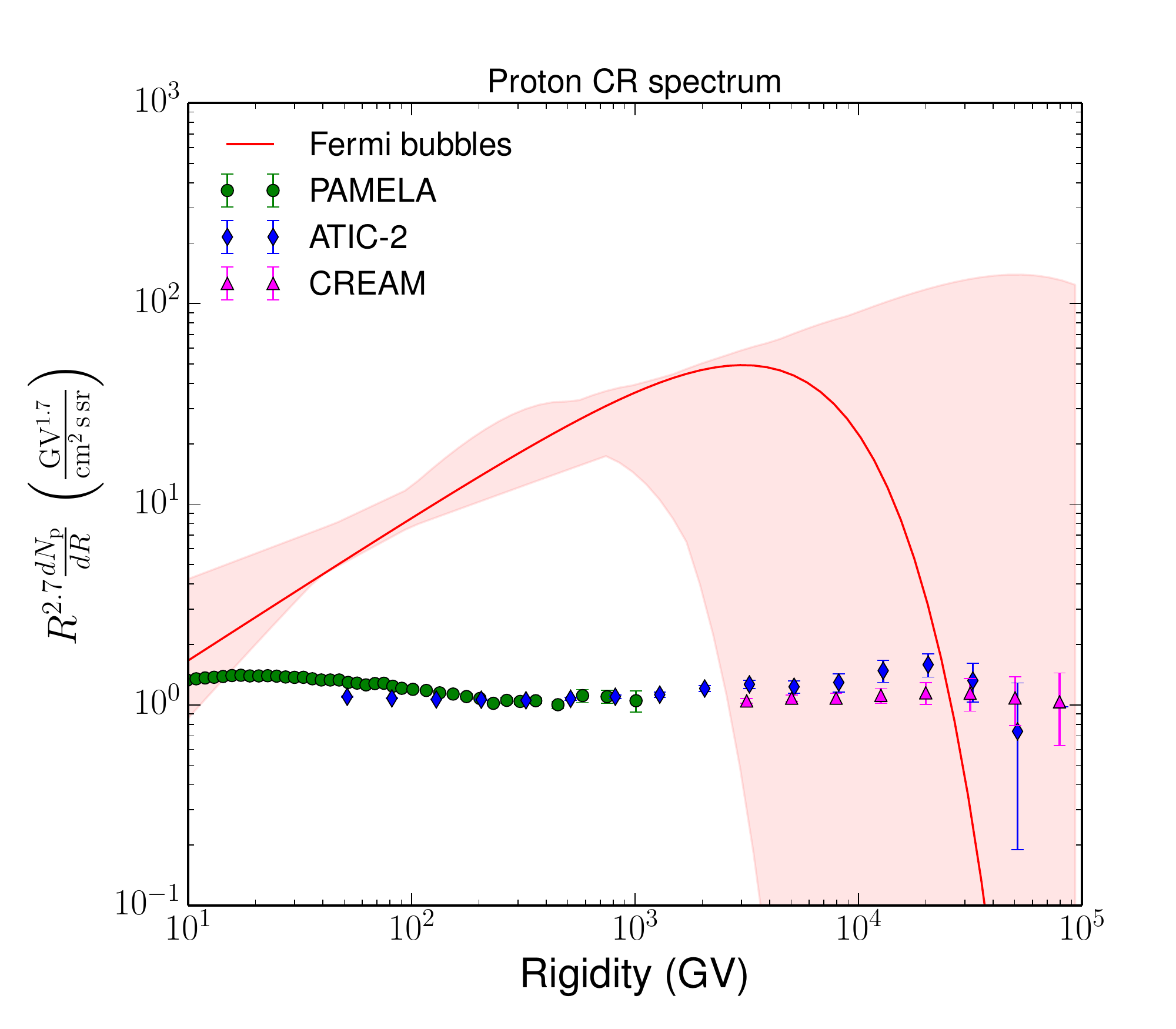, scale=\twopic}
\noindent
\caption{\small 
Left: comparison of leptonic CR spectra in the IC model of the {\Fermi} bubbles
to the local CR spectra: 
\PAMELA electron only spectrum~\citep{2011PhRvL.106t1101A}, HESS~2008~\citep{2008PhRvL.101z1104A},
HESS~2009~\citep{2009A&A...508..561A},
and {\Fermi}~LAT~2010~\citep{2010PhRvD..82i2004A}.
Right: comparison of proton CR spectra in the hadronic model of the {\Fermi} bubbles
to the local proton CR spectrum: \PAMELA \citep{2011Sci...332...69A}, 
ATIC-2 \citep{2006astro.ph.12377P},
CREAM \citep{2011ApJ...728..122Y}.
In both cases, the band represents an envelope of the CR spectra fitted to the gamma-ray spectra of the 
bubbles for different models of the foreground emission and definitions of the bubbles templates. We do the comparison at energies above 10 GeV because this is the energy range relevant for the production of the gamma rays from the bubbles and the microwave haze. Below $\sim10$ GeV the local CR spectra are affected by solar modulation.
}
\label{fig:99localCR}
\end{center}
\vspace{1mm}
\end{figure}

In Figure \ref{fig:99localCR} we compare the CR energy density in leptonic and hadronic models of the {\Fermi} bubbles
to the local CR energy density.
In hadronic model of the bubbles, the energy density of CR is about an order of magnitude larger than the 
energy density of local CR.
In the leptonic model, the energy density of leptonic CR inside the bubbles
is comparable to the local energy density of leptonic CR.
The spectrum of leptonic CR in the bubbles is harder. 
As a result, above (below) 100 GeV the spectrum of leptonic CR inside the bubbles is larger (smaller) than the local CR spectrum.

